%% file: phys_rept_main.tex
\documentclass[11pt]{elsarticle}

\usepackage{amssymb}
\usepackage{amsmath}
\usepackage{mathtools}
\usepackage{upgreek}
\usepackage[letterpaper, left=1in, right=1in, top=1in, bottom=1in]{geometry}
\usepackage[figuresright]{rotating}
\usepackage[dvipsnames]{xcolor}
\usepackage[colorlinks = true,
            linkcolor = Maroon,
            urlcolor  = Maroon,
            citecolor = Maroon]{hyperref}
\usepackage{soul}
\usepackage{tikz}
\usetikzlibrary{decorations.pathmorphing}
\usetikzlibrary{decorations.markings}
\usetikzlibrary{arrows}
\tikzset{snake it/.style={decorate, decoration=snake}}
\usepackage{bm}
\usepackage{dsfont}
\usepackage{microtype}
\usepackage{scalerel}
\usepackage{makecell}
\usepackage{cancel}
\usepackage{multirow}
\usepackage{nicematrix}

\biboptions{numbers,sort&compress}

\makeatletter
\hypersetup{colorlinks=true}
\AtBeginDocument{\@ifpackageloaded{hyperref}
  {\def\@linkcolor{Maroon}
   \def\@anchorcolor{Maroon}
   \def\@citecolor{Maroon}
   \def\@filecolor{Maroon}
   \def\@urlcolor{Maroon}
   \def\@menucolor{Maroon}
   \def\@pagecolor{Maroon}
\begingroup
  \@makeother\`%
  \@makeother\=%
  \edef\x{%
    \edef\noexpand\x{%
      \endgroup
      \noexpand\toks@{%
        \catcode 96=\noexpand\the\catcode`\noexpand\`\relax
        \catcode 61=\noexpand\the\catcode`\noexpand\=\relax
      }%
    }%
    \noexpand\x
  }%
\x
\@makeother\`
\@makeother\=
}{}}
\makeatother

\newcommand{\be}{\begin{equation}}
\newcommand{\ee}{\end{equation}}
\newcommand{\la}{\langle}
\newcommand{\ra}{\rangle}

\newcommand*{\paral}{\stretchrel*{\parallel}{\perp}}

\newcommand{\spacy}{\mkern2mu}

\newcolumntype{L}[1]{>{\raggedright\let\newline\\\arraybackslash\hspace{0pt}}m{#1}}
\newcolumntype{C}[1]{>{\centering\let\newline\\\arraybackslash\hspace{0pt}}m{#1}}
\newcolumntype{R}[1]{>{\raggedleft\let\newline\\\arraybackslash\hspace{0pt}}m{#1}}

\numberwithin{equation}{section}

\begin{document}


\vspace{1in}

\begin{frontmatter}

\title{Effective Field Theories for Material Media}

\author{Angelo Esposito}
\address{Dipartimento di Fisica, Sapienza Universit\`a di Roma and INFN Sezione di Roma,  I-00185 Rome, Italy}

\author{Alberto Nicolis}
\address{Center for Theoretical Physics, Department of Physics,  Columbia University, New York, NY 10027, USA}

\author{Riccardo Penco}
\address{Department of Physics, Carnegie Mellon University, Pittsburgh, PA 15213, USA}

\begin{abstract}
We review recent progress in understanding certain aspects of condensed matter systems from a high energy theory perspective. We discuss effective field theories that describe collective bulk and localized excitations in a variety of solid and fluid systems. Particular emphasis is placed on the role played by spacetime symmetries and their spontaneous breaking. The resulting Goldstone dynamics can be seen as underlying a wide variety of phenomena. We attempt to bridge the language gap between subfields while underscoring the numerous conceptual similarities.
\end{abstract}





\end{frontmatter}


\newpage

\tableofcontents

\include{introduction}

\include{CM_theory}

\include{solids}

\include{fluids}

\include{superfluids}

\include{classification}

\include{education}

\include{conclusions}

\include{app_strain_tensor}

\include{app_nonlinear_Lorentz}

\include{Routhian}

\include{app_feynman_rules}



\bibliographystyle{elsarticle-num}
\bibliography{biblio}

\end{document}

%% file: introduction.tex
\section{Introduction}
\label{sec: introduction}

\noindent In the last several decades it has become clear that the connection between condensed matter\footnote{We use “condensed matter” loosely, to denote systems at finite density with some, perhaps approximate, translational invariance. Then, according to this definition, solids, fluids, cold atom clouds, etc.,~all count as condensed matter. This is a substantially broader class than what  condensed matter physicists usually call ``condensed matter”.}  and particle physics goes well beyond the ancient (correct) idea that matter is ultimately made up of individual particles. In fact, we might even say that from the modern, field theoretical viewpoint, that extremely concrete and physical connection is, in many situations, irrelevant, both in the colloquial and in the technical sense of the term.
The reason is that when we deal with macroscopic matter we are not necessarily interested in following the evolution of the individual constituent particles, and, even if we  are, the physical properties of such particles inside matter can be so different from those they have in vacuum, that we might as well call them different particles. Sometimes, the constituent particles don’t even propagate as nearly free excitations, but are instead grouped together with others into collective excitations or bound states.
Rather, from the modern viewpoint, what unites condensed matter with particle physics is the language of field theory, with  special emphasis on symmetries, renormalization group ideas, and effective field theories. 

We now understand that under very general conditions many physical systems can be described by field theories~\cite{Weinberg:1995mt}. Particle physics describes one system in this class, that whose ground state is empty space, which is invariant under spacetime translations, rotations, and, crucially, Lorentz boosts. We can think of different condensed matter systems as being other systems in this class, with  different ground states, symmetries, and dynamics. At finite temperature, instead of ground states we will have to talk about equilibrium density matrices. In many cases, these ground states or density matrices will still be approximately translational invariant, at least at large enough distances, and in some cases they might even be rotationally invariant, again perhaps after some spatial averaging. From this viewpoint, what really distinguishes condensed matter from particle physics is the absence of  boost-invariance for the former. We can even take this as the symmetry-based defining feature of condensed matter~\cite{Nicolis:2015sra}.

In many cases, the breaking of boost invariance can be considered as ``explicit", that is, never there to begin with. In other cases, if one is really considering the full dynamics of the medium making up the system, including its mechanical deformations, it is useful to remember that boost invariance is there at the level of the fundamental laws of physics, but it is broken by the state that the system is in. {In other words, any state of matter is nothing but a (very complicated) aggregate of elementary particles. The laws obeyed by these particles are boost invariant, but the aggregate itself is not.} In such cases one can use the tools of spontaneous symmetry breaking for spacetime symmetries, with phonon-like excitations or other collective modes playing the role of the associated Goldstone bosons.  

Given the characterization of many condensed matter systems in terms of symmetries and not much more, and the relative insensitivity of the low-energy dynamics to the microscopic constituents' physical properties, it is clear that effective field theory ideas and techniques {can be an important addition to} our toolbox for addressing questions in condensed matter physics.

\vspace{1em}

The last two decades have seen much progress in this direction, with an ongoing interchange of ideas between high energy and condensed matter physics. This interchange is, in fact, too vast a phenomenon for us to be able to cover it in its entirety here. What we want to do, instead, is to focus on the general role that spacetime symmetries and their spontaneous breaking play for condensed matter systems. As we will see with a number of examples and results, such an approach is powerful, in that many phenomena in condensed matter can be understood in terms of the dynamics of the Goldstone excitations associated with spontaneously broken spacetime symmetries. Much of this is still work in progress, and so our Report is meant to be pedagogical and illustrative, and by no means exhaustive. We do, however, present some of these topics in a new way, which we hope the reader will find systematic and unified---examples of this are our framing of the nonrelativistic limit as a scaling limit in Section~\ref{sec: nonrelativistic limit} and our discussion on the Schwinger--Keldysh formalism in Section~\ref{sec: Schwinger-Keldysh}. In the process of doing that, we also derive some new results, such as the thermodynamical interpretation of the effective theory for solids and fluids of Sections~\ref{sec: thermodynamics of solids} and \ref{sec: perfect fluids}, as well as amend a few discrepancies found in the literature, such as the roton--phonon scattering rate of Section~\ref{sec:rotons}. Moreover, we introduce a certain degree of redundancy between the sections devoted to different media, in order to make each of them self-contained. This way, the reader interested in one particular medium (solids, fluids, superfluids, and so on), can directly skip to that section and still get the gist of it.

Our goal is not only to review these ideas for the high energy physics crowd, but also to present them to the condensed matter physics community. We will do that in a way that highlights the fact that, when looking at the low energy dynamics of a given state of matter, the differences between the high energy formulation and the more traditional one are nothing but linguistic: a different framework for the same physical concepts. 

In this respect, we believe that two aspects deserve particular emphasis: the role of relativistic versus nonrelativistic boost invariance, and the connection between the typical observables of interest. As far as the former aspect is concerned, while we will build most of our effective field theories assuming Lorentz invariance, we will explain how to consistently take the nonrelativistic limit, to recover results that are more suitable to condensed matter systems realized in a laboratory. As far as observables are concerned, instead, for each medium we consider we will present at least one example on how to employ the effective field theories to compute specific observable, such as dispersion relations, scattering amplitudes, and so on.

It is also important to stress that the fields of high energy and condensed matter physics are often characterized by two rather different viewpoints on a given physical problem. In high energy physics one usually strives towards results that are as universal as possible, common to a large class of microscopic models and independent on their details. In condensed matter physics, instead, the details of these short distance descriptions are often exactly what one tries to understand, with the hope of identifying special systems with desired, maybe fine tuned, properties. The effective field theories we present here fall in the first class of descriptions. This, however, does not mean that they should not be of interest to someone who leans more towards the study of specific microscopic models. Indeed, when applicable, effective field theories generally provide a powerful {\it computational tool}, which allows to considerably simplify the problem at hand. A notable example is when the microscopic interactions are strong, making an ab initio calculation demanding. Even in this context, a low energy effective theory for the Goldstone modes is always weakly coupled, hence allowing to work in perturbation theory. The only information about the complicated short distance physics is encoded in a few effective parameters, which can be extracted once and for all for a given system, and then used in all contexts of interest. 

Finally, as of today, quantum field theory is arguably the most successful tool for describing Nature at its smallest scales, with the Standard Model of particle physics being the most notable incarnation. The effective field theories presented here provide simple, real world examples of quantum field theories with properties that are more ``exotic'' than those typically encountered in high energy physics. For this reason, understanding these systems could also be an important test field for the high energy physicist who is interested in exploring what quantum field theory (if any) could describe physics beyond the Standard Model. 

\vspace{1em}

\noindent \emph{Conventions:} We will work in units such that $\hbar = k_{\rm B} = 1$ and, unless otherwise specified, $c = 1$. When discussing relativistic systems, we adopt a ``mostly plus'' signature for the Minkowski metric, $\eta_{\mu\nu} = {\rm diag}(-,+,+,+)$. The overall sign of the Levi--Civita symbol is such that $\epsilon^{0123} = +1$. Moreover, we will use Einstein's summation convention, in which repeated indices are summed over. 

%% file: CM_theory.tex
\section{Bridging the Gap between High Energy and Condensed Matter Theory}

\noindent We start by delineating the boundaries of this Report. We focus solely on those excitations whose existence is guaranteed by the symmetries, in particular the spontaneously broken ones. Moreover, since any condensed matter system breaks at least some spacetime symmetries, our emphasis will be on the role of these. 
As an example, a crystalline solid breaks Lorentz boosts, spatial translations, and rotations. The mechanical deformations of the solid, which correspond to the phonon degrees of freedom, can be thought of as the associated Goldstone excitations. Their dynamics are highly constrained by those broken symmetries, and we want to elucidate this connection, and the analogous ones for other phases of matter.

And so, in particular, our Report will not deal with the standard questions that occupy much of the condensed matter community, such as what are the electronic properties of a given material, what is the correct theory of high-$T_c$ superconductivity, and so on. We will be dealing instead  only with what we may call the mechanical degrees of freedom of the medium at hand. In a sense, we will present effective field theories for the {\em underlying media} that provide the arena where much of the standard condensed matter action takes place. This is very much analogous, conceptually and technically, to general relativity's being a theory for the dynamics of spacetime, which in turn provides an arena for all non-gravitational physical processes, such as particle physics ones. 

Considering again the above example of the solid, we will deal with phonon degrees of freedom, which  usually play only a tangential role (for example, in BCS theory) in modern condensed matter research. To be fair, there is a good reason why to a first approximation phonons can sometimes be neglected: like all Goldstone excitations, they are derivatively coupled{---i.e., the strength of their interactions is dictated by their momentum---}and thus decouple at low enough energies. However, thanks to symmetries, they are also gapless, and so for some questions, such as thermodynamic ones, they cannot be neglected, {as they can be excited with an arbitrarily small amount of energy.}

These considerations bring us to an important aspect of our Report: the role of boost invariance. If one neglects the phonon dynamics of the underlying medium in a given condensed matter process---say the scattering of two electrons in a metal---boost invariance plays no role. The reason is that boost invariance is (spontaneously) broken by the medium itself, and it is only thanks to the phonons that it is realized, nonlinearly, in the dynamics of the system. So, if the phonons are neglected, there is no boost invariance to talk about. In the standard field theory language, it is ``explicitly broken." The flip side of this is that when  one instead {\em is} interested in the dynamics of phonons, boost invariance cannot be neglected. This is good, because, like all symmetries, it constrains the structure of the theory, and thus makes the theory more predictive.

Finally, on whether one should insist on Galilei rather than Lorentz boost invariance: as far as we know, Lorentz invariance is a fundamental symmetry of physics, and so there is no harm in insisting on it. On the other hand, Galilei invariance can be a very good and very useful {\em approximation} to Lorentz invariance, but it is never more precise or fundamental. So, our approach will be to use a relativistic framework and Lorentz invariance whenever possible. When appropriate and convenient, we will take the nonrelativistic limit of our results, perhaps to connect to more familiar concepts, but at least at the level of constructing the effective field theories, we will try to be as general as possible and keep a relativistic viewpoint. This also has the advantage of being often technically convenient, because writing down Lorentz invariant terms is just a matter of ``contracting the indices'', whereas ensuring Galilean invariance can be more subtle and less systematic.

\subsection{Continuous Media} \label{sec:continuousmedia}

\noindent Consider a space-filling medium. Let us first try to define precisely what we mean by ``medium". If we  consider familiar examples of physical media, such as solids, fluids, and superfluids, we can think of what common features they share, especially in terms of symmetries, and propose a possible operational definition: a medium is a stationary state of matter that has a preferred reference frame (the local rest frame), and, at least at long enough distances, is homogeneous.
Some media, like ordinary fluids, are inherently thermal systems, but  for now we want to consider only zero-temperature systems. 

So,  in QFT language, we can  define a generic zero-temperature medium as a 
\begin{center}
{\em Boost breaking, space- and time-translationally invariant state}
\end{center}
 in a generic  quantum field theory. As we will see, the translational invariance part of this definition must be qualified in most cases, but, ignoring this subtlety for the time being, this simple characterization has far reaching implications.
 
The most important of these implications is the Goldstone phenomenon: since boosts are broken by the state of the system rather than by the dynamics of the fundamental QFT that describes the system, they still constrain the dynamics and, crucially, the spectrum of the theory. In particular, they imply the existence of gapless excitations. The low-energy, long-distance dynamics of a medium can thus be thought of as the dynamics of these Goldstone excitations.

\begin{figure}
	\centering
	\resizebox{0.95\textwidth}{!}{
		\begin{tikzpicture}
			\draw[->, gray, line width = 0.07mm] (-0.1,0) -- (2.15,0);
			\draw[->, gray, line width = 0.07mm] (-0.1,0) -- (-0.1,0.5);
			\draw[] (0,0) sin (0.5,0.25) cos (1,0) sin (1.5,-0.25) cos (2,0);
			\draw[dashed, thin, dash pattern = on 1pt off 1pt] (0,0) -- (2,0);
			
			\draw[->, Maroon, semithick] (0.25, 0) -- (0.25, 0.165);
			\draw[->, Maroon, semithick] (0.5, 0) -- (0.5, 0.24);
			\draw[->, Maroon, semithick] (0.75, 0) -- (0.75, 0.165);
			\draw[->, Maroon, semithick] (1.25, 0) -- (1.25, -0.165);
			\draw[->, Maroon, semithick] (1.5, 0) -- (1.5, -0.24);
			\draw[->, Maroon, semithick] (1.75, 0) -- (1.75, -0.165);
		
			\node[scale=0.4] at (2.25,0) {\color{gray}{$x$}};
			\node[scale=0.4] at (-0.1,0.58) {\color{gray}{$y$}};
			\node[scale=0.4] at (0.54,0.39) {$\delta y(x)$};
			
			\draw[->, gray, line width = 0.07mm] (2.9,0) -- (5.15,0);
			\draw[->, gray, line width = 0.07mm] (2.9,0) -- (2.9,0.5);
			\draw[] (3,0) sin (3.5,0.25) cos (4,0) sin (4.5,-0.25) cos (5,0);
			\draw[dashed, thin, dash pattern = on 1pt off 1pt] (3,0) -- (5,0);
			
			\node[scale=0.4] at (5.25,0) {\color{gray}{$x$}};
			\node[scale=0.4] at (2.9,0.58) {\color{gray}{$y$}};
			\node[scale=0.4] at (3.54,0.39) {$\delta y(x)$};
			
			\draw[->, Maroon, semithick] (3.25,0) arc (180:160:0.5);
			\draw[->, Maroon, semithick] (3.5,0) arc (180:145:0.4);
			\draw[->, Maroon, semithick] (3.75,0) arc (180:138:0.2);
			\draw[->, Maroon, semithick] (4.25,0) arc (0:-42:0.2);
			\draw[->, Maroon, semithick] (4.5,0) arc (0:-35:0.4);
			\draw[->, Maroon, semithick] (4.75,0) arc (0:-20:0.5);
		\end{tikzpicture}
	}
	\caption{The equilibrium configuration (dashed line) of a string in two spatial dimensions spontaneously breaks three symmetries: boosts along $y$, translations along $y$ and rotations in the $xy$-plane. A given perturbation (solid line) can be seen as generated by the action of locally modulated translations along $y$ ({\bf left panel}), locally modulated rotations in the $xy$-plane ({\bf right panel}) or locally modulated boosts (not shown). As a consequence, despite the breaking of these three symmetries, the perturbation can be described by a single Goldstone mode, $\delta y(x)$, corresponding to the vertical displacement of the string.} \label{fig:string}
\end{figure}
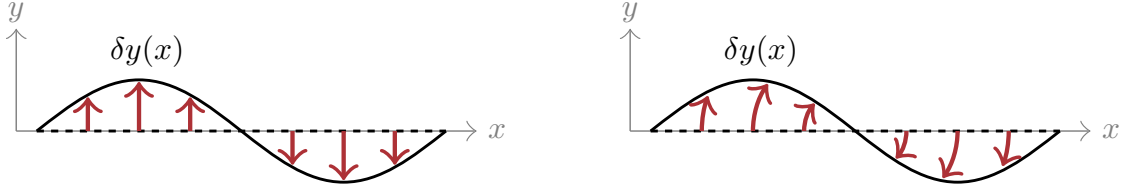

Before we proceed with presenting the effective field theories for the different kinds of media one can have, it is important to stress a number of subtleties behind this logic:
\begin{enumerate}
\item
nonrelativistic Goldstone theorems, that is, Goldstone theorems for systems that break boosts either explicitly or spontaneously, are less powerful than relativistic ones. Formally, they guarantee the existence of gapless excitations only for $\vec k \to 0$, but say nothing about the spectrum at finite $\vec k$~\cite[e.g.,][]{Strocchi:2021abr}. As a physically relevant example, a phonon in superfluid ${^4}{\rm He}$ has a decay rate scaling like $k^5$~\cite[e.g.,][]{maris1977phonon}. It does become more and more stable as $\vec k$ goes to zero, but at any finite $\vec k$ a single-phonon state is not a sharply defined excitation{---it is not an exact eigenstate of the Hamiltonian, but only  an approximate one.}
\item
In fact, the Goldstone theorem for  spontaneously broken boosts presents further subtleties compared to those for more standard symmetries~\cite{Morchio:1985ti, Ojima:1985is, Requardt:2008xx, Alberte:2020eil}, and it is not yet clear in how many unconventional ways it can be obeyed. For example, in a Fermi liquid the role of Goldstone excitations is played by the particle-hole continuum rather than by single-particle states.
\item
Even ignoring the subtleties above, when spacetime symmetries are spontaneously broken the counting of Goldstone bosons does not work in the same way as in particle physics. In particular, the number of broken symmetries is always an {\em upper bound} on the number of (independent) Goldstone fields. Put another way, the same Goldstone field can be associated with more than one broken symmetry, and so  in general one needs fewer than usual Goldstone fields to realize all the symmetries. {See Figure~\ref{fig:string} for an example.} This general phenomenon goes under the somewhat confusing name of ``inverse Higgs constraints''~\cite[e.g.,][]{Ivanov:1975zq,Low:2001bw,McArthur:2010zm,Nicolis:2013sga}. An alternative way of explaining this is by noticing that, in the presence of spontaneously broken spacetime symmetries, the associated Noether currents are typically linearly related to each other. As explained in~\cite{Watanabe:2013iia}, for this reason different Noether currents create the same Goldstone.
\item
Finally, there is a systematic technique---the {\em coset construction}---that allows one to write down the most general Goldstone effective field theory associated with a given symmetry breaking pattern. It has the advantage of being foolproof, but at the expense of having to rely on a somewhat heavy and opaque formalism. In this work, we use some physical intuition and hindsight to construct our EFTs ``by hand". The reader interested in the details of the coset construction, for both unbroken and broken spacetime symmetries, can refer to~\cite[e.g.,][]{Coleman:1969sm,Callan:1969sn,ogievetsky1974nonlinear,Ivanov:1975zq,Delacretaz:2014oxa}. 

\end{enumerate}

\subsection{Nonrelativistic Limit} \label{sec: nonrelativistic limit}

\noindent There is no doubt that, for what concerns any condensed matter system realized on Earth, the nonrelativistic approximation is an excellent one. This is because the typical speeds of the elementary constituents, as well as the phase and group velocities of collective excitations, are typically much smaller than the speed of light, $c$. This, in turn, can be traced back to the weakness of electromagnetic interactions and the fact that the boundary conditions we are able to impose in a laboratory setting correspond to temperatures $T$, pressures $p$ and number densities $n$ such that
\begin{align}
	T , \ p/n \ll m \spacy c^2 \,,
\end{align}
where $m$ can be either the electron or the nucleus mass, depending on which phenomena one is interested in.

On the one hand, the presence of characteristic speeds other than the speed of light invalidates the use of dimensional analysis, which cannot constrain how observable quantities will depend on dimensionless ratios of speeds. On the other hand, it opens up the possibility of taking a \emph{scaling limit} of any relativistic theory such that the ratios $v/c \to 0$, where $v$ stands for any characteristic speed. Operationally, this scaling limit is implemented in three steps:
\begin{enumerate}
	\item Starting from a relativistic theory defined in natural units, i.e. with $c=1$, reintroduce all powers of $c$ explicitly via the replacements 
\begin{align}
	d^4 x \to c \ dt d^3 x \,,  \qquad \qquad \partial_\mu = \left( \frac{1}{c}\frac{\partial}{\partial t} , \vec \nabla\right) \,.
\end{align}
\item Rescale all the fields so that they have a canonical kinetic term, i.e. such that its coefficient is exactly $1/2$. 
\item Determine the scaling of all the parameters in the EFT by demanding that the action remains finite and contains the largest possible number of terms as we let $c \to \infty$.
\end{enumerate}
This procedure defines unambiguously what we mean by nonrelativistic limit, and yields effective actions that are invariant under Galilean transformations. We will show explicitly how this scaling limit works for various systems in the following sections. However, our approach will be to always introduce first the relativistic description of any given system, because it is more fundamental and, usually, relativistic invariants are easier to write down.

%% file: solids.tex
\section{Solids} \label{sec:solids}

\noindent We start our investigation of different condensed matter systems with solids. As it turns out, their long-distance EFT is not the simplest there is. However, it is arguably the most intuitive one, and it can be easily understood by the reader who is familiar with the standard treatment of elasticity~\cite[e.g.,][]{LL_elasticity}. For this reason, we choose it as our entry point, to pave the way for the EFT of other systems, which can be either less intuitive or more subtle, or both.

We usually think of solids as being made up of a lattice of atoms, such as a cubic lattice. At large distances, however, the discreteness of the lattice becomes invisible, and we are left with a continuous material. (This is, in fact, how homo sapiens directly perceives solids.) From a symmetry viewpoint, this creates a little puzzle. On the one hand, the lattice structure breaks spatial translations down to a discrete subgroup. We thus expect to have Goldstone excitations, transforming nonlinearly under generic translations. On the other hand, in the long-wavelength limit the unbroken discrete subgroup becomes indistinguishable from the full group of continuous translations, and so we might be tempted to say that translations are unbroken at large distances. So, do the Goldstone excitations survive at long-wavelengths, and how do they transform under translations?

The resolution of this puzzle is that a long-distance EFT for solids must involve {\em two} different continuous translation groups. Indeed, we can consider a field theory that is invariant under Poincar\'e symmetry as well as under some internal symmetries, in a state that breaks boosts, spatial translations and the internal symmetries, but such that there are unbroken symmetries that play the role of unbroken translations at long distances.

Concretely, we can think of the configuration space of a continuous solid in the following way.
Each volume element comes with three {\em comoving} coordinates $(\phi^1, \phi^2, \phi^3)$, associated with some arbitrary internal coordinate system. If the solid gets deformed and volume elements move around, their comoving coordinates just follow them. They are ``labels" attached to the individual volume elements. So, the configuration space of a continuous solid
can be parametrized in terms of three scalar fields~\cite[e.g.,][]{Soper:1976bb}
\begin{align} \label{solid fields}
    \phi^I(\vec x, t) \,, \qquad \text{ with } \qquad I =1,2,3 \, ,
\end{align}
which tell us the comoving coordinates corresponding to a volume element that, at time $t$, occupies physical position $\vec x$. One could alternatively use the equivalent description in terms of $\vec x \left(\phi^I, t\right)$---the physical position at time $t$ of the volume element with labels $\phi^I$---but the one above is more convenient as far as the spacetime symmetries go: the $\phi^I$'s are three Poincar\'e scalar fields, as functions of $(\vec x, t) = x^\mu$.\footnote{The descriptions in terms of $\phi^I(\vec{x},t)$ and $\vec{x}\left(\phi^I,t\right)$ typically go under the names of ``Eulerian'' and ''Lagrangian'' descriptions, respectively.} 

\begin{figure}
	\centering
	\resizebox{0.85\textwidth}{!}{
        \begin{tikzpicture}
			\draw[->, gray] (0,0) -- (2,0);
			\draw[->, gray] (0,0) -- (0,2);
			\draw[->, gray] (0,0) -- (-1,-1);

            		\draw[->, semithick, Maroon] (0,0) -- (1.25,0);
			\draw[->, semithick, Maroon] (0,0) -- (0,1.25);
			\draw[->, semithick, Maroon] (0,0) -- (-0.55,-0.55);
			
			\node[scale = 0.5, gray] at (2.1,0) {$y$};
			\node[scale = 0.5, gray] at (-1.08,-1.08) {$x$};
			\node[scale = 0.5, gray] at (0,2.1) {$z$};

           		\node[scale = 0.5, Maroon] at (1.35,0.17) {$\phi^2$};
			\node[scale = 0.5, Maroon] at (-0.55,-0.3) {$\phi^1$};
			\node[scale = 0.5, Maroon] at (0.15,1.35) {$\phi^3$};
									
			\draw[gray, fill = gray, opacity = 0.1] (0,0) rectangle (1.75,1.75);

            		\draw[gray, fill = gray, opacity = 0.1] (-0.75,-0.75) rectangle (1,1);

            		\draw[gray, fill = gray, opacity = 0.1] (1,-0.75) -- (1.75,0) -- (1.75,1.75) -- (1,1);

            		\draw[gray, fill = gray, opacity = 0.1] (0,0) -- (-0.75,-0.75) -- (-0.75,1) -- (0,1.75);

            		\draw[gray, fill = gray, opacity = 0.1] (0,1.75) -- (1.75,1.75) -- (1,1) -- (-0.75,1);

            		\draw[gray, fill = gray, opacity = 0.1] (0,0) -- (1.75,0) -- (1,-0.75) -- (-0.75,-0.75);

            		\node[scale = 0.5, Maroon, align = center] at (0.5, 2.6) {Solid at equilibrium \\[0.3em] $\langle \phi^I(x) \rangle \propto x^I$};

            		\draw[->, gray] (5,0) -- (2+5,0);
			\draw[->, gray] (5,0) -- (5,2);
			\draw[->, gray] (5,0) -- (-1+5,-1);

            		\draw[->, semithick, Maroon] (5,0) .. controls (5.5,0.3) and (5.7,0) .. (1.25+5,-0.04);
			\draw[->, semithick, Maroon] (5,0) .. controls (5.23,0.5) and (4.95,1) .. (4.93,1.25);
			\draw[->, semithick, Maroon] (5,0) .. controls (4.95,-0.35) and (4.65,-0.25) .. (4.45,-0.38);
			
			\node[scale = 0.5, gray] at (2.1+5,0) {$y$};
			\node[scale = 0.5, gray] at (-1.08+5,-1.08) {$x$};
			\node[scale = 0.5, gray] at (5,2.1) {$z$};

            		\node[scale = 0.5, Maroon] at (1.35+5,0.15) {$\phi^2$};
			\node[scale = 0.5, Maroon] at (4.4,-0.22) {$\phi^1$};
			\node[scale = 0.5, Maroon] at (4.8,1.35) {$\phi^3$};

            		\draw[gray, fill = gray, opacity = 0.1] (5,0) .. controls (5.5,0.35) and (6,-0.2) .. (6.75,0) .. controls (6.55, 0.5) and (7.05, 1.2) .. (6.75,1.75) .. controls (6.2,1.55) and (5.2,1.7) .. (5,1.75) .. controls (4.7,1.4) and (5.3,0.5) .. (5,0);

            		\draw[gray, fill = gray, opacity = 0.1] (5,0) .. controls (5.5,0.35) and (6,-0.2) .. (6.75,0) .. controls (6.45, 0.1) and (6.45, -0.8) .. (6,-0.75) .. controls (5.6,-0.95) and (4.7,-0.6) .. (4.25,-0.75) .. controls (4.3,-0.1) and (4.9,-0.5) .. (5,0);

            		\draw[gray, fill = gray, opacity = 0.1] (6.75,0) .. controls (6.45, 0.1) and (6.45, -0.8) .. (6,-0.75) .. controls (6.1,-0.1) and (5.7,0.4) .. (6,1) .. controls (6.2,1.2) and (6.65,1.4) .. (6.75,1.75) .. controls (7.05, 1.2) and (6.55, 0.5) .. (6.75,0);

            		\draw[ gray, fill = gray, opacity = 0.1] (6,-0.75) .. controls (5.6,-0.95) and (4.7,-0.6) .. (4.25,-0.75) .. controls (4.05,-0.5) and (4.45,0.85) .. (4.25,1) .. controls (4.6,1.2) and (5.7,0.8) .. (6,1) .. controls (5.7,0.4) and (6.1,-0.1) .. (6,-0.75);

            		\draw[gray, fill = gray, opacity = 0.1] (5,0) .. controls (4.9,-0.5) and (4.3,-0.1) .. (4.25,-0.75) .. controls (4.05,-0.5) and (4.45,0.85) .. (4.25,1) .. controls (4.6,1.5) .. (5,1.75) .. controls (4.7,1.4) and (5.3,0.5) .. (5,0);

            		\draw[gray, fill = gray, opacity = 0.1] (5,1.75) .. controls (5.2,1.7) and (6.2,1.55) .. (6.75,1.75) .. controls (6.65,1.4) and (6.2,1.2) .. (6,1) .. controls (5.7,0.8) and (4.6,1.2) .. (4.25,1) .. controls (4.6,1.5) .. (5,1.75);

            		\node[scale = 0.5, Maroon, align = center] at (0.5+5, 2.6) {Solid away from equilibrium \\[0.3em] $\phi^I(x) \propto x^I + \pi^I(x)$};
		\end{tikzpicture}
	}
	\caption{\textbf{Left panel:} Relation between physical and comoving coordinates for a solid at equilibrium, Eq.~\eqref{solid equilibrium}. \textbf{Right panel:} Same relation but for a solid slightly away from equilibrium, Eq.~\eqref{eq: solid phonon field def}.} \label{fig: solid comoving coordinates}
\end{figure}
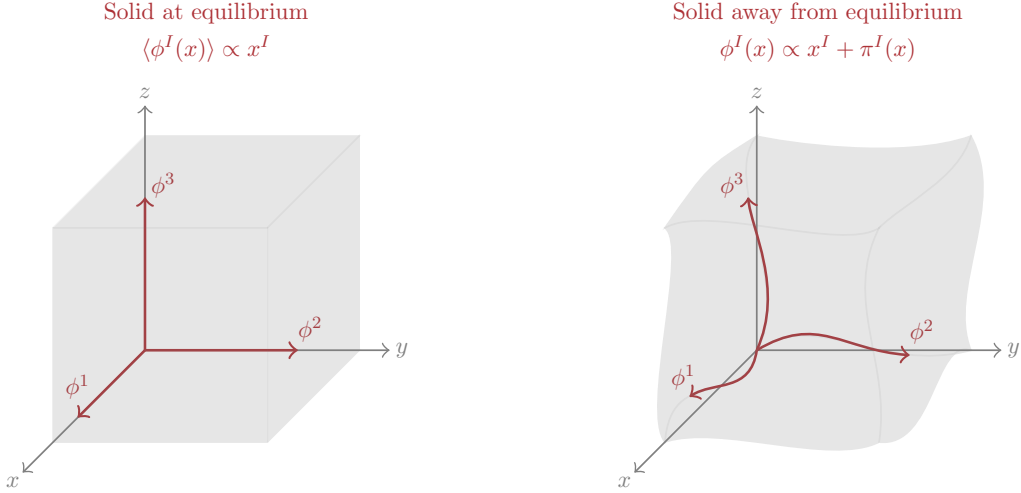

This characterization is still too general though. To make progress, we have to choose the internal coordinate system wisely. A particularly convenient choice corresponds to requiring that for a solid in static equilibrium at some given (possibly vanishing) external pressure, the internal coordinate system is aligned to that of our lab, 
\begin{align} \label{solid equilibrium}
    \left\langle \phi^I(x) \right\rangle = \alpha \spacy x^I \,,
\end{align}
where we indicate the equilibrium configuration with $\langle{\cdots}\rangle$, anticipating its field theory meaning of ground state expectation value. See Figure~\ref{fig: solid comoving coordinates} for a pictorial representation.
Different values of the dimensionless parameter $\alpha$ correspond to physically distinct solid configurations that can be achieved by varying the external pressure.  The above field configuration breaks a number of spacetime symmetries: Lorentz boosts, rotations, and, most  importantly, spatial translations. On the other hand, we would like the equilibrium state of our continuous solid to be homogeneous, that is, translationally invariant. 

The only way out seems to be to postulate the existence of internal translational symmetries, that is, shift symmetries
\begin{align} \label{solid shifts}
    \phi^I(x) \to \phi^I(x) + a^I \, , \qquad \text{ with } \qquad a^I = {\rm constant} \, .
\end{align}
In this way, even though the equilibrium configuration above breaks {\em both} spatial translations {\em and} these shift symmetries, it is invariant under their combined action. One can first perform a spatial translation, and then a suitably chosen internal shift to cancel it out. It is this unbroken linear combination of spatial and internal symmetries that plays the role of the unbroken translational symmetry of a solid in the continuum limit. 

One might question pushing the homogeneous continuous solid analysis too far: if the unbroken translations are  secretly  discrete because of the microscopic underlying lattice, then one can expect higher derivative corrections to break the continuous ``unbroken" translations. In other words, perhaps more familiar to the particle physicists among our readers: if the unbroken continuous translations are in fact  accidental symmetries at low energies, akin to baryon number in the Standard Model, then one generically expects higher dimensional operators to break them. However, it can be proved that in the long-distance effective theory there is no difference between discrete translations with microscopic spacing and continuous translations {\em to all orders in the derivative expansion} \cite{Esposito:2020wsn}. From the viewpoint of the derivative expansion, the difference is non-perturbative.

We are now in a position to write down the effective theory for the mechanical deformations of a homogeneous continuous solid. Invariance under the internal shift symmetries, Eq.~\eqref{solid shifts}, requires for the comoving coordinates to always appear derived, $\partial_\mu\phi^I$. Imposing Lorentz invariance is now simply a matter of contracting free Lorentz indices. 
To lowest orders in the derivative expansion, this isolates the Lorentz scalar matrix
\begin{align} \label{B solid}
    B^{IJ} \equiv \partial_\mu \phi^I \partial^\mu \phi^J \,,
\end{align}
as the only independent invariant. We now have to decide what to do regarding the internal $I,J $ indices.

\subsection{Rotational Symmetries}

\noindent Different solids in general have different symmetries. At the microscopic level, the symmetries of a given crystal make up a group that involves both discrete translations and discrete rotations (and possibly inversion symmetries). At the macroscopic  level---that is, in the continuum limit discussed here---all translational symmetries become continuous. Still, the underlying lattice structure selects some preferred directions, which survive at long distances. And so, at equilibrium, our solid will only be invariant under a discrete subgroup of rotations, $G \subset SO(3)$.  We refer the reader to \cite{Landau:1980mil}, where all this is reviewed in detail. 

For our purposes here, it is enough to understand how to make the equilibrium configuration in Eq.~\eqref{solid equilibrium} invariant under a discrete subgroup of rotations. Clearly, the right hand side breaks rotations. So, following the example of what we did for translations, we can postulate that there is an {\em internal} symmetry group $G$ acting on our fields,
\begin{align} \label{solid rotations}
    \phi^I(x) \to O^{I} {}_J \, \phi^J(x) \, , \qquad \text{ with }  \qquad O \in G \subset SO(3) \, ,
\end{align}
so that the configuration \eqref{solid equilibrium} breaks both spatial translations and the internal $G$, but is invariant under a combined action of a {\em spatial} $G$ transformation and an identical internal one. It is this unbroken combination that corresponds to the $G$ rotational symmetry of the underlying crystal at equilibrium.

It is useful to keep in mind the isotropic solid limit. This corresponds to choosing $G = SO(3)$. Clearly, no crystal has full rotational symmetry, and so the isotropic limit is at best a toy model or, perhaps more interestingly, a long-distance approximation: in many cases, the orientation of the underlying crystal structure does not survive over very large distances,  for instance because of how the material solidified in the first place. If we call $\ell$ the lattice spacing, and $L \gg \ell$ the typical size of the domain over which the crystal structure is regular, and assuming that at much larger scales the orientations of such domains are random, we have two qualitatively different regimes, both in the continuum limit. At intermediate distances, $\ell \ll \lambda \ll L$, we can focus on a single domain, and we can use an effective field theory for our fields, $\phi^I$, with the same symmetry group, $G$, as the underlying lattice. On the other hand, at much larger distances,  $\lambda \gg L$, many domains with random orientations are averaged over, and we recover full rotational invariance, in which case our solid will behave as if $G = SO(3)$.
The same considerations apply to {\em amorphous} solids, which have no underlying crystal structure even at microscopic distances \cite{Landau:1980mil}. In that case, the relative locations of the atoms making up solid is random, and statistically isotropic. Averaging over large enough volumes is equivalent to statistical averaging, and one ends up with an isotropic solid at large distances.

Finally, notice that not all discrete subgroups of rotations are symmetries of possible crystal structures. For example, there is no periodic crystal whose symmetry group is the icosahedral group~\cite{aroyo2016international}.\footnote{In that particular case, there is a {\em quasi}-crystal with that symmetry group, both as a theoretical construct and as a solid realized in nature~\cite{levine1984quasicrystals}.}
This is because a crystal is invariant under certain discrete translations as well, and not all discrete subgroups of rotations can be augmented with discrete translations to form a discrete subgroup of the Euclidean group.

\subsection{The Effective Action}

\noindent To summarize: a solid in the continuum limit can be described by three scalar fields $\phi^I(x)$ acted upon by the shift symmetries in Eq.~\eqref{solid shifts}, and by some rotational symmetry group $G$, as in Eq.~\eqref{solid rotations}. To lowest order in the derivative expansion, the effective Lagrangian must be a $G$-invariant function of the Lorentz- and shift-invariant combination $B^{IJ}$ in Eq.~\eqref{B solid},
\begin{align} \label{general solid}
    S = -\int d^4 x \, F \left( B^{IJ} \right) + \text{higher } \partial\text{'s} \, ,
\end{align}
with $F$ invariant under $B \to O \cdot B \cdot O^T$, for any $O \in G$.

For example, in the isotropic limit, when $G = SO(3)$, $F$ must be a rotationally invariant function of the matrix $B$, which means that it can only depend on its determinant, its trace, the trace of $B^2$, and so on. In particular, since in four spacetime dimensions $B$ is a $3 \times 3$ symmetric matrix, there are only three independent invariants, which we could take to be the traces of $B$, $B^2$, and $B^3$. However, we will see that there is a different basis of invariants that it more convenient to use on physical grounds, namely
\begin{align} \label{eq: isotropic solid invariants}
	b \equiv \sqrt{\det B} \, , \qquad Y \equiv \frac{[B^2]}{{[B]}^2} \, , \qquad Z \equiv \frac{[B^3]}{{[B]}^3} \, ,
\end{align}
where we have denoted the trace operation with $[{\dots}]$.\footnote{The two parametrizations are completely equivalent, since all these invariants are regular around the background in Eq.~\eqref{solid equilibrium}, and the determinant of a $3\times3$ symmetric matrix can be expressed as
\begin{align}
	\det B = \tfrac{1}{6} \left( {[B]}^3 -3[B][B^2]+2 [B^3] \right) \, .
\end{align}
} The advantage of this basis is that local, isotropic compressions or dilutions are now solely capture by the invariant $b$, which in fact is equal to the Jacobian of the change from comoving to spatial coordinates: 
\begin{align} \label{eq: Jacobian comoving to spatial}
	dt d^3 \phi = b \, d^4 x \ .
\end{align}
As we will see in the next section, $b$ admits a simple thermodynamic interpretation. The quantities $Y$ and $Z$, instead, are invariant under rescalings of $B^{IJ}$ and capture anisotropic local deformations---a choice that proves to be convenient from a thermodynamic perspective, and also for cosmological applications~\cite{Endlich:2012pz}. In this case, the Lagrangian takes the form: 
\begin{align} \label{eq: solid effective Lagrangian}
F = F\big( b, Y, Z \big) \, , \qquad \qquad \qquad (\, G = SO(3) \,) \,.
\end{align}

For discrete subgroups of $SO(3)$, it is not immediate how to write down general nonlinear invariants in the same fashion. However, the process is systematic at any given order in a Taylor expansion about the equilibrium configuration \eqref{solid equilibrium}. We refer the reader to the literature~\cite[e.g.,][]{Kang:2015uha,Kang:2018bqc,Nicolis:2020rqz,Esposito:2020wsn} for a few examples of this procedure.

Before proceeding further with an analysis of our effective theory, we want to characterize the function $F$ that defines the action and the invariants it depends on in physical terms.

\subsection{The Thermodynamics of Solids} \label{sec: thermodynamics of solids}

\noindent One can easily generalize the standard thermodynamic relations of gases and more general fluids to solids, provided one takes into account that, if the solid is strained, equilibrium configurations will be characterized by general anisotropic stresses rather than just an isotropic pressure. As a consequence, the first law of thermodynamics, usually written as $d E = T dS -p \, dV + \upmu \, dN$, gets generalized to
\cite{LL_elasticity}
\begin{align}
    d E = T dS - V \, T_{ij} \, d u_{ij} + \upmu \, dN\, ,
\end{align}
where $T_{ij}$ is the stress tensor, $d u_{ij}$ is a suitably defined variation of the strain tensor (see \ref{sec: strain}), and of course $E$, $T$, $S$, $N$ and $V$ are the  energy, temperature, entropy, particle number and volume of the solid's portion under consideration. As usual, it is convenient to factor out the volume from the extensive quantities,
\begin{align}
    E = \rho \spacy V \, , \qquad\qquad  S = s \spacy V \, ,  \qquad\qquad  N = n \spacy V \,,
\end{align}
so that the first law becomes
\begin{align}
    V d\rho + \rho \, dV = V T ds + T s \, dV - V \, T_{ij} \, d u_{ij} + V \upmu \,
    d n + \upmu \, n \, dV\, .
\end{align}
Using that $V d u_{ii} = dV$ \cite{LL_elasticity}, 
and decomposing the stress tensor into its trace and traceless parts,
\begin{align} \label{eq:Tijsolid}
    T_{ij}  \equiv p \, \delta_{ij} + {\cal T}_{ij} \, , \quad \text{ with } \quad {\cal T}_{ii} = 0 \, , 
\end{align}
this becomes
\begin{align}
    (\rho + p -Ts - \upmu \, n) \, d V +  (d \rho - T  ds + {\cal T}_{ij} \,  du_{ij} - \upmu \, dn) \, V  = 0 \, .
\end{align}
As usual, this is a convenient way to rewrite the first law, because the terms in parentheses only involve {\em intensive} quantities, that is, local physical quantities that do not depend on the volume of the solid's portion being considered. Since the extensive quantity $dV$ can be varied independently from the intensive differentials $d \rho$, $ds$, and $du_{ij}$, the two terms in parentheses must vanish separately.
We thus find the two independent relationships
\begin{align} \label{thermo eqs solid}
    \rho + p = T s + \upmu \, n \, ,\qquad\qquad   d \rho= T  ds - {\cal T}_{ij} \,  du_{ij} + \upmu \, dn \, .
\end{align}
We see that the standard thermodynamic relationships are here modified by the presence of the last term in the second equation, which is nonzero for strained solids featuring anisotropic stresses.

How do we recover such  relationships from our field theory, and what do the thermodynamic quantities above map to in terms of our field variables? We start from the effective action in Eq.~\eqref{general solid} and derive the energy--momentum tensor, either through the Noether procedure~\cite[e.g.,][]{Weinberg:1995mt}, or by varying with respect to the metric~\cite[e.g.,][]{Weinberg:1972kfs}, upon the replacements $d^4 x \to d^4x \spacy \sqrt{-g}$ and $B^{IJ} \to g^{\mu\nu}\partial_\mu \phi^I \partial_\nu \phi^J$. We get,
\begin{align} \label{eq: solid stress energy tensor}
    T_{\mu\nu} = 2 \frac{\partial F}{\partial B^{IJ}} \spacy {\partial_\mu \phi^I \partial_\nu \phi^J} - F \spacy \eta_{\mu\nu} \,.
\end{align}
To match our EFT to the standard thermodynamics of solids at equilibrium, we restrict ourselves to a generic {\it static} configuration, $\phi^I(\vec x)$. Even in the homogeneous limit, a static configuration is actually more general than what we presented earlier, in Eq.~\eqref{solid equilibrium}. In particular, it can also feature some strain, that is, a spatially modulated combination of a dilation (or compression) and a shear deformation. Accounting for this as well, the most general {\it homogeneous} and {\it static} configuration of our solid is described by a field profile that is invariant under the combined action of spatial translations and internal shifts, which is (up to an irrelevant additive constant vector),
\begin{align}
    \phi^I (\vec x) = A_i{}^I \spacy x^i  \, ,
\end{align}
where $A$ is a constant (invertible) matrix. The equilibrium configuration in Eq.~\eqref{solid equilibrium} is obtained as a special case of this, $A_i{}^I = \alpha \spacy \delta^I_i$.

Let us now give up the assumption of homogeneity, while still working in the static limit. In this instance, the entries of the energy--momentum tensor reduce to,
\begin{align} \label{Tmn solid}
    T_{00} = F \, , \qquad T_{0i} = 0 \, , \qquad T_{ij} = 2 \frac{\partial F}{\partial B^{IJ}} J_i{}^I \spacy J_j{}^J  - F \spacy \delta_{ij} \,,
\end{align}
where $J$ is the Jacobian matrix (the generalization of the $A$ matrix above),
\begin{align} \label{eq: def Jacobian solid}
    J_i{}^I (\vec x \, ) \equiv \frac{\partial \phi^I(\vec x \,)}{\partial x^i} \,.
\end{align}
From $T_{00}$ and by comparing $T_{ij}$ with Eq.~\eqref{eq:Tijsolid}, we can immediately read off $\rho$ and $p$  in terms of our Lagrangian and field theory variables,
\begin{align} \label{eq: rho and p solid}
    \rho = F \, , \qquad p = \frac{2}{3} \frac{\partial F}{\partial B^{IJ}} \spacy B^{IJ} - F \, ,
\end{align}
where we used the fact that, for static configurations, $B^{IJ} = J_i{}^I \spacy J_i{}^J$. In order to obey the first of the thermodynamics identities in Eq.~\eqref{thermo eqs solid}, we need
\begin{align} \label{eq: thermo identification solid}
\frac23 \frac{\partial F}{\partial B^{IJ}} \spacy B^{IJ} = b \spacy \frac{\partial F}{\partial b} \bigg|_{Y,Z} = b \spacy \frac{\partial \rho}{\partial b} \bigg|_{Y,Z} = T \spacy s + \upmu \spacy n \, ,
\end{align}
where we explicitly indicated that, when deriving with respect to $b$, $Y$ and $Z$ are kept constant.

In light of the thermodynamic relations $T = \partial\rho/\partial s|_{n,u}$ and $\mu = \partial\rho/\partial n|_{s,u}$ (see Eq.~\eqref{thermo eqs solid}), the last equality can be read as
\begin{equation}
b\,\frac{\partial\rho}{\partial b}\bigg|_{Y,Z}  = s\,\frac{\partial\rho}{\partial s}\bigg|_{n,u} + n\,\frac{\partial\rho}{\partial n}\bigg|_{s,u}\,.
\end{equation}
In other words, the variation of $\rho$ induced by a compression $b\,\partial_b$ coincides with its variation along the \emph{radial} direction $(s\,\partial_s + n\,\partial_n)$ in the $(n,s)$ plane. The EFT therefore has access to changes in $\rho$ only along this single, fixed direction---the ray emanating from the origin through the point $(n,s)$---rather than to the two independent directions $\partial_s$ and $\partial_n$ separately. Matching the two requires $n$ and $s$ to vary proportionally to $b$, i.e. $n \propto s \propto b$, so that the ratio is held fixed and we are restricted to the \emph{isentropic regime},
\begin{equation} \label{eq: isentropic condition}
\sigma \equiv s/n = \text{constant}\,.
\end{equation}
This makes sense: Eq.~\eqref{eq: Jacobian comoving to spatial} shows that, loosely speaking, $dV \propto 1/b$, so $b$ acts as an overall scaling of the comoving volume. The EFT can thus describe changes in the densities of extensive quantities only through local mechanical compressions or dilations, which rescale $n$ and $s$ together and leave $\sigma$ untouched. The isentropic condition \eqref{eq: isentropic condition} is more restrictive than the \emph{adiabatic} condition, $u^\mu \partial_\mu (s / n) = 0$, which only demands that $s / n$ remain constant along the direction of motion.\footnote{It is unfortunately confusing that cosmological perturbations satisfying the isentropic condition are referred to as ``adiabatic'' in the literature~\cite{Weinberg:2008zzc}.} As we will discuss in Section~\ref{sec: hydro modes fluid}, in order to describe more general adiabatic motion we need to consider an effective theory with an additional degree of freedom and an additional symmetry. 

It is convenient to set the proportionality constant between $n$ and $b$ equal to one, which  can be done without loss of generality by rescaling the fields $\phi^I$. Thus, we identify  
\begin{align} \label{eq: solid thermo identifications}
    b = n  \, ,\qquad\qquad \frac{\partial F}{\partial b} \bigg|_{Y,Z} = \left.\frac{\partial \rho}{\partial n} \right|_{\sigma, u} = T\sigma + \upmu\, .
\end{align}
Note that, because of the first equation in \eqref{thermo eqs solid}, the derivative $\partial F/\partial b|_{Y,Z}$ is also equal to the specific enthalpy $h = (\rho + p)/n$.  The identification of $b$ with $n$ was, to the best of our knowledge, first put forward in~\cite{Soper:1976bb}, and it is natural from a physical standpoint: it guarantees that our EFT admits a smooth $T \to 0$ limit. Indeed, $b$ is the geometric density built out of the comoving field configuration, and it remains finite for any solid at equilibrium irrespective of its temperature. Tying it to the conserved charge density $n$---which generically stays finite as $T \to 0$---rather than to the entropy density $s$---which generically vanishes in that limit by the third law of thermodynamics---keeps $b$ finite throughout, so the EFT remains well-defined down to zero temperature. This is in contrast to fluids, where, as we will see in Section~\ref{sec:Eulerianfluids}, the $T \to 0$ limit is most likely unphysical and the natural choice is instead $b = s$.

The identifications \eqref{eq: solid thermo identifications} ensure that the second thermodynamic relationship in Eq.~\eqref{thermo eqs solid} is also satisfied, as we now check. To this end, we notice that
\begin{align}
    dn = \frac{ds}{\sigma} = d b = \frac12 b \; [B^{-1} \cdot d B] \, ,
\end{align}
and therefore
\begin{align}
    d \rho - T ds - \upmu \, dn = \left[ \frac{\partial F}{\partial B} \cdot dB \right] - \frac{1}{3} \left[ \frac{\partial F}{\partial B} \cdot B \right] \left[ B^{-1} \cdot dB \right]  \, ,
\end{align}
where we have once again used the notation $[{\cdots}]$ to denote the trace.  On the other hand, the  strain tensor and our $B = J^t \cdot J$ are related by (see \ref{sec: strain}),
\begin{align}
    d B = - 2 \spacy J^t \cdot du \cdot J \, ,
\end{align}
and so, using the cyclicity of the trace, we get,
\begin{align}
    \begin{split}
        d \rho - T ds - \upmu \, dn ={}& -2 \left[ J \cdot \frac{\partial F}{\partial B} \cdot J^t \cdot du - \frac13 \left[ J \cdot \frac{\partial F}{\partial B} \cdot J^t \right]   du \right] = - {\cal T}_{ij} \, d u_{ij} \, ,
    \end{split}
\end{align}
where we used the fact that ${\cal T}_{ij}$ is precisely the traceless part of $T_{ij}$ in Eq.~\eqref{Tmn solid}.

In summary, the thermodynamic relationships \eqref{thermo eqs solid} are automatically obeyed by our solid EFT upon the identifications,
\begin{align}
    \rho = F(B^{IJ}) \, , \qquad\qquad  n = \sqrt{\det{B}} \, , \qquad\qquad T\sigma + \upmu = \frac{\partial F\left(B^{IJ}\right)}{\partial \sqrt{\det B}} \bigg|_{Y,Z} \, .
\end{align}
Furthermore, we show in \ref{sec: strain} that the relationship between the strain tensor and our $B$ matrix is purely geometrical, and simple in comoving space. Explicitly, we find
\begin{align}
    \big(B^{-1}\big)_{IJ} = \delta_{IJ } + 2 \spacy u_{IJ}  \, ,
\end{align}
but we refer the reader to that appendix for details about how to interpret this relationship for nonlinear deformations.

We notice in passing that our $B^{IJ}$ is  Poincar\'e invariant, invariant under the solid's internal shifts, and transforms linearly under the internal rotational symmetry group $G \subset SO(3)$. On the other hand, the strain tensor that is usually introduced in the literature~\cite{LL_elasticity} has more subtle transformation properties, and so it is not obvious  how to use it as a  building block beyond linear order. (We elaborate on this in \ref{sec: strain}.) Thus, we believe that one of the advantages of our solid effective theory, with its emphasis on symmetries, is to provide a nonlinear generalization of the strain tensor that is the {\em most general} shift invariant, Poincar\'e invariant building block with tensorial transformation properties under the solid's rotation symmetry group.

Although in most of this section we have restricted our attention to static configurations in order to make contact with the standard thermodynamic description of solids in equilibrium, the effective theory also captures all dynamical phenomena taking place at scales much larger than the interatomic distance. For instance, for a solid in motion, we can define its four-velocity  field $u^\mu(x)$ by demanding that the energy flux be $u^\mu \spacy T_{\mu\nu} = - \rho \spacy u_\nu = -F \spacy u_\nu$. Based on the form of the stress energy tensor in Eq. \eqref{eq: solid stress energy tensor}, this relation is verified if only if $u^\mu \partial_\mu \phi^I = 0$, which matches an alternative definition of the four-velocity field---a vector field along which the comoving coordinates do not change. Keeping in mind the normalization condition $u^\mu u_\mu = -1$, this implies \footnote{The fact that $u^\mu \partial_\mu \phi^I = 0$ follows because the $I$-type indices can only take three values. When computing $u^\mu \partial_\mu \phi^I = 0$, two of them will then take the same value. At the same time, the $\epsilon^{\mu\nu\lambda\sigma}$ factor anti-symmetrizes all the factors, making the expression vanish.} 
\begin{align} \label{eq:umu for solids}
	u^\mu =  \frac{1}{3!} \frac{1}{b} \, \epsilon^{\mu\nu\lambda\sigma} \epsilon_{IJK} \partial_\nu \phi^I \partial_\lambda \phi^J \partial_\sigma \phi^K  \, .
\end{align}
Incidentally, this also shows that the particle number current is conserved:
\begin{align} \label{eq: conservation particle number current}
	\partial_\mu (n u^\mu) =  \partial_\mu (b u^\mu) = \tfrac{1}{2!} \,  \epsilon^{\mu\nu\lambda\sigma} \epsilon_{IJK} \partial_\mu \partial_\nu \phi^I \partial_\lambda \phi^J \partial_\sigma \phi^K  = 0 \, .
\end{align}
It is interesting to note that conservation of the particle number current is an identity that holds regardless of the solid equation of state—that is, for any functional form of $F$. Furthermore, in the isentropic regime, Eq. \eqref{eq: conservation particle number current} also implies conservation of the entropy current, which holds at lowest order in the derivative expansion where dissipative effects are neglected.\footnote{In fact, the isentropic regime is no longer well-defined in the presence of dissipation since, unlike entropy, particle number remains exactly conserved.}

\subsection{Phonons} \label{sec:solidphonons}

\noindent Phonons correspond to small fluctuations of the solid around the equilibrium configuration in Eq.~\eqref{solid equilibrium}. They are Goldstone bosons associated with the symmetries that are spontaneously broken by this background solution~\cite[e.g.,][]{Leutwyler:1996er}. Therefore, they are derivatively coupled and their interactions can be treated perturbatively at small energy and momenta. They can be parametrized as small fluctuations of the fields $\phi^I$ around their equilibrium configuration:
\begin{align} \label{eq: solid phonon field def}
	 \phi^I(x) = \alpha \left( x^I + \pi^I (x) \right) \,,
\end{align}
as again represented pictorially in Figure~\ref{fig: solid comoving coordinates}.
Focusing for simplicity on the isotropic case, $G=SO(3)$, we can plug this expression into the effective Lagrangian \eqref{eq: solid effective Lagrangian} and expand up to cubic order in powers of the phonon field  $\pi^I (x)$, to find
\begin{align} \label{eq: cubic action phonon solids}
    \begin{split}
        S = & \int d t \spacy d^3 x  \, \left(\bar b \, F_b \right) \bigg\{\frac{1}{2} \spacy \dot{\vec{\pi}}^2-\frac{c_L^2-c_T^2}{2}  \left[\nabla \pi \right]^2 -\frac{c_T^2}{2} \left[\nabla \pi \, \nabla \pi^t\right]  \\
        & \qquad \qquad \qquad \qquad \; + g_1\left[\nabla \pi\right]^3 + g_2\left[\nabla \pi\right]\left[\nabla \pi^2\right] +g_3 \left[\nabla \pi\right]\left[\nabla \pi \, \nabla \pi^t\right] \\
        & \qquad \qquad \qquad \qquad \quad +g_4\left[\nabla \pi^2 \, \nabla \pi^t\right]+g_5\left[\nabla \pi\right] \dot{\vec{\pi}}^2+g_6\spacy \dot{\vec{\pi}} \cdot \nabla \pi \cdot \dot{\vec{\pi}} + \dots \bigg\} \, ,
    \end{split}
\end{align}
where we have defined the matrix $(\nabla \pi)_{ij} \equiv \nabla_i \pi_j$, the background value of $b$ is $\bar b \equiv \alpha^3$, and we have parametrized the quadratic part of the action in terms of the sound speeds for longitudinal ($c_L$) and transverse ($c_T$) phonons. This is confirmed by splitting the phonon field into its longitudinal and transverse components: 
\begin{align}
        \vec{\pi} = \vec{\pi}_L + \vec{\pi}_T \,, \qquad\qquad \text{ with } \qquad\qquad \vec{\nabla}\cdot\vec{\pi}_T = \vec{\nabla}\times\vec{\pi}_L = 0 \,.   
\end{align}
The linear equations of motions for these two components are indeed given by,
\begin{align}
    \ddot{\vec{\pi}}_L - c_L^2 \vec{\nabla}\big(\vec{\nabla}\cdot\vec{\pi}_L\big) = 0  \,, \qquad\qquad \ddot{\vec{\pi}}_T - c_T^2 \nabla^2 \vec{\pi}_T = 0 \,.
\end{align}
Finally, we have denoted with $F_b$ the derivative $\partial F / \partial b$ evaluated on the background solution. Adopting a similar notation for the other derivatives of $F$, the explicit expressions for the sound speeds and the cubic coefficients are,
\begin{subequations} \label{eq:solidcoefficients}
\begin{align}
    c_T^2 ={}& \frac{4}{9} \frac{F_Y+F_Z}{\bar b F_b} \,, \\
    c_L^2 ={}& \frac{\bar b F_{bb}}{F_b}+\frac{16}{27} \frac{F_Y+F_Z}{\bar b F_b} \,, \label{eq: longitudinal speed solids} \\
    g_1={}& \frac{2}{9} c_T^2-\frac{1}{2} c_L^2-\frac{7}{243} \frac{F_Z}{\bar bF_b}+\frac{4}{27}\frac{F_{bY}+F_{bZ}}{F_b}-\frac{1}{6} \frac{{\bar b}^2 F_{bbb}}{F_b} \,, \\
    g_2={}& \frac{1}{2} c_L^2+\frac{1}{27} \frac{F_Z}{\bar b F_b}-\frac{2}{9}\frac{F_{bY}+F_{bZ}}{F_b} \,, \\
    g_3={}&c_T^2+\frac{4}{27} \frac{F_Z}{\bar bF_b}-\frac{2}{9}\frac{F_{bY}+F_{bZ}}{F_b} \,, \\
    g_4={}&-c_T^2-\frac{2}{9} \frac{F_Z}{\bar bF_b} \,, \\
    g_5={}&\frac{1}{2}+\frac{1}{2} c_L^2- c_T^2 \,, \\
    g_6={}&-1+c_T^2 \,.
\end{align}
\end{subequations}
Note that in Eq.~\eqref{eq: cubic action phonon solids} we have extracted a factor of $\bar b F_b$, which is now the only dimensionful parameter that appears in the Lagrangian and, on naturalness grounds, it is the scale that suppresses phonon self-interactions. Based on Eqs.~\eqref{thermo eqs solid} and \eqref{eq: thermo identification solid}, this prefactor is equal to the background value of the enthalpy density, $\bar \rho + \bar p$. Thus, absence of ghosts is guaranteed provided the null energy condition is satisfied, i.e. $\bar \rho + \bar p>0$~\cite{Kontou:2020bta}, which is of course true for any ordinary solid. 

Note that, up to cubic order there are eight effective coefficients: the two sound speeds, and the six cubic coefficients $g_i$. Nonetheless, only five independent quantities appear in Eqs.~\eqref{eq:solidcoefficients}: 
\begin{align} \label{eq: solid 5 cubic parameters}
	c_L^2 \,, \qquad c_T^2 \,, \qquad \frac{F_Z}{\bar bF_b} \,, \qquad \frac{F_{bY}+F_{bZ}}{F_b} \,, \qquad \frac{{\bar b}^2 F_{bbb}}{F_b} \,.
\end{align}
Indeed, the eight effective coefficients are not all independent, as they satisfy the following three relations between cubic and quadratic coefficients,
\begin{align} \label{eq: solid cubic constraints}
    2 g_2 - g_3 - g_4 = c_L^2 - c_T^2 \,, \qquad 2g_5 + g_6 = c_L^2 - c_T^2 \,, \qquad g_6 = -1 + c_T^2 \,.
\end{align}
These constraints all follow from invariance under the original Lorentz group (boosts and rotations), which acts nonlinearly on the Goldstone fields. These are nontrivial predictions of the effective theory. 
We show how this comes about in full detail in~\ref{app:nonlinearLorentz}.

As a simple application of the cubic action \eqref{eq: cubic action phonon solids}, we can calculate perturbatively the corrections to the phonon dispersion relations in a slightly stressed sample. Such configuration can be modeled by giving the Goldstone fields a small background value, $\langle\pi^I\rangle = \gamma^I{}_J x^J$, with $\gamma^I{}_J \ll 1$. As one can check, this is a solution to the equations of motion for the $\vec{\pi}$'s, and changes the background value of the comoving coordinates to,
\begin{align}
    \langle \phi^I(x) \rangle = \left( \alpha \spacy \delta^I{}_J + \gamma^I{}_J \right) x^J \,.
\end{align}
In other words, at equilibrium, there is now a mismatch between the physical and comoving coordinates, because the latter have been compressed, dilated, or sheared with respect to the former. At the level of the action, this amounts to taking Eq.~\eqref{eq: cubic action phonon solids} and replacing $\pi^I \to \gamma^I{}_J \spacy x^J + \pi^I$.
In this regime, the leading corrections to the quadratic action come from replacing one of the Goldstone field in the cubic part of the action with its background value. A similar logic was employed in~\cite{Endlich:2013jia} to derive consistency relations in a model of the early universe where inflation is driven by a solid~\cite{Endlich:2012pz}. It was also used in~\cite{Pavaskar:2021pfo} to calculate the corrections to the dispersion relations of magnons in a stressed sample coming from their interactions with phonons.

It is convenient to decompose the matrix $\gamma_{IJ}$ into spin-0, spin-1, and spin-2 representations: 
\begin{align}
	\gamma_{IJ} = \frac{1}{3} \delta_{IJ} \spacy \gamma  + \epsilon_{IJK} \spacy \theta^K \ + \sigma_{IJ} \,,
\end{align}
with $\sigma_{IJ}$ symmetric and traceless, and $\gamma$ the trace of $\gamma_{IJ}$. The term proportional to $\gamma$ corresponds to a uniform compression of the solid that changes $\alpha \to \alpha + \gamma / 3$. Thus, the dispersion relations remain qualitatively of the same form, with the sounds speeds evaluated at a slightly different value of the entropy density. At first order in $\gamma$, the new dispersion relations are obtained by replacing
\begin{align}
	c_{L,T}^2 \to c_{L,T}^2 + \frac{\gamma}{\alpha} \frac{\partial c_{L,T}^2}{\partial \log s} \, .
\end{align}
The term proportional to $\theta^K$, instead, describes an infinitesimal spatial rotation of the solid. Since this is a symmetry transformation, it cannot modify the dispersion relations. Finally, $\sigma_{IJ}$ describes shear deformations---the only ones that can introduce anisotropies in the dispersion relations of an otherwise isotropic solid. At linear order in $\sigma_{IJ}$, the quadratic action for the phonons in the presence of a shear deformation becomes 
\begin{align}
\begin{aligned} \label{eq: cubic action phonon solids sheared}
    S^{(2)}_{\rm sheared} = & \int d t d^3 x  \, (\bar b F_b )\left\{\frac{1}{2} \dot{\vec{\pi}}^2-\frac{c_L^2-c_T^2}{2} [\nabla \pi]^2 -\frac{c_T^2}{2} \left[\nabla \pi \, \nabla \pi^t\right] + 2 (g_2 + g_3)[\nabla \pi] [ \sigma \nabla \pi] \right.\\
    & \qquad \qquad \qquad \qquad  +g_4\left[\sigma \nabla \pi \, \nabla \pi^t\right] +g_4\left[\sigma \nabla \pi^t \, \nabla \pi\right] +g_4\left[\sigma \nabla \pi^2 \right]+g_6\spacy\dot{\vec{\pi}}\cdot\sigma \cdot\dot{\vec{\pi}} \bigg\} \,.
\end{aligned}
\end{align}
The wave equation that follows from this action is, in Fourier space,
\begin{align}
    \begin{split}
        & \Big\{  \big(\omega^2 - c_T^2 \spacy k^2 + 2 g_4 \spacy \vec{k}\cdot\sigma\cdot\vec{k} \big) \delta_{IJ} - \left(c_L^2 -c_T^2\right) k_I k_J + 2 \left(g_6 \omega^2 +  g_4 k^2 \right)\sigma_{IJ} \\ 
        & \qquad \qquad \qquad \qquad \qquad \qquad \qquad + \left(2g_2 + 2 g_3 + g_4\right) \left(k_I \sigma_{JL} k^L + k_J \sigma_{IL} k^L \right)\Big\} \pi^J (\omega, \vec k) = 0 \,.
    \end{split}
\end{align}
The dispersion relations are obtained by setting to zero the determinant of the matrix multiplying $\pi^J (\omega, \vec k)$ and solving for $\omega^2$. In fact, it is convenient to study longitudinal and transverse modes separately. The linear correction to the dispersion relation of the longitudinal mode can be easily obtained by projecting the matrix along $\hat k^I \hat k^J$. At leading order in $\sigma_{IJ}$ one finds
\begin{align}
	\omega_L^2 = \tilde c_L^2 (\hat k) \spacy k^2 \,, \qquad \qquad \tilde  c_L^2 (\hat k) \simeq c_L^2 - 2 (g_6 c_L^2 +2 g_2 + 2g_3 + 3g_4) \hat \sigma \,, 
\end{align}
where we have defined $\hat \sigma \equiv \hat k \cdot \sigma \cdot \hat k$. 

To study the transverse modes, we project instead the kinetic matrix in the transverse direction using the projector $P_{IJ} = \delta_{IJ} - \hat k_I \hat k_J$. Then, the relevant $2 \times 2$ matrix is
\begin{align} \label{eq: kinetic matrix M transverse}
	M_{ab} = \left(\omega^2 - c_T^2 \spacy k^2 + 2 g_4 \spacy \hat \sigma \spacy k^2 \right) \delta_{ab} + 2 \left(g_6 \spacy \omega^2 + g_4 \spacy k^2\right) \sigma_{ab} \,, 
\end{align}
where $a,b$ are indices along the two directions orthogonal to $\hat k$, and $\sigma_{ab} \equiv P_a{}^I \sigma_{IJ} P^J{}_b$.  
Now, if $\lambda_\pm$ are the eigenvalues of $\sigma_{ab}$, the eigenvalues of the matrix above are simply,
\begin{align}
    M_\pm = \omega^2 - c_T^2 k^2 + 2 g_4 k^2 \hat{\sigma} + 2(g_4 \spacy k^2 + g_6 \spacy \omega^2) \lambda_\pm \,.
\end{align}
The condition $\det M = 0$ is then satisfied if either of these vanish. Again, at leading order in $\sigma_{ab}$, this implies the following dispersion relations for the transverse modes,
\begin{align}
    \omega_{T,\pm}^2 \simeq \tilde{c}_{T,\pm}^2(\hat k) \spacy k^2 \,,
\end{align}
with,
\begin{align} \label{eq:sound speeds of tranverse modes}
    \begin{split}
        \tilde{c}_{T,\pm}^2(\hat k) ={}& c_T^2 - 2 g_4 \spacy \hat \sigma - 2 (g_4 + g_6 \spacy c_T^2) \lambda_\pm \\
        ={}& c_T^2 - 2 g_4 \spacy \hat \sigma - (g_4 + g_6 \spacy c_T^2) \left[ \sigma_{aa} \pm \sqrt{\sigma_{aa}^2 - 4 \det \sigma_{ab}} \right] \\
        ={}& c_T^2 - 2 g_4 \spacy \hat \sigma + (g_4 + g_6 \spacy c_T^2) \left[ \hat\sigma \mp \sqrt{\hat{\sigma}^2 + 2 [\sigma^2] - 4 \hat k \cdot \sigma^2 \cdot \hat k} \spacy\right] \,,
    \end{split}
\end{align}
where in the second line we used the known expression for the eigenvalues of a $2\times 2$ matrix, and in the last one we related them to scalar quantities involving  $\sigma_{IJ}$ and $\hat k_I$. In a strained solid, therefore, the two transverse modes are not degenerate anymore. They have different sound speeds, and each of them depends on the direction of propagation.

Interestingly, however, there exist special directions $\hat k$ along which the transverse modes remain degenerate.\footnote{An analogous result holds for the propagation of electromagnetic waves in media with an anisotropic dielectric tensor~\cite{landau2013electrodynamics,Born:1999ory}.} These directions can be derived by working in a frame where $\sigma_{IJ}$ is diagonal, with the three eigenvalues such that $\sigma_1 + \sigma_2 + \sigma_3 = 0$ because $\sigma_{IJ}$ is traceless. If we label the eigenvalues in such a way that $\sigma_1 \geqslant \sigma_2 \geqslant \sigma_3$ and choose a direction of propagation in the plane corresponding to the smallest and largest eigenvalues, i.e. $\hat k = (\sin \theta, 0, \cos \theta)$, then the square root on the last line of \eqref{eq:sound speeds of tranverse modes} reduces to $| \sigma_2 - \sigma_1 \cos^2 \theta - \sigma_3 \sin^2 \theta|$, which vanishes when
\begin{align} \label{angle sigma}
	\cos^2 \theta = \frac{\sigma_2 - \sigma_3}{\sigma_1 - \sigma_3} \, .
\end{align}
This condition admits two solutions, which generically correspond to two independent axis of propagation. It is only in the special case where two eigenvalues of $\sigma_{IJ}$  are degenerate that the two solutions yield a single axis.\footnote{Notice that, given the ordering that we chose, the denominator in \eqref{angle sigma} cannot vanish unless the numerator also does. In fact, our choice of ordering implies the standard inequalities for a squared cosine,
\be
0 \le \cos^2 \theta \le 1 \; .
\ee
}

\subsection{Nonrelativistic Limit} \label{sec: non-relativistic limit of solids}

\noindent In order to derive the effective description of nonrelativistic solids, we will follow the strategy outlined in Section~\ref{sec: nonrelativistic limit}. More specifically, we will reintroduce powers of $c$ by replacing $dt \to c \spacy dt$ and $\partial_t \to \partial_t / c$, and then express the action \eqref{eq: cubic action phonon solids} in terms of the canonically normalized fields $\vec \pi_c= \sqrt{\bar b F_b/c} \, \vec \pi $. This leads to
\begin{align}
	S = & \int d t d^3 x  \, \left\{\frac{1}{2} \dot{\vec{\pi}}_c^2-\frac{c_L^2-c_T^2}{2} [\nabla \pi_c]^2 -\frac{c_T^2}{2} \left[\nabla \pi_c \nabla \pi_c^t\right]\right. \nonumber \\
& \qquad \qquad \qquad  + \hat{g}_1[\nabla \pi_c]^3+\hat{g}_2[\nabla \pi_c]\left[\nabla \pi_c^2\right] +\hat{g}_3[\nabla \pi_c]\left[\nabla \pi_c \nabla \pi_c^t\right] \\
& \qquad \qquad \qquad \qquad \quad +\hat{g}_4\left[\nabla \pi_c^2 \nabla \pi_c^t\right]+\hat{g}_5[\nabla \pi_c] \dot{\vec{\pi}}_c^2+\hat{g}_6\dot{\vec{\pi}}_c \cdot \nabla \pi_c \cdot \dot{\vec{\pi}}_c + \dots \bigg\} \, , \nonumber
\end{align}
where we have absorbed two factors of $c$ in the definition of the sounds speed, which now are
\begin{align} \label{eq: sound speeds solid}
	c_T^2 = c^2 \left( \frac{4}{9} \frac{F_Y+F_Z}{\bar b F_b} \right) \,, \qquad \qquad c_L^2 = c^2 \left( \frac{\bar b F_{bb}}{F_b}+\frac{16}{27} \frac{F_Y+F_Z}{\bar b F_b} \right) \, ,
\end{align}
while the remaining coefficients in the effective action read
\begin{subequations} \label{eq: non-relativistic self-interactions}
\begin{align}
    & \hat{g}_1= \frac{1}{\sqrt{\bar b F_b/c}} \left( \frac{2}{9} c_T^2 -\frac{1}{2} c_L^2 -\frac{7}{243} \frac{c^2 F_Z}{\bar bF_b}+\frac{4}{27}c^2 \frac{F_{bY}+F_{bZ}}{F_b}-\frac{1}{6} \frac{c^2 {\bar b}^2 F_{bbb}}{F_b} \right) \,, \\
    & \hat{g}_2= \frac{1}{\sqrt{\bar b F_b/c}} \left( \frac{1}{2} c_T^2+\frac{1}{27} \frac{c^2 F_Z}{\bar b F_b}-\frac{2}{9}c^2 \frac{F_{bY}+F_{bZ}}{F_b} \right) \,, \\
    & \hat{g}_3=\frac{1}{\sqrt{\bar b F_b/c}}\left(c_T^2+\frac{4}{27} \frac{c^2 F_Z}{\bar bF_b}-\frac{2}{9} c^2 \frac{F_{bY}+F_{bZ}}{F_b} \right) \,, \\
    & \hat{g}_4= \frac{1}{\sqrt{\bar b F_b/c}}\left(-c_T^2-\frac{2}{9} \frac{c^2 F_Z}{\bar bF_b} \right) \,, \\
    & \hat{g}_5= \frac{1}{\sqrt{\bar b F_b/c}}\left(\frac{1}{2}+\frac{1}{2} \frac{c_L^2}{c^2}- \frac{c_T^2}{c^2} \right) \,, \\
    & \hat{g}_6= \frac{1}{\sqrt{\bar b F_b/c}}\left(-1+\frac{c_T^2}{c^2} \right)\,.
\end{align}
\end{subequations}
Demanding now that the sound speeds remain finite and that the largest number of interactions survive as we take the formal $c \to \infty$ limit for arbitrary values of the background parameter $\alpha$, we conclude that the derivatives of $F$ must scale with powers of $c$ as follows:
\begin{align}
	\bar b \spacy F_b \sim c \ , \qquad \qquad F_Y\,, F_Z\,, \bar b \spacy F_{b Y} \,, \bar b \spacy F_{b Z}\,, {\bar b}^2 F_{b b} \,, {\bar b}^3 F_{b b b} \,, {\dots}  \sim 1/c \, .
\end{align}
These scalings are realized provided $F(b, Y, Z)$ has the form
\begin{align} \label{eq: nonrelativistic solid F}
	c F(b, Y, Z) = \bar \rho_m c^2 b + U (b, Y, Z) \, ,
\end{align}
where all derivatives of the function $U$ are of $\mathcal{O}(1)$, and $\bar \rho_m$ is a constant with the units of mass density---hence our notation. 
In fact, in the nonrelativistic limit, $\bar \rho_m$  is exactly the mass density when the compressibility parameter is $\alpha = 1$; equivalently, it is the {\em comoving} mass density. Note that $\bar \rho_m$ is also the scale that determines the strength of phonon self-interactions when we let $c \to \infty$---see Eqs.~\eqref{eq: non-relativistic self-interactions}. It is interesting to see how the notion of mass density arises naturally from the requirement that the nonrelativistic limit be well defined, and does not necessitate breaking up the energy density into a sum of rest mass and kinetic contributions as an additional input. 

Eq. \eqref{eq: nonrelativistic solid F} allows us to derive the full nonlinear form of the effective action for a nonrelativistic solid. Restoring powers of $c$ in the definition of $B^{IJ}$,
\begin{align}
	B^{IJ} =- \frac{1}{c^2} \dot \phi^I \dot \phi^J + \partial_k \phi^I  \partial^k \phi^J = -\frac{1}{c^2} \dot \phi^I \dot \phi^J + \left(J^t \cdot J\right)^{IJ} \, ,
\end{align}
with $J_k{}^I$ being the Jacobian introduced in Eq.~\eqref{eq: def Jacobian solid}. Plugging this expression into the definition of the invariants in Eq.~\eqref{eq: isotropic solid invariants}, and expanding in inverse powers of $c$, we obtain
\begin{subequations}
\begin{align}
	b &= \det J  \left\{ 1 - \frac{1}{2 c^2} \dot \phi^I \left(J^{-1}\right)_I^{\;\;k} \spacy \dot \phi^J \left(J^{-1}\right)_{Jk} +  \mathcal{O}\left(1/c^4\right) \right\} \,, \label{eq: non-rel limit b}\\
    Y &= \frac{\left[J^t \cdot J \cdot J^t \cdot J\right]}{\left[J^t \cdot J\right]^2} + \mathcal{O}\left(1/c^2\right) \equiv Y_{\rm nr}  + \mathcal{O}\left(1/c^2\right) \,, \\
    Z &= \frac{\left[J^t \cdot J \cdot J^t \cdot J \cdot J^t \cdot J\right]}{\left[J^t \cdot J\right]^3} + \mathcal{O}\left(1/c^2\right) \equiv Z_{\rm nr}  + \mathcal{O}\left(1/c^2\right) \,.
\end{align}
\end{subequations}
Taking then the $c \to \infty$ limit of the effective action yields\footnote{Note that the term proportional to $c^2$ drops out because $$\det J = \tfrac{1}{3!} \epsilon_{IJK}\epsilon^{ijk} \partial_i\phi^I\partial_j\phi^J\partial_k\phi^K= \partial_i \left( \tfrac{1}{3!} \epsilon_{IJK}\epsilon^{ijk}\phi^I\partial_j\phi^J\partial_k\phi^K\right)$$ is a total derivative.}
\begin{align} \label{eq: nonrelativistic action solids}
    \begin{split}
        S ={}& - \int dt \spacy d^3 x \spacy c F(b, Y, Z) \\
        \to{}& \int dt \spacy d^3 x \left\{ \frac{\bar \rho_m \det J}{2} \dot \phi^I \left(J^{-1}\right)_I^{\;\;k} \dot \phi^J \left(J^{-1}\right)_{Jk} - U (\det J \,,  Y_{\rm nr} \,,  Z_{\rm nr}) \right\} \,.
    \end{split}
\end{align}
Interpreting this nonrelativistic Lagrangian as the difference between a kinetic and a potential terms, we are lead to identify the mass density of the solid with $\rho_m = \bar \rho_m \det J$, its local velocity with $v^k = -\dot \phi^I \left(J^{-1}\right)_I^{\;\;k}$, and the internal energy associated with  deformations with $U (\rho_m  / \bar \rho_m,  Y_{\rm nr},  Z_{\rm nr})$. Our expression for the local mass density $\rho_m$ is consistent with the fact that $\det J$ is the nonrelativistic limit of $b$---see Eq. \eqref{eq: non-rel limit b}---which in turn was identified with the number density $n$ in Eq. \eqref{eq: solid thermo identifications}. 

With some tedious algebra, one can check that a variation of the action with respect to the fields $\phi^I$ yields the familiar Euler's equation for solids---see Section~\ref{sec:edu}---where mass density and velocity are precisely given by the combinations of fields above. Furthermore, these identifications are also such that the familiar continuity equation arises as an identity (i.e., it is satisfied even for field configurations that do not obey the equations of motion):
\begin{align}
	\partial_t \rho_m + \partial_i \left(\rho_m v^i\right) = 0 \, .
\end{align}
In fact, this is simply the nonrelativistic limit of Eq.~\eqref{eq: conservation particle number current}.

\subsection{Further Readings}

\noindent The EFT presented in this section can be extended in a number of directions that we have not attempted to cover. We collect here some entry points to the literature, organized by theme.

\paragraph{Other symmetry-breaking patterns related to solids} Crystals lacking inversion symmetry can host phonons carrying angular momentum, and active or driven elastic media admit parity-odd elastic moduli (``odd elasticity''~\cite{scheibner2020odd}); these are natural extensions of the solid EFT to broken discrete symmetries. Supersolids---phases that simultaneously break translations and a $U(1)$ associated with particle-number conservation---were originally proposed in~\cite{andreev1969quantum,chester1970speculations,PhysRevLett.25.1543} and given an EFT treatment in~\cite{Son:2005ak}; we devote Section~\ref{sec:supersolids} to them. Liquid crystals interpolate between fluids and solids by breaking only a subset of translational and rotational symmetries, and the EFT viewpoint cleanly identifies which Goldstones survive and how to systematically incorporate anisotropies~\cite[e.g.,][]{Brauner:2024juy}. Quasicrystals carry additional Goldstone-like degrees of freedom---phasons---that have no analog in periodic crystals~\cite{levine1984quasicrystals,levine1986quasicrystals,socolar1986quasicrystals}; their EFT description requires enlarging the field content beyond the three $\phi^I$'s used in the main body~\cite{Baggioli:2020haa}. Amorphous solids and glasses, which lack long-range crystalline order but still display elastic response, can be accommodated within a unified topological-field-theory framework that interpolates between liquids, solids, and glasses~\cite{Baggioli:2021ntj}. Active matter introduces driven, non-equilibrium phases; an EFT for active nematics, built within the Schwinger--Keldysh formalism---to be covered in detail in the context of fluids in Section~\ref{sec: Schwinger-Keldysh}---has been formulated in~\cite{landry2023active}. 

\paragraph{Other collective excitations in solids} Beyond the acoustic phonons we have focused on, real solids host a wealth of additional low-energy degrees of freedom. Electrons near a Fermi surface admit their own Wilsonian description~\cite{Polchinski:1992ed,Shankar:1993pf}; the resulting EFT has not yet been combined, to our knowledge, with the relativistic solid EFT discussed here, although the corresponding nonrelativistic electron-phonon couplings have been extensively studied (see~\cite{giustino2017electron} for a review). When the underlying crystal is magnetically ordered, low-energy spin waves (magnons) couple to phonons through magneto-elastic interactions, recently revisited from an EFT perspective in~\cite{Pavaskar:2021pfo,Li:2023onq}. Solids also support a heat-transport mode---second sound---originally explored in the 1960s~\cite{chester1963second,guyer1964dispersion,gurzhi65ab,Pitaevski__1968} and recast in an effective non-equilibrium framework in~\cite{Landry:2020ire}. Beyond the acoustic phonons we have studied, real solids generically host gapped (optical) modes whose proper EFT treatment requires distinguishing intra- from inter-cell symmetries~\cite{Vallone:2019gxx,Esposito:2020hwq}. Topological defects of the lattice---dislocations and disclinations---can also be incorporated as dynamical fields: the classic gauge-field formulation is due to Kleinert~\cite{kleinert1989gauge}, while a modern EFT description of dislocation modes and their sensitivity to the lattice structure has been developed in~\cite{Lin:2022rvx}. Disclinations have been argued to be dual to fracton excitations, which display restricted mobility and have attracted considerable attention as a novel phase of matter~\cite{pretko2018fracton,Pretko:2020cko}.

\paragraph{External fields: electromagnetism, gravity, and dark matter} Coupling the solid EFT to external sources opens up a rich set of applications. The cleanest derivation of the (small but nonzero) gravitational mass carried by sound waves comes from coupling the solid to an external gravitational field~\cite{Esposito:2018sdc}. Solids minimally coupled to gravity can also drive cosmological dynamics: in particular, they support a phase of ``solid inflation''~\cite{Endlich:2012pz,Kang:2015uha,Bartolo:2015qvr,Meszaros:2023ipu}, whose distinctive non-Gaussian signatures stem from the breaking of spatial diffeomorphisms by the inflationary sector. Along similar lines, a solid coupled to gravity has been proposed as a dark matter candidate~\cite{Bucher:1998mh}. At much shorter scales, the coupling of dark matter to phonons in laboratory solids is being actively pursued as a target for direct detection of sub-GeV dark matter~\cite{Cox:2019cod,Trickle:2019nya,Trickle:2020oki,Mitridate:2023izi}.

\paragraph{Scale invariance and holography} An interesting subclass of solids is the one endowed with an additional scaling symmetry. Conformal solids and their holographic duals were first studied in~\cite{Esposito:2017qpj}, while~\cite{Baggioli:2019elg,Baggioli:2020qdg} examined scale-invariant solids and their nonlinear elastic response from a complementary perspective. More generally, holographic constructions provide a complementary handle on solids in the strongly coupled regime, with explicit bulk realizations of phonons and their transport properties developed in~\cite{Alberte:2017oqx,Amoretti:2019cef,Baggioli:2019abx,Armas:2019sbe,Amoretti:2019kuf,Xia:2025ewv}.

\paragraph{Formal aspects} On the formal side, stability considerations have been used to derive nontrivial bounds on the Wilson coefficients of the solid EFT~\cite{Alberte:2018doe}. Explicit weakly-coupled and Lorentz-invariant UV completions of solids  have been constructed in~\cite{Musso:2018wbv,Musso:2019kii,Esposito:2020wsn}. Finally, the breaking of Lorentz invariance in the solid ground state leads to scattering amplitudes for phonons that obey unusual soft theorems, including fractional soft limits~\cite{Brauner:2022ymm,Cheung:2023qwn}.

%% file: fluids.tex
\section{Fluids} \label{fluids}

\noindent In this section we introduce the action-based approach to hydrodynamics.  Hydrodynamics, in general, is an effective theory organized in an expansion in powers of $\lambda\spacy\partial_\mu$, with $\lambda$  the mean free path of the elementary constituents. Our goal is to identify the relevant degrees of freedom and appropriate symmetries that, \emph{without the need for additional assumptions}, lead to an effective action formulation of the problem.
This should be contrasted with the standard approach based directly on the conservation equations, which we will now proceed to summarize for completeness. We will adopt the notation of~\cite{Kovtun:2012rj}, and refer the reader to~\cite{Kovtun:2012rj,Jeon:2015dfa} for more extensive reviews of relativistic hydrodynamics.

\subsection{Relativistic Hydrodynamics} \label{sec: relativistic hydro}

\noindent The first step in the standard treatment of hydrodynamics consists of decomposing $T^{\mu\nu}$ and any additional conserved current $J^\mu$ along directions parallel and perpendicular to the 4-velocity~$u^\mu$:
\begin{subequations} \label{eq: T and J decomposition}
\begin{align}
	T^{\mu\nu} &= \mathcal{E} u^\mu u^\nu + \mathcal{P} \Delta^{\mu\nu} + (q^\mu u^\nu + u^\mu q^\nu) +t^{\mu\nu} \, , \\
	J^\mu &= \mathcal{N} u^\mu + j^\mu \, ,
\end{align}
\end{subequations}
where we have introduced the projector $\Delta^{\mu\nu} = \eta^{\mu\nu} + u^\mu u^\nu$, the vectors $q^\mu$ and $j^\mu$ are transverse to $u^\mu$, and the symmetric tensor $t^{\mu\nu}$ is traceless and transverse. The fundamental assumption of hydrodynamics is that the quantities $\mathcal{E}, \mathcal{P}, \mathcal{N}, q^\mu, j^\mu,$ and $t^{\mu\nu}$ should be organized in an expansion in powers of derivatives of $u^\mu, T$ and $\upmu$. For instance, we have
\begin{align} \label{eq: derivative expansion of P}
	\mathcal{P} =  p(T,\upmu) +  p_1(T,\upmu) \spacy u^\nu \partial_\nu T + p_2(T,\upmu) \spacy u^\nu \partial_\nu \upmu + p_3(T,\upmu) \spacy \partial_\nu u^\nu + \mathcal{O}\left(\partial^2\right) \, ,
\end{align}
with higher derivative corrections suppressed by the mean free path. The motivation behind this assumption is that, at the microscopic level, our system is described in an arbitrary frame by the density matrix,
\begin{align}
	\hat \rho = \frac{e^{u^\nu P_\nu / T}}{Z} \, .
\end{align}
Hydrodynamic excitations are captured by allowing $T$ and $u^\mu$ to vary over length scales much larger than the mean free path of the microscopic constituents. In the presence of an additional conserved charge $Q$, the appropriate density matrix becomes $\hat \rho \sim e^{(u^\nu P_\nu + \upmu Q)/T}$, with $\upmu$ the associated chemical potential, which in the hydrodynamic regime becomes an additional quantity that varies smoothly at large scales. The extension to multiple conserved charges is straightforward.

Eq. \eqref{eq: derivative expansion of P} makes it apparent that additional functions of $T$ and $\upmu$ will appear at any new order in the derivative expansion. However, not all these functions are physically independent. In fact, we are always free to redefine our hydrodynamic quantities, 
\begin{subequations}
\begin{align}
	u^\mu &\to u^\mu + \delta u^\mu \,, \label{eq: u gauge transformation}\\
	T &\to T + \delta T \,, \label{eq: T gauge transformation} \\
	\upmu &\to \upmu + \delta \upmu \label{eq: mu gauge transformation}
\end{align}
\end{subequations}
(with $\delta u$, $\delta T$, and $\delta \upmu$ of first order in gradients) and, by doing so, change the form of all the quantities we have introduced in \eqref{eq: T and J decomposition} except $t^{\mu\nu}$~\cite{Kovtun:2012rj}. These transformations correspond to  the usual freedom of performing field redefinition to remove redundant operators from EFTs~\cite{Glorioso:2017fpd}. For instance, it is conventional to exploit the transformations \eqref{eq: T gauge transformation} and \eqref{eq: mu gauge transformation} to zero out higher derivative corrections to the energy and number density, so that 
\begin{subequations} \label{eq: energy and number density constitutive relations}
\begin{align}
	\mathcal{E} &= \rho(T,\upmu) \,, \\
	\mathcal{N} &= n(T,\upmu) \, ,
\end{align}
\end{subequations}
where the energy density $\rho (T, \mu)$ should not be confused with the density matrix discussed above. Similarly, we can always perform a transformation of the form \eqref{eq: u gauge transformation} (with $u^\mu \delta u_\mu = 0$ to preserve the normalization of the 4-velocity) to set $j^\mu = 0$---a condition that specifies the so-called \emph{Eckart frame}~\cite{Eckart:1940te}---or to work in the \emph{Landau frame} by setting $q^\mu = 0$~\cite{landau:1987bo}. More general choices of frame are also possible~\cite{Kovtun:2012rj}.

Furthermore, the form of first derivative corrections can be simplified using conservation equations at zeroth order in derivatives. For example, the continuity equations $\partial_\lambda J^\lambda = 0$ at zeroth order in derivatives (where $j^\lambda = 0$) is 
\begin{align}
	\frac{\partial n}{\partial T} \spacy u^\lambda \spacy\partial_\lambda T + \frac{\partial n}{\partial \upmu} \spacy u^\lambda \spacy \partial_\lambda \upmu + n \spacy \partial_\lambda u^\lambda = \mathcal{O}\left(\partial^2\right) \, .
\end{align}
We can use this relation to, say, eliminate $u^\lambda \spacy \partial_\lambda T$ from the righthand side of equation \eqref{eq: derivative expansion of P}. Combining this result with $u_\nu \spacy \partial_\mu T^{\mu\nu} = 0$ allows us to eliminate also the term proportional to $u^\lambda \spacy\partial_\lambda \upmu$, so that the derivative expansion for the pressure, up to first order, reduces to 
\begin{align} \label{eq: constitutive relation pressure}
	\mathcal{P} =  p(T,\upmu) - \zeta(T,\upmu) \spacy \partial_\nu u^\nu + \mathcal{O}\left(\partial^2\right) \, , 
\end{align}
with $\zeta (T,\upmu)$ the usual \emph{bulk viscosity} coefficient. In a similar fashion, the transverse part of the conservation equation of energy and momentum, $\Delta_{\lambda\nu}\spacy \partial_\mu T^{\mu\nu} = 0$, allows us to reabsorb one of the arbitrary functions appearing in the derivative expansion of $q^\mu$ or $j^\mu$, depending on the frame one chooses to work in. In the case of the Landau frame, where $q^\mu = 0$, we can parametrize for instance\footnote{If the $J^\mu$ current is anomalous, $j^\mu$ contains an additional parity-odd, non-dissipative transport term proportional to $ \epsilon^{\mu\nu\lambda\sigma}u_\nu \partial_\lambda u_\sigma$~\cite{Son:2009tf}.}
\begin{align} \label{eq: constitutive relation j Landau frame}
	j^\mu = - \sigma_c (T, \upmu) \spacy \Delta^{\mu\lambda} \spacy \partial_\lambda \upmu + \chi (T, \upmu) \spacy \Delta^{\mu\lambda} \spacy \partial_\lambda T + \mathcal{O}\left(\partial^2\right) \, ,
\end{align}
where $\sigma_c(T, \upmu)$ is the familiar \emph{conductivity} parameter.\footnote{We have added a subscript ``$c$'' to avoid potential confusion with the ratio $\sigma = s/n$ introduced in Eq. \eqref{eq: isentropic condition}.} Finally, at first order in derivatives there exists a unique symmetric, traceless, and transverse tensor, and thus
\begin{align} \label{eq: constitutive relation t}
	t^{\mu\nu} = - \eta(T, \upmu) \spacy \Delta^{\mu\alpha} \spacy \Delta^{\nu\beta} \spacy\left(\partial_\alpha u_\beta + \partial_\beta u_\alpha - \tfrac{2}{3} \eta_{\alpha\beta}\partial_\lambda u^\lambda\right) + \mathcal{O}\left(\partial^2\right) \, ,
\end{align}
with the $\eta (T, \upmu)$ the \emph{shear viscosity} coefficient.

At this stage we have implemented all the constraints that follow from Lorentz invariance, and exhausted all the freedom associated with possible field redefinitions. Nonetheless, the constitutive relations we have arrived at---Eqs. \eqref{eq: energy and number density constitutive relations}, \eqref{eq: constitutive relation pressure}, \eqref{eq: constitutive relation j Landau frame}, and \eqref{eq: constitutive relation t}---are still too general and must be further restricted based on physical considerations. The entropy current is of the form
\begin{align}\label{eq: entropy current}
	S^\mu = s(T, \upmu) \spacy u^\mu + \mathcal{O}\left(\partial\right) \, ,
\end{align}
with 
\begin{align}
	s(T, \upmu) = \frac{1}{T} \big[\rho(T,\upmu) + p(T,\upmu) - \upmu\spacy n(T,\upmu) \big] \, .
\end{align}
Multiplying this last equation by $u^\mu$ and using the fact that $q^\mu = 0$ in the Landau frame, we obtain $s u^\mu = \left[ - T^{\mu\nu} u_{\nu} + p \spacy u^\mu - \upmu \spacy n \spacy u^\mu \right]/T$.
This equation is only valid at lowest order in the derivative expansion, but can be bootstrapped to the following all-order expression~\cite{ISRAEL1981204}:
\begin{align} \label{eq: entropy current fluid}
	S^\mu = \frac{1}{T} \big[ - T^{\mu\nu} \spacy u_{\nu} + \mathcal{P} \spacy u^\mu - \upmu \spacy J^\mu \big] \, .
\end{align}
This equation determines the higher order derivative corrections on the right-hand side of Eq. \eqref{eq: entropy current} in terms of the higher derivative corrections appearing in the constitutive relations for $T^{\mu\nu}$ and $J^\mu$. Then, imposing that $\partial_\mu S^\mu \geqslant 0$ we are led to the conclusion that
\begin{align} \label{eq: entropy constraints on transport coefficients}
	\zeta >0 \,, \qquad \qquad \sigma_c = \frac{T}{\upmu} \chi > 0 \,, \qquad \qquad  \eta >0 \, .
\end{align}
Equations Eqs. \eqref{eq: energy and number density constitutive relations}, \eqref{eq: constitutive relation pressure}, \eqref{eq: constitutive relation j Landau frame}, and \eqref{eq: constitutive relation t}, supplemented by the constraints  \eqref{eq: entropy constraints on transport coefficients}, are the first order constitutive relations of hydrodynamics in the Landau frame. We refer the reader to \cite{Kovtun:2012rj} for analogous expressions in the Eckart frame. This discussion highlights the main conceptual downside of the traditional approach to hydrodynamics, which is that the second law of thermodynamics must be imposed by hand and does not follow from more fundamental principles.

\subsection{Perfect Fluids} \label{sec: perfect fluids}

\noindent We will now turn our attention to the EFT approach to hydrodynamics. In this section we will focus on the perfect fluid limit, which from the EFT perspective is just the leading term in a derivative expansion of the effective action. We will restrict our attention to conservation of energy and momentum, postponing a discussion of additional conserved charges to Section~\ref{sec: fluid conserved charges}. We will introduce several equivalent formulations that have been put forward in the literature, and explain how they are connected to one another.

\subsubsection{Eulerian Formulation with 3 Fields} \label{sec:Eulerianfluids}

\noindent The EFT approach to perfect fluids discussed, e.g., in~\cite{Soper:1976bb,Dubovsky:2005xd} is arguably the most minimal one, as the only degrees of freedom it involves are the comoving coordinates $\phi^I$ of the fluid elements. In the Eulerian picture of hydrodynamics, an arbitrary fluid configuration can be described by specifying which fluid element $\phi^I$ can be found at the position $\vec x$ at time $t$. This assignment defines three scalar fields, $\phi^I (\vec x,t)$, i.e. exactly the same degrees of freedom we used to describe a solid in section \ref{sec:solids}. Compared to solids, however, fluids are characterized by a large symmetry group: as long as the volume of the fluid element we are considering remains unchanged one can traslate it, rotate it, shear it, and so on. Consequently, the low energy effective action for our fields $\phi^I(x)$ must be invariant under a very large set of global symmetries, namely, the \emph{volume preserving diffeomorphisms}
\begin{equation} \label{eq: volume preserving diffs}
	\phi^I \to \xi^I (\phi) \, , \qquad \qquad \quad \det \frac{\partial \xi^I(\phi)}{\partial \phi^J} = 1 \, . 
\end{equation}
The requirement that these transformation have unit determinant encodes the fact that compressions of the comoving volume elements are the only transformations that produce a physically distinct fluid configuration. The simplest possible such transformation are rigid shifts of all comoving coordinates, i.e. $\phi^I (x) \to \phi^I(x) + c^I$, with $c^I$ constant parameters. Clearly, the action will be invariant under these transformations provided each field appears with at least one derivative, $\partial_\mu$, acting on it, very much analogously to what happened for solids in the previous pages. Thus, at lowest order in a derivative expansion, our effective action must be a function of single derivatives $\partial_\mu \phi^I$. Furthermore, Lorentz invariance demands that all the indices carried by derivatives be contracted pairwise, so that our action will actually depend once again on the combination $B^{IJ} \equiv \partial_\mu \phi^I \partial^\mu \phi^J$. So far, everything is identical to the case of solids. The last ingredient discriminating the latter from fluids is the requirement of invariance under volume-preserving diffeomorphisms. This, in fact, singles out the determinant of $B^{IJ}$ as the only possible invariant quantity at this order, so that
\begin{align} \label{eq: U(B) Lagrangian fluids}
    S[\phi] = -\int d^4 x \, F(b) \, , \qquad \text{ with } \qquad b \equiv \sqrt{\det B} \,.
\end{align}
Notice that, from this viewpoint, a fluid is nothing but a very special solid---one tuned at a point of enhanced symmetry.\footnote{In fact, the reader will find some derivations in this section almot identical to the ones in Section~\ref{sec: thermodynamics of solids}. We repeat them here to keep this section self-contained.} The Lagrangian above, in fact, is the same as that in Eq.~\eqref{eq: solid effective Lagrangian}, where the dependence on $Y$ and $Z$ has been dropped because they are not invariant under volume preserving diffeomorphisms.

To make contact with the standard treatment of relativistic hydrodynamic discussed in the previous section, one can derive the energy--momentum tensor associated with this action and find that it takes the perfect fluid form shown in Eq.~\eqref{eq: superfluid current and stress energy tensor}, with 
\begin{align} \label{eq:umu fluids}
	\rho = F\,, \qquad p = b \spacy F' -F\,, \qquad u^\mu = \frac{1}{3!} \frac{1}{b} \epsilon^{\mu\nu\lambda\sigma} \epsilon_{IJK} \partial_\nu \phi^I \partial_\lambda  \phi^J \partial_\sigma \phi^K \, . 
\end{align}
These are nothing but Eqs.~\eqref{eq: rho and p solid} and \eqref{eq:umu for solids} for solids, taken in the limit where the Lagrangian solely depends on $b$.
It is easy to show that the equations of motion for the fields $\phi^I$ are equivalent to the conservation of the energy--momentum tensor~\cite{Soper:1976bb,Dubovsky:2005xd}. 

Because the Lagrangian depends on a single invariant $b$, the same argument that we ran for solids applies here as well: this EFT can describe a fluid only in the \emph{isentropic regime}, $\sigma \equiv s/n = \text{constant}$, in which the entropy and charge densities are not independent but vary together in proportion to $b$. The only difference with respect to solids is that here we have no $Y$ and $Z$ at our disposal. As a result, $b$ is the unique extensive density carried by the EFT, and whether we call it the entropy density or the conserved charge density is purely a matter of convention: the two are locked together by the fixed value of $\sigma$. Indeed, a constant rescaling of the fields $\phi^I \to \phi^I/\sigma^{1/3}$ merely rescales $b$, and can be used to trade the identification $b = s$ for $b = n$ without changing any physics. In order to match the existing literature on EFT of fluid more closely, we will choose here to identify $b$ with the entropy density,
\begin{align} \label{b def}
	b = s \, , \qquad\qquad \frac{dF}{db} = \left. \frac{\partial \rho}{\partial s} \right|_{\sigma, u} = T + \frac{\upmu}{\sigma} \, ,
\end{align}
where the second relation follows from the Euler relation $\rho + p = b \spacy F'$ together with the first law along the isentropic ray. This is the analogue, in the $b = s$ normalization, of the solid identification $\partial F/\partial b|_{Y,Z} = T\sigma + \upmu$ in Eq.~\eqref{eq: solid thermo identifications}; the two differ precisely by the rescaling of $b$ that trades $b = n$ for $b = s$.

The choice of normalization is largely a matter of convenience, but it is worth pausing on the physical distinction between solids and fluids. For solids, the identification $b = n$ in Eq.~\eqref{eq: solid thermo identifications} is natural: we typically want our EFT to admit a smooth $T \to 0$ limit, in which the solid retains its rigidity and the conserved charge density $n$ remains the meaningful extensive quantity. For regular fluids, by contrast, the $T \to 0$ limit is most likely unphysical---see however~\cite{Endlich:2010hf,Dersy:2022kjd, Cuomo:2024ekf,Goldberger:2025mgb} for recent attempts to make sense of quantum perfect fluids at zero temperature. The  $\upmu \to 0$ limit, on the other hand, is certainly relevant---a gas of massless particles being the obvious example. In fact, in the rest of this section we will set 
\begin{align} \label{eq: upmu to zero fluid}
	\upmu = 0 \, ,
\end{align}
in order to streamline the equations that follow and to align with the literature. For an isentropic fluid, this is formally equivalent to letting the constant $\sigma \to \infty$. In this limit, the conserved charge plays no dynamical role and $dF/db$ is unambiguously the temperature. The identically satisfied conservation law $\partial_\mu (b u^\mu) =0$ now captures the fact that the entropy current is conserved in a perfect fluid.\footnote{Because this result is an identity, it holds at all orders in the derivative expansion. However, $s \neq b$ beyond leading order, which is why the entropy current is no longer conserved in the presence of dissipation.} We will reinstate an independent chemical potential, and with it a genuinely independent number density, when we discuss fluids with conserved charges in Section~\ref{sec: fluid conserved charges}.

A few comments are in order at this point. Based on Noether's theorem, every global symmetry of the action should correspond to a conserved quantity. The volume-preserving diffeomorphisms \eqref{eq: volume preserving diffs} are in effect an infinite number of global symmetries, since 
\begin{align}
	\xi^I(\phi) = c^I + c^{IJ} \phi_J + c^{IJK}\phi_J\phi_K + \dots \,,
\end{align}
with the $c$'s are an infinite number of transformation parameters that are constrained only by the requirement that $\det \partial\xi/ \partial \phi = 1$. This implies the existence of an infinite number of conserved quantities for perfect fluids: what could they be? Kelvin's circulation theorem gives us a hint by stating that, in a nonrelativistic perfect fluid, the line integral of the velocity field along \emph{any} closed loop $\mathcal{C}$ that is comoving with the fluid is conserved. The relativistic version of this statement can be derived by applying the standard arguments used to derive Noether currents~\cite{Weinberg:1996kr} to find~\cite{Soper:1976bb,Dubovsky:2005xd,endlich2013effective} 
\begin{align} \label{eq: relativistic Kelvin's theorem}
	\frac{dQ_{\mathcal{C}}}{dt} = \frac{d}{dt} \oint_\mathcal{C} d \vec x \cdot \vec u \, \left[\frac{dF}{db}\right] = 0 \, .
\end{align}
In order to compare this result with the usual nonrelativistic Kelvin's theorem, we need to momentarily back off from the infinite $\sigma$ limit to take the nonrelativistic limit. Then, as was the case for solids, the effective Lagrangian contains a term linear in $b$ proportional to $c^2$, i.e. $F(b) = \rho_m c^2
b /\sigma + U(b)$.\footnote{The additional $1/\sigma$ factor in the term proportional to $c^2$ compared to Eq.~\eqref{eq: nonrelativistic solid F} for solids because here we set $b = s = \sigma n$.} Therefore, $dF/db$ becomes dominated by a large, constant piece in the $c \to \infty$ limit, and \eqref{eq: relativistic Kelvin's theorem} reduces to the statement that the line integral of $\vec u$ along any closed loop $\mathcal{C}$ must be conserved.

\subsubsection{Eulerian Formulation with 4 Fields}\label{euler 4}

\noindent The EFT approach discussed in the previous subsection can be reformulated by introducing an auxiliary field and extending the internal symmetry group.  This approach turns out to have a few conceptual advantages, and can be more easily translated to the Lagrangian picture or extended to imperfect fluids. In this section, we will follow the logic first discussed in a set of unpublished notes by Sergei Sibiryakov.

The starting point is the introduction of an additional field, $\kappa(x)$, that transforms under a generic change of comoving coordinates as follows:
\begin{align} \label{eq: sigma transformation rule}
	\kappa \to \kappa' = \kappa \det \spacy \frac{\partial \phi^I }{ \partial \xi^J} \, .
\end{align}
This means that the product $\kappa b$ is invariant under all possible changes of comoving coordinates $\phi^I$, and so would be the action
\begin{align}
	S[\phi, \kappa] = - \int d^4 x \, F(\kappa b) \, , \label{eq: sibiryakov's action without constraint}
\end{align}
The energy--momentum tensor associated with this action would still have the perfect fluid form, with $b \to \kappa b$. In particular, we would have that
\begin{align}
	\rho + p =  \kappa b \ \frac{d F}{d(\kappa b)} \, .
\end{align}
However, if we now want to interpret $\kappa b$ as the entropy density, we need to restrict the dynamics associated with the action above by imposing the additional constraint $\partial_\mu ( \kappa b u^\mu) = 0$. We can do so by introducing a Lagrange multiplier, $\tau$:
\begin{align}
	S[\phi, \kappa, \tau] = - \int d^4 x \big[ F(\kappa b) + \tau  \partial_\mu ( \kappa b u^\mu) \big] \,. \label{eq: sibiryakov's action with constraint}
\end{align}
Interestingly, by doing so we have arrived at an action that is not only invariant under generic diffeomorphisms of the comoving coordinates, but also under a $\phi^I$-dependent shift symmetry acting on $\tau$, which has been dubbed ``chemical shift" \cite{Dubovsky:2011sj}. The full set of internal symmetries of the action \eqref{eq: sibiryakov's action with constraint} is
\begin{align} \label{eq: target reparametrizations of fluid}
	 \phi^I \to \xi^I (\phi) \, , \qquad \qquad \quad  \tau \to \tau + g(\phi) \, ,
\end{align}
with $\xi^I$ and $g$ arbitrary functions of the comoving coordinates.

We can check explicitly that our new action \eqref{eq: sibiryakov's action with constraint} is equivalent to the one we started with in Eq. \eqref{eq: U(B) Lagrangian fluids}. Indeed, varying with respect to $\tau$ we get the entropy conservation equation
\begin{align}
\partial_\mu ( \kappa b u^\mu) = b u^\mu \partial_\mu \kappa = 0 \, ,
\end{align}
which implies that $\kappa(x) = \kappa(\phi^I(x))$.\footnote{To show this, recall that $\partial_\mu \phi^I u^\mu = 0$---see discussion above Eq.~\eqref{eq:umu for solids}.} By performing an appropriate change of comoving coordinates as in \eqref{eq: sigma transformation rule} we can then always set $\kappa(x) = 1$. By plugging this choice into the action \eqref{eq: sibiryakov's action with constraint}, the second term in the action is identically zero, and we get back the original action  \eqref{eq: U(B) Lagrangian fluids} for the $\phi^I$ fields only.

However, rather than solving for the equation of motion of $\tau$, we can choose to solve the equation of motion of $\kappa$,
\begin{align}
u^\mu \partial_\mu \tau = \frac{d F}{d(\kappa b)} \label{eq: eoms sibiryakov 2} \, ,
\end{align}
to integrate $\kappa$ out while keeping a nontrivial dependence of the action on $\tau$. Note that the equation above allows us to interpret the Lagrangian in Eq.~\eqref{eq: sibiryakov's action with constraint} as a regular Legendre transform of the Lagrangian in \eqref{eq: U(B) Lagrangian fluids} (up to an integration by parts). By inverting Eq. \eqref{eq: eoms sibiryakov 2}, we find that $\sigma b$ can be expressed as a function of $u^\mu \partial_\mu \tau$. This implies that, after integrating out $\kappa$, the effective action will acquire the form
\begin{align}
	S[\phi, \tau] = \int d^4 x \ P(u^\mu \partial_\mu \tau) \label{eq: sibiryakov's action} \, ,
\end{align}
for a suitable $P$, which is related to the original function $F$ by Legendre transform (up to integration by parts).
The energy--momentum tensor associated with this new action is
\begin{align}
	T_{\mu\nu} = \eta_{\mu\nu} \left(P-P'z\right)+P'z\left(B^{-1}\right)_{IJ}\partial_\mu \phi^I \partial_\nu \phi^J \, , \label{Tmunu Euler 4 fields}
\end{align}
where we have defined $z \equiv u^\mu \partial_\mu \tau$. Because $u^\mu \left(B^{-1}\right)_{IJ}\partial_\mu \phi^I \partial_\nu \phi^J = 0$ and $B^{-1}_{IJ} B^{IJ} = 3$, we have that $\left(B^{-1}\right)_{IJ}\spacy\partial_\mu \phi^I \partial_\nu \phi^J = \eta_{\mu\nu} + u_\mu u_\nu$, and therefore $T_{\mu\nu}$ has again the perfect fluid form with
\begin{align}
	\rho = P' z - P \, , \qquad \qquad \quad p = P \, .
\end{align}

In order to obtain a thermodynamic interpretation of $z$, we notice that the equation of motion for $\tau$ is $\partial_\mu (P' u^\mu) = 0$, which suggests that we should identify
\begin{align} \label{eq: thermodynamic interpretation P(z)}
	s \equiv P'(z) \, , \qquad \qquad \quad T \equiv z \, .
\end{align}
This identification is consistent with the thermodynamic relation $s = dP/dT$. Interestingly, in this new approach the entropy current is conserved on-shell rather than identically.

\subsubsection{Lagrangian Formulation with 4 Fields} \label{sec: Lagrangian Formulation with 4 Fields}

\noindent Starting from the 4-field formulation discussed in the previous subsection, we can switch from the Eulerian to the Lagrangian viewpoint by inverting our fields:
\begin{align} \label{eq: fluid reparametrizations}
	\left\{ \tau(x), \phi^I(x) \right\} \quad \to \quad x^\mu = X^\mu \left(\tau, \phi^I\right) \, .
\end{align}
Now, describing a fluid configuration amounts to specifying the spacetime position of every fluid element $\phi$ for any value of the ``internal time'' $\tau$. Poincar\'e invariance now plays the role of an internal symmetry acting on $X^\mu$, while the redefinitions of $\tau$ and $\phi^I$ shown in Eq.~\eqref{eq: target reparametrizations of fluid} are a restricted form of reparametrization invariance. This viewpoint is similar to the one we adopt when writing down, say, the relativistic action for a string. There, the embedding coordinates $X^\mu$ are a function of only two coordinates on the worldsheet, whereas here they are a function of four coordinates because the fluid fills the entire space. Furthermore, in the string case we know that we can distinguish between different types of strings by imposing different types of reparametrization invariance. For instance, the fundamental string is invariant under full reparametrizations of the worldsheet coordinates, while a material string (in unitary gauge) would only be invariant under a restricted set of reparametrizations~\cite{Horn:2015zna}. Invariance under the reparametrizations \eqref{eq: target reparametrizations of fluid} could be taken as the definition of a fluid in our formulation~\cite{Glorioso:2018wxw}.

Poincaré invariance dictates that the effective action in the Lagrangian picture, at lowest order in derivatives, can only depend on the induced metric
\begin{align} \label{eq: induced metric fluid Lagrangian}
	h_{ab} (\sigma) = \partial_a X^\mu (\sigma) \spacy \partial_b X_\mu (\sigma) \, ,
\end{align}
where, from now on, we will denote with $\sigma^a = (\tau, \phi^I)$ the spacetime comoving coordinates. In order to write down an invariant effective action, it is convenient to decompose $h_{ab}$ as follows~\cite{Glorioso:2018wxw}:
\begin{align} 
	h_{ab} (\sigma) \spacy d\sigma^a \spacy d\sigma^b = - \beta^2 (\sigma)\spacy \left(d\tau - v_I (\sigma)d\phi^I\right)^2 + \gamma_{IJ} (\sigma)\spacy d\phi^I \spacy d\phi^J \, .
\end{align}
This is reminiscent of, but not quite the same as, an ADM decomposition~\cite{Arnowitt:1962hi}. In terms of the components of the induced metric, $h_{ab}$, these new quantities are defined as follows:
\begin{align} \label{eq: ADM-like breaking of induced metric fluid}
	\beta^2 \equiv - h_{\tau\tau} \, , \qquad \quad v_I \equiv - \frac{h_{\tau_I}}{h_{\tau\tau}} \, , \qquad \quad \gamma_{IJ} \equiv h_{IJ} - \frac{h_{\tau I}h_{\tau J}}{h_{\tau\tau}} \, .
\end{align}
Under diffeomorphisms of the spatial comoving coordinates $\phi^I$, the quantities  $\beta, v_I$ and $\gamma_{IJ}$ transform respectively as a scalar, a vector, and a rank-2 tensor; under chemical shifts, they transform instead as, 
\begin{subequations}
\begin{align}
	\beta'(\sigma') &=\beta(\sigma) \, , \\
	v_I'(\sigma') &= v_I(\sigma) + \frac{\partial g(\phi) }{\partial \phi^I} \, , \label{eq: transformation of v^I chemical shifts}\\
	\gamma_{IJ}'(\sigma') &= \gamma_{IJ}(\sigma) \, .
\end{align}
\end{subequations}
In light of these transformation properties, the effective action at lowest order in derivatives takes the form
\begin{align} \label{eq: action fluid Lagrangian no charge}
	S[X] = \int d^4 \sigma \spacy \sqrt{\gamma} \spacy G (\beta) \, .
\end{align}
The arbitrary function $G$ is related to the one in the action \eqref{eq: sibiryakov's action} by $G(\beta) \equiv \beta P(\beta^{-1})$. This can be seen using the relations
\begin{align} \label{eq: relation Lagrangian to Eulerian}
	\sqrt{-h} = \beta \sqrt{\gamma} \, , \qquad \qquad \quad u^\mu \partial_\mu \tau = 1/\beta \, .
\end{align}
Because we are only imposing a restricted form of reparametrization invariance, this action contains an arbitrary function $P(\beta^{-1})$ at lowest order in derivatives, unlike the Nambu--Goto action for a fundamental string which must be equal to $\sqrt{-h}$ (up to an overall normalization) to ensure full reparametrization invariance. The second equation in \eqref{eq: relation Lagrangian to Eulerian} also justifies the notation we have introduced: according to Eq. \eqref{eq: thermodynamic interpretation P(z)}, and recalling that $z=u^\mu \partial_\mu \tau$, the quantity $\beta$ should indeed be interpreted as the inverse temperature: $T = \beta^{-1}$. Finally, the stress--energy tensor associated with the action \eqref{eq: action fluid Lagrangian no charge} has once again a perfect fluid form~with
\begin{align} \label{eq: rho p mu in Lagrangian variables}
	\rho = \frac{P'\left(\beta^{-1}\right)}{\beta} - P\left(\beta^{-1}\right)  \, , \qquad \quad p = P\left(\beta^{-1}\right) \,, \qquad \quad u^\mu = \frac{\partial_\tau X^\mu}{\beta} \, .
\end{align}
Given the definitions \eqref{eq: induced metric fluid Lagrangian} and \eqref{eq: ADM-like breaking of induced metric fluid}, the expression for $u^\mu$ reproduces the usual 4-velocity of a fluid element.

\subsubsection{Lagrangian Formulation with 3 Fields}\label{LF3fields}

\noindent An other useful rewriting of the perfect fluid effective theory is the standard Lagrangian one, in which one uses just the fields $\vec X (t, \phi^I)$, where $t = X^0$ is the standard Minkowski time. To derive it, it is convenient to start from the $\{\tau(x),\phi^I(x)\}$ description of Section~\ref{euler 4}, and change the spatial coordinates from the physical ones ($\vec x \,$) to the comoving ones ($\phi^I$). We then have four fields, $\tau(t,\phi^I)$ and $\vec X(t, \phi^I) = \vec x$. Using the relations~\eqref{eq: induced metric fluid Lagrangian}, \eqref{eq: ADM-like breaking of induced metric fluid} and \eqref{eq: relation Lagrangian to Eulerian}, the action \eqref{eq: sibiryakov's action} written in terms of these new fields becomes
\begin{align}
    S\big[\tau, X \big] = \int d^3 \phi \, dt \,  \left| \frac{\partial X}{\partial \phi}\right| P \big(\gamma \, \dot \tau\big) \, , \qquad \text{ with } \qquad \gamma \equiv \frac{1}{\sqrt{1 - \dot X^i \dot X^i }} \, ,
\end{align}
where $\big| \partial X/\partial \phi\big|$ is the determinant of the spatial Jacobian $\partial X^i/\partial \phi^J$ accounting for the comoving-to-physical volume ratio, and $\gamma$ is the standard Lorentz time-dilation factor for a particle moving with velocity $\dot{\vec X}$. Note that one might be tempted to confuse $\gamma$ with the determinant of the matrix appearing in Eq.~\eqref{eq: ADM-like breaking of induced metric fluid}. Please don't.

Then, one notices that $\tau$ only appears in the action through its time derivative. This implies that its conjugate momentum density,
\begin{align} \label{p_tau}
    \pi_\tau \equiv  \left| \frac{\partial X}{\partial \phi}\right| \spacy P' \big(\gamma \, \dot \tau\big) \spacy \gamma  \, ,
\end{align}
is constant in time {\em at any point} $ \phi^I$~\cite{Dubovsky:2011sj},
that is, it is only a function of the comoving coordinates:
\begin{align}
    \pi_\tau =  \pi_\tau(\phi^I) \, .
\end{align}
Because of the Jacobian factor in \eqref{p_tau}, $\pi_\tau$ transforms as a scalar density under $\phi^I$ diffeomorphisms; we can therefore set it to one at $t=0$ by performing a diffeomorphism with a suitable determinant, and its conservation will imply that it will be one at all times. Once this is arranged, the theory remains invariant only under the residual subgroup of volume-preserving diffeomorphisms. 

The conservation law of $\pi_\tau$ is a direct consequence of the chemical shift symmetry \eqref{eq: target reparametrizations of fluid} acting on $\tau$, and 
we can use it to integrate out $\tau$. To do so, in order to be consistent with the variational principle, we have to use the Routhian \cite{landau1976mechanics}---the Legendre transform of the Lagrangian with respect to $\dot \tau$,
\begin{align} \label{eq: routhian}
    R\Big(\vec X, \dot {\vec X}, \pi_\tau \Big) = \left| \frac{\partial X}{\partial \phi}\right| P \big(\gamma \, \dot \tau\big)  - \dot \tau \, \pi_\tau \, , 
\end{align}
with $\dot \tau$ expressed in terms of $\pi_\tau$ and everything else through \eqref{p_tau}.\footnote{There is a subtlety in the variational problem: when integrating out a cyclic variable ($\tau$, in our case) using the associated conservation law, one usually fixes the value of the conserved charge, which is {\em not} the same as fixing the boundary conditions for the field being integrated out. Using the Routhian takes care of this subtlety \cite{landau1976mechanics}. See 
\ref{app: routhian} for a more detailed discussion of this point. \label{foot: routhian}}
Choosing $\pi_\tau = 1$ as mentioned above, we  have
\begin{align} \label{eq: solution for tau dot}
    \dot \tau = \gamma^{-1} \, f \left( \, \left| \frac{\partial X}{\partial \phi} \right|^{-1}  \gamma^{-1} \right) \, ,
\end{align}
where the function $f$ is the inverse of $P'$. Plugging this into the Routhian, we finally get the effective action for the $\vec X(t, \phi^I)$ fields,
\begin{align} \label{action LF3fields}
    S\big[X \big] = - \int d^3 \phi \, dt \,  \left| \frac{\partial X}{\partial \phi} \right| \, F \bigg(  \left| \frac{\partial X}{\partial \phi}\right| ^{-1} \sqrt{1 - \dot X^i \dot X^i} \,\bigg) \, ,
\end{align}
where $F$ happens to be the same function we started with in Eq.~\eqref{eq: U(B) Lagrangian fluids}, and thus shares the same thermodynamic intepretation as the energy density: $F = \rho$. (Recall that $P$ was related to $F$ by a Legendre transform.) In fact, the above action can be obtained by going to comoving coordinates directly  in \eqref{eq: U(B) Lagrangian fluids}  \cite{Endlich:2010hf,Endlich:2013dma}. Furthermore, the argument of $F$ in the final action \eqref{action LF3fields} is precisely equal to $b = \sqrt{\det B}$ evaluated in comoving coordinates:
\begin{align} \label{eq: b in LF3fields}
	b = \left| \frac{\partial X}{\partial \phi}\right| ^{-1} \sqrt{1 - \dot X^i \dot X^i} \, ,
\end{align}
and thus corresponds to the entropy density $s$. The temperature then follows from $T = dF/db$. We therefore notice a pattern in the thermodynamic interpretation of the four equivalent formulations of perfect fluids discussed so far: formulations involving three fields feature a combination of (derivatives of) fields that can be interpreted as the entropy density, while formulations with four fields describe more directly the local temperature.

\subsubsection{Clebsch parametrization}

\noindent A completely different parametrization of perfect fluid dynamics is provided by the so-called Clebsch variables. For nonrelativistic fluids, these  are three scalar fields $\chi$, $\alpha$, and $\beta$ that characterize the velocity field as \cite{Jackiw:2004nm}
\begin{align} \label{NR Clebsch}
    \vec v =  \vec \nabla \chi + \alpha \vec \nabla \beta \, .
\end{align}
The vorticity then is simply
\begin{align}
    \vec \omega = \vec \nabla \alpha \times \vec \nabla \beta \, ,
\end{align}
which makes the Clebsch parametrization particularly useful to study vorticose fluid flows.

For relativistic fluids, since the vorticity field that enters Kelvin's theorem is not purely kinematical---that is, its definition involves not only the velocity field but also thermodynamic quantities, and thus depends on the Lagrangian---one has to work a little harder. Following~\cite{Cuomo:2024ekf}, we can start with the Eulerian formulation of Section~\ref{sec:Eulerianfluids} and consider the Legendre transform
\begin{align} \label{cuomo transform}
    S[\xi, \phi] = \int d^{d+1} x \, \left[ P(-\xi^\mu \xi_\mu) - \xi_\mu J^\mu \right] \, ,
\end{align}
where $J^\mu$ is the identically conserved current
\begin{align}
    J^\mu = \frac{1}{d!}  \epsilon^{\mu \, \nu_1 \dots \nu_d} \, \epsilon_{I_1 \dots I_d} \, \partial_{\nu_1} \phi^{I_1} \dots  \partial_{\nu_d} \phi^{I_d} \, ,
\end{align}
and $P$ is, for the moment, a generic function. We are also keeping the spatial dimensionality $d$ generic for the time being, for reasons that will soon be clear.  Varying with respect to $\xi_\mu$ yields the corresponding equation of motion,
\begin{align} \label{eom xi}
    2 P' (-\xi^2) \, \xi^\mu = -J^\mu  \,, 
\end{align}
which one can solve for $\xi^\mu$ and end up with our original action \eqref{eq: U(B) Lagrangian fluids}, since $J_\mu J^\mu = - \det B$. We thus see that the action \eqref{cuomo transform} is equivalent to our original action, and the functions $P$ and $F$ are related by a Legendre transform. In particular, $P$ has the interpretation of the pressure, as we will confirm below.

We can then take Eq.~\eqref{cuomo transform} and try to integrate out our original  $\phi^I$ fields. The details of this procedure depend quite nontrivially on the spatial dimensionality $d$. It is shown in~\cite{Cuomo:2024ekf} that, in two and three spatial dimensions, the end result is an action for three scalar fields,
\begin{align} \label{Clebsch}
    S[\chi, \alpha, \beta] = \int d^{d+1} x \, P( - \xi^\mu \xi_\mu ) \, , \qquad \xi_\mu \equiv \partial_\mu \chi + \frac{1}{2} \left( \alpha \spacy \partial_\mu \beta - \beta \spacy \partial_\mu \alpha \right) \; \qquad (d=2,3) \, .
\end{align}
Notice that the $\chi$, $\alpha$, $\beta$ fields appearing here play similar roles to those appearing in \eqref{NR Clebsch}, but do not reduce
exactly to those in the nonrelativistic limit.

Upon straightforward manipulations, the equations of motion deriving from this action can be written as
\begin{align} \label{eom Clebsch}
    \partial_\mu\big( P' \, \xi^\mu) = 0 \,, \qquad \text{ and } \qquad  \xi^\mu \partial_\mu \alpha = \xi^\mu \partial_\mu \beta = 0 \,.
\end{align}
On the other hand, the stress--energy tensor is
\begin{align}
    T_{\mu\nu} = -2 P' \xi_\mu \xi_\nu + P \, \eta_{\mu\nu} \, ,
\end{align}
and the current associated with the shift symmetry on $\chi$ is
\begin{align}
    J^\mu = - 2 P' \xi^\mu \, ,
\end{align}
which matches the $J^\mu$ in Eq.~\eqref{eom xi}.
Notice that these take a form very reminiscent of the conserved currents for a superfluid, which we will see in Section~\ref{sec:superfluids}. Contrary to that case, however, here $\xi_\mu$ is not controlled just by the gradient of a scalar. 

Using $\rho+p = T s$, we can match these expression to a perfect fluid's stress--energy tensor ($(\rho+p) u_\mu u_\nu + p \, \eta_{\mu\nu}$) and entropy current ($s u^\mu$) by the identifications
\begin{align}
    P = p \, , \qquad\qquad  \xi^\mu = T u^\mu \, .
\end{align}
Then the equations of motion \eqref{eom Clebsch} are nothing but the conservation of the entropy current and the statement that $\alpha$ and $\beta$, which characterize the vorticity two-form, 
\begin{align}
    \omega_{\mu\nu} = T \, \partial_{[\mu} u_{\nu]} = \partial_{[\mu} \alpha \, \partial_{\nu]} \beta \, ,
\end{align}
are conserved along the flow.

We refer the reader to \cite{Cuomo:2024ekf} and references therein for a more extensive analysis of the properties of the Clebsch parametrization---including its vast symmetry structure---and for potential applications. We stress again that the Clebsch parametrization crucially depends  on the spacetime dimensionality. The one briefly reviewed here is the appropriate one in 2+1 and 3+1 dimensions.

\subsection{Perfect Fluids with Conserved Charges} \label{sec: fluid conserved charges}

\noindent All the formulations of Section~\ref{sec: perfect fluids}---the 3- and 4-field Eulerian descriptions as well as their Lagrangian counterparts---rest on a single thermodynamic invariant. As a consequence, the entropy and charge densities are locked together by a fixed ratio $\sigma = s/n$, confining the fluid to the isentropic ray: the conserved charge carries no independent dynamical role and the chemical potential is not an independent variable (which is why we could set $\upmu = 0$ there without loss of generality). To move away from this limit we must let $\sigma$ vary from point to point, which requires a second extensive density, and hence a second invariant, in the EFT. The Eulerian formulation with four fields discussed in Section~\ref{euler 4} provides the blueprint for doing so. In that context, we learned that
\begin{enumerate}
	\item the field $\tau$ is the conjugate variable to the entropy density $s$;
	\item the entropy current $s u^\mu$ is conserved and aligned with the energy current $-u_\mu T^{\mu\nu} = \rho u^\mu$.
\end{enumerate}
This suggests that, in the presence of an additional conserved current $n u^\mu$, we should introduce another field, $\psi$, and our low energy effective Lagrangian should become a function of two quantities, 
\begin{align}
	S[\phi, \tau,\psi] = \int d^4 x \spacy P(u^\mu \partial_\mu \tau, u^\mu \partial_\mu \psi) \, . \label{eq: action Euler 4 fields + charge}
\end{align}
This action is invariant under an additional chemical shift of $\psi$, 
\begin{align} \label{eq: chemical shift conserved charge}
	\psi \to \psi + f (\phi) \, .
\end{align}
The energy--momentum tensor that follows from the action \eqref{eq: action Euler 4 fields + charge} is a straightforward generalization of the one in Eq.~\eqref{Tmunu Euler 4 fields}, namely
\begin{align}
	T_{\mu\nu} = \eta_{\mu\nu} \left(P- z \partial_z P - y \partial_y P \right)+ (z \partial_z P + y \partial_y P)(\eta_{\mu\nu} + u_\mu u_\nu) \, ,
\end{align}
with 
\be
z \equiv u^\mu \partial_\mu \tau, \qquad y \equiv u^\mu \partial_\mu \psi \; .
\ee
From this expression, we deduce that 
\begin{align}
	\rho = z \partial_z P + y \partial_y P - P \, , \qquad \qquad \quad p = P \, .
\end{align}
Furthermore, the equation of motion for $\psi$ takes indeed the form of a conservation equation,
\begin{align}
	\partial_\mu (\partial_y P u^\mu) = 0 \, .
\end{align}
In complete analogy with the entropy sector of Section~\ref{euler 4}---where we identified $s = \partial_z P$ and $T = z$, see Eq.~\eqref{eq: thermodynamic interpretation P(z)}---we identify the charge density with $\partial_y P$ and the quantity $y$ with its associated chemical potential:
\begin{align} \label{eq: n and mu fluid conserved charge}
	n = \partial_y P \, , \qquad \qquad \qquad y = \upmu \, . 
\end{align}
As in the pure-entropy case of Section~\ref{euler 4}, both currents are conserved on-shell rather than identically: the equation of motion for $\tau$ gives $\partial_\mu(s u^\mu) = 0$, while that for $\psi$ gives $\partial_\mu(n u^\mu) = 0$. Because $s$ and $n$ are now separately conserved densities, the ratio $s/n$ is no longer constant, as promised. Instead, the weaker, adiabatic condition is satisfied:
\begin{align} \label{eq: adiabatic condition fluid}
	u^\mu \partial_\mu (s/n) = \frac{s}{n} \left( \frac{u^\mu \partial_\mu s}{s}  - \frac{u^\mu \partial_\mu n}{n} \right) = \frac{s}{n} (\partial_\mu u^\mu - \partial_\mu u^\mu) = 0 \, ,
\end{align}
where in the last step we use the conservation equations of $s$ and $n$.
 
Extending this approach to account for multiple conserved charges is straightforward. It is also easy to reverse the steps that took us from the 3-field Eulerian description to Eq. \eqref{eq: sibiryakov's action} and obtain a 4-field Eulerian description of a perfect fluid with a conserved charge based on the action~\cite{Dubovsky:2011sj}
\begin{align} \label{U(b,y)}
	S[\phi,\psi] = - \int d^4 x \ F(b,y) \, ,
\end{align}
where $F$ is the Legendre transform of $P$ with respect to $z$, of which $b$ is the conjugate variable.  Notice that the adiabatic condition \eqref{eq: adiabatic condition fluid}---the conservation of $s/n$ along the flow---is now simply a consequence of the shift invariance on $\psi$. Indeed, the associated Noether current is
\begin{align} \label{J_mu_psi}
    J^\mu_\psi = u^\mu F_y = J^\mu \, \frac{F_y}{b} \, ,
\end{align}
and since $J^\mu$ is identically conserved, the conservation of $J^\mu_\psi$ reduces to
\begin{align}
    J^\mu  \partial_\mu \big( n/s \big) = 0 \,, 
\end{align}
where we used that $b =s$ and $F_y = n$, the latter being a direct consequence of \eqref{J_mu_psi}.

The adiabatic condition can be used to integrate out $y$, up to some initial conditions, which specify the initial value of $n/s$ throughout the fluid. Explicitly, the fact that the ratio $F_y/b$ is conserved along the flow implies that it must a function of the comoving coordinates $\phi^I$, so that
\begin{align} \label{N(phi)}
    F_y (b,y) = b / \sigma(\phi)  \qquad \qquad \mbox{(on shell)} \; ,
\end{align}
where $\sigma$ is a generic function, which has the interpretation of the inverse number density in $\phi$ space~\cite{Dubovsky:2011sj}.\footnote{This should not be confused with the comoving coordinates defined below Eq.~\eqref{eq: induced metric fluid Lagrangian}.} A nontrivial $\phi$-dependence in $\sigma$ corresponds to a different value of $n/s$ in different points of the fluid. In the isentropic case, $\sigma$ is a constant throughout the fluid
\begin{align}
    \sigma(\phi) = \sigma_0 = {\rm constant} \qquad \qquad \mbox{(isentropic case)} \,,
\end{align}
and inverting \eqref{N(phi)} one can eliminate algebraically $y$ in favor of $b$
\begin{align}
    y(x) \equiv y_{\sigma_0}(b(x)) \, ,
\end{align}
where we are displaying the $x$-dependence to emphasize that the same function $y_{\sigma_0} ({\cdots})$ relates $y$ to $b$ at any spacetime point $x$.
 
When eliminating variables using conservation laws, the correct variational problem is reproduced by plugging back the solutions for such variables not into the Lagrangian, but into the Routhian~\cite{landau1976mechanics}, as we already did in Section~\ref{LF3fields}. In our case, the effective action for the remaining variables is~\cite{Dubovsky:2011sj}
\begin{align}
    S_{\sigma_0}[\phi] = - \int d^4 x \, \left[ F(b, y_{\sigma_0}(b)) - \frac{b}{\sigma_0} \, y_{\sigma_0}(b) \right] \equiv -  \int d^4 x \,  \tilde F_{\sigma_0}(b) \, ,
\end{align}
where we are keeping $\sigma_0$ at the subscript to remind the reader and ourselves that this is the correct action for an isentropic fluid with a specific value for $s/n=\sigma_0$; different $\sigma_0$'s yield different effective Lagrangians $\tilde F$. We have now gone full circle, back to the effective action of Section~\ref{sec:Eulerianfluids} for an isentropic fluid parametrized in terms of the three $\phi^I$ fields.


\subsection{Hydrodynamical Modes} \label{sec: hydro modes fluid}

\noindent So far our treatment of perfect fluids has been somewhat abstract. We now want to make contact with the real world, and highlight possible applications. It should be emphasized that our formulation of perfect fluid dynamics is completely equivalent to the standard one based on conservation laws---in fact, our equations of motion are precisely those conservation laws. And so, when it comes to having to solve those equations of motion for a  complicated fluid flow, our approach is not necessarily adding much compared to more standard ones. What it offers, instead, is the possibility of using the know-how, viewpoint, and techniques of (quantum) field theory in situations where these can be useful, such as for perturbative computations. 

Let's first consider a perfect fluid without conserved charges, in its three-field Eulerian version of Section~\ref{sec:Eulerianfluids}. Small perturbations about a homogeneous equilibrium configuration are parametrized as,
\begin{align} \label{perturbed fluid}
    \phi^I(x) = \bar b^{\spacy 1/3} \big( x^I + \pi^I(x) \big) \, ,
\end{align}
very much analogous to what done for solids in Eq.~\eqref{eq: solid phonon field def}.
Recall that the $\phi^I$'s are the comoving coordinates of the fluid elements, and  a homogeneous equilibrium configuration corresponds to having the comoving coordinates aligned with the physical ones. The constant prefactor $\bar b^{\spacy 1/3}$ measures the compression/dilation level of such an equilibrium configuration, with $\bar b$ being  the value of $b$ in Eq.~\eqref{b def} at equilibrium. The three fields $\pi^I(x)$ describe generic perturbations of our $\phi^I$ fields about equilibrium. 

Plugging  \eqref{perturbed fluid} into the action \eqref{eq: U(B) Lagrangian fluids} and expanding to quadratic order in perturbations, one finds \cite{Dubovsky:2005xd},
\begin{align} \label{quadratic fluid action}
    S = \bar w \int d^4 x \, \frac12 \left[ \dot{\vec \pi} \,^2 - c_s^2 \big( \vec \nabla \cdot \vec \pi \big)^2 + \dots \right]  \, ,
\end{align}
where $\bar w = \bar \rho +\bar p = \bar b F_b$ is the equilibrium enthalpy density, and $c_s$ is the speed of sound:
\begin{align}
    c_s^2 = \frac{\bar b \spacy F_{bb}}{F_b} = \frac{d p}{d \rho}\Big|_{\bar b} \, .
\end{align}
Notice that all derivatives are evaluated on the equilibrium configuration with $b=\bar b$, and so the speed of sound in general depends on $\bar b$. The expression above is nothing but the longitudinal sound speed for a solid, Eq.~\eqref{eq: longitudinal speed solids}, when the Lagrangian is independent on $Y$ and $Z$, as expected.

We immediately see from the quadratic action above that, contrary to what happens in solids, the longitudinal and transverse excitations of a fluid behave in qualitatively different ways. The longitudinal ones obey a standard wave equation, with propagation speed $c_s$,
\begin{align}
    \ddot \pi_L - c_s^2 \spacy \nabla^2 \pi_L = 0 \, .
\end{align}
Clearly, these are the sound modes, just like the solid's longitudinal phonons. On the other hand, the transverse ones obey a free particle-like equation
\begin{align}
    \ddot {\vec \pi}_T = 0 \, ,
\end{align}
whose general solution is
\begin{align} \label{transverse mode solutions}
    \vec \pi_T = \vec g(\vec x) + \vec h(\vec x) \, t \, ,  
\end{align}
where $\vec g$ and $\vec h$ are two arbitrary transverse (i.e., divergence-free) vector functions of $\vec x$. 

These peculiar behavior of the transverse excitations can be traced back directly to our volume-preserving diffeomorphisms symmetry. At the infinitesimal level, a volume preserving diffeomorphism in 
$\phi^I$ space acts as
\begin{align}
    \phi^I \to \phi^I + \epsilon^I (\phi) \, , \qquad \text{with} \qquad \partial_I \epsilon^I = 0  \, .
\end{align}
Applying this to the perturbed configuration in Eq.~\eqref{perturbed fluid} is equivalent, to lowest order, to shifting $\vec \pi$ by
\begin{align}
    \vec \pi(\vec x, t) \to \vec \pi(\vec x, t) + \vec \epsilon \,  (\vec x) \, , 
\qquad \text{with} \qquad \vec \nabla \cdot \vec \epsilon = 0 \, ,
\end{align}
where we are implicitly changing the overall normalization of $\vec \epsilon$ by a $\bar b^{1/3}$ factor. We see that such a transformation is equivalent to adding to $\vec \pi$ a {\em generic time-independent} transverse perturbation. Since this is a symmetry, all this means is  that a time independent transverse perturbation cannot contribute to the action. As a consequence, there cannot be gradient energy terms for transverse perturbations, because such terms would survive for time-independent perturbations.\footnote{Notice that, in this argument, we did use that $\vec \pi$ is small and that we are working to lowest order in it, but we did not use anywhere that we are working to lowest order in the derivative expansion. We thus reach the conclusion that, to {\em all} orders in the derivative expansion, 
the quadratic action for $\vec \pi$ cannot contain terms that only involve spatial-derivatives for transverse excitation. As a consequence, to all orders in the derivative expansion, transverse modes only feature zero frequency solutions of the form \eqref{transverse mode solutions}. Beyond lowest order in the derivative expansion though, such a conclusion is somewhat academic, in that dissipative effects modify things in a substantial way---see Section~\ref{sec: Schwinger-Keldysh}.}

What do the peculiar transverse mode solutions \eqref{transverse mode solutions} correspond to, physically? They are the lineared version
of vorticose solutions. Indeed, to lowest order in $\vec \pi$, Eq.~\eqref{eq:umu fluids} implies that the fluid velocity field is
\begin{align} \label{eq: u hydro modes}
    \vec u(\vec x, t) \simeq - \dot {\vec \pi} (\vec x, t) \, ,
\end{align}
and so the vorticity is
\begin{align}
    \vec \omega \equiv \vec \nabla \times \vec u \simeq - \vec \nabla  \times \dot {\vec \pi} \, .
\end{align}
Not surprisingly, the longitudinal modes do not contribute to vorticity, but the transverse ones do: for the solution in \eqref{transverse mode solutions}, we get
\begin{align}
    \vec \omega(\vec x, t) \simeq - \vec \nabla  \times \vec h(\vec x) \, .
\end{align}
So, our transverse modes correspond to solutions with {\em time-independent} vorticity. 

Clearly, the vast majority of vorticose solutions in hydrodynamics are not in this class, but rather feature a highly time-dependent vorticity, even in the perfect fluid limit. The reason our small perturbation analysis completely misses those is that the dynamics of vorticose solutions is intrinsically nonlinear. We can already get a glimpse of this fact from the linearized solution \eqref{transverse mode solutions} itself: our perturbation grows linearly in time, at a rate that is related to vorticity itself. The higher the vorticity, the sooner our perturbation will exit the regime of validity of the linearized approximation. In fact, one can rewrite the nonlinear hydrodynamical equations of motion  as a time-evolution equation for vorticity. Restricting for simplicity to a nonrelativistic fluid in the incompressible limit, one finds
\begin{align}
\dot {\vec \omega} = \vec \nabla \times (\vec u \times \vec \omega) \, . 
\end{align}
The term on the right hand side is quadratic in perturbations. If one neglects it, there is no time evolution for $\vec \omega$. That is, the dynamics of vorticity are completely dominated by nonlinearities, and this is why they are so complicated and so difficult to solve in general. In particular,
the problem of turbulence is not amenable to a perturbative treatment.

On the other hand, compressional modes, i.e., sound waves, can be dealt with in perturbation theory. In particular, for fluid motions that are slow compared to the speed of sound, all fluids behave as nearly incompressible, which is just another way to say that in those situations it is difficult to excite compressional modes. This simple observation suggests an interesting application of our field-theoretical approach: in the near incompressible limit, we can separate  our degrees of freedom into vorticose  and compressional ones. We can treat the former as (arbitrarily nonlinear) sources for the latter; the latter, we can deal with in perturbation theory. Without attempting to solve for the time-evolution of the vorticose degrees of freedom, one can compute in this way interesting observables related to them, such as the sound emitted by vorticose regions as well as the vortex--vortex long-distance interaction energy mediated by the exchange of compressional modes. This program was initiated in \cite{Endlich:2013dma}. There it was shown, for example, that the power radiated by a vorticose, possibly turbulent, fluid configuration, $\vec v(x)$, due to the emission of sound modes is given by,
\begin{align} \label{eq: vortex radiated power}
    P = \frac{\bar w}{\pi c_s^5} \left[ \frac{1}{4} \big\langle \ddot Q^2 \big\rangle \left( 1 - 3 \frac{c_s^2}{c^2} + \frac{9}{4} \frac{c_s^4}{c^4} \right) + \frac{1}{30} \big\langle \ddot Q_{ij}^2 \big\rangle \right] \,,
\end{align}
where $\langle{\cdots}\rangle$ indicates time averages and,
\begin{align}
    Q_{ij}(t) \equiv \int d^3x  \left( v_i v_j - \frac{1}{3} v^2 \delta_{ij} \right) \,, \qquad \text{ and } \qquad Q(t) \equiv \int d^3x \spacy \frac{1}{3}v^2 \,.
\end{align}
This result generalizes the classic result by Lighthill (e.g.,~\cite{landau:1987bo}) to relativistic fluids, and it reduces to it in the nonrelativistic limit ($c_s \ll c$). 
Notice in particular that Eq.~\eqref{eq: vortex radiated power} is derived in the near incompressible limit, $v \ll c_s$, and it holds to lowest nontrivial order in $v/c_s$, but the equation of state of the fluid can be arbitrarily relativistic, in particular with $c_s \sim c$. The relativistic corrections above are exact to all orders in $c_s/c$.

Going back to the spectrum of small perturbations, one can easily extend the analysis to fluids carrying a conserved charge, such as particle number \cite{Dubovsky:2011sj}. We can use the parametrization of Eq.~\eqref{U(b,y)} and perturb the fields as
\begin{align} \label{eq: background phi^I and tau}
    \phi^I(x) = \bar b^{\spacy 1/3} \big( x^I + \pi^I(x) \big) \, , \qquad \psi(x) = \bar y \spacy (t + \pi^0(x)) \, , 
\end{align}
where $\bar b$ and $\bar y$ are, respectively, the entropy density and chemical potential at equilibrium. To quadratic order in perturbations, we then find the Lagrangian~\cite{Nicolis:2011cs}
\begin{align} \label{eq: L fluids with charge}
    \mathcal{L} = \frac12  \Big[ \big(F_b \bar b - F_y \bar y \big) \spacy \dot{\vec \pi}^{\spacy 2} - F_{bb} \bar b^2  \, (\vec \nabla \cdot \vec \pi)^2
    - F_{yy} \bar y^2 \, (\dot \pi^0)^2 + 2 \big( F_y \bar y - F_{by} \bar b \bar y \big) \, \dot \pi^0 (\vec \nabla \cdot \vec \pi) \Big] \, ,
\end{align}
where the subscripts denote derivatives with respect to the corresponding variables, all implicitly evaluated at equilibrium.

One notices right away that the  transverse modes of $\vec \pi$ do not feature a gradient energy, precisely as for the fluid without charges, and  precisely for the same reason. Also, they do not mix with $\pi^0$, thanks to the unbroken rotational symmetry. Their general solution is thus still Eq.~\eqref{transverse mode solutions}, and the above discussion about their relationship with general vorticose solutions applies unaltered. On the other hand, the longitudinal mode of $\vec \pi$ mixes with $\pi^0$ thanks the last term in Eq.~\eqref{eq: L fluids with charge}, and so it is not immediately obvious what the general solution for $\pi_L$ and $\pi^0$ look like. To figure that out, it is convenient to perform the (nonlocal) field redefinition \cite{Nicolis:2011cs}
\begin{align} \label{pi0 redefinition}
    \dot \pi^0 = \dot {\tilde \pi} + \frac{F_y - F_{by} \bar y}{F_{yy} \bar y} \spacy (\vec \nabla \cdot \vec \pi) \, , 
\end{align}
upon which the Lagrangian becomes diagonal:
\begin{align}
    \mathcal{L} = \frac12  \bigg[ \big(F_b \bar b - F_y \bar y \big) \dot{\vec \pi} \, ^2 - \frac{F_{bb} F_{yy} \bar b^2 - \big(F_y-F_{by} \bar b\big)^2}{F_{yy}}  \, \big(\vec \nabla \cdot \vec \pi\big)^2
    - F_{yy} \bar y^2 \, \dot {\tilde \pi}^2 \bigg] \, .
\end{align}
We thus see that the longitudinal $\vec \pi$ obeys a standard wave equation with propagation speed
\begin{align}
    c_s^2 = \frac{F_{bb} F_{yy} \bar b^2 - \big(F_y-F_{by} \bar b \big)^2}{F_{yy} \big(F_b \bar b - F_y \bar y \big)} \,.
\end{align}
Upon using thermodynamic identities, one can show that this matches the standard expression \cite{Nicolis:2011cs}
\begin{align}
    c_s^2 = \frac{\partial p}{\partial \rho} \Big|_{S,N} \, ,
\end{align}
where $S$ and $N$ are the total entropy and charge of the fluid.
On the other hand, $\tilde \pi$ has dynamics reminiscent of those of the transverse $\vec \pi$ modes: it does not feature a gradient energy and its general solution is
\begin{align} \label{pi tilde solution}
    \tilde \pi(x) = f(\vec x) + g(\vec x) \, t \, ,
\end{align}
for arbitrary $f$ and $g$. It is easy to interpret these solutions, and they happen to be less problematic (or interesting) than the analogous ones for $\vec \pi_T$.
Because of \eqref{pi0 redefinition}, $\pi^0$ in general will be a linear combination of the wave solutions of $\vec \pi$ and of the ``drifting" ones of $\tilde \pi$. Let's for simplicity set $\vec \pi$ to zero, and analyze only the $\tilde \pi$ contribution to $\pi^0$. The latter only enters the thermodynamic quantities through the variable $y$, which, for vanishing $\vec \pi$, simply reads
\begin{align}
    y = \bar y \big(1 + \dot \pi^0\big) = \bar y \big(1 + g(\vec x)\big) \, ,
\end{align}
where in the last step we used the general solution for $\tilde \pi$. In particular, from Eq.~\eqref{eq: n and mu fluid conserved charge}, we get a first order correction to the number density
\begin{align} \label{delta n}
    \delta n \propto g(\vec x) \, .
\end{align}
On the other hand, the fluid velocity field and its entropy density depend only on the $\phi^I$ variables, and are thus unperturbed if $\vec \pi$ is zero:
\begin{align}
    u^\mu = (1, \vec 0 \, ) \, , \qquad \delta s = 0 \, .
\end{align}
However, a perfect fluid carrying a conserved charge satisfies the adiabatic condition \eqref{eq: adiabatic condition fluid} which, for an unperturbed $u^\mu$ and $s$, simply reduces to
\begin{align}
    \delta \dot n = 0 \, ,
\end{align}
whose general solution is precisely  \eqref{delta n}. In other words, the solutions \eqref{pi tilde solution} for $\tilde \pi$  correspond simply to the local  conservation of $n/s$ along the fluid flow: one can give arbitrarily modulated initial conditions for $n/s$ all over the volume of the fluid, and these will be transported along the fluid flow, but will have  no other dynamics. This is a consequence of the fact that charge diffusion, like other dissipative phenomena, only appears at higher orders in the derivative expansion.

\subsection{Nonrelativistic Limit} \label{sec: nonrelativistic fluid}

In order to describe the nonrelativistic of an isentropic fluid, we can simply mirror our treatment of the nonrelativistic limit of solids in Section~\ref{sec: non-relativistic limit of solids}. We leave it to the reader to read off the result from there, simply dropping the dependence on the $Y$ and $Z$ invariants everywhere. Instead, here we will consider the non-relativistic limit of the EFT for a non-isentropic fluid discussed in Section~\ref{sec: fluid conserved charges}. We thus begin from the four-field Eulerian action of Eq.~\eqref{U(b,y)},
\begin{align} \label{eq: NR starting action fluid}
	S = \int d^4 x \spacy F(b,y) \, ,
\end{align}
and apply the scaling limit laid out in Section~\ref{sec: nonrelativistic limit}. The first step is to write $b$ and $y$ in non-covariant notation and reintroduce all factors of $c$ recalling that $\partial_0 = \tfrac{1}{c} \partial_t$ and $d^4 x = c \spacy dt \spacy d^3x$. Then, parametrizing $\psi = \bar \upmu \spacy (t + \pi^0)$ and expanding in inverse powers of the speed of light, we obtain: 
\begin{subequations}
\begin{align}
    b &= \det J \spacy \sqrt{\det \Big( \delta^{mn} - \tfrac{1}{c^2} \dot{\phi}^I {(J^{-1})}_I^{\;\;m} \spacy \dot{\phi}^J {(J^{-1})}_J^{\;\;n} \Big)}  = \det J + \mathcal{O} (1/c^2) \, , \\
    y &= \frac{\det J}{c \spacy b} \left[  \dot{\psi} - \spacy \dot{\phi}^I {(J^{-1})}_I^{\;\;k} \partial_k \psi \right] \,, \nonumber \\
    \begin{split}
        &= \frac{\bar \upmu}{c} \bigg[ 1 + \dot{\pi}^0  - \spacy \dot{\phi}^I {(J^{-1})}_I^{\;\;k} \partial_k \pi^0 \\
        & \qquad \quad \; + \frac{1
        }{2c ^2} \left[1 + \dot{\pi}^0  - \spacy \dot{\phi}^L {(J^{-1})}_L^{\;\;k} \partial_k \pi^0 \right] \dot{\phi}^I {(J^{-1})}_I^{\;\;m} \spacy \dot{\phi}^J {(J^{-1})}_{J m} + \mathcal{O} (1/c^4) \bigg] \, , 
    \end{split}
\end{align}
\end{subequations}
where the matrix $J$ was introduced in Eq.~\eqref{eq: def Jacobian solid}. 

It is worth pausing to discuss how our thermodynamic intepretation influences the scaling of derivatives of $F$ with powers of $c$. In contrast to solids---where the well-defined nonrelativistic limit forced the appearance of a term linear in $b$ proportional to $c^2$, namely $c F = \bar \rho_m c^2 b + U$ in Eq.~\eqref{eq: nonrelativistic solid F}---here we must omit any such term and demand that $F(b,y)$ contain no piece linear in $b$ that scales like $c^2$. The reason is that, in the nonrelativistic limit, the mass density is $\rho_m = m \spacy n$, so that the number density $n$ can be read off from the nonrelativistic kinetic term---it appears as the coefficient of the $\tfrac{m}{2} \dot{\phi}^I {(J^{-1})}_I^{\;\;k} \spacy \dot{\phi}^J {(J^{-1})}_{J k}$ piece in the expansion of $y$. Generically, however, the density read off in this way is a linear combination of $b$ and $F_y$, and the same is true of the entropy density $s$; the clean identifications $b = s$ and $y$ conjugate to $n = F_y$ correspond to a particular choice that diagonalizes this mixing. A nonzero $c^2 b$ term is precisely what spoils it: it would feed the large rest energy into the entropy channel as well, so that $n$ would emerge as a combination of $b$ and $F_y$ rather than $F_y$ alone. Consistency with the thermodynamic interpretation $b = s$ therefore requires this term to vanish, and we set it to zero from the outset. 

Then, proceeding as we did for the solid in section \ref{sec: non-relativistic limit of solids}, we arrive at the conclusion that Eq.~\eqref{eq: NR starting action fluid} will admit a nonrelativistic limit if the derivatives of $F(b,y)$ scale with $c$ as
\begin{align} \label{eq: nonrelativistic scaling fluids conserved charge}
	 \bar b^n \bar y^m \left. \frac{\partial^{n+m} \spacy F}{\partial b^n \partial y^m} \right|_{b = \bar b, y = \bar y} \sim c^{2m-1}  \ .
\end{align}
In particular, the scaling for the first derivative of $F$ with respect to $y$ naturally introduces a quantity with the dimensions of mass density:
\begin{align}
    \bar y \left. F_y \right|_{b = \bar b, y = \bar y} \equiv \bar \rho_m c	
\end{align}
Since the background value of the number density, which must remain finite in the non-relativistic limit, is $\bar n = \left. F_y \right|_{b = \bar b, y = \bar y}$, we can define the mass $m = \lim_{c \to \infty} \bar \rho_m / \bar n = \lim_{c \to \infty} \bar \upmu / c^2 $, and the non-relativistic chemical potential at equilibrium as $\bar \upmu_{\rm nr} = \bar \upmu - m c^2$. We stress that these definitions, although they agree with our physical intuition, they arise naturally  from the requirement that the non-relativistic limit is a well defined scaling limit of the relativistic theory.

Galilean invariance organizes the expansion of $y$ in a transparent way. Collecting the $\mathcal{O}(c^0)$ terms in Eq.~\eqref{eq: NR starting action fluid}, and using $v^k = - \dot{\phi}^I (J^{-1})_I^{\;\;k}$ for the nonrelativistic velocity (we made the same identification for solids below Eq.~\eqref{eq: nonrelativistic action solids}), the dynamical part of $y$ can be written compactly as
\begin{align} \label{eq: y_NR}
	y_{\rm nr} = D_t \psi_{\rm nr} + \frac{m}{2} \spacy \vec v^{\spacy 2} \, , \qquad \qquad D_t \equiv \partial_t + \vec v \cdot \vec \nabla \, ,
\end{align}
where $D_t$ is the usual convective (material) derivative, and we have defined $\psi_{\rm nr} = \bar \upmu_{\rm nr} t + \pi^0$. One can check that $y_{\rm nr}$ is invariant under Galilean boosts. Under $\vec x \to \vec x + \vec v_0 \spacy t$ the velocity transforms by the usual addition law $\vec v \to \vec v + \vec v_0$, while the charge field picks up an inhomogeneous shift,
\begin{align} \label{eq: psi galilean shift}
	\psi \to \psi + m \left( \tfrac{1}{2} \vec v_0^{\spacy 2} \spacy t - \vec v_0 \cdot \vec x \right) \, .
\end{align}
The two terms in \eqref{eq: y_NR} then shift separately: the velocity addition changes $\tfrac{m}{2} \vec v^{\spacy 2}$ by $m \spacy \vec v_0 \cdot \vec v + \tfrac{m}{2} \vec v_0^{\spacy 2}$ and, together with \eqref{eq: psi galilean shift}, changes $D_t \psi$ by exactly the opposite amount so that the combination $y_{\rm nr}$ transforms as a genuine scalar. The nonrelativistic effective action is therefore
\begin{align} \label{eq: action NR fluid}
	S = \int dt d^3 x \spacy F_{\rm nr}(\det J, y_{\rm nr}) \, ,
\end{align}
with $c F (b,y) = F_{\rm nr}(b, c y)$, is constructed so that all of its derivatives are $\mathcal{O}(1)$ in the $c \to \infty$ limit; in particular, $\partial F_{\rm nr}/\partial y_{\rm nr} = F_y$. Expanding $F_{\rm nr}$ around the equilibrium values $\det J = \bar s$ and $y_{\rm nr} = \bar \upmu_{\rm nr}$, the term proportional to $\partial F_{\rm nr}/\partial y_{\rm nr}$ contains the convective derivative $D_t \pi^0$, which is a total derivative and can be dropped, together with the kinetic piece $\tfrac{m}{2} \vec v^{\spacy 2}$, which is invariant up to a total derivative under Galilean boosts. The latter generates a nonrelativistic kinetic term
\begin{align}
	\frac{m \spacy F_y}{2} \spacy \vec v^{\spacy 2} \, ,
\end{align}
in which the coefficient $m \spacy F_y$ is the mass density, consistent with the thermodynamic interpretation~$F_y = n$. It is also straightforward to verify that the action \eqref{eq: action NR fluid} reproduces the equations of nonrelativistic hydrodynamics. Variation with respect to $\psi$ yields the continuity equation, whereas variation with respect to the $\phi^I$'s yields the familiar Euler's equation, 
\begin{align} \label{eq: NR momentum conservation}
	\partial_t \big( \rho_m \spacy v_i \big) + \partial_k \big( \rho_m \spacy v_i \spacy v^k + p \spacy \delta_i^k \big) = 0 \, , 
\end{align}
where 
\begin{align} \label{eq: NR dictionary}
	p = \det J \frac{\partial F_{\rm nr}}{\partial \det J} - F_{\rm nr} \, .
\end{align}

\subsection{Imperfect Fluids and the Schwinger--Keldysh Formalism} \label{sec: Schwinger-Keldysh}

\noindent The effective actions for fluids presented up to this point are appropriate at lowest order in the derivative expansion, capturing the behavior of  perfect fluids. However, for many applications it is important to move away from the perfect fluid limit and include dissipative effects encoded by the leading derivative corrections. As discussed in Section~\eqref{sec: relativistic hydro}, these are corrections to the equations of motion of perfect fluids that contain a single additional derivative. As consistent for dissipative corrections, they cannot be captured by amending any of the effective actions discussed so far in this chapter, since all one-derivative corrections are either total derivatives or removable by field redefinition \cite{Dubovsky:2011sj}. Indeed, in order to capture these additional effects, effective theories of hydrodynamics should be formulated within a broader framework known as \emph{Schwinger-Keldysh formalism}~\cite{10.1063/1.1703727,keldysh1965diagram}. 

This formalism provides a systematic way to construct effective field theories for systems in an arbitrary, possibly mixed state, and is essential for capturing the irreversible, entropy-producing processes that characterize dissipative hydrodynamics. Indeed, unlike the scattering amplitudes that are the traditional focus of particle physics, the central objects of interest in hydrodynamics are expectation values of conserved currents in a thermal state and their time evolution. This requires a path integral approach that is defined on a doubled time contour (see, e.g.,~\cite{Niemi:1983nf}), rather than a single time contour connecting asymptotic in-states in the far past and out-states in the far future.

Considerable work has gone in the last decade into understanding how the EFT knowhow developed in high energy physics can be ported over to the Schwinger-Keldysh formalism, with particular emphasis on hydrodynamic applications---see Sec. \ref{sec: further readings fluids} for an entry point into the literature. We will start by reviewing the main principles of Schwinger--Keldysh EFTs and how they apply to fluids with no additional conserved charges.  Afterwards, we will  reproduce the lowest-order, perfect fluid action, and finally show how dissipative corrections are captured within this framework. Along the way, we will also obtain a first-principle treatment of thermal fluctuations in hydrodynamics. In doing all this, we will take a practical approach. Rather than rederiving the whole Schwinger--Keldysh formalism from scratch, as it can be found at length in the literature, we will present the reader with the rules of the game and show, simply, that they work.

\subsubsection{Principles of Schwinger-Keldysh EFTs} \label{sec: principles of SK EFTs}

\noindent We will find it convenient to work with the Lagrangian formulation with 4 fields discussed in Section~\ref{sec: Lagrangian Formulation with 4 Fields}. In this context, the rules to build a Schwinger--Keldysh effective action for fluids up to arbitrary orders in the derivative expansion are the following:
\begin{enumerate}
    \item Compared to the effective actions we have considered so far, \emph{Schwinger--Keldysh effective actions always involve two copies of each degree of freedom.} Traditionally, these two copies are thought of as the values that each degree of freedom takes on the two branches of a doubled time contour. However, from an EFT perspective, the effective action is defined on the usual time axis and simply features a larger field content. In the case of a fluid, we will denote the two copies of our fields as $X^\mu_+(\sigma)$ and $X^\mu_-(\sigma)$. Note that \emph{these two fields must be expanded around the same physical equilibrium configuration.}\footnote{For some applications, the relevant saddle points may be complex. In this case, the imaginary parts of $X^\mu_+(\sigma)$ and $X^\mu_-(\sigma)$ can differ from one another. For a pedagogical discussion of this point, see Section~4.4 of \cite{kamenev2011field}.}
    \item Unlike in standard EFTs, where unitarity demands that the effective action be real, \emph{Schwinger--Keldysh effective actions can be complex}. However, \emph{the real part of the effective action must be odd under interchange of + and $-$ fields, while the imaginary part must be even and must vanish when these two types of fields are equal to each other.}
    \item \emph{The Schwinger--Keldysh effective action for a closed system}\footnote{Our operational definition of closed system is that it is a system whose time evolution is governed by the von Neumann equation $i \partial_t \rho = [H,  \rho]$. It is worth stressing that, although it is customary to think of a thermal state as being reached due to interactions with an environment, this doesn't mean that a system in a thermal state must be an open system. In fact, the thermal state is a stationary solution of the von Neumann equation.} \emph{enjoys two copies of each internal continuous symmetry.} A fluid should be regarded as a closed system since, as we reviewed at the beginning of Section~\ref{fluids}, the starting point of fluid dynamics is the validity of the conservation equation $\partial_\mu T^{\mu\nu} = 0$, which would not hold as an exact statement in an open system \cite{Sieberer:2015svu}. Moreover, in Lagrangian space, Poincar\'e symmetries behave like internal symmetries because they do not act on the coordinates $\sigma^a$. Therefore, the effective action must be invariant under two copies of spacetime translations and Lorentz transformations acting independently on the + and $-$ variables: $X^\mu_\pm(\sigma) \to \left(\Lambda_\pm\right)^\mu{}_{\nu} X^\nu_\pm(\sigma) + a_\pm^\mu$. This implies that our effective action must be built out of the induced metrics $h_{ab}^\pm = \partial_a X^\mu_\pm \spacy \partial_b X_\mu^\pm$ and their derivatives. As we did in Section~\ref{sec: Lagrangian Formulation with 4 Fields}, it will be convenient to parametrize these metrics using the quantities $\beta_\pm, v_I^\pm, \gamma_{IJ}^\pm$, which have simpler transformation properties under the subgroup of diffeomorphisms of $\sigma^a$ that our action is invariant under. Note in particular that there are five quantities that transform as scalars under chemical shifts: $\beta_\pm,  \gamma_{IJ}^\pm, v^+_I - v^-_I$.
    \item \emph{The Schwinger--Keldysh effective action may have additional symmetries depending on the state of the system.} For instance, in the case of a thermal state, the action must be invariant under the discrete \emph{KMS transformations}:\footnote{This is not the only way to implement the discrete KMS transformations---see e.g.~\cite{Glorioso:2018wxw} for a more general version of these transformations. One of the advantages of our implementation is that it lends itself to a simple mnemonic rule: under a KMS transformation, the temporal argument of $X_\pm$ shifts by $\pm i/2$.}
    \begin{align} \label{eq: KMS transformation}
	    \tilde X_+^\mu (\tau, \phi^I) =  X_+^\mu (- \tau + i /2, \phi^I) \, , \qquad \qquad 
        \tilde X_-^\mu (\tau, \phi^I) =  X_-^\mu (- \tau - i /2, \phi^I) \, .
    \end{align}
    To make the connection between this KMS transformation and the usual notion that fields should have a period equal to the inverse temperature along the imaginary time direction, let's consider the equilibrium configuration of a homogeneous fluid at rest: $X^\mu_+ = X^\mu_- \equiv \bar{X}^\mu$, with $\bar{X}^0 = \tau / T$ and $\bar{X}^i = \ell \spacy \delta^i_I \spacy \phi^I$, with $\ell$  and $T$ dimensionful constants with units of length and inverse length respectively.
    It is necessary to introduce these dimensionful quantities because the coordinates $\tau, \phi^I$ are dimensionless, while the $X^\mu$'s are not. Evaluating the decomposition \eqref{eq: ADM-like breaking of induced metric fluid} on this background, one finds
    \begin{align}
         \bar \beta_+ = \bar  \beta_- = 1/T \, ,
    \end{align}
    and this allows us to identify $T$ with the background value of the temperature. After performing the KMS transformation \eqref{eq: KMS transformation}, it is no longer true that the background values of + and $-$ fields coincide, because
    \begin{align}
        \bar{\tilde{X}}_+^0 - \bar{\tilde{X}}_-^0 = \frac{-\tau + i/2}{T} - \frac{-\tau - i/2}{T} = i/T \,. 
    \end{align}
    If the transformation \eqref{eq: KMS transformation} is a symmetry, this new background configuration must be physically equivalent to the original one, for which $\bar X^0_+ - \bar X^0_-=0$. This is only possible if the shift by $i/T$ in physical Euclidean time is an identification, i.e.\ if imaginary time is periodic with period $1/T$.
    \item \emph{In general, the Schwinger--Keldysh effective action is local only at energies and momenta much smaller than the scales that characterize the state of the system.} For instance, in the case of a thermal state, the KMS transformation~\eqref{eq: KMS transformation} involves evaluating the fields at imaginary-shifted values of $\tau$ and is therefore nonlocal. Any effective action invariant under it must be nonlocal as well. To recover a local description, we can Taylor-expand in small time derivatives:
    \begin{align} \label{eq: expanded KMS transformation}
        \begin{split}
            \tilde X_\pm^\mu (\tau, \phi^I) ={}& e^{\mp \frac{i}{2} \partial_\tau} X_\pm^\mu(-\tau, \phi^I) \\
            ={}& X_\pm^\mu(-\tau, \phi^I) \mp \tfrac{i}{2} \partial_\tau X_\pm^\mu(-\tau, \phi^I) - \tfrac{1}{8} \partial_\tau^2 X_\pm^\mu(-\tau, \phi^I)+ \dots \,.
        \end{split}
    \end{align}
    On the equilibrium background, $\partial_\tau = \frac{1}{T} \partial_t$, where $t$ is physical time. Each additional power of $\partial_\tau$ in the expansion therefore corresponds to a factor of $\partial_t / T$, and the expansion is controlled by the dimensionless ratio $\omega/T$, with $\omega$ the characteristic frequency of the field configuration. This ratio is small precisely in the hydrodynamic regime, $\omega \ll T$. In this limit, the KMS transformation becomes local order-by-order in the derivative expansion, and so does the effective action. At zeroth order in $\partial_\tau$, the KMS transformation reduces to time reversal, $\tau \to - \tau$. Information about the thermal nature of the state enters at first and higher orders. 
\end{enumerate}
In what follows, we will apply these principles first to rederive the effective action for a perfect fluid from a Schwinger--Keldysh perspective, and then to reproduce the usual first order dissipative corrections.

\subsubsection{Schwinger--Keldysh Description of a Perfect Fluid}

\noindent To streamline our notation, we will denote our +/$-$ building blocks collectively as $\varphi_\pm = \{\beta_\pm,  v^\pm_I, \gamma_{IJ}^\pm\}$.  At lowest order in the derivative expansion, the effective action depends exclusively on $\varphi_\pm$, without any additional derivative. KMS transformations act on $\varphi_\pm$ as in Eq. \eqref{eq: expanded KMS transformation},\footnote{To be precise, $v_I^\pm$ picks up an additional overall minus sign, i.e. $\tilde v_I^\pm (\tau) = - v_I^\pm (- \tau) \pm \tfrac{i}{2} \partial_\tau v_I^\pm (- \tau) + \dots$. This is because $v_I^\pm = - h^\pm_{\tau I}/h^\pm_{\tau\tau}$, and 
$$\tilde h^\pm_{\tau I}(\tau) = \partial_\tau \tilde X_\mu^\pm (\tau) \partial_I \tilde X^\mu_\pm(\tau) = \partial_\tau X_\mu^\pm (- \tau) \partial_I  X^\mu_\pm(- \tau) + {\dots} = - (\partial_\tau X_\mu^\pm) (- \tau) (\partial_I  X^\mu_\pm) (- \tau) + {\dots} = -  h^\pm_{\tau I}(- \tau) + {\dots}\, .$$
Since the lowest order effective action will turn out to be independent of $v_\pm^I$, we gloss over this point to avoid complicating the notation further. However, we will need to take this into account when discussing invariance of the next-to-leading order terms in Section~\ref{sec: fluids NLO}. \label{foot: KMS transformation of v_I}} and the action must be invariant order-by-order in the derivative expansion. At first order, the change in the action is
\begin{align}
	\delta_1 S = \delta_1 \int d^4 \sigma \spacy \mathcal{L}(\varphi_\pm) = \int d^4 \sigma \left( - \frac{i}{2} \frac{\partial \mathcal{L}}{\partial \varphi_+} \partial_\tau \varphi_+ +\frac{i}{2} \frac{\partial \mathcal{L}}{\partial \varphi_-} \partial_\tau \varphi_- \right) \, .
\end{align}
Because of the relative sign, the two terms on the right-hand side do not combine to form a total derivative. Therefore, the only way to ensure that the action is invariant is to demand that both terms are separately total derivatives, i.e. that
\begin{align}
	\frac{\partial \mathcal{L}}{\partial \varphi_+} = \frac{d f_1(\varphi_+)}{d \varphi_+} \, , \qquad \qquad \qquad  \frac{\partial \mathcal{L}}{\partial \varphi_-} = \frac{d f_2(\varphi_-)}{d \varphi_-} \, .
\end{align}
This implies that the Lagrangian must have the separable form $\mathcal L (\varphi_\pm) = f_1 (\varphi_+) + f_2(\varphi_-)$.\footnote{More precisely, such a Lagrangian leads to a path integral with an integrand that is factorized: $e^{iS} = e^{i S_+} e^{-i S_-}$.} From this, we can already conclude that the action can only depend on $\beta_\pm$ and $\gamma^\pm_{IJ}$ at this order. The only other combination that is invariant under chemical shifts would be $v^+_I - v^-_I$, but invariance under spatial diffeomorphisms prevents it from appearing linearly in the action, and thus the action would not be separable.

Furthermore, such a separable action can only satisfy the unitarity requirements listed in point 2. of the previous section if $\text{Re}f_1 = - \text{Re}f_2$ and $\text{Im}f_1 = \text{Im}f_2 = 0$. Note that, because of this, the whole action must vanish for $\varphi_+ = \varphi_-$, at this order. Finally, invariance under spatial diffeomorphisms completely fixes the dependence on $\gamma_{IJ}^\pm$, so that the lowest order action is
\begin{align} \label{eq: SK perfect fluid action}
	S = \int d^4 \sigma \bigg\{ \sqrt{\gamma_+} \spacy G(\beta_+) -   \sqrt{\gamma_-} \spacy G(\beta_-) \bigg\} \, .
\end{align}
In other words, the lowest order action is the difference between two copies of the perfect fluid action in Eq.~\eqref{eq: action fluid Lagrangian no charge}. 

It is easy to verify that this factorized action remains invariant under KMS transformations also at higher orders in the derivative expansion. For instance, setting $f(\varphi) = \sqrt{\gamma} \spacy G(\beta)$, the second order variation of the action is
\begin{align}
	\delta_2 S &= \int d^4 \sigma \left[ - \frac{1}{8} \frac{d f(\varphi_+)}{d \varphi_+} \partial_\tau^2 \varphi_+ +\frac{1}{8} \frac{d f(\varphi_-)}{d \varphi_-} \partial_\tau^2 \varphi_- + \frac{1}{2} \left(-\frac{i}{2}\right)^{\!\!\spacy 2} \frac{d^2 f}{d \varphi_+^2} (\partial_\tau \varphi_+)^2 - \frac{1}{2} \left(\frac{i}{2}\right)^{\!\!\spacy 2} \frac{d^2 f}{d \varphi_-^2} (\partial_\tau \varphi_-)^2 \right] \nonumber \\
    &= \int d^4 \sigma \left[ - \frac{1}{8} \partial_\tau \left(  \frac{d f(\varphi_+)}{d \varphi_+} \partial_\tau \varphi_+ \right) + \frac{1}{8} \partial_\tau \left(  \frac{d f(\varphi_-)}{d \varphi_-} \partial_\tau \varphi_- \right) \right] = 0 \, .
\end{align}
As we will see, the effective action will no longer be separable at next-to-leading order in the derivative expansion.

\subsubsection{The Keldysh Basis}

\noindent Before extending the perfect fluid action \eqref{eq: SK perfect fluid action} to next-to-leading order in the derivative expansion, it is helpful to pause for a moment to introduce an alternative combination of fields known in general as the \emph{Keldysh basis}. Starting from the collective variables $\varphi_\pm$ introduced in the previous section, we can switch to a new basis $\varphi_{a, r}$ defined by
\begin{align} \label{eq: keldysh basis}
	\varphi_r \equiv \tfrac{1}{2} (\varphi_+ + \varphi_-) \, , \qquad \qquad \qquad \varphi_a \equiv \varphi_+ - \varphi_- \, .
\end{align}
The advantage of these new combinations of fields is that now only $\varphi_r$ acquires a large classical background when $\bar \varphi_+ = \bar \varphi_- \equiv \bar \varphi$, whereas $\varphi_a$ always has a vanishing background. Because of this, instead of keeping all nonlinearities in both $\varphi_+$ and $\varphi_-$, we can organize the effective action by expanding it in powers of $\varphi_a$ while keeping all nonlinearities in $\varphi_r$. This expansion would need to be put on firmer ground by assigning definite power counting rules to the fields $\varphi_{a,r}$, and we'll address this in the next section. For now, we can invert the definitions \eqref{eq: keldysh basis}, plug the resulting expressions for $\varphi_\pm$ in the perfect fluid action \eqref{eq: SK perfect fluid action}, and expand in powers of $\varphi_a$ fields to obtain
\begin{align} \label{eq: SK action fluid expanded}
	S = \int d^4 \sigma \left[ \frac{d f(\varphi_r)}{d \varphi_r} \varphi_a + \frac{1}{4!} \frac{d^3 f(\varphi_r)}{d \varphi_r^3} \varphi_a^3 + \mathcal{O} \left(\varphi_a^5\right) \right] \, , 
\end{align}
where, once again, $f(\varphi) = \sqrt{\gamma} \spacy G(\beta)$. This action contains only odd powers of $\varphi_a$ because of the unitarity constraints discussed in Section~\ref{sec: principles of SK EFTs}. More in general, \emph{the real part of a Schwinger--Keldysh effective action is always an odd function of $a$-type fields, while the imaginary part is always even.} 

We could also switch to a Keldysh basis at the level of our fundamental fields $X_\pm^\mu$, introducing
\begin{equation}
    X_r^\mu = \tfrac{1}{2}\left(X^\mu_+ + X^\mu_-\right) \, , \qquad \qquad \qquad X_a^\mu = X_+^\mu - X_-^\mu \, .
\end{equation}
Because $X_+^\mu$ and $X^\mu_-$ transform under two different copies of the Poincar\'e group, $X^\mu_r$ transforms as a set of coordinates only under the diagonal subgroup. The other combination, $X^\mu_a$, transforms as a 4-vector under diagonal Lorentz transformations but is invariant under diagonal translations. The action of off-diagonal transformations is, instead, more complicated and mixes $X_r^\mu$ and $X^\mu_a$. Consequently, an expansion in powers of $X^\mu_a$ would explicitly break the off-diagonal symmetries, obscuring the existence of additional Ward identities. The KMS transformations also mix the $X_{r,a}^\mu$ fields, and in the small frequency limit where the transformations \eqref{eq: expanded KMS transformation} are local, they reduce to\footnote{After deriving the power counting rules, it will become clear that the term $-\tfrac{i}{4} \partial_\tau  X_a^\mu (-\tau, \phi)$ is subleading and can be consistently dropped at lowest order in the derivative expansion.}
\begin{subequations} \label{eq: KMS Keldysh basis}
\begin{align}
    \tilde X_r^\mu (\tau, \phi) &=  X_r^\mu (-\tau, \phi) -\tfrac{i}{4} \partial_\tau  X_a^\mu (-\tau, \phi) + \dots \, , \label{eq: KMS Keldysh basis a} \\
    \tilde X_a^\mu (\tau, \phi) &=  X_a^\mu (-\tau, \phi)- i \partial_\tau  X_r^\mu (-\tau, \phi) + \dots \, . \label{eq: KMS Keldysh basis b}
\end{align}
\end{subequations}

Although symmetries act in a more complicated way on the Keldysh basis of fields $X_{r,a}^\mu$, these variables are still convenient to introduce because they lend themselves to a more direct physical interpretation. In fact, varying the action \eqref{eq: SK action fluid expanded} with respect to $X^\mu_a$ and then setting it to zero yields the classical equations for the fluid coordinates $X^\mu_r$:
\begin{align}
	\begin{split}
   \left. \frac{\delta S}{\delta X_a^\mu} \right|_{X_a^\mu = 0} &= \int d^4 \sigma \, \frac{d f(\varphi_r)}{d \varphi_r} \left. \frac{\delta \varphi_a}{\delta X^\mu_a}\right|_{X_a^\mu = 0} \\
   &= \int d^4 \sigma \left. \frac{d f(\varphi_r)}{d \varphi_r} \right|_{X_a^\mu = 0} \left[ \frac{1}{2}  \frac{\delta \varphi_+}{\delta X^\mu_+} +  \frac{1}{2}  \frac{\delta \varphi_-}{\delta X^\mu_-}\right]_{X_+^\mu = X_-^\mu = X_r^\mu} \\
   &= \int d^4 \sigma \left. \frac{d f(\varphi_+)}{d \varphi_+}  \frac{\delta \varphi_+}{\delta X^\mu_+} \right|_{X_+^\mu = X_r^\mu} \,. 
   \end{split}
\end{align}
The final result is exactly what we would have obtained by varying the single-field effective action for a perfect fluid with $X^\mu \to X^\mu_r$. This is consistent with our previous observation that $r$-type fields have a classical expectation value. Instead, $a$-type fields should be interpreted as parametrizing (in our case, thermal) fluctuations around the classical solution. Notice also that only the term linear in $\varphi_a$ ultimately contributes to the classical equations of motion after setting $X_a^\mu = 0$. This structure makes contact with the Martin--Siggia--Rose (MSR) formalism for stochastic systems~\cite{Martin:1973zz}, where the role of the $a$-type field is played by the response field. In both cases, the auxiliary field couples linearly to the equations of motion and generates retarded response functions. The Schwinger--Keldysh framework is, however, the more systematic and fundamental construction where effective actions follow systematically from symmetry considerations rather than phenomenological considerations, and  noise can be introduced consistently at fully nonlinear level.

\subsubsection{Covariant Derivatives}

\noindent Before considering higher derivative corrections to the perfect fluid effective action \eqref{eq: SK action fluid expanded}, we need to introduce derivatives that are covariant under the coordinate transformations \eqref{eq: target reparametrizations of fluid} and have definite ``parity'' under interchange of + and $-$ fields. The latter requirement will make it easier to enforce unitarity constraints, and will be implemented more easily by working in the Keldysh basis, $\varphi_{a,r}$. 

Let us start by considering chemical shifts, under which regular derivatives transform as follows:
\begin{align}
	\frac{\partial}{\partial \tau'} = \frac{\partial}{\partial \tau} \, , \qquad \qquad \qquad \frac{\partial}{\partial \phi^{I \prime}} = \frac{\partial}{\partial \phi^{I}} - \frac{\partial g (\phi)}{\partial \phi^I} \frac{\partial}{\partial \tau} \, .
\end{align}
In order for the derivatives with respect to $\phi^I$ to transform covariantly, we need to supplement them with an additional term transforming in such a way as to cancel the piece proportional to $\partial g / \partial \phi^I$. Because the combinations $v^\pm_I$ shift by precisely this quantity under chemical shifts---see Eq. \eqref{eq: transformation of v^I chemical shifts}---and regular derivatives are even under $+/-$ interchange, it is easy to conclude that the combination of partial derivatives we are after is
\begin{align}
	\hat \partial_I = \partial_I + v_I^r \partial_\tau \, .
\end{align}

We can now ensure covariance also under the diffeomorphisms $\phi^I \to \xi^I (\phi)$ by supplementing $\hat \partial_I$ with a connection. The fields $\gamma_{IJ}^{r,a}$ both transform like metrics, and would both give rise to a Christoffel connection that is even under $+/-$ interchange. However, the connection for $\gamma^a_{IJ}$ becomes singular when we expand around a classical background $\gamma^+_{IJ} = \gamma^-_{IJ} = \bar \gamma_{IJ}$, since $\gamma^a_{IJ}$ vanishes. For this reason, we must work with the Christoffel connection $\gamma_{IJ}^{r}$, 
\begin{align} \label{eq: NLO fluids connection}
	\hat \Gamma^I{}_{JK} = \tfrac{1}{2} \gamma_r^{IL} \left(\hat \partial_J \gamma^r_{LK} + \hat \partial_K \gamma^r_{LJ} - \hat \partial_K \gamma^r_{JK} \right) \, ,
\end{align}
and use the covariant derivative $\hat \nabla = \hat \partial + \hat \Gamma$.

\subsubsection{Power Counting} \label{sec: power counting}

\noindent In order to organize the effective action in a systematic expansion, we need to assign definite power counting rules to the fields $\varphi_{a,r}$ and to derivatives. Recall that the comoving coordinates $\sigma^a = (\tau, \phi^I)$ are dimensionless, and so are derivatives with respect to them. As discussed in item 5 of Section~\ref{sec: principles of SK EFTs}, on the equilibrium background one has $\partial_\tau = \bar\beta \, \partial_t$, so that $\partial_\tau$ acting on a slowly varying field configuration produces a factor of order $\omega / T$, with $\omega$ the characteristic frequency. Similarly, from the background configuration $\bar X^i = \ell \, \delta^i_I \phi^I$ we have $\partial_I = \ell \, \partial_i$, so that spatial derivatives produce factors of order $k \ell$, with $k$ the characteristic wavenumber.\footnote{The length scale $\ell$ is related to the equilibrium entropy density by $\bar b = 1/\ell^3$. In general, $\ell$ is an independent scale from the inverse temperature $1/T$. However, the hydrodynamic regime requires both $\omega/T \ll 1$ and $k \ell \ll 1$, and both conditions are met when the characteristic spacetime scales of the fluid configuration are much larger than the mean free path.} In principle, the two small parameters $\omega/T$ and $k\ell$ could scale differently, leading to an independent power counting for temporal and spatial derivatives. However, assigning a definite scaling to the covariant derivative $\hat\nabla_I$ introduced in the previous section requires that we treat $\partial_\tau$ and $\partial_I$ on the same footing. Indeed, recall that the chemical-shift-covariant partial derivative is $\hat\partial_I = \partial_I + v^r_I \partial_\tau$. Since $v^r_I$ acquires a background value of order one, this combination has a definite power counting only if $\partial_\tau \sim \partial_I$, i.e.\ only if $\omega/T \sim k \ell$. This is the regime in which the chemical shift symmetry places nontrivial constraints on the allowed combinations of derivatives in the effective action. We will therefore treat each derivative, $\partial_\tau$ or $\hat\nabla_I$, as $\mathcal{O}(\partial)$ in a single, unified expansion in this small parameter.
 
The $r$-type fields $\varphi_r = \{ \beta_r, v_I^r, \gamma_{IJ}^r \}$ acquire nonzero background values and should be counted as $\mathcal{O}(1)$. What about the $a$-type fields? Their background values vanish because $\bar \varphi_+ = \bar \varphi_-$, but this does not by itself fix their scaling---one needs an additional argument. The key observation is that the KMS transformation must impose nontrivial constraints on the effective action at each order in the derivative expansion. Inspecting the local form of the KMS transformation of $X_a^\mu$ in the Keldysh basis, Eq.~\eqref{eq: KMS Keldysh basis b}, we see that it relates $X_a^\mu$ to $\partial_\tau X_r^\mu$. If the latter term were subleading with respect to the former, KMS would end up being simply time reversal, at lowest order. In order for it to be more nontrivial transformation, it must be $X_a^\mu \sim \partial X_r^\mu \sim \mathcal{O}(1)$. Here we used the fact that $X_r^\mu$ always appears with at least one derivative, and we can thus start our power counting directly from $\partial X_r^\mu$. (As a mnemonic rule, $X_r^\mu \sim \mathcal{O}(1/\partial)$.)
The scaling we just discussed, implies,\footnote{If we instead treated $\varphi_a$ as $\mathcal{O}(1)$, the KMS transformation would only relate operators at different orders in the derivative expansion. In that case, KMS would fail to impose any constraint at a given order, and the information about the thermal nature of the state would be lost order by order. Assigning $\varphi_a \sim \mathcal{O}(\partial)$ is therefore necessary for the KMS symmetry to have any constraining power.}
\begin{align} \label{eq: power counting}
	\varphi_r \sim \mathcal{O}(1) \, , \qquad \qquad \qquad \varphi_a \sim \mathcal{O}(\partial) \, .
\end{align}
With this power counting we can now consistently truncate the KMS transformations~\eqref{eq: KMS Keldysh basis} at any desired order. The leading term in Eq.~\eqref{eq: KMS Keldysh basis a} is $X_r^\mu(-\tau, \phi)$, while the correction $-\tfrac{i}{4} \partial_\tau X_a^\mu$ is $\mathcal{O}(\partial)$, since it involves a derivative acting on an $a$-type field. Similarly, the two terms shown explicitly in Eq.~\eqref{eq: KMS Keldysh basis b} are both $\mathcal{O}(1)$, while the next correction is $\mathcal{O}(\partial^2)$. Therefore, up to and including next-to-leading order, the KMS transformations can be consistently truncated to
\begin{subequations} \label{eq: KMS truncated NLO}
\begin{align}
    \tilde X_r^\mu (\tau, \phi) &=  X_r^\mu (-\tau, \phi) + \mathcal{O}(\partial)\, , \label{eq: KMS truncated NLO a}\\
    \tilde X_a^\mu (\tau, \phi) &=  X_a^\mu (-\tau, \phi)- i \partial_\tau  X_r^\mu (-\tau, \phi) + \mathcal{O}(\partial^2) \, . \label{eq: KMS truncated NLO b}
\end{align}
\end{subequations}
That is, at this order the KMS transformation acts as time reversal on $r$-type fields, while on $a$-type fields it acts as time reversal supplemented by a shift proportional to $\partial_\tau X_r^\mu$. It is the latter piece that encodes information about the thermal state and, as we will see, will be responsible for relating dissipative and noise terms in the effective action.
 
It is important to appreciate that the expansion parameter $\omega/T$ controlling the locality of the KMS transformation is \emph{kinematic} in nature: it arises from the Taylor expansion of the imaginary-time shift $e^{\mp \frac{i}{2}\partial_\tau}$ in Eq.~\eqref{eq: expanded KMS transformation}, and reflects only the structure of the thermal density matrix $\rho \propto e^{-H/T}$, independently of the microscopic dynamics. By contrast, the physical expansion parameter governing the hydrodynamic gradient expansion---i.e.\ the ratio of dissipative corrections to ideal fluid terms---is $\omega t_{\rm therm}$, where $t_{\rm therm}$ is the thermalization time of the microscopic constituents. Based on the uncertainty principle, we expect $t_{\rm therm} \gtrsim 1/T$---a relation referred to in some contexts as {\it Planckian bound}~\cite{Hartnoll:2021ydi}. At strong coupling this bound is saturated and the two expansion parameters coincide. At weak coupling, however, there exists a parametric hierarchy between them: for instance, in a weakly coupled gauge theory at temperature $T$, $t_{\rm therm}$ is set by the rate of nearly forward scattering with small momentum transfer $q \sim g T$~\cite{Arnold:2002zm}, so that $1/t_{\rm therm} \sim n\,\sigma \sim T^3 \cdot g^4 \log  (1/g) /T^2 \sim g^4 T \log (1/g)$, with $\log (1/g)$ the usual Coulomb logarithm. Therefore, in this case $\omega/T \ll \omega t_{\rm therm} \ll 1$, and the two quantities are parametrically separated.
 
These two expansions are reconciled by the fact that the Wilson coefficients appearing in the next-to-leading order effective action are generically smaller than $T t_{\rm therm}$,
and are therefore not $\mathcal{O}(1)$ in units of $T$ at weak coupling. As a result, while the formal derivative counting assigns $S_{\rm NLO} \sim \mathcal{O}(\partial^2)$ and $S_{\rm LO} \sim \mathcal{O}(\partial)$ in units of $\omega/T$, the physical suppression of next-to-leading order (NLO) corrections relative to leading order (LO) is
\begin{align} \label{eq: physical suppression NLO}
\frac{S_{\rm NLO}}{S_{\rm LO}} \quad \lesssim \underbrace{\omega/T}_{\text{derivative counting}} \times \underbrace{Tt_{\rm therm}}_{\text{Wilson coefficients}} \sim \quad \omega t_{\rm therm} \, ,
\end{align}
as expected on physical grounds. In other words, the $\omega/T$ power counting is a \emph{structural} organizing principle that determines which operators KMS relates to one another at each order, but the actual physical size of those operators is set by the Wilson coefficients, which encode the microscopic dynamics and carry the information about $t_{\rm therm}$. The scheme is nonetheless consistent, because $\omega t_{\rm therm} \ll 1$ always implies $\omega/T \ll 1$, so the KMS locality expansion converges whenever the hydrodynamic gradient expansion does. Furthermore, the constraints that KMS imposes---such as the fluctuation-dissipation relations between dissipative and noise terms---hold regardless of the size of the Wilson coefficients, and are therefore equally valid at weak and strong coupling.

Note that the dependence of the Wilson coefficients on the UV coupling runs counter to the intuition from standard particle physics effective theories. There, one integrates out a heavy particle of mass $M$ coupled to the light fields with strength $g$, and the resulting Wilson coefficients scale as positive powers of $g$: weaker coupling means a smaller imprint of the heavy state, and the Wilson coefficients \emph{shrink} as $g \to 0$. The crucial difference is that the cutoff---the scale $M$ at which new physics enters---is the particle mass, and is \emph{independent} of the coupling. In hydrodynamics, by contrast, the cutoff is the inverse thermalization rate $1/t_{\rm therm} \sim n\,\sigma$, which is itself controlled by the microscopic coupling. As the UV physics becomes weakly coupled, $\sigma$ decreases, $t_{\rm therm}$ grows, and the NLO Wilson coefficients---which scale as $T\, t_{\rm therm}$---grow with it. Weak coupling therefore produces \emph{larger} Wilson coefficients, not smaller ones, and the matching delivers couplings in the denominator precisely because the cutoff scales with the coupling rather than being fixed independently of it.

\subsubsection{Fluids at Next-to-Leading Order: Viscosity and Non-dissipative Transport} \label{sec: fluids NLO}

\noindent With these power counting rules in hand, the structure of the effective action organizes itself as follows. The explicit form of the LO action is obtained by looking at Eq.~\eqref{eq: SK action fluid expanded}, and explicitly spelling out the fact that the $\varphi_{a,r}$ field on which the Lagrangian depends are $\beta_{a,r}$ and $\gamma_{a,r}$ (and summing over their derivatives). The result is
\begin{align} \label{eq: LO SK action fluids}
    S_{\rm LO} = \int d^4 \sigma \left( \beta_a \frac{\partial}{\partial \beta_r} + \gamma_{IJ}^a \frac{\partial}{\partial \gamma_{IJ}^r} \right) \sqrt{\gamma_r} \spacy G(\beta_r) \, ,
\end{align}
is $\mathcal{O}(\partial)$ because it is linear in $a$-type quantities without any additional derivatives. This expression details the first term in the action \eqref{eq: SK action fluid expanded}, which we have previously argued reproduces the perfect fluid equations. The NLO action is $\mathcal{O}(\partial^2)$, and includes two types of contributions:
\begin{enumerate}
\item \emph{Terms linear in $\varphi_a$ with one derivative.} These are real (because the real part of the action is odd in $\varphi_a$) and will give rise to dissipative corrections to the equations of motion upon variation with respect to $X_a^\mu$.
\item \emph{Terms quadratic in $\varphi_a$ with no derivatives.} These are purely imaginary (because the imaginary part of the action is even in $\varphi_a$) and encode thermal fluctuations. They will generate stochastic noise terms in a Langevin-type description of the fluid.
\end{enumerate}
As we will see, KMS invariance relates the coefficients of these two types of terms to one another, leading to a fluctuation-dissipation relation. Once the dust settles, we will see the familiar bulk and shear viscosity terms as well as the parity-odd non-dissipative transport term arise purely from symmetry consideration. Furthermore, the second law of thermodynamics will follow directly from unitarity constraints~\cite{Glorioso:2016gsa}.

Let's start by classifying all possible $\mathcal{O}(\partial^2)$ terms that can appear in the imaginary part of the action. To this end, it is convenient to break up $\gamma_{IJ}^a$ into a pure trace and a traceless part with respect to $\gamma_{IJ}^r$:
\begin{align} \label{eq: decomponsition gamma a}
    \theta_a \equiv \gamma^{IJ}_r \gamma_{IJ}^a \,, \qquad \qquad \qquad \Gamma_{IJ}^a \equiv \gamma_{IJ}^a - \tfrac{1}{3} \gamma_{IJ}^r \theta_a \, .
\end{align}
Then, the most general linear combination of $\mathcal O(\partial^2)$ operators that are quadratic in $a$-type fields is  built out only of terms of the type discussed in item 2 above, i.e., operators quadratic in the $\varphi_a$ fields. This reads,
\begin{align} \label{eq: NLO Im S fluids}
    \begin{split}
        \text{Im} \spacy S_{\rm NLO} ={}& \int d^4 \sigma \spacy \sqrt{\gamma_r} \spacy \beta_r \bigg(c_{\beta \beta}(\beta_r) \spacy \beta_a^2 + 2 c_{\beta \theta}(\beta_r) \spacy \beta_a \theta_a + c_{\theta\theta}(\beta_r) \spacy \theta_a^2 + c_{vv} (\beta_r) \spacy v^I_a v_I^a \\
        {}& \qquad \qquad \qquad + c_{\Gamma\Gamma} (\beta_r)  \spacy \Gamma^{IJ}_a \Gamma_{IJ}^a \bigg) \,,
    \end{split}
\end{align}
where we indicated explicitly that the coefficient of this expansion can depend on $\beta_r$, which is $\mathcal{O}(1)$, and we have extracted an overall factor of $\beta_r$ for later convenience.
Unitarity requires that $\text{Im} \spacy S \geqslant 0$ for all field configurations or, alternatively, for the quadratic kernel in the equation above to be a positive semi-definite matrix. Sylvester's criterion then implies the following positivity constraints on the otherwise arbitrary functions of $\beta_r$ appearing in Eq.~\eqref{eq: NLO Im S fluids}:
\begin{align} \label{eq: positivity NLO fluids}
    c_{\beta\beta} \geqslant 0 \, , \qquad \qquad c_{\beta \beta} c_{\theta \theta} - c_{\beta \theta}^2 \geqslant 0 \, , \qquad \qquad  c_{vv} \geqslant 0 \, , \qquad \qquad  c_{\Gamma\Gamma}\geqslant 0 \, .
\end{align}
We should also point out that a term of the form $(v^I_a v_I^r)^2$ would not be invariant under chemical shifts because of the transformation properties of $v_I^r$.\footnote{One might also be tempted to write $\left( \gamma_r^{IJ} \Gamma^a_{JI} \right)^2$ and $\gamma_r^{IJ} \Gamma^a_{JK} \gamma_r^{KN} \Gamma^a_{NI}$, which are also quadratic in $a$-type fields. However, recalling that we are using $\gamma_r$ to raise and lower indices, the first combination vanishes, and the second is precisely $\Gamma_a^{IJ} \Gamma^a_{IJ}$, which is already included in Eq.~\eqref{eq: NLO Im S fluids}.}

Let us now turn our attention to the real NLO part of the effective action, which is linear in the $a$-type fields and contains a single additional derivative compared to the LO action \eqref{eq: LO SK action fluids}. Without loss of generality, we can always perform integrations by parts to ensure that the derivative acts only on $r$-type fields. Thus, every NLO operator must be proportional to one of the following quantities:
\begin{align}
    \partial_\tau \beta_r \,, \;\;\, \partial_\tau v_I^r \,, \;\;\, \Theta_r =  \gamma^{IJ}_r\partial_\tau \gamma_{IJ}^r \,, \;\;\, K_{IJ}^r = \partial_\tau \gamma_{IJ}^r - \tfrac{1}{3} \gamma_{IJ}^r  \Theta_r \,, \;\;\, \hat \nabla_I \beta_r \,, \;\;\, F_{IJ}^r = \hat \partial_I v^r_J - \hat \partial_J v^r_I \,,
\end{align}
where we have decomposed $\partial_\tau \gamma_{IJ}^r$ into its trace and traceless part, similarly to what we did for $\gamma_{IJ}^a$ in Eq.~\eqref{eq: decomponsition gamma a}. The spatial covariant derivative $\hat \nabla_K \gamma_{IJ}^r$ does not appear in this list because it vanishes, as the covariant derivative of the metric tensor should~\cite[e.g.,][]{Weinberg:1972kfs}. Finally, the time derivative $\partial_\tau v_I^r$ is invariant under chemical shifts because they do not depend on $\tau$, but this symmetry forces the spatial derivatives of $v_I^r$ to enter the action via the field strength $\hat F_{IJ}^r$. Invariance under chemical shifts of $\tau$ and reparametrizations of the spatial coordinates $\phi^I$ constrains the real part of the NLO action to have the following form: 
\begin{align}\label{eq: NLO Re S fluids}
    \text{Re} \spacy S_{\rm NLO} &= \int d^4 \sigma \spacy \sqrt{\gamma_r} \spacy \beta_r \bigg( a_{\beta\beta} (\beta_r) \spacy \beta_a \partial_\tau \beta_r + a_{\beta\theta} (\beta_r) \spacy \theta_a \partial_\tau \beta_r + a_{\beta\Theta} (\beta_r) \spacy \beta_a \Theta_r + a_{\theta\Theta} (\beta_r) \spacy \theta_a \Theta_r  \\ 
    & \qquad \qquad  + a_{vv}(\beta_r) \spacy v^I_a \partial_\tau v_I^r  + a_{v\beta}(\beta_r) \spacy v^I_a  \hat \nabla_I \beta_r + a_{\Gamma K}(\beta_r) \spacy \Gamma^{IJ}_a K_{IJ}^r + a_{vF} (\beta_r) \varepsilon^{IJK} v_I^a F_{JK}^r \bigg) \,, \nonumber
\end{align}
where $\varepsilon^{IJK} = \epsilon^{IJK} / \sqrt{\gamma_r}$, with $\epsilon^{IJK}$ the Levi--Civita symbol, and we have once again extracted an overall factor of $\beta_r$.

Let's now impose invariance under the KMS symmetry, which will relate the arbitrary functions of $\beta_r$ appearing in the real and imaginary parts of $S_{\rm NLO}$. At the order relevant for $S_{\rm NLO}$, the KMS transformations of $X_{a,r}^\mu$ are given in Eq. \eqref{eq: KMS truncated NLO}. The crucial step is to determine what transformations they induce on the composite building blocks appearing in $S_{\rm NLO}$. A direct computation yields these remarkably simple results:
\begin{subequations} \label{eq: KMS shift composite fields}
\begin{align}
    \tilde \beta_a (\tau, \phi) &= \beta_a(-\tau, \phi) - i \, \partial_\tau \beta_r (-\tau, \phi) + \mathcal{O}(\partial^3) \, , \label{eq: KMS beta_a} \\
    \tilde \theta_a (\tau, \phi) &= \theta_a(-\tau, \phi) - i \, \Theta_r (-\tau, \phi) + \mathcal{O}(\partial^3)  \, , \label{eq: KMS theta_a} \\
    \tilde \Gamma^a_{IJ} (\tau, \phi) &= \Gamma^a_{IJ}(-\tau, \phi) - i \, K^r_{IJ} (-\tau, \phi) + \mathcal{O}(\partial^3)  \, , \label{eq: KMS Gamma_a} \\
    \tilde v^a_I (\tau, \phi) &= -v^a_I(-\tau, \phi) + i \, \partial_\tau v^r_I (-\tau, \phi) + \mathcal{O}(\partial^3)  \, , \label{eq: KMS v_a}
\end{align}
\end{subequations}
while all $r$-type fields simply undergo time reversal, e.g. \ $\tilde \beta_r (\tau, \phi) = \beta_r(-\tau, \phi) + \mathcal{O}(\partial^2),$ $\tilde v_I^r (\tau, \phi) = - v_I^r(-\tau, \phi) + \mathcal{O}(\partial^2),$ and so on. Note that the fields $v^{a,r}_I$ pick up an additional minus sign under KMS transformations for the reasons we already discussed in footnote \ref{foot: KMS transformation of v_I}. Under the transformations~\eqref{eq: KMS shift composite fields}, the real part in Eq. \eqref{eq: NLO Re S fluids} becomes
\begin{align}
    \text{Re} \spacy \tilde S_{\rm NLO} &= - \left. \text{Re} \spacy S_{\rm NLO}\right|_{a_{vF} \to - a_{vF}} - i \int d^4 \sigma \spacy \sqrt{\gamma_r} \spacy \beta_r \bigg[ a_{\beta\beta} (\partial_\tau \beta_r)^2 + (a_{\beta\theta} + a_{\beta\Theta}) \Theta_r \partial_\tau \beta_r \\
    & \qquad \qquad \qquad \qquad \qquad + a_{\theta\Theta} \Theta_r^2 + a_{vv} \gamma_r^{IJ} \partial_\tau v_I^r \partial_\tau v_J^r + a_{v\beta} \gamma_r^{IJ} \partial_\tau v_I^r \hat \nabla_J \beta_r + a_{\Gamma K} K^{IJ}_r K_{IJ}^r \bigg] \,. \nonumber 
\end{align}
To obtain this result, we changed the integration time variable $\tau \to - \tau$, so that all fields are evaluated at $(\tau, \phi^I)$. The imaginary part of $S_{\rm NLO}$ transforms, instead, as follows:
\begin{align}
    \begin{split}
    i \spacy \text{Im} \spacy \tilde S_{\rm NLO} ={}& i \spacy \text{Im} \spacy S_{\rm NLO}  - 2 \int d^4 \sigma \spacy \sqrt{\gamma_r} \spacy \beta_r \bigg[ c_{\beta \beta} \beta_a \partial_\tau \beta_r + c_{\beta \theta} (\theta_a  \partial_\tau \beta_r + \beta_a \Theta_r ) + c_{\theta\theta} \theta_a \Theta_r \\ 
    & \quad \qquad \qquad \qquad \qquad \qquad \qquad  \qquad \qquad \qquad \qquad  + c_{vv} v^I_a \partial_\tau v_I^r + c_{\Gamma\Gamma} \Gamma^{IJ}_a  K_{IJ}^r \bigg] \\
    & \qquad \qquad \ - i \int d^4 \sigma \spacy \sqrt{\gamma_r} \spacy \beta_r \bigg[ c_{\beta \beta} (\partial_\tau \beta_r)^2 + 2 c_{\beta \theta} \Theta_r  \partial_\tau \beta_r + c_{\theta\theta} \Theta_r^2 \\
    & \qquad \qquad \qquad \qquad \qquad  \qquad \qquad \qquad \qquad + c_{vv} \gamma_r^{IJ}\partial_\tau v_I^r \partial_\tau v_J^r + c_{\Gamma\Gamma} K^{IJ}_r  K_{IJ}^r \bigg] \, ,
    \end{split}
\end{align}
where, once again, we have the integration time variable $\tau \to - \tau$, ensuring that all fields on the right-hand side are evaluated at $(\tau, \phi^I)$. As one can see from the two equations above, a KMS transformation mixes the action's real and imaginary parts. Requiring now that $\tilde S_{\rm NLO} = S_{\rm NLO}$, we obtain the \emph{fluctuation-dissipation relations}:
\begin{align} \label{eq: FDR fluid}
    a_{\beta\beta} = -c_{\beta\beta} \, , \quad  a_{\beta\theta} = a_{\beta\Theta} = -c_{\beta\theta} \, , \quad a_{\theta\Theta} = -c_{\theta\theta} \, , \quad a_{vv} = -c_{vv} \, , \quad a_{\Gamma K} = -c_{\Gamma\Gamma} \, ,
\end{align}
along with the constraint $a_{v\beta} = 0$. Interestingly, the parity-odd coefficient $a_8$ receives no constraint from the KMS symmetry: the shifts in Eq.~\eqref{eq: KMS shift composite fields} cannot generate a real parity-odd term starting from Im$\spacy S_{\rm NLO}$, while the $r$-only imaginary term coming from Re$\spacy S_{\rm NLO}$ is a total derivative (it has the structure of a Chern--Simons form). We thus see that KMS reduces the $8+5 = 13$ arbitrary functions to a total of $5+1 = 6$ independent ones: five parity-even noise coefficients $c_{\beta\beta}, c_{\beta\theta}, c_{\theta\theta}, c_{vv}, c_{\Gamma\Gamma}$ plus the parity-odd transport coefficient $a_{vF}$. In conclusion, after imposing the fluctuation-dissipation relations \eqref{eq: FDR fluid}, Re$\spacy S_{\rm NLO}$ reduces to:
\begin{align} \label{eq: Re S NLO final} 
	\begin{split}
    \text{Re} \spacy S_{\rm NLO} &= - \int d^4 \sigma \spacy \sqrt{\gamma_r} \spacy \beta_r \bigg( c_{\beta\beta} \, \beta_a \partial_\tau \beta_r + c_{\beta\theta} (\beta_a \Theta_r + \theta_a \partial_\tau \beta_r) + c_{\theta\theta} \, \theta_a \Theta_r  \\
    & \qquad \qquad \qquad \qquad \qquad \qquad \qquad+ c_{vv} \, v^I_a \partial_\tau v_I^r + c_{\Gamma\Gamma} \, \Gamma^{IJ}_a K_{IJ}^r - a_{vF} \, \varepsilon^{IJK} v_I^a  F^r_{JK} \bigg) \, . 
    \end{split}
\end{align}
In the next subsection, we will show how this action reproduces the usual bulk and shear viscosity contributions to the stress--energy tensor and, in addition, a conservative parity-odd term (proportional to $a_{vF}$).

\subsubsection{Stress--Energy Tensor and Transport Coefficients}

\noindent Before extracting the NLO transport coefficients from the action~\eqref{eq: Re S NLO final}, it is instructive to illustrate the procedure at leading order, where the result must reproduce the perfect fluid stress--energy tensor of Section~\ref{sec: Lagrangian Formulation with 4 Fields}. The key idea is to rewrite the LO action \eqref{eq: LO SK action fluids} in terms of spacetime coordinates $x^\mu \equiv X_r^\mu(\sigma)$, rather than the internal ones $\sigma^a$, and the ``response field'' $X_a^\mu (x) \equiv X_a^\mu\left(\sigma(x)\right)$, and then read off the stress--energy tensor from the coefficient of $\partial_\mu X^a_\nu$. 

To start, we express the $a$-type building blocks appearing in the LO action~\eqref{eq: LO SK action fluids} in terms of $X_{r,a}^\mu$, and expand them up to leading order in $X_a^\mu$ to obtain
\begin{align} \label{eq: a-fields linearized text}
    \beta_a = -  \frac{1}{\beta} \spacy \partial_\tau X_r^\mu \partial_\tau X^a_\mu + \mathcal{O} (\partial^3) \, , \qquad \qquad
    \gamma^a_{IJ} = \hat \partial_{I} X_{r}^\mu \, \hat \partial_{J} X_{a\mu} + \hat \partial_{J} X_{r}^\mu \, \hat \partial_{I} X_{a\mu}+ \mathcal{O} (\partial^3)  \, ,
\end{align}
with $\beta$ defined as the leading part of $\beta_r$:
\begin{align} \label{eq: beta def}
    \beta_r = \sqrt{- \partial_\tau X_r^\mu \partial_\tau X^r_\mu} + \mathcal{O}(\partial^2) \equiv \beta +  \mathcal{O}(\partial^2) \, .
\end{align}
We then change integration variables from $\sigma^a = (\tau, \phi^I)$ to physical spacetime coordinates $x^\mu = X_r^\mu(\sigma)$. The measure of integration in the action becomes $d^4 x = d^4 \sigma \spacy\big|\!\det(\partial X_r / \partial \sigma)\big| = d^4 \sigma \spacy \sqrt{-h_r} = d^4 \sigma \spacy \sqrt{\gamma_r} \spacy \beta$. Furthermore, we can trade the comoving derivatives for spacetime ones using $\partial_\tau = \partial_\tau X^\mu_r \partial_\mu$ and $\hat \partial_I = \hat \partial_I X_r^\mu \partial_\mu$. After some algebra, the LO action becomes
\begin{align} \label{eq: S LO spacetime form}
    S_{\rm LO} = \int d^4 x \,  \left\{ - G'(\beta) \spacy u^\mu u^\nu + \frac{1}{\beta}\spacy \Delta^{\mu\nu} G (\beta) \right\}\spacy \partial_\mu X_{a\nu} + \mathcal{O} (\partial^3) \, ,
\end{align}
where we have defined 
\begin{align} \label{eq: u mu def}
    u^\mu \equiv \frac{1}{\beta} \spacy \partial_\tau X_r^\mu \, ,
\end{align}
and used the fact that $\gamma_r^{IJ} \hat{\partial}_I X^\mu_r \hat{\partial}_J X^\nu_r = \eta^{\mu\nu} + u^\mu u^\nu + O(\partial^2) \equiv \Delta^{\mu\nu} + O(\partial^2)$.\footnote{This result can be derived by noticing that $$u_\mu \hat \partial_I X^{\mu}_r = \partial_\tau  X_{\mu}^{r} \partial_I X^{\mu r} + v_I^r \partial_\tau  X_{\mu}^{r} \partial_\tau  X^{\mu r} = \partial_\tau  X_{\mu}^{r} \partial_I X^{\mu r} -  \partial_\tau  X_{\mu}^{r} \partial_\tau  X^{\mu r} h_{\tau I}^r / h_{\tau\tau}^r + \mathcal O(\partial^2) = O(\partial^2)\,.$$ This also provides a geometric interpretation of the covariant derivatives $\hat \partial_I$ as those that yield a basis of 4-vectors orthogonal to $u^\mu$ at leading order.} The term in curly brackets is the stress--energy tensor for a perfect fluid  
with energy density $\rho = - G'(\beta)$ and pressure  $p = G(\beta)/\beta$. These identifications agree with the ones in Eq.~\eqref{eq: rho p mu in Lagrangian variables}.

The form of Eq.~\eqref{eq: S LO spacetime form} makes the physical content of the LO action transparent. The equation of motion for $X_a^\mu$---obtained by varying~\eqref{eq: S LO spacetime form} and then setting $X_a^\mu = 0$---is simply $\partial_\mu T^{\mu\nu} = 0$. In other words, the conservation of the stress--energy tensor is the classical equation of motion for the response field. Note also that the classical stress--energy tensor has a natural interpretation as the leading contribution to the canonical Noether currents associated with the off-diagonal translations acting as $X_a^\mu \to X_a^\mu + \epsilon^\mu$. More generally, \emph{classical currents are always the leading contribution to the Noether currents of the off-diagonal symmetries}. 

With the LO procedure established, we can now apply the same strategy to the NLO action~\eqref{eq: Re S NLO final} to read off the transport coefficients. In addition to the linearized $a$-fields in Eq.~\eqref{eq: a-fields linearized text}, we will also need
\begin{align}
    v^a_I = \frac{1}{\beta^2} \spacy \!\big(\partial_\tau X_{r\mu}\, \hat \partial_I X_a^\mu + \hat \partial_I X_{r\mu}\,\partial_\tau X_a^\mu\big) + \mathcal{O}(\partial^3) \, , \qquad \qquad \theta_a = 2\,\Delta^{\mu\nu} \,\partial_\nu X^a_\mu + \mathcal{O}(\partial^3) \, ,
\end{align}
as well as the following derivatives of $r$-type quantities:
\begin{subequations}
\begin{align} \label{eq: dictionary text}
    \partial_\tau \beta_r &= \beta \spacy u^\mu \partial_\mu \beta + \mathcal O (\partial^2) \, , \\
    \qquad \Theta_r &= 2 \beta \spacy \partial_\mu u^\mu + \mathcal O (\partial^2) \, , \\
    \hat \partial^I X_\mu^r \partial_\tau v_I^r &=  u^\nu \partial_\nu u_\mu - \frac{1}{\beta}\spacy\Delta_{\mu\nu} \partial^\nu \beta \, ,\\
    K_r^{IJ} \hat \partial_I X_r^\mu \hat \partial_J X_r^\nu &= \beta \spacy \Delta^{\mu\alpha} \spacy \Delta^{\nu\beta} \spacy\left(\partial_\alpha u_\beta + \partial_\beta u_\alpha - \tfrac{2}{3} \eta_{\alpha\beta}\partial_\lambda u^\lambda\right) + \mathcal O (\partial^2) \equiv 2 \beta \spacy \sigma^{\mu\nu}  + \mathcal O (\partial^2) \, , \\
    F_r^{IJ} \hat \partial_I X_r^\mu \hat \partial_J X_r^\nu &= \frac{1}{\beta}\spacy\Delta^{\mu\alpha}\Delta^{\nu\beta}(\partial_\alpha u_\beta - \partial_\beta u_\alpha) + \mathcal O (\partial^2) \equiv \frac{2}{\beta}\spacy \omega^{\mu\nu}  + \mathcal O (\partial^2) \, .
\end{align}
\end{subequations}
Note the last two expressions are proportional to the shear and vorticity tensors respectively, yielding a physical interpretation to the covariant quantities $F_r^{IJ}$ and $K_r^{IJ}$. 

After the same sequence of steps that led to~\eqref{eq: S LO spacetime form}---linearizing the $a$-fields in $X_a^\mu$, changing integration variables to physical coordinates $x^\mu = X_r^\mu(\sigma)$, and converting all comoving building blocks to spacetime quantities using the results above---the real part of the NLO action takes the form
\begin{align}
    \text{Re} \, S_{\rm NLO} &= \int d^4 x \spacy \bigg\{ c_{\beta\beta} \, \beta^2 u^\mu u^\nu u^\lambda \partial_\lambda \beta  - 2 c_{\beta\theta} (\beta \Delta^{\mu\nu} u^\lambda \partial_\lambda \beta -\beta^2 u^\mu u^\nu \partial_\lambda u^\lambda )  \nonumber  \\
    & \qquad \qquad \qquad - 4 c_{\theta\theta} \, \beta  \Delta^{\mu\nu} \partial_\lambda u^\lambda - \frac{c_{vv}}{\beta} \spacy \left( u^\mu u^\lambda \partial_\lambda u^\nu - u^\mu \Delta^{\nu \lambda} \partial_\lambda \beta / \beta + \mu \leftrightarrow \nu  \right) \label{eq: Re SNLO after KMS} \\
    & \qquad \qquad \qquad  - 4 c_{\Gamma\Gamma} \, \beta \spacy \sigma^{\mu\nu}  + 2 \spacy \frac{a_{vF}}{\beta^2} \spacy \epsilon_{\alpha\beta\gamma\delta}\left( u^\mu \Delta^{\alpha\nu} +  u^\nu \Delta^{\alpha\mu} \right) \omega^{\beta\gamma}u^\delta \bigg\} \, \partial_\nu X_\mu^a + \mathcal{O}(\partial^3) \, . \nonumber
\end{align}
To rewrite the last term in the action, we used the result $\varepsilon^{IJK} =  \epsilon_{\mu\nu\lambda\rho} \hat \partial^I X_r^\mu \hat \partial^J X_r^\nu \hat \partial^K X_r^\lambda u^\rho$.

The terms in the curly brackets are the NLO correction to the stress--energy tensor. By comparing this expression with the general parametrization of the stress--energy tensor in Eq. \eqref{eq: T and J decomposition}, we conclude that 
\begin{subequations} \label{eq: matching to NLO Tmunu}
\begin{align}
    \mathcal{E} &= \rho(\beta) + c_{\beta\beta} (\beta) \,\beta^2\, u^\lambda\partial_\lambda \beta + 2\, c_{\beta\theta}(\beta) \,\beta^2\,\partial_\lambda u^\lambda \,, \\[4pt]
    \mathcal{P} &= p(\beta) -\,2\, c_{\beta\theta}(\beta) \,\beta\, u^\lambda\partial_\lambda \beta - 4\, c_{\theta\theta}(\beta) \,\beta\,\partial_\lambda u^\lambda \,, \\[4pt]
    q^\mu &= -\,\frac{c_{vv}(\beta)}{\beta}\!\Delta^{\mu\sigma} \left(u^\lambda\partial_\lambda u_\sigma - \frac{\partial_\sigma \beta}{\beta}\right) + \frac{2\, a_{vF}(\beta)}{\beta^2}\,\epsilon_{\alpha\beta\gamma\delta}\,\Delta^{\alpha\mu}\,\omega^{\beta\gamma}\,u^\delta \label{eq: Schwinger-Keldysh q mu} \,, \\[4pt]
    t^{\mu\nu} &= -\,4\, c_{\Gamma\Gamma}(\beta)\,\beta\,\sigma^{\mu\nu} \,.
\end{align}
\end{subequations}
As we discussed in Section~\ref{sec: relativistic hydro}, we can redefine the local temperature and 4-velocity as in Eqs. \eqref{eq: u gauge transformation} and \eqref{eq: T gauge transformation} to set $\cal E = \rho$ and eliminate the vector part of $q^\mu$, i.e. the first term on the right-hand side of \eqref{eq: Schwinger-Keldysh q mu}, and use conservation of $T^{\mu\nu}$ at LO to simplify the correction to $\mathcal P$. More accurately, these procedures should be thought of as redefinitions of the fundamental fields $X^\mu_{r,a}$~\cite{Glorioso:2017fpd}, as we now discuss.

\subsubsection{Field Redefinitions of $X_{a,r}^\mu$ and Frame Fixing} \label{sec: SK frame fixing}

\noindent The NLO constitutive relations \eqref{eq: Schwinger-Keldysh q mu} contain more arbitrary functions of $\beta$ than the physical transport coefficients, $\zeta$, $\eta$, and $a_8$, we expect to retain in the end. In the standard treatment of Section~\ref{sec: relativistic hydro}, this excess of arbitrary functions was removed by means of two distinct operations: \emph{(i)} the redefinitions \eqref{eq: u gauge transformation} and \eqref{eq: T gauge transformation} of the local temperature and 4-velocity, which can be used to fix a hydrodynamic frame (such as the Landau frame, where $q^\mu = 0$ and $\mathcal{E} = \rho$); and \emph{(ii)} the use of the LO conservation equations $\partial_\mu T^{\mu\nu}_{\rm LO} = 0$ to impose linear relations among the independent first-derivative covariants of $\beta$ and $u^\mu$ (e.g.\ relating the scalar $u^\mu \partial_\mu \beta$ to $\partial_\mu u^\mu$, and the vector $u^\lambda \partial_\lambda u^\mu$ to $\Delta^{\mu\lambda} \partial_\lambda \beta$). We will now show that in the Schwinger--Keldysh framework these two operations admit a unified interpretation: they both descend from field redefinitions of the fundamental fields $X_{a,r}^\mu$ at NLO. As such, they require no additional input beyond what is already built into the construction of the effective action.

The use of the LO equations of motion corresponds to a field redefinition of $X_a^\mu$ at  NLO. The most general form of such redefinition is\footnote{We are not including terms of $\mathcal{O}(X_a^2)$ because, although of the same order, they would modify the imaginary part of $S_{\rm NLO}$, which is not the focus of our discussion. Nonetheless, these additional redefinitions are important to preserve the KMS invariance of the effective action---something that can be achieved in the Landau frame~\cite{Glorioso:2017fpd}, but not necessarily in all hydrodynamic frames~\cite{Jain:2023obu}.}
\begin{align}
    \begin{split}
        X_a^\mu \to{}& X_a^\mu + \big[ F_1(\beta) u^\mu u^\nu u^\rho + F_2(\beta) u^\mu  \Delta^{\nu\rho}  + F_3(\beta)  u^\rho \Delta^{\mu\nu} \\[4pt]
        {}& \qquad\quad + F_4(\beta) u^\nu \Delta^{\mu\rho}+ F_5(\beta) \epsilon^{\mu\nu\rho\sigma} u_\sigma \big] \partial_\nu X^a_\rho \,.
    \end{split}
\end{align}
As a result, the LO part of the action changes as follows:
\begin{align} \label{eq: first delta SLO}
    \delta S_{\rm LO} 
    &= - \int d^4 x \, \partial_\mu T^{\mu\nu}_{\rm LO} 
    \big( F_1\, u_\nu u_\lambda u_\rho + F_2\, u_\nu \Delta_{\lambda\rho} 
    + F_3\, \Delta_{\nu\lambda} u_\rho + F_4\, \Delta_{\nu\rho} u_\lambda 
    + F_5\, \epsilon_{\nu\lambda\rho\sigma} u^\sigma \big)
    \partial^\lambda X^\rho_a \nonumber \\
    \begin{split}
        &= \int d^4x\, \Big\{
        F_1\big[u^\lambda\partial_\lambda\rho + (\rho+p)\,\partial_\lambda u^\lambda\big]\, u^\mu u^\nu \\[-2pt]
        & \qquad\qquad\qquad
        + F_2\big[u^\lambda\partial_\lambda\rho + (\rho+p)\,\partial_\lambda u^\lambda\big]\, \Delta^{\mu\nu} \\[2pt]
        & \qquad\qquad\qquad \qquad 
        - F_3\big[(\rho+p)\,u^\lambda\partial_\lambda u_\sigma + \partial_\sigma p\big]\, \Delta^{\sigma\mu}\, u^\nu \\[2pt]
        & \qquad\qquad\qquad\qquad \qquad 
        - F_4\big[(\rho+p)\,u^\lambda\partial_\lambda u_\sigma + \partial_\sigma p\big]\, \Delta^{\sigma\nu}\, u^\mu \\[2pt]
        & \qquad\qquad\qquad\qquad \qquad \qquad 
        - F_5\big[(\rho+p)\,u^\lambda\partial_\lambda u_\sigma + \partial_\sigma p\big]\, \epsilon^{\sigma\mu\nu\tau} u_\tau 
        \Big\}\, \partial_\mu X^a_\nu\,.
    \end{split}
\end{align}
Adding this expression to the one for Re$\spacy S_{\rm NLO}$ in Eq. \eqref{eq: Re SNLO after KMS} effectively changes the NLO part of the stress--energy tensor. If we want this to remain symmetric, we must restrict ourselves to field redefinitions such that $F_4 = F_3$ and $F_5 = 0$. Then, our field redefinition amounts to the following shifts of $\mathcal E, \mathcal P$. and $q^\mu$:
\begin{subequations} \label{eq: NLO transformations 1}
\begin{align}
    \mathcal{E} &\to \mathcal{E} + F_1\bigg[\rho' u^\lambda\partial_\lambda\beta + (\rho+p)\,\partial_\lambda u^\lambda\bigg] \, , \\
    \mathcal{P} &\to \mathcal{P} + F_2\bigg[\rho' u^\lambda\partial_\lambda\beta + (\rho+p)\,\partial_\lambda u^\lambda\bigg] \, , \\
    q^\mu &\to q^\mu - F_3 (\rho+p) \, \Delta^{\mu\sigma} \left(u^\lambda\partial_\lambda u_\sigma - \frac{\partial_\sigma \beta}{\beta} \right)  \, ,
\end{align}
\end{subequations}
where we used the thermodynamic relation $p' = - (\rho+p) /\beta$ to simplify the last equation. By a judicious choice of the functions $F_{1,2,3}$ we can set to zero some of the terms on the right-hand side of Eqs. \eqref{eq: matching to NLO Tmunu}. Before doing so, however, we should also consider a similar redefinition of $X^\mu_r$. To be precise, we are going to consider a field redefinition of the form 
\begin{align}
    X_r^\mu (\sigma) \to X_r^\mu  (\sigma)  + \xi^\mu (\sigma) \, ,
\end{align}
which induces the following infinitesimal transformations on $\beta$ and $u^\mu$:
\begin{align} \label{eq: induced delta beta delta u}
    \delta \beta = -u^\mu \partial_\tau \xi_{\mu}  \, , \qquad \qquad
    \delta u^\mu =  \frac{1}{\beta} \spacy \Delta^{\mu\nu} \, \partial_\tau \xi_{\nu} \, .
\end{align}
This induces a further change in the LO effective action \eqref{eq: S LO spacetime form}, on top of the one in Eq.~\eqref{eq: first delta SLO}. This additional change is
\begin{align}
    \delta S_{\rm LO} = \int d^4 x \bigg\{  \rho'(\beta) \, \delta\beta \, u^\mu u^\nu + p'(\beta) \, \delta\beta \, \Delta^{\mu\nu} + (\rho+p)(u^\mu \delta u^\nu + u^\nu \delta u^\mu)\bigg\} \partial_\mu X^a_\nu \, ,
\end{align}
after dropping a total derivative arising from the infinitesimal change of  coordinates $x^\mu \to x^\mu + \xi^\mu(x)$. 
Note in particular that $\delta u^\mu$ is automatically transverse, $u_\mu \delta u^\mu = 0$, as required to preserve the normalization $u^\mu u_\mu = -1$ at this order. In fact, by an appropriate choice of $\partial_\tau \xi_\mu$ we can generate arbitrary shifts $\delta u^\mu$  and $\delta \beta$. Since $\partial X^\mu_r \sim \mathcal{O}(1)$, a subleading redefinition of $X^\mu_r$ requires $\partial_\tau \xi_\mu \sim \mathcal{O}(\partial)$. The most general such redefinition takes the form
\begin{align} \label{eq: partial_tau xi_mu}
	\begin{split}
    \partial_\tau \xi_\mu ={}& u_\mu \big[ F_6(\beta) u^\nu \partial_\nu \beta + F_7(\beta) \partial_\nu u^\nu \big] +\Delta_{\mu\nu} \big[F_8(\beta) \partial^\nu \beta + F_9(\beta) u^\lambda \partial_\lambda u^\nu \big] \\
    & + F_{10}(\beta) \epsilon_{\mu\nu\lambda\sigma} \omega^{\nu\lambda}u^\sigma\, ,
    \end{split}
\end{align}
which we can integrate in $\tau$ to obtain a transformation for $X^\mu_r$ that, unlike the one for $X^\mu_a$, is generically nonlocal. Indeed, the shift in $X^\mu_r$ would be local only for the very special choice $F_6 (\beta) = F_9'(\beta) = f'(\beta), \, F_7(\beta) = F_8(\beta) = F_{10}(\beta) = 0$, in which case $\xi_\mu = f(\beta) u_\mu$. There is however no need to restrict ourselves to this local form, since the more general form Eq.~\eqref{eq: partial_tau xi_mu} still leads to a local change in the stress--energy tensor---and this is, ultimately, the observable quantity~\cite{Glorioso:2017fpd}. The transformation \eqref{eq: partial_tau xi_mu} modifies $\mathcal{E}, \mathcal{P}$ and $q^\mu$ at NLO without altering $t^{\mu\nu}$: 
\begin{subequations} \label{eq: NLO transformations 2}
\begin{align}
    \mathcal{E} &\to \mathcal{E} +  \rho' (F_6 u^\nu \partial_\nu \beta + F_7 \partial_\nu u^\nu ) \, , \\
    \mathcal{P} &\to \mathcal{P} + p' (F_6 u^\nu \partial_\nu \beta + F_7 \partial_\nu u^\nu ) \, , \\
    q^\mu &\to q^\mu + \frac{ \rho+p}{\beta} \left[ \Delta^{\mu\nu} ( F_8 \partial_\nu \beta + F_9 u^\lambda \partial_\lambda u_\nu) + F_{10}(\beta) \epsilon_{\mu\nu\lambda\sigma} \omega^{\nu\lambda}u^\sigma \right]\,  \, ,
\end{align}
\end{subequations}

The transformations \eqref{eq: NLO transformations 1} and \eqref{eq: NLO transformations 2} allow enough freedom to impose the Landau frame conditions, i.e. to set $\mathcal{E} = \rho$, $q^\mu = 0$, and demand that the part of $\mathcal{P}$ proportional to $u^\lambda \partial_\lambda \beta$ vanish. Specifically, this can be achieved by choosing:
\begin{subequations}
\begin{align} \label{eq: fluid frame fixing}
    F_2 &= \frac{2 c_{\beta \theta} \beta}{\rho'} + c_{\beta\beta} \frac{\beta^2 p'}{\rho^{\prime 2} } + F_1 \frac{p'}{\rho'} \, , \\
    F_6 &= - c_{\beta\beta} \frac{\beta^2}{\rho'} -F_1 \, , \\
    F_7 &= -2 c_{\beta\theta} \frac{\beta^2}{\rho'} - \frac{(\rho+p)}{\rho'} F_1 \, , \\
    F_8 &= - \frac{c_{vv}}{\beta (\rho+p)} - F_3\, , \\
    F_9 &= \frac{c_{vv}}{(\rho+p)} + \beta F_3 \, , \\
    F_{10} &= - \frac{2 a_{vF}}{\beta (\rho + p)} \, . 
\end{align}
\end{subequations}
In summary, in the Landau frame we have
\begin{align}
    \mathcal{E} = \rho \,, \qquad\quad \mathcal{P} = p - \zeta \, \partial_\lambda u^\lambda \,, \qquad\quad 
    q^\mu = 0 \,, \qquad\quad  t^{\mu\nu} = - \eta\, \sigma^{\mu\nu}\, ,
\end{align}
where $\zeta$ and $\eta$ are the familiar bulk and shear viscosity coefficients that are now determined by the EFT coefficients:
\begin{align} \label{eq: transport coefficients fluids}
    \zeta  &=  \beta^3 \,c_s^4 \, c_{\beta\beta}  +  4 \beta^2 \,c_s^2 \, c_{\beta\theta}  +  4 \beta\, c_{\theta\theta}\, , \qquad \quad
    \eta =  4 \beta \, c_{\Gamma\Gamma}\, .
\end{align}
These results are obtained by using once again the thermodynamic relation $p' = - (\rho+p) /\beta$, as well as $c_s^2 = p' / \rho'$.

A few comments are in order. First, the result \eqref{eq: transport coefficients fluids} maps the six KMS-surviving Wilson coefficients $\{c_{\beta\beta},\, c_{\beta\theta},\, c_{\theta\theta},\, c_{vv}, \,c_{\Gamma\Gamma},\, a_{vF}\}$ onto the two physical transport coefficients $\{\zeta,\, \eta\}$. Although the relations \eqref{eq: fluid frame fixing} do not completely fix all possible field redefinitions, the physical transport coefficients \eqref{eq: transport coefficients fluids} are invariant under the residual transformations parametrized by $F_1$ and $F_3$. Note also that the vector noise coefficient $c_{vv}$ does not appear at all in \eqref{eq: transport coefficients fluids}: this is the EFT counterpart of the fact that, on shell, the parity-even part of the heat flux \eqref{eq: Schwinger-Keldysh q mu} is proportional to the LO transverse equation of motion and therefore vanishes. 

Second, the positivity conditions on $\text{Im}\,S_{\rm NLO}$ stated in \eqref{eq: positivity NLO fluids}---themselves a consequence of unitarity---immediately imply the positivity of the bulk and shear viscosity. In particular, the expression for $\zeta$ in Eq. \eqref{eq: transport coefficients fluids} can be rewritten as the manifestly positive quadratic form
\begin{align} \label{eq: zeta positivity}
    \zeta \,=\, \beta\,
    \begin{pmatrix} \beta c_s^2 & 2 \end{pmatrix}\!\!
    \begin{pmatrix} c_{\beta\beta} & c_{\beta\theta} \\ c_{\beta\theta} & c_{\theta\theta} \end{pmatrix}\!\!
    \begin{pmatrix} \beta c_s^2 \\ 2 \end{pmatrix} \geqslant 0  \,,
\end{align}
which is non-negative because unitarity implies that $c_{\beta\beta}\geqslant 0$ and  $c_{\beta\beta} c_{\theta\theta} - c_{\beta\theta}^2 \geqslant 0$. This is the EFT incarnation of the second law of thermodynamics, which here is automatic rather than an extra assumption (compare with Section~\ref{sec: relativistic hydro}). 

Finally, we stress that we never imposed parity as a symmetry, yet every parity-violating term was removed by the frame-fixing redefinitions of $X_{a,r}^\mu$, so that the final stress--energy tensor is automatically parity-even. This is no longer the case once an additional conserved charge is present: there, the residual frame freedom is too narrow to eliminate the parity-odd vortical term, which now resides in the charge current $J^\mu$. A non-dissipative one-derivative correction---the chiral vortical effect, with a coefficient fixed by the underlying anomalies---therefore survives as a genuine, frame-independent piece of the constitutive relations~\cite{Son:2009tf}.

\subsection{Further Readings} \label{sec: further readings fluids}

\noindent Hydrodynamics plays a central role across physics, with applications ranging from the quark--gluon plasma to neutron-star astrophysics, condensed-matter transport, cosmological perturbation theory, and biological active matter. No review can do justice to this enormous range, and we make no attempt at one here. Instead, in keeping with the rest of this paper, we collect some entry points to the literature on EFT approaches to hydrodynamics---as opposed to the more traditional formulation based directly on the conservation equations---organized by theme. Other topics, applications and viewpoints on relativistic hydrodynamics can be found in countless review articles such as, for example,~\cite{Gourgoulhon:2006bn,Andersson:2006nr,Font:2008fka,Banerjee:2011tg,Jaiswal:2016hex,Santos:2022fpf,Baggioli:2022pyb,Basar:2024srd}.

\paragraph{Generalized fluids} Beyond the standard setup of an energy--momentum tensor and a $U(1)$ charge current, a number of generalized hydrodynamic frameworks have been developed that extend the EFT presented in this section. The constraints imposed by quantum anomalies on hydrodynamic transport were systematically derived in~\cite{Son:2009tf} and shortly thereafter reformulated using equilibrium partition functions in~\cite{Banerjee:2012iz,Jensen:2012jh,Jensen:2012kj}. Relativistic magnetohydrodynamics admits an elegant reformulation as the hydrodynamics of a $U(1)$ one-form (``higher-form'') global symmetry~\cite{Grozdanov:2016tdf}, with non-dissipative and dissipative EFTs for the associated Goldstones developed in~\cite{Glorioso:2018kcp,Armas:2018zbe,Vardhan:2024qdi}; the relation to conventional MHD coupled to a dynamical photon is worked out in~\cite{Hernandez:2017mch}. Spin hydrodynamics promotes the spin tensor to an independent hydrodynamic variable~\cite{Florkowski:2017ruc,Hattori:2019lfp,Hongo:2021ona,Gallegos:2022jow}, motivated by polarization measurements of $\Lambda$ hyperons in heavy-ion collisions~\cite{Becattini:2016gvu}; see~\cite{Becattini:2024uha} for a recent review of spin polarization phenomena in this context. In a separate direction, integrable many-body systems support infinitely many local conservation laws and require a generalization of hydrodynamics tailored to this enlarged set of currents, ``generalized hydrodynamics''~\cite{Castro-Alvaredo:2016cdj,Bertini:2016tmj}, with~\cite{DeNardis:2018omc} extending the framework to diffusive transport; see~\cite{Doyon:2019nhl} for a pedagogical review.
 
\paragraph{Lagrangian and group-theoretic formulations} A complementary line of work, initiated by Jackiw, Nair and collaborators, develops fluid dynamics from a Lagrangian/symplectic standpoint based on the Clebsch parametrization and its non-abelian generalizations~\cite{Jackiw:2000cd,Bistrovic:2002jx,Jackiw:2002tw,Jackiw:2004nm}, with a subsequent systematic group-theoretic incorporation of anomalies and chiral magnetic/vortical effects in~\cite{Nair:2011mk}.
 
\paragraph{Schwinger--Keldysh, KMS symmetry, and superspace} The Schwinger--Keldysh effective action for dissipative hydrodynamics introduced in Section~\ref{sec: Schwinger-Keldysh} has been refined and extended along several largely parallel lines. The Crossley--Glorioso--Liu (CGL) approach implements the constraints of unitarity and the second law as a (dynamical) KMS $\mathbb{Z}_2$ symmetry acting on doubled fields~\cite{Crossley:2015evo,Glorioso:2016gsa,Glorioso:2017fpd}, with the unitarity and causality structure of the resulting EFTs further analyzed in~\cite{Gao:2018bxz}; see~\cite{Glorioso:2018wxw,Delacretaz:2026owo} for pedagogical lectures. The viewpoint we adopted in this review is close in spirit to this approach. By contrast, the Haehl--Loganayagam--Rangamani (HLR) program provides a systematic ``eightfold'' classification of dissipative and adiabatic transport~\cite{Haehl:2015pja}, repackages the corresponding constraints as a topological BRST symmetry and an emergent thermal equivariant cohomology in superspace~\cite{Haehl:2015foa,Haehl:2015uoc,Haehl:2016pec,Haehl:2016uah}, and culminates in an explicit superspace effective action implementing the local second law via an entropy-inflow mechanism~\cite{Haehl:2018lcu}; an explicit comparison between the CGL and HLR approaches is given in~\cite{Haehl:2017zac}. A closely related thread implements the SK contour directly in superspace at the level of the effective action~\cite{Jensen:2017kzi,Jensen:2018hhx}, with the resulting classification of constitutive relations and Schwinger--Keldysh positivity constraints worked out in~\cite{Jensen:2018hse}. A holographic dual realization, obtained by gluing a Lorentzian eternal AdS black hole to its Euclidean counterpart, was developed in~\cite{deBoer:2018qqm}. Analogous SK effective actions for the transport of internal (non-spacetime) charges, built directly using the coset construction, have been developed in~\cite{Akyuz:2023lsm} and extended to non-abelian symmetries and to all orders in $\hbar\omega/T$ in~\cite{Firat:2025upx}.

\paragraph{Classical mechanics} Several structures underlying the
Schwinger--Keldysh effective action have counterparts already in the classical
mechanics of finitely many degrees of freedom. The path-integral formulation of
classical Hamiltonian dynamics of~\cite{Gozzi:1989bf}, obtained by weighting
histories so as to localize onto classical solutions, carries a hidden BRST
invariance with anticommuting ghosts---the classical avatar of the ghost fields
and BRST symmetry of the CGL effective action~\cite{Crossley:2015evo}, which
likewise implement the functional determinant of the equations of motion and
relate it to the doubled bosonic sector. The closed-time-path doubling itself
appears in the variational formulation of nonconservative dynamics
of~\cite{Galley:2012hx}, where Hamilton's principle is rendered compatible with
initial data by duplicating the trajectory and identifying the two copies at the
final time. Both dissipation and nonholonomic constraints can in turn be
captured without doubling of degrees of freedom, but rather by coupling explicitly the system to an effective action for its environment (see e.g.~\cite{Besharat:2023ylg} for a modern discussion).
 
\paragraph{Holography} Holography supplies a powerful complementary handle on strongly coupled fluids. Building on the holographic shear-viscosity computation of~\cite{Policastro:2001yc}, the celebrated KSS bound $\eta/s \geqslant 1/(4\pi)$ was conjectured in~\cite{Kovtun:2004de}; it is saturated by holographic theories with two-derivative (Einstein) gravity duals. The full nonlinear fluid/gravity correspondence, mapping long-wavelength solutions of Einstein's equations in AdS to relativistic hydrodynamic flows on the boundary, was developed in~\cite{Bhattacharyya:2008jc}; see~\cite{Rangamani:2009xk} for a pedagogical review. Motivated in part by holographic results---notably on anomaly-induced transport---the equilibrium partition-function approach~\cite{Banerjee:2012iz,Jensen:2012jh,Jensen:2012kj}, anticipated by the 2+1d analysis of~\cite{Jensen:2011xb}, reproduces and extends the constraints from the entropy-current analysis without invoking a phenomenological entropy current.
 
\paragraph{Cosmological fluids} Fluids feature prominently in cosmology, both as the matter content sourcing the expansion of the universe and as the effective description of the inhomogeneities that grow into large-scale structure. The fluid EFT introduced in this section, with its characteristic invariance under internal volume-preserving diffeomorphisms, has been applied directly to cosmological perturbation theory in~\cite{Ballesteros:2012kv}. On nonlinear scales, the long-wavelength dynamics of cold dark matter is described as a viscous effective fluid coupled to gravity, with Wilson coefficients matched to short-distance dynamics~\cite{Baumann:2010tm,Carrasco:2012cv}; subsequent developments have addressed renormalization, IR safety, the Lagrangian-space formulation, and IR resummation~\cite{Pajer:2013jj,Carrasco:2013sva,Porto:2013qua,Senatore:2014via}. This ``EFT of large-scale structure'' is now widely applied to galaxy-survey data analyses, with accessible reviews provided in~\cite{Cabass:2022avo,Ivanov:2022mrd}.
 
\paragraph{Formal aspects} Several formal aspects of the hydrodynamic gradient expansion have attracted considerable interest in recent years. The expansion is generically factorially divergent~\cite{Heller:2013fn}---as remains the case once the high symmetry of Bjorken flow is relaxed~\cite{Heller:2021oxl}---but its Borel resummation reveals an underlying hydrodynamic-attractor structure governed by short-lived (non-hydrodynamic) modes~\cite{Heller:2015dha}, with analogous resurgent behavior identified in non-conformal holographic plasmas~\cite{Buchel:2016cbj}; see~\cite{Florkowski:2017olj,Jankowski:2023fdz} for reviews. At the linearized level, the convergence radius of the dispersion-relation series is set by level-crossings between hydrodynamic and gapped modes in the complex \emph{momentum} plane~\cite{Grozdanov:2019uhi,Heller:2020uuy}, and microscopic causality has been used to derive sharp two-sided bounds on transport coefficients themselves~\cite{Heller:2022ejw,Heller:2023jtd}. Closely related is the BDNK program~\cite{Bemfica:2017wps,Kovtun:2019hdm,Bemfica:2019knx,Hoult:2020eho,Hoult:2021gnb}, which exploits the freedom in defining the fluid frame to write down first-order relativistic hydrodynamics that is causal, stable, and locally well-posed---providing a simpler alternative to the M\"uller--Israel--Stewart formulation. A Schwinger--Keldysh EFT that incorporates statistical fluctuations on top of M\"uller--Israel--Stewart-type stable causal hydrodynamics was constructed in~\cite{Jain:2023obu}. More broadly, the effort to put hydrodynamics on a systematic EFT footing has been a major driver of progress on the formal structure of real-time effective field theories more generally; see~\cite{Haehl:2024pqu} for a recent overview of Schwinger--Keldysh techniques in QFT and holography.

%% file: superfluids.tex
\section{Superfluids} \label{sec:superfluids}

\noindent So far we have focused our attention to solids and fluids, and their corresponding EFTs. This has been done for pedagogical reasons, as the phenomenology of these systems is experienced every day. In our view, it is then possible to be guided by intuition when building their EFTs. Nonetheless, from the purely field-theoretical viewpoint, the simplest medium one can consider is arguably a zero-temperature superfluid. Historically, superfluids have been characterized in terms of the phenomenon of Bose--Einstein condensation and of their vanishing viscosity. In this context, one considers a gas of weakly repulsive bosons, whose interactions are assumed to be suppressed by $a_{\rm s} \bar n^{1/3} \ll 1$, where $a_{\rm s}$ is the boson-boson scattering length and $\bar n$ their number density. The repulsion between particles gives rise to a gapless collective mode, the superfluid phonon, whose density of states rapidly vanishes at low momenta, thus forbidding the superfluid from dissipating energy by emitting excitations. This is ultimately the reason for its being inviscid. For a standard textbook treatment see, for example,~\cite{landau1980statistical,pethick2002bose}.

From a modern field-theoretical viewpoint, however, it is more useful to adopt a different characterization. The reason is that Bose--Einstein condensation is well defined only in a free theory, and as soon as there are interactions among the particles making up the superfluid, it is not clear how to generalize the idea that a nonzero fraction of them sit in a zero-momentum ``condensate": at finite coupling and at finite density, multiparticle states with definite momenta for the individual particles are not eigenstates of the Hamiltonian. Moreover, the weak coupling assumption breaks down in phenomenologically relevant instances, most notably for superfluid ${^4}{\rm He}$. The latter is characterized by a scattering length $a_{\rm s} \simeq 3$~\AA~\cite{PhysRevLett.26.735}, and a number density, $\bar n \simeq 2 \times 10^{22}$~cm$^{-3}$~\cite{abraham1970velocity}, leading to $a_{\rm s} \bar n^{1/3} \sim 1$.

To provide a more general definition, we will now follow the path that we so far only hinted at: rather than being guided by phenomenological intuition, we will rely on symmetries as our guiding principles. The application of this approach to superfluids has been pioneered in~\cite{Greiter:1989qb,Son:2002zn}, and later developed in a number of further works~\cite[e.g.,][]{Son:2005rv,Escobedo:2010uv,Nicolis:2011pv,Nicolis:2013lma,Berezhiani:2018oxf}.
Specifically, for a generic relativistic QFT with a $U(1)$ conserved charge, a superfluid state is a state $| \Psi \ra $ with both of these properties:
\begin{enumerate}
    \item It has a nonzero density for the $U(1)$ charge;
    \item It spontaneously breaks the corresponding $U(1)$ symmetry. 
\end{enumerate}
These two properties are independent, and we have examples of systems that feature neither, or just either one of them~\cite{Nicolis:2023pye}. A superfluid features both.
In particular, property 2 can be thought of as the generalization of Bose-Einstein condensation to interacting theories; property 1 then qualifies that we want the symmetry to be broken in a finite density system, rather in the Poincar\'e { (or Galilei)} invariant vacuum of the theory.

An interesting aspect of this characterization is the interplay with spacetime symmetries. Property 1 implies that Lorentz boosts are spontaneously broken, because the charge density is the zeroth component of a 4-vector operator, $J^\mu$, and so its having a nonzero expectation value necessarily breaks boosts. Perhaps less obviously, property 2 generically implies that {\em time translations} are also spontaneously broken. To see this, consider the way we usually talk about systems at finite charge density: we introduce a chemical potential $\bar\upmu$ (the meaning of the bar will be clear later) and the modified Hamiltonian $H_\upmu \equiv H - \bar\upmu Q$. We then use this modified Hamiltonian in the partition function, or, at zero temperature, we look for its ground state. But, if our $| \Psi \ra$ is an eigenstate of $H_\upmu$---in particular, its ground state---and if it is {\em not} an eigenstate of $Q$---because $Q$ is spontaneously broken---then it cannot be an eigenstate of $H$ either. That is, the time translational symmetry generated by the original Poincar\'e generator $H = P^0$ is spontaneously broken \cite{Nicolis:2011pv}.
 
Notice that we are characterizing spontaneous breaking of a given (Hermitian) generator, say, $T$, in terms of the state of the system not being an eigenstate of $T$. This is a more precise characterization than the usual one in terms of the state not being {\em annihilated} by $T$, which only refers to eigenstates with zero eigenvalue. The reason is that the physical implication of spontaneous symmetry breaking is the existence of a local operator $O(x)$ (the order parameter) whose expectation value is not invariant under the symmetry,
\begin{align}
    \la \Psi | [T, O(x)] | \Psi \ra \neq 0 \,.
\end{align}
However, it is immediate to see that if $| \Psi \ra$ is an eigenstate of $T$ this cannot happen, for any $O(x)$~\cite{Nicolis:2011pv}.

So, the Poincar\'e time-translational invariance is spontaneously broken, but there is a new definition of time-translational invariance, generated by $H_\upmu$, that is unbroken. It is in this sense that a superfluid is a stationary medium.

Having concluded our preamble, we are now in a position to construct the low-energy effective field theory for a superfluid state and its excitations. The simplest way to do so it to start by considering the implications of property 2 above for a relativistic QFT, construct the corresponding relativistic Goldstone effective theory, and consider property 1 as picking a state in this effective theory. This is mostly a technical trick, because nothing guarantees that, for a given superfluid, spontaneous symmetry breaking survives at zero density---for example, ${^4}{\rm He}$ spontaneously breaks the particle number $U(1)$ symmetry only if there are helium atoms around. However, as shown by means of the coset construction~\cite{Nicolis:2013lma}, one gets the same final result by considering properties 1 and 2 simultaneously.

Property 2 implies the existence of a field, $\psi(x)$, which transforms as a scalar field under the Poincar\'e group. As far as the spontaneously broken $U(1)$ symmetry goes, up to field redefinitions, it can be taken to transform by a constant shift,
\begin{align} \label{shift}
    U(1): \quad \psi(x) \to \psi(x) + a \, , \qquad \text{with} \qquad a= {\rm constant} \,.
\end{align}
As an explicit example, one can consider a $U(1)$ symmetry broken by the expectation value of a complex scalar field $\Phi(x)$. Then, one can conveniently parametrize the perturbations of $\Phi$ about its vacuum expectation value by a radial mode and a phase,
\begin{align} \label{eq: first time Phi}
    \Phi(x) = \frac{1}{\sqrt 2} \rho(x) e^{- i \psi(x)} \, .
\end{align}
The phase field $\psi$, which is our Goldstone field, shifts under the original $U(1)$ symmetry. 

Shift invariance requires for the field to always appear acted on by derivatives, while Lorentz invariance mandates that indices must be contracted. Consequently, to lowest order in a derivative expansion, there is only one independent invariant,
\begin{align} \label{X}
    X \equiv \sqrt {-\partial_\mu \psi \, \partial^\mu \psi } \, ,
\end{align}
(the square root is conventional, and simplifies some of the manipulations that follow,)
and so the low-energy effective Lagrangian can be a generic function $P$ of $X$ \cite[e.g.,][]{Son:2002zn},
\begin{align} \label{superfluid action}
    S = \int d^4 x \, P(X) + \mbox{higher $\partial$'s} \, . 
\end{align}

A note about the derivative expansion: if we were interested to use this effective theory to study particle physics processes in the Poincar\'e invariant vacuum, we should apply the usual power counting rules of relativistic effective theories~\cite[e.g.,][]{Rothstein:2003mp,Kaplan:2005es,Manohar:2018aog,Penco:2020kvy}. In particular, $X$ itself involves derivatives, and so higher powers of $X$ should be considered as higher derivative terms. Here however, as we will now see, we will be expanding this Lagrangian around {\em large} but constant values of $X$, and consider the dynamics of the corresponding perturbations. (Exactly as it was constructively done for solids in Section~\ref{sec:solids}.) The correct power counting rules to use in this case are that $X$ and  more in general $\partial_\mu \psi$ are of {\em zero-th} order in derivatives, while further derivatives increase the order in the usual way. So, for example, any power of $X$ is of zero-th order, but $\Box X$ is of second order. This is a consistent power counting scheme thanks to shift invariance~\cite{Son:2005rv,Son:2005ak}.
For the time being, we will ignore all higher derivative terms.

We can now impose property 1 above. In order to do so, we first compute the $U(1)$ Noether current for our effective theory,
\begin{align} \label{superfluid current}
    J^\mu = \frac{\partial{\cal L}}{\partial \, \partial_\mu \psi} = -\frac{P'(X)}{X} \partial^\mu \psi \, .
\end{align}
Then, we require that our state have a nonzero expectation value for the charge density, which means that $\psi$ must have a nonzero time-derivative. This is how  the spontaneous breaking of time translations manifest itself in the low-energy effective theory. Luckily, we can still retain a high degree of symmetry if we pick a $\psi(x)$ with constant derivatives. In particular, if we want spatial rotations to be unbroken, we can take vanishing spatial derivatives. We can then consider the background configuration  
\begin{align} \label{superfluid background}
    \bar \psi(x) = \bar \upmu \, t \, , \qquad \text{ with } \qquad \bar \upmu = {\rm constant} \, ,
\end{align}
where from now on barred quantities refer to the ground state, and, as we will soon see, the constant $\bar \upmu$ has the interpretation of the ground state's chemical potential.  Such a background configuration is not invariant under the $U(1)$ shifts generated by $Q$, or under the time translations generated by $H = P^0$, but only under the ``diagonal" transformation generated by  $H_\upmu = \bar P^0 = H - \bar\upmu Q$, as desired.

Notice that {\em any} configuration with constant $\partial_\mu \psi$ is a solution of the field equation of our effective theory, which is in fact nothing but the conservation of the $U(1)$ current,
\begin{align}
    \partial _\mu J^\mu = -\partial_\mu \bigg( \frac{ P'(X)}{X} \partial^\mu \psi \bigg) =  0 \, .
\end{align}
Our background solution above corresponds to having a nonzero, constant, purely time-like current:
\begin{align}
    \bar J \,^0 = P' (\bar \upmu)  \; , \qquad \bar J \, ^i = 0 \, .
\end{align}

We can also compute the energy--momentum tensor associated with our effective theory \eqref{superfluid action}, either as the Noether current for spacetime translations~\cite{Weinberg:1995mt}, or by minimally coupling our system to curved spacetime metric~\cite{Weinberg:1972kfs},
\begin{align}
    d^4 x \to \sqrt{-g} \, d^4 x \, , \qquad X \to \sqrt{-g^{\mu\nu} \spacy \partial_\mu \psi \partial_\nu \psi} \, .
\end{align}
The result is
\begin{align} \label{superfluid Tmn}
    T_{\mu\nu} = \frac{P'(X)}{X} \partial_\mu \psi \partial_\nu \psi + P(X) \eta_{\mu\nu} \, ,
\end{align}
so that on the background solution \eqref{superfluid background} we have 
\begin{align} \label{superfluid rho and p}
    \bar \rho = P'(\bar \upmu)  \bar \upmu - P(\bar \upmu) \, , \qquad \bar p =  P(\bar \upmu) \, .
\end{align}
as, respectively, the energy density and pressure. These obey the zero-temperature thermodynamical identities
\begin{align}  \label{thermo}
    \bar \rho + \bar p = \bar \upmu \, \bar n \, , \qquad d \bar p = \bar n \, d \bar \upmu  \,,
\end{align}
where $\bar n = \bar J^0$ is the number density,  thus confirming that  our constant $\bar \upmu$ is nothing but the chemical potential associated with the background solution~\eqref{superfluid background}.

In fact, when we consider configurations more general than the background \eqref{superfluid background}, by inspecting the functional and tensor form of the current  \eqref{superfluid current} and of the stress-tensor \eqref{superfluid Tmn}, and matching them to those appropriate for a superfluid (or perfect fluid),
\begin{align} \label{eq: superfluid current and stress energy tensor}
    J^\mu = n \, u^\mu \,, \qquad \qquad \quad T_{\mu\nu} = (\rho + p ) u_\mu u_\nu + p \, \eta_{\mu\nu} \, ,
\end{align}
we discover that the same relationships as above work, without any reference to background quantities,
\begin{align} \label{eq:Pprime}
    \upmu & = X \, ,  \qquad p = P(X) \, , \qquad  n = P'(X) \, , \qquad \rho  = n \, \upmu - p 
\end{align}
with the qualification that $\rho$, $n$, and $\upmu$ are all thermodynamic variables defined in the local rest frame of the superfluid, 
\begin{align}
    n = -J^\mu u_\mu \, , \qquad \rho = T_{\mu\nu} \, u^\mu u^\nu \, , \qquad \upmu = \partial_\mu \psi \, u^\mu \, ,
\end{align}
whose four-velocity is
\begin{align} \label{eq:superfluidu}
    u_\mu =- \frac {\partial_\mu \psi}{X} \, .
\end{align}
In particular, we see that the so far arbitrary function $P(X)$ is in fact completely determined by the equation of state: it is the pressure as a function of the chemical potential.

\subsection{Phonons} \label{sec:phonon}

\noindent One can use the above effective Lagrangian to study the dynamics of a generic superfluid to lowest order in the derivative expansion. As we just saw, the only input needed is the equation of state $p(\upmu)$, which gives us directly the function $P(X)$ entering the effective action. As an example, we can consider small perturbations about the background configuration \eqref{superfluid background},
\begin{align} \label{eq:superfluidgoldstone}
    \psi (x) = \bar \upmu \big( t + \pi(x) \big) \, ,
\end{align}
where $\pi$ can be thought of as the phonon field, and we find it useful to normalize it with units of time. Our building block~\eqref{X} becomes
\begin{align} \label{X expansion} 
    X = \bar \upmu \sqrt{1 + 2 \dot \pi + \dot \pi^2 - \big( \vec \nabla \pi\big)^2} \, ,
\end{align}
and expanding the effective action in powers of $\pi$, and neglecting total derivatives, we get the phonon action,
\begin{align} \label{eq:Sintsuperfluid}
    S \quad \to \quad S_{(2)} + S_{(3)} + S_{(4)} + \dots \, ,
\end{align}
where the subscripts on the right hand side denote the order of the term in the expansion. 

The quadratic action, which describes the propagation of free phonons, is
\begin{align} \label{eq:S2superfluid}
    S_{(2)} = \frac{\bar \rho + \bar p}{c_s^2} \int d^4 x \, \frac{1}{2} \Big[\dot \pi^2 - c_s^2 \big( \vec \nabla \pi\big)^2 \Big] \, ,
\end{align}
where $\bar \rho$ and $\bar p$ are the background energy density and pressure \eqref{superfluid rho and p}, and the sound speed $c_s$ is given by
\begin{align} \label{sf cs}
    c_s^2 = \frac{P'(X)}{P''(X) X } \bigg|_{X = \bar \upmu} = \frac{d p }{d \rho} \bigg|_{X = \bar \upmu} \, ,
\end{align}
as expected. Indeed, varying $S_{(2)}$ with respect to $\pi$ yields the wave equation,
\begin{align}
    \ddot \pi - c_s^2 \, \nabla^2 \pi = 0 \, ,
\end{align}
which describes sound waves propagating at speed $c_s$.

The higher order $S_{(n)}$'s describe interactions among phonons:
\begin{subequations} \label{eq:S3and4superfluid}
\begin{align}
    S_{(3)} & = \frac{\bar \rho + \bar p}{c_s^2} \int d^4 x \, \bigg[ \frac{g_3}{3!} \, \dot \pi^3 - \frac{1- c_s^2}{2!} \, \dot \pi \big( \vec \nabla \pi\big)^2 \bigg] \,, \label{eq:S3superfluid} \\
    S_{(4)} & = \frac{\bar \rho + \bar p}{c_s^2} \int d^4 x \, \bigg[ \frac{g_4}{4!} \, \dot \pi^4 
    + \frac{2\left(1-c_s^2\right) -g_3}{(2!)^2} \, \dot \pi^2 \big( \vec \nabla \pi\big)^2 
    +  \frac{1-c_s^2}{2 (2!)^2} \, \left( \big( \vec \nabla \pi\big)^2 \right)^2 \bigg] \,, \label{eq:S4superfluid}
\end{align}
\end{subequations}
and so on. The dimensionless couplings, $g_n$, are given by suitably normalized derivatives of $P(X)$, evaluated on the background:
\begin{align} \label{eq:gn}
    g_n \equiv \frac{P^{(n)}(\bar \upmu) \bar \upmu^{n-1} \, c_s^2 }{P'(\bar \upmu)} \,.
\end{align}
A few comments about the expanded action:
\begin{itemize}
\item
Since $P$ is nothing but the pressure, and the sound speed can be expressed as in \eqref{sf cs}, all the $g_n$'s can be expressed in terms of the sound speed and its derivatives with respect to the chemical potential. For example,
\begin{align} \label{g3}
    g_3 = \frac{1-c_s^2}{c_s^2} - \frac{d \log c_s^2}{d \log \bar \upmu} \, ,
\end{align}
and so on.

\item
At each new order in the phonon field, $n$, there is only one new independent coupling $g_n$ compared to the previous orders, and only one term weighed by that coupling. Conceptually, this is a consequence of the nonlinearly realized symmetries, which relate different orders in the expansion in powers of $\pi$, analogously to what happens for solids---see Eq.~\eqref{eq: solid cubic constraints}. In practice, this comes from the structure of our building block $X$ in Eq.~\eqref{X expansion}: at order $n$, the  $\dot \pi^n$ term will be multiplied by $P^{(n)}(\bar \upmu) \propto g_n$, whereas terms of the form $\dot \pi^{n- 2m} (\nabla \pi)^{2 m}$, with positive $m$, will be multiplied by lower order derivatives of $P$, that is, lower order $g_n$'s.

\item
For each $S_{(n)}$, including the quadratic one, we have pulled out the same constant factor,
\begin{align}
    Z \equiv \frac{\bar \rho + \bar p}{c_s^2} \, .
\end{align}
For classical computations, since the overall normalization of the action does not matter, one can simply ignore it. For perturbative quantum mechanical computations, such as an $S$-matrix element,  the rules to keep track of $Z$ are quite simple: there is a factor of $Z$ for each vertex, a factor of $1/Z$ for each propagator, and a factor of $1/\sqrt{Z}$ for each external line in a scattering amplitude. {Alternatively, one could normalize the phonon field ``canonically'', by defining $\pi_c \equiv \sqrt{Z} \pi$. This way, the factor of $Z$ disappears from the kinetic term of the latter field, and gets reshuffled into the nonlinear interaction terms. This factor $\sqrt{Z}$ then controls the strength of the phonon self-interactions; it appears to be larger than that for a normal fluid's sound modes
by a factor of $1/c_s$ (compare Eqs.~\eqref{eq:S2superfluid} and~\eqref{quadratic fluid action}), but this is because of the different structure---spatial vs.~time-derivatives---of interactions in the two cases.}

\item
The factorial prefactors in the interaction actions are those that make the combinatorics of Feynman rules easy. They are the analogs of the $1/4!$ for $\lambda \phi^4$ theory.
\end{itemize}

The expanded action can be used to compute phonon processes in perturbation theory, using standard diagrammatic rules. {These are widely used in both condensed matter and high energy physics, although they often appear in a slightly different fashion. The procedure to obtain the diagramatic rules associate to our theory is reviewed in detail in~\ref{app:feynmanrules}. At this stage, we stress that,} even though Lorentz boosts are spontaneously broken, it is consistent (and convenient) to use the so-called relativistic normalization for single phonon states,
\begin{align}
    \big\la \vec k | \vec k^{\,\prime} \big\ra = (2 \omega_{\vec k} )\, (2 \pi)^3 \delta^3\big(\vec k - \vec k^{\,\prime}\big) \, , \qquad \text{ with } \qquad \omega_{\vec k} = c_s k \, ,
\end{align}
and for the associated relativistic Feynman rules.\footnote{{The relation between the matrix element obtained with a relativistic normalization, $\mathcal{M}_{\rm r}$, and that obtained with a nonrelativistic one, $\mathcal{M}_{\rm nr}$, is $\mathcal{M}_{\rm r} = \prod_i \sqrt{2\omega_i} \mathcal{M}_{\rm nr}$, where the $\omega_i$'s are the energies of the particles involved in the external legs. Similarly, the infinitesimal phase space element for plane waves in the relativistic normalization are given by $d^3k / [(2\pi)^3 2\omega_{\vec k}]$, while those in the nonrelativistic one is $d^3k/(2\pi)^3$.}} 
As an example, consider the decay of a phonon of momentum $\vec k$ and energy $\omega$ into two phonons of momenta $\vec k_1$ and $\vec k_2$ and energies $\omega_1$ and $\omega_2$:

\begin{figure}[h!]
	\centering
	\resizebox{0.31\textwidth}{!}{
		\begin{tikzpicture}
			\draw[->, snake it, semithick] (-1,0) -- (-0.02,0);
			\draw[->, snake it, semithick] (0,0) -- (0.75,0.75);
			\draw[->, snake it, semithick] (0,0) -- (0.75,-0.75);
			
			\node at (0,0) [circle,fill,inner sep=0.8pt]{};
			
			\node[scale=0.8] at (-1.5,0.05) {$(\omega, \vec k)$};
			\node[scale=0.8] at (1.4,0.75) {$(\omega_1, \vec k_1)$};
			\node[scale=0.8] at (1.4,-0.75) {$(\omega_2, \vec k_2)$};
		\end{tikzpicture}
	}
\end{figure}

Provided a certain condition about higher derivative corrections is met \cite[e.g.,][]{maris1977phonon}, such a process is kinematically allowed. The tree level amplitude is simply {(see again~\ref{app:feynmanrules})},
\begin{align}
    i {\cal M}_{(3)} = \frac{c_s}{\sqrt{\bar \rho + \bar p}}  \left[g_3 \, \omega \omega_1 \omega_2 
    -(1-c_s^2)\big( \omega \vec k_1 \cdot \vec k_2 + \omega_1 \vec k_2 \cdot \vec k +\omega_2 \vec k \cdot \vec k_1 \big)  \right] \, ,
\end{align}
leading to the total decay rate
\begin{align}
	\begin{split}
    \Gamma & = \frac12 \int \frac{1}{2 \omega} |{\cal M}_{(3)}|^2 (2\pi)^4 \delta\big(\omega - (\omega_1 + \omega_2)\big)
    \delta^3\big(\vec k - (\vec k_1 + \vec k_2)\big)
    \frac{d^3 k_1}{(2\pi)^3 \, 2\omega_1} \frac{d^3 k_2}{(2\pi)^3 \, 2\omega_2} \\
    & = \frac{c_s^4}{960 \pi(\bar \rho + \bar p)} \left[\frac{2(1-c_s^2)}{c_s^2}  +  \frac{d \log c_s^2}{d \log \bar \upmu} \right]^2 k^5 \, ,
    \end{split}
\end{align}
where we used the expression \eqref{g3} for $g_3$.
This result is completely general, and, in particular, fully relativistic. 

For a nonrelativistic superfluid, we can approximate $\bar \rho + \bar p$ with the background mass density, $\bar \rho_m$, and express $d \log \bar \upmu$ as
\begin{align}
    d \log \bar \upmu = \frac{d \bar \upmu}{\bar \upmu} = \frac{d \bar p}{\bar \rho + \bar p} \simeq c_s^2 \, d \log \bar \rho_m \, ,
\end{align}
where we used the thermodynamic relations \eqref{thermo}, and the relationship $c_s^2 = dp / d \rho$.
Keeping the lowest order in $c_s \ll 1$, we thus get
\begin{align} \label{eq:GammaNR}
    \Gamma \simeq \frac{1}{240 \pi \, \bar \rho_m} \bigg( 1+ \frac{d \log c_s}{d \log \bar \rho_m} \bigg)^2 \, k^5 \;, \qquad 
    \qquad \qquad \mbox{(nonrelativistic limit)} \,, 
\end{align}
which matches the standard result obtained from hydrodynamical theory~\cite[e.g.,][]{maris1977phonon}.

We can also consider the opposite limit, that of an ultra-relativistic superfluid in a conformal field theory, which is expected to be the universal behavior of a superfluid at such high values of the chemical potential that the underlying microscopic theory can be approximated by its UV fixed point (assuming it has one). In this case, we can replace $\bar p$ with $\bar \rho/3$, and $c_s^2$ with $1/3$, thus getting
\begin{align}
    \Gamma \simeq \frac{1}{720 \pi \, \bar \rho} \,  k^5 \;, \qquad \qquad \qquad \mbox{(Conformal limit)} \,, 
\end{align}

One can of course use the action above to compute more complicated amplitudes and rates, such as those for generic phonon scattering processes~\cite[e.g.,][]{Caputo:2019xum}. One can also push perturbation theory to any desired order in the loop expansion. Notice however that this has to be done in a way consistent with the derivative expansion, as usual for non-renormalizable effective field theories~\cite[e.g.,][]{donoghue1994dynamics,Son:2005rv}.

\vspace{1em}

{Finally, we can use Eq.~\eqref{eq:GammaNR} to determine the criterion of validity of the EFT presented above. To do so in a way that encompasses both the relativistic case and non-relativistic one, let us temporarily step away from natural units. In particular, since the system now features an additional parameter with dimensions of speed, $c_s$, which can also be widely different from the speed of light, $c$, let us refrain from setting the latter to unity, but let us still work with $\hbar = 1$. In this units, length and mass scales are now distinguished, while time scales are related to the previous two by $\text{time} = \text{mass} \times \text{length}^2$. A nonrelativistic superfluid features two independent scales, one with dimensions of mass and one with dimensions of length. They are given by the typical mass of the microscopic constituents, $m$, and the typical inter-particle distance, $\ell$. With this at hand, the natural energy scale of the microscopic system is given by $(\ell^2 m)^{-1}$.\footnote{Recall that, in this units, $\text{energy} = (\text{time})^{-1} = (\text{mass})^{-1} \times (\text{length})^{-2}$.} Now, in order for the EFT to be weakly coupled, the width of a phonon with momentum $k$ must be much smaller than its energy, $\omega_{\vec k}$, meaning that the phonon has to be a proper, quasi-stable excitation of the medium. This means that, 
\begin{align}
	\frac{\Gamma(k)}{\omega_{\vec k}} \sim  \frac{k^4}{240 \pi \bar \rho_m c_s} \sim \frac{1}{240\pi} \left(k \ell \right)^4 \ll 1 \, ,
\end{align}
where we used the fact that, by dimensional analysis, the equilibrium mass density scales as $\bar \rho_m \sim m / \ell^3$, while the speed of sounds scales as $c_s \sim ( m \ell )^{-1}$. We also used the fact that the logaritmic derivative in Eq.~\eqref{eq:GammaNR} is typically of order unity.
Therefore, modulo obvious phase space suppressions, the EFT is valid in the regime where momenta are much smaller than the inverse inter-particle separation, $k  \ll \ell^{-1}$, as one would expect. The frequencies must instead be much smaller than the typical microscopic frequency, $\omega \ll (\ell^2 m)^{-1}$. We will get back to this in a more quantitative way in Section~\ref{sec: effective coefficients}. For a relativistic superfluid, the speed of sound is comparable to the speed of light, implying that $m 
\sim (c\ell)^{-1}$, and the expansion in both momenta and energies are controlled by the length scale alone, $k \ell \sim \omega \ell / c \ll 1$.
}

\subsection{Nonrelativistic Limit} \label{sec:NRsuperfluid}

\noindent We will now proceed to take the nonrelativistic limit of the superfluid effective action \eqref{superfluid action}, following the general strategy outlined in Section~\ref{sec: nonrelativistic limit} and what already done for solids and fluids. This will allow to make contact with the nonrelativistic treatment of superfluid phonons. The first step is to reintroduce all the powers of $c$ explicitly, so that the superfluid action becomes  
\begin{align}
	S = \int dt d^3 x  \, c \, P(X) \, , \qquad \quad X = \sqrt{\dot{\psi}^2 / c^2 - (\vec\nabla \psi)^2} \, , \qquad \quad \psi(x)  = \bar{\upmu} (t + \pi(x) )\, .
\end{align}
Note that at this stage we actually don't need to make any assumptions about the dimensions of $\psi(x)$, and therefore of $\bar{\upmu}$ and $\pi(x)$. Expanding the effective action in powers of the phonon field now yields
\begin{align}
	S =& \int dt d^3 x  \, c \, \bigg\{ P(\bar{\upmu}/c) + \bar{\upmu} P'(\bar{\upmu}/c) \left[ \frac{\dot \pi}{c} - \frac{c}{2} (\vec\nabla \pi)^2 + \frac{c}{2} \dot \pi (\vec\nabla \pi)^2 + \mathcal{O}(\pi^4) \right] \\
	&  \qquad\qquad\quad   + \frac{\bar{\upmu}^2}{2} P''(\bar{\upmu}/c) \left[ \frac{\dot{ \pi}^2}{c^2} - \dot \pi (\vec\nabla \pi)^2  + \mathcal{O}(\pi^4) \right] + \frac{\bar{\upmu}^3}{3!} P'''(\bar{\upmu}/c) \left[ \frac{\dot \pi^3}{c^3}  + \mathcal{O}(\pi^4) \right] +   \dots \bigg\} \, \notag .
\end{align}
The field $\pi$ does not have a canonical normalization because the term $\frac{1}{2} \dot {\pi}^2$ appears in the Lagrangian with a nontrivial coefficient. This can be remedied by switching to a new, canonically normalized field defined as: 
\begin{align}
	\pi_c (x) \equiv \sqrt{\frac{\bar{\upmu}^2 P''(\bar{\upmu}/c)}{c}} \, \pi(x) \, ,
\end{align}
When expressed in terms of this field, the Lagrangian now reads
\begin{align}
	\begin{split}
		S =& \int dt d^3 x  \, \bigg\{ c \, P + \frac{\dot{\pi}^2_c}{2} - \frac{c^3}{2 \bar{\upmu}} \frac{P'}{P''} \, (\vec\nabla \pi_c)^2  \\ 
		& \qquad \qquad \; + \left(\frac{c}{P''}\right)^{3/2} \left[ \left( \frac{c^2 P'}{2 \bar{\upmu}^2} - \frac{c P''}{2 \bar{\upmu}} \right)  \, \dot{\pi}_c (\vec\nabla \pi_c)^2 + \frac{P'''}{6 c^2} \, \dot \pi_c^3 \right] + \mathcal{O}(\pi_c^4)	 \bigg\} \, ,
	\end{split}
\end{align}
where we have omitted the argument of $P$ and its derivatives to streamline the notation. 
To implement the nonrelativistic limit, we want to keep the sound speed squared, 
\begin{align}
	c_s^2 \equiv \frac{c^3}{\bar{\upmu}} \frac{P'}{P''} \, ,
\end{align}
constant while we let $c^2 \to \infty$. Furthermore, demanding that the cubic interactions survive in this limit implies the following scalings
\begin{align} \label{eq: superfluid P scalings nonrelativistic}
	\frac{\bar{\upmu}}{c} P'(\bar{\upmu}/c)\sim c \, , \qquad \quad \frac{\bar{\upmu}^2}{c^2} P''(\bar{\upmu}/c) \sim c^3 \, , \qquad \quad \frac{\bar{\upmu}^3}{c^3} P'''(\bar{\upmu}/c) \sim c^5 \, .
\end{align}
It can be checked that this pattern extends to all orders, i.e. that the higher derivatives of $P$ must scale like $(\bar{\upmu}/c)^n P^{(n)} (\bar{\upmu}/c) \sim c^{2n-1}$.

To make Eqs. \eqref{eq: superfluid P scalings nonrelativistic} more precise from a dimensional viewpoint, we must introduce a quantity that remains finite in the limit $c \to \infty$ and has  dimensions of mass density, $\bar \rho_m$, so that the first two scalings become
\begin{align} \label{eq: nr superfluid introduction of rho_m}
	\frac{\bar{\upmu}}{c} P'(\bar{\upmu}/c) \equiv \bar \rho_m c \, , \qquad \qquad  \frac{\bar{\upmu}^2}{c^2} P''(\bar{\upmu}/c) = \frac{c^2}{c_s^2} \frac{\bar{\upmu}}{c} P'(\bar{\upmu}/c) = \frac{c^2}{c_s^2} \bar \rho_m c \, . 
\end{align}
Once again, our implementation of the nonrelativistic limit has naturally led to the introduction of a mass density, without the need to assume \emph{a priori} that the chemical potential $\bar{\upmu}$ should be split on physical grounds into a rest mass contribution and a nonrelativistic chemical potential, i.e. $\bar{\upmu} = m c^2 + \bar{\upmu}_{\rm nr}$. In light of Eqs. \eqref{eq: superfluid P scalings nonrelativistic} and \eqref{eq: nr superfluid introduction of rho_m}, a natural parametrization for higher derivatives of $P$ in terms of dimensionless coefficients is
\begin{align} \label{eq:lambdan}
	\frac{\bar{\upmu}^n}{c^n} P^{(n)}(\bar{\upmu}/c) \equiv \lambda_n \left(\frac{c}{c_s} \right)^{2n-2} \bar \rho_m c \, .
\end{align}
After these manipulations, the nonrelativistic action we are left with is
\begin{align} \label{eq: Snr cubic superfluid phonon}
	S_{\rm nr} = \int dt d^3x \left\{ \frac{\dot \pi_c^2 }{2} - \frac{c_s^2}{2} \big( \vec \nabla \pi_c\big)^2 + \frac{\lambda_3}{3!} \frac{\dot \pi_c^3}{\sqrt{\bar \rho_m} \, c_s} - \frac{1}{2} \frac{c_s}{\sqrt{\bar \rho_m}} \, \dot \pi_c \big( \vec \nabla \pi_c \big)^2 + \mathcal{O}(\pi_c^4) \right\} \, . 
\end{align}
This effective action can be resummed to all orders in (first derivatives of) $\pi_c$ and expressed as (see~\cite{Greiter:1989qb} for an equivalent result):
\begin{align}
	S_{\rm nr} = \int dt d^3 x \spacy \mathcal{P} (\mathcal{X}) \, , \qquad \qquad \mathcal{X} = \bar{\mathcal{X}} + \dot \pi_c -\frac{1}{2} \frac{c_s}{\sqrt{\bar{\rho}_m}} (\vec\nabla \pi_c)^2 \, ,
\end{align}
with
\begin{align} \label{eq:Xderivatives}
	\mathcal{P}' = \sqrt{\bar{\rho}_m c_s^2} \, , \qquad\quad   \mathcal{P}'' = 1 \, , \qquad\quad  \mathcal{P}^{(n)} =  \frac{\lambda_n}{(\bar \rho_m c_s^2 )^{n/2-1}} \, .
\end{align}
Note that the combination of time and spatial derivatives of $\pi_c$ appearing in $\mathcal{X}$ is invariant under Galilean boosts. This can be made more manifest by an appropriate rescaling of $\pi_c$.

\subsection{Weak Coupling} \label{sec:weak coupling}

\noindent The EFT we just presented is completely general, as it can describe {\it any} s-wave superfluid at sufficiently low energies. All the information about the details of the microscopic physics giving rise to the superfluid at hand is encoded in the specific functional form of the Lagrangian, $P(X)$. Different systems will correspond to different functions and, thus, different effective couplings, but the structure of the theory will be the same.

In this respect, it is instructive to consider an explicit weakly-coupled model whose low-energy physics is captured by the effective action \eqref{superfluid action}, and for which the $P(X)$ function can be computed explicitly. As anticipated in Eq.~\eqref{eq: first time Phi}, arguably the simplest possibility is a model with a single, self-interacting complex scalar field described by the action~\cite{Kapusta:2006pm},
\begin{align} \label{eq: action complex relativistic scalar quartic}
	S = \int d^4 x \left( - |\partial_\mu \Phi|^2 - m^2 |\Phi|^2 - \lambda |\Phi|^4 \right) \, ,
\end{align}
with $m^2 > 0$.
This action is invariant under $U(1)$ transformations, $\Phi \to e^{i \alpha} \Phi$, and the corresponding conserved Noether current is 
\begin{align}
	J_\mu = i (\Phi^* \partial_\mu \Phi - \Phi \partial_\mu \Phi^*) \, .
\end{align}
In order for this model to be in a superfluid phase, we need the $U(1)$ symmetry to be spontaneously broken and the associated charge density to be nonzero. Equivalently, both $\Phi$ and $J_0$ must have a nonzero expectation value, and this can be achieved in two ways:
\begin{enumerate}
	\item If $m^2>0$, we can turn on a chemical potential $\bar \upmu^2 > m^2$:  in this case, spontaneous symmetry breaking occurs if and only if there is a nonzero charge density~\cite{Nicolis:2023pye}.
	\item  If $m^2 <0$, the $U(1)$ symmetry is spontaneously broken even at zero density, and the spectrum already contains a massless Goldstone boson. Turning on any $\bar \upmu^2 > 0$ will yield a finite charge density.
\end{enumerate} 
It is customary to introduce a chemical potential by modifying the time derivatives in the action: $\partial_0 \Phi \to \partial_0 \Phi - i \bar \upmu \Phi$ (see~\cite{Kapusta:1981aa} for a derivation of this prescription). It is easy to see that, when $\Phi$ develops a nonzero expectation value, this is equivalent to considering a time-dependent solution of the form~\cite{Nicolis:2011pv}
\begin{align} \label{eq: vev complex scalar superfluid}
	\bar \Phi  = \frac{v}{\sqrt{2}} \, e^{- i \bar \upmu t} \, . 
\end{align}
The conceptual advantage of this second viewpoint is that it makes manifest the fact that Lorentz invariance is also spontaneously (rather than explicitly) broken by the finite-density state.

It is easy to verify that the field configuration \eqref{eq: vev complex scalar superfluid} satisfies the equation of motion provided $v^2 = (\bar \upmu^2 - m^2)/\lambda$, in which case $ \bar J_0  = \bar \upmu v^2$. Fluctuations around this solution can be parametrized~as
\begin{align}
	\Phi (x)  = \frac{1}{\sqrt{2}} \big( v+\sigma(x) \big)  e^{- i \psi (x)} \ , \qquad \quad \psi(x) =  \bar \upmu \left( t + \pi(x) \right) \, .
\end{align}
We can derive the full spectrum of this model by plugging these expressions in the action \eqref{eq: action complex relativistic scalar quartic} and focusing on the quadratic terms:
\begin{align}
	S_{(2)} = \int d^4 x \left\{ -\frac{1}{2} (\partial_\mu \sigma)^2 - \frac{1}{2} (\partial_\mu \pi_c)^2 + 2 \bar \upmu \dot \pi_c \sigma - \lambda v^2 \sigma^2 \right\} \, ,
\end{align}
where we have introduced the canonically normalized field $\pi_c \equiv \bar \upmu v \pi$. Setting to zero the determinant of the kinetic matrix in Fourier space we obtain one gapless and one gapped dispersion relation:
\begin{align} \label{eq: U(1) relativistic superfluid dispersion relations}
	\omega_{\vec k}^2 = c_s^2 k^2 \left[ 1 + {(1-c_s^2)}^2 \frac{k^2}{c_s^2 M^2} + \dots \right]  \, , \qquad \qquad \omega_{\vec k}^2 = M^2 + (2 - c_s^2) k^2  + \dots \, , 
\end{align}
where $M^2 \equiv 4 \bar \upmu^2 + 2 \lambda v^2$ and $c_s^2 \equiv 2 \lambda v^2 /M^2 = 1 - 4 \bar \upmu^2 /M^2$.  Note that nonlinear corrections to the gapless dispersion relation are suppressed by the scale $c_s M$, which for nonrelativistic sound speeds is parametrically smaller than the gap $M$. We are only showing the small-$k$ limit of these dispersion relations, but in fact one can trust the full nonlinear expressions since this model is weakly coupled and UV complete.

At energy lower than the gap $M$, we can integrate out the field $\sigma$ to obtain a low-energy effective action for the gapless mode. At tree-level, this amounts to solving the full equations of motion for $\sigma$ in a derivative expansion~\cite{Burgess:2007pt} to find
\begin{align} \label{eq: sol eom sigma}
	v + \sigma = \sqrt{\frac{X^2 - m^2}{\lambda}} + \frac{1}{2 (X^2 - m^2)} \spacy \square \spacy \sqrt{\frac{X^2 - m^2}{\lambda}} + \dots \, , \qquad \text{with} \qquad X \equiv \sqrt{- \partial_\mu \psi \partial^\mu \psi} \, , 
\end{align}
and then plugging this solution back into \eqref{eq: action complex relativistic scalar quartic} to obtain an action that only depends on $\psi$ \cite{Babichev:2018twg, Creminelli:2019kjy, Joyce:2022ydd}: 
\begin{align} \label{eq: effective action superfluid U(1) relativistic model}
	S = \int d^4 x \, \frac{m^4}{4 \lambda} \left(1 - \frac{X^2}{m^2} \right)^2 + \mbox{higher $\partial$'s} \, . 
\end{align}
This action is precisely of the form \eqref{superfluid action} for a particular choice of $P(X)$, determined by the UV completion we are considering. This example illustrates how the characteristic scale entering $P(X)$ is in general not the gap $M$, the symmetry breaking scale $v$, or the cubic root of the density $J_0^{1/3}$---all of which would depend on the background value of $\psi$. Instead, this characteristic scale is independent of $\bar \upmu$ and is fixed by the underlying microscopic physics---which, in our particular case, leaves $m$ (possibly multiplied by some powers of the dimensionless coupling $\lambda$) as the only option.

Eq. \eqref{eq: sol eom sigma} shows instead that higher derivative correction in the action \eqref{eq: effective action superfluid U(1) relativistic model} are suppressed at tree level by the combination $ 2(X^2 - m^2)$, which on the background is equal to $c_s^2 M^2$. This is because these corrections are needed to reproduce the nonlinear terms in the gapless dispersion relation \eqref{eq: U(1) relativistic superfluid dispersion relations}, which are suppressed precisely by this scale.
Note that this agrees with the standard expectation for an effective theory describing the low-energy dynamics of a weakly coupled sector. The scale suppressing higher derivative corrections, in fact, correspond to the scale at which new degrees of freedom emerge:
\begin{align} \label{eq: LambdaUV}
    \Lambda_{\rm UV} \sim c_s M \sim \sqrt{\bar\upmu^2 - m^2} \,.
\end{align}
The strong coupling scale, $\Lambda_{\rm strong}$, instead, is the scale at which the Goldstone dynamics becomes strongly coupled. This can be estimated by expanding the lowest order action~\eqref{eq: effective action superfluid U(1) relativistic model}. One can then canonically normalize the Goldstone field, and compare, for example, the cubic term with the quadratic one. In doing that one finds, schematically,
\begin{align}
    \frac{\mathcal{L}_{(3)}}{\mathcal{L}_{(2)}} \sim \frac{\sqrt{\lambda} \spacy \bar{\upmu}}{M^3} \spacy \partial \pi \sim \frac{\sqrt{\lambda}}{\bar{\upmu}^2} \spacy E^2 \,,
\end{align}
where we used the fact that, for all regimes described by our theory $M \sim \bar{\upmu}$, and that the typical amplitude of a canonically normalized field is of order of the energy involved in the process under consideration, $E$. The strong coupling scale is then $\Lambda_{\rm strong} \sim \bar{\upmu}/\lambda^{1/4}$. Comparing this with Eq.~\eqref{eq: LambdaUV} we deduce that,
\begin{align}
    \Lambda_{\rm UV} \ll \Lambda_{\rm strong} \,,
\end{align}
just like it happens when one integrates out weakly coupled gapped states, as in many particle physics contexts.

Once again, one can contrast this with what happens in the hydrodynamical theory of fluids, as discussed in Section~\ref{sec: power counting}. In that case, the cutoff of the effective theory is $\Lambda_{\rm UV} \sim n \spacy \sigma$, where $n$ is the number density and $\sigma$ the scattering cross section of the microscopic constituents. In this case, not only does $\Lambda_{\rm UV}$ depend on the coupling of the UV theory, but it vanishes in the weakly coupled limit, making the effective theory very {\it ineffective}. The reason is that in the standard hydrodynamic description of a fluid, the effective theory is not obtained by integrating out gapped degrees of freedom. Rather, it is a coarse graining over distances much larger than the mean free path. At very weak coupling, the latter becomes very large, relegating the regime of validity of the hydrodynamical description only to very large distances. 
This makes the relationship between ordinary hydrodynamics and the dynamics of superfluids somewhat subtle:
while the leading order theories look the same, their relationships to their UV completions are qualitatively different in the two instances.

\subsection{Gross--Pitaevskii Model and Bogoliubov Spectrum} \label{sec:GP}

\noindent In condensed matter applications, the scale $m$ in the previous model is usually very large (e.g., atomic mass) compared to the typical energies of the collective excitations one is interested in. For this reason, one can work with the nonrelativistic limit of the action \eqref{eq: action complex relativistic scalar quartic}. More accurately, we introduce an effective theory organized in an expansion in powers of $\omega / m$ and $k / m$, and work at lowest order in such an expansion. The relevant nonrelativistic degree of freedom $\varphi$ is related to the complex field $\Phi$ in the previous section by
\begin{align}
	\Phi = \frac{e^{-i m t}}{\sqrt{2m}} \, \varphi \ .
\end{align}
Plugging this definition in the relativistic action \eqref{eq: action complex relativistic scalar quartic} we obtain
\begin{align} \label{eq: U(1) scalar relativistic in terms of nonrelativistic field}
	S = \int dt d^3 x \left\{ i \varphi^\dag \partial_t \varphi - \frac{|\partial_i \varphi|^2}{2m} - \frac{\lambda}{4 m^2} |\varphi|^4 + \frac{| \partial_t \varphi|^2}{2m} 
	 \right\} \, ,
\end{align}
where, as we'll see in a moment, we have ordered the terms in the action according to their scaling with $k/m$. In the absence of a chemical potential, the first two terms yield the usual nonrelativistic dispersion relation $\omega_{\vec k} = k^2 / 2 m$. Imposing that their contribution to the action is of $\mathcal{O}(1)$, and using the uncertainty principle to estimate the scaling $d t d^3 x \sim \omega^{-1} k^{-3} \sim m / k^5$, we conclude that the field must scale like $\varphi \sim k^{3/2}$. According to these scalings, the quartic interaction is of $\mathcal{O}(k/m)$ while the term with two time derivatives is of $\mathcal{O}(k^2/m^2)$. It is customary to drop the last term in Eq. \eqref{eq: U(1) scalar relativistic in terms of nonrelativistic field} but keep the quartic one since, despite being an irrelevant operator according to the nonrelativistic power counting, it provides the leading interactions (the same approach underlies for instance Fermi's theory of weak interactions).

At finite density, it is once again helpful to extract the rest mass from the relativistic chemical potential, i.e. $ \bar \upmu = m + \bar \upmu_{\rm nr}$. In the nonrelativistic limit we have $ \bar \upmu_{\rm nr} \ll m $, and thus the expectation value of the current reduces to $\bar J_0 = 2 \bar \upmu_{\rm nr} \spacy m^2 /\lambda$, while the background value of $\varphi$ corresponding to the solution \eqref{eq: vev complex scalar superfluid} is
\begin{align}
	\bar \varphi = \sqrt{\bar J_0} \, e^{- i \bar \upmu_{\rm nr} t} \, .
\end{align}
In the condensed matter literature it is customary to work instead with a field $\phi \equiv e^{i \bar \upmu_{\rm nr} t} \varphi $ that has a constant expectation value $\sqrt{\bar J_0}$, and is described by the nonrelativistic action
\begin{align} \label{eq: action Gross Pitaevskii}
	S = \int dt d^3 x \left\{ i \phi^\dag \partial_t \phi - \frac{|\partial_i \phi|^2}{2m}+ \bar \upmu_{\rm nr} |\phi|^2 - \frac{\lambda}{4 m^2} |\phi|^4 \right\} \, . 
\end{align}
This action defines the celebrated \emph{Gross--Pitaevskii model}, corresponding to the mean-field description of a system of weakly repulsive bosons. For a textbook treatment see, for example,~\cite{pethick2002bose,leggett2008quantum}, while for a review dedicated to this model see~\cite[e.g.,][]{Andersen:2003qj}. (For further references, we refer the reader to Section~\ref{sec: further superfluids}.) In this model, turning on a nonrelativistic chemical potential manifestly gives rise to an instability that prompts the field $\phi$ to develop a nonzero expectation value. 

When $\bar \upmu_{\rm nr} \ll m$, the gapped mode in the relativistic model has a gap $M \simeq 2m$ that is of the same order as the cutoff of the Gross--Pitaevskii model, and thus it is outside its regime of validity. In fact, the action \eqref{eq: action Gross Pitaevskii} describes a single gapless degree of freedom with dispersion relation
\begin{align} \label{eq: Bogoliubov spectrum}
	\omega_{\vec k}^2 = c_s^2 k^2 \left( 1+ \frac{k^2}{4 m^2 c_s^2} \right) \, ,
\end{align}
with $c_s^2 = \bar \upmu_{\rm nr} / m \ll 1$. This nonlinear dispersion relation is known as \emph{Bogoliubov spectrum}, and could have also been derived by taking the limit $\bar \upmu_{\rm nr} \ll m$ of the gapless dispersion relation in Eq. \eqref{eq: U(1) relativistic superfluid dispersion relations}. This spectrum interpolates between the linear dispersion relation of phonons and the quadratic one of free particles with mass $m$ at the crossover scale $k_\star = 2 m c_s = 2 \bar \upmu_{\rm nr} $. Because this scale is well within the regime of validity of the Gross--Pitaevskii model, i.e. $k_\star \ll m$, and loop corrections are suppressed by powers of $\lambda \ll 1$, we can trust the full nonlinear structure of the Bogoliubov spectrum.

\subsection{Effective Coefficients} \label{sec: effective coefficients}

\noindent Let us now pause the formal developments for a moment. Given its simplicity, we want to use the EFT for superfluids developed so far to show how one can match it to experimental data to extract the effective coefficients. Indeed, in order to employ the EFTs to extract quantitative results, with controlled theoretical uncertainties, one must typically find a way to determine, (a) the effective coefficients appearing in the Lagrangian, and (b) the momentum scale at which the theory breaks down. While both of them can be estimated by means of simple dimensional analysis, as we did in Section~\ref{sec:phonon}, it is crucial to be able to extract them from data. This is even more so given that
many systems realized in the lab have some of their properties finely tuned, in which case na\"ive dimensional analysis might fail. Since the most common phases of matter are highly nonrelativistic, in this section we always work in that limit, unless otherwise specified.

As a paradigmatic example, we look into the superfluid phase of $^4{\rm He}$, at temperatures much smaller than the critical one, $T \ll T_c \simeq 2.17 \text{ K}$, where the system is well described by its superfluid component only~\cite[e.g.,][]{landau1980statistical,leggett2008quantum}. First, we estimate the effective parameters using dimensional analysis, to then compare them with a proper determination. At lowest order in the derivative expansion, the theory is completely determined once we assign a value to the number density, $\bar n$, to the couplings $\lambda_n$ in Eq.~\eqref{eq:lambdan}, as well as to the {\it momentum} cutoff, $\Lambda$.\footnote{When boosts are broken, the momentum and energy cutoffs don't coincide anymore. The energy cutoff is obtained from the momentum one from the dispersion relation. In this case, $\Lambda_\omega = c_s \Lambda$.} The latter corresponds to the momentum beyond which our theory ceases to be valid or, in other words, when the higher derivative corrections neglected in Eq.~\eqref{superfluid action} become as relevant as the lower order ones and the derivative expansion breaks down.

As already anticipated at the end of Section~\ref{sec:phonon}, a nonrelativistic superfluid features two independent scales:  the typical mass of its microscopical constituents and the typical inter-particle separation. For the case of $^{4}{\rm He}$, the first one is $m \simeq 6.8 \times 10^{-24} \text{ g} \simeq 3.8 \text{ GeV}$~\cite{ParticleDataGroup:2022pth}. As far as the inter-particle separation is concerned, it can be estimated from the equilibrium number density, $\bar n \simeq 2.2 \times 10^{22} \text{ cm}^{-3}$~\cite{abraham1970velocity}, as $\ell \sim \bar n^{-1/3} \simeq 3.6 \text{ \AA} \simeq 1.8 \text{ keV}^{-1}$. Once this is given, dimensional analysis suggests the following values for the effective parameters,
\begin{subequations}
	\begin{align}
		&{\rm dim}[c_s] = (\text{mass} \times \text{length})^{-1} \qquad \quad\;\;\, \Rightarrow \qquad c_s \sim (m \ell)^{-1} \sim 1.5 \times 10^{-7} c \simeq 44 \text{ m/s} \,, \\
		&{\rm dim}[\lambda_n] = 1 \qquad\qquad\qquad\qquad\qquad\;\;\; \Rightarrow \qquad \lambda_n \sim 1 \,, \\
		&{\rm dim}[\Lambda] = (\text{length})^{-1} \qquad \qquad\qquad\quad\; \Rightarrow \qquad \Lambda \sim 1/\ell \sim 0.5 \text{ keV} \,.
	\end{align}
\end{subequations}
To extract the above quantities properly from data, we use the nonrelativistic version of the thermodynamical relation~\eqref{thermo}, $d\bar p = \bar n \, d\bar\upmu_{\rm nr}$, together with the definition of the speed of sound, $c_s^2 = d\bar p / d\bar \rho_m$. These allow us to rewrite the effective coefficients in a form that is more amenable to experimental determination, i.e., in terms of derivatives of the mass density with respect to pressure. In particular, the cubic coefficient appearing in the nonrelativistic phonon action, Eq.~\eqref{eq: Snr cubic superfluid phonon}, can be rewritten as,
\begin{align}
	\lambda_3 = 1 + \bar \rho_m c_s^4 \frac{d^2 \bar \rho_m}{d\bar p^2} \,.
\end{align}
The dependence of the mass density on the pressure, up to the third derivative, has been determined experimentally~\cite{caupin2008static}\footnote{We notice that the parameter $b$ in the Appendix of~\cite{caupin2008static} has a typo in its units, which should be ${\rm MPa}^{1/2} {\rm m}^{9/2} {\rm kg}^{-3/2}$.} and, applied to our case, it returns the following values for the effective coefficients at atmospheric pressure,
\begin{align}
	c_s \simeq 247 \text{ m/s} \simeq 8.2 \times 10^{-7} \, c \,,  \qquad  \lambda_3 \simeq - 4.5 \,.
\end{align}

\begin{figure}[t]
	\centering
	\includegraphics[width=0.65\textwidth]{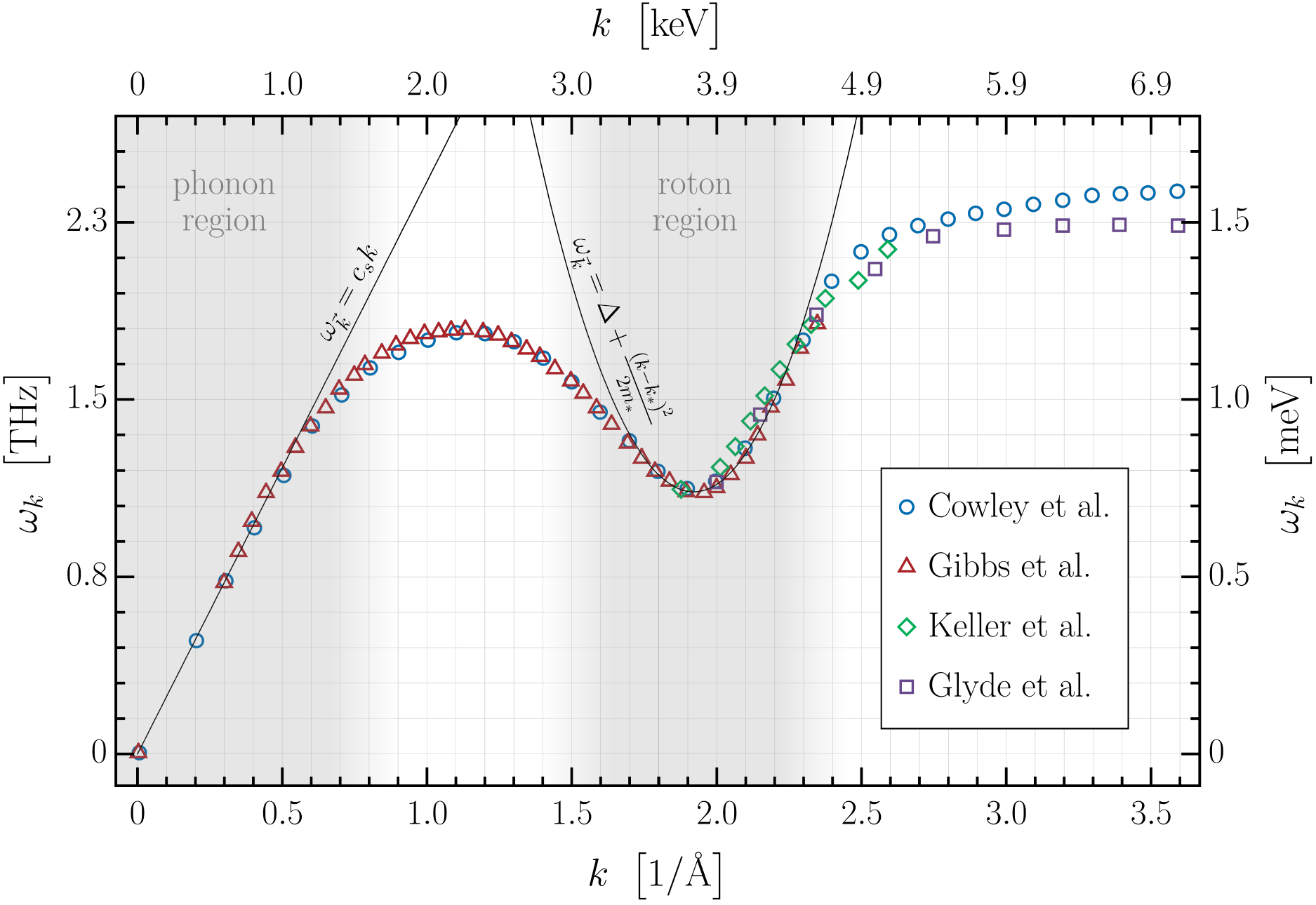}
	\caption{Measured spectrum of collective modes in superfluid ${^4}{\rm He}$, as report in~\cite{campbell2015dynamic}. In particular, blues circles are taken from~\cite{cowley1971inelastic}, red triangles from~\cite{gibbs1999collective}, green diamonds from~\cite{keller2004roton}, and purple squares from~\cite{ma1998excitations}. The phonon region corresponds to the lowest momentum one, $k \lesssim \Lambda_{\rm UV} \simeq 1.5 \text{ keV}$, where the lowest order linear dispersion relation is also plotted.} \label{fig:heliumspectrum}
\end{figure}

Finally, we need to determine the momentum scale at which the theory breaks down. One possible way to do that is by looking at the dispersion relation of the medium, and determine for which momentum it deviates appreciably from the linear behavior expected at very low momenta. In Figure~\ref{fig:heliumspectrum} we report the spectrum of $^{4}{\rm He}$ as measured by different experiments. As one can see, order one deviations from the linear behavior, $\omega_{\vec k} = c_sk$, happen for momenta around $\Lambda \simeq 0.75 \text{ \AA}^{-1} \simeq 1.5 \text{ keV}$.
As one can see from the results above, depending on the application one is interested in, na\"ive dimensional analysis should be taken with a grain of salt.


\subsection{Finite Temperature Effects and the Two-Fluid Model}

\noindent It is interesting to consider the effects of a nonzero temperature for a superfluid. On general grounds, one expects  a finite temperature superfluid to behave, at low frequencies and at long distances, as a mixture of two fluids: a superfluid like the one we have been describing so far, and a normal fluid, whose dynamics has been discussed at length in Section~\ref{fluids}. The reason is simple to understand if one starts with a zero temperature superfluid and imagines turning on a very small temperature, much below the typical energy scales of the superfluid, such as its chemical potential. In that case, only the gapless degrees of freedom---the phonons---will get thermally excited. The phonons only have derivative interactions, and so at very low temperatures their interactions are very weak. 

One can thus  think of the equilibrium state of a low-temperature superfluid as a near ideal thermal gas of phonons living in a zero temperature superfluid. Consider now very long-wavelength deviations from equilibrium, such a spatially modulated temperature. Despite the weakness of the thermal phonons' interactions, there will be a critical length scale, the thermal phonons' mean free path, above which they cannot be neglected. Above that scale, the thermal phonon gas will evolve in time like any other thermal gas, according to the laws of hydrodynamics. It particular, it will sustain sound waves, which go under the name of ``second sound." So, second sound is the compressional mode of a thermal gas of phonons, which are themselves the quanta of another compressional mode, ``first sound," that of the underlying, zero-temperature superfluid. For these reasons, first sound is usually thought of as a mass-density wave, whereas second sound is thought of as a temperature wave. For a standard treatment of this see, for example,~\cite{Landau:1941lul}.

It is useful to quantify the above considerations in terms of some relevant length and frequency scales. Considering for simplicity only the nonrelativistic limit, we have that at low temperatures ($T$) the thermal phonons' number density and typical momentum are
\begin{align}
    n_{\rm ph} \sim \frac{T^3}{c_s^3} \, , \qquad k_{\rm ph} \sim \frac{T}{c_s} \, .
\end{align}
Their typical scattering cross section can be estimated using the Feynman rules given above, and is~\cite[e.g.,][]{Nicolis:2017eqo}
\begin{align}
    \sigma \sim \frac{k_{\rm ph}^6}{c_s^2 \bar \rho_m^{\, 2}} \sim \frac{T^6}{c_s^8 \bar \rho_m^{\, 2}} \, ,
\end{align}
where, we remind the reader, $\bar \rho_m$ stands for the underlying superfluid's mass density. Their mean free path thus is
\begin{align}
    \ell_{\rm mfp} \sim \frac{1}{\sigma \, n_{\rm ph}} \sim \frac{c_s^{11} \bar \rho_m^{\, 2}}{T^9} \, .
\end{align}
Related to this, there is a mean free time
\begin{align}
    \tau = \frac{\ell_{\rm mfp}}{c_s} \, .
\end{align}

Notice that the thermal phonons' mean free path and mean free time increase very rapidly for $T$ approaching zero, as a result of the phonons' density going to zero as $T^3$ and their typical cross section going to zero as $T^6$.
Regardless, if our system is large enough to accommodate perturbations with wavelengths $\lambda$ much larger than $\ell_{\rm mfp}$, in the presence of these the thermal phonon gas will behave as a normal fluid, coupled to the underlying superfluid. There are thus a number of different kinematical regimes:
\begin{enumerate}
\item Long distances, low frequencies: $\lambda \gg \ell_{\rm mfp}$, $\omega \ll 1/\tau$. In this case we have a normal fluid of thermal phonons interacting with a zero-temperature superfluid. This is the celebrated two-fluid model, whose effective field theory we will describe shortly.
\item Intermediate distances and frequencies: $a \ll \lambda \ll \ell_{\rm mfp}$, $1/\tau \ll \omega \ll c_s/a$, where $a$ is the UV cutoff of the underlying superfluid's effective theory---for instance, the inter-particle separation for the microscopic constituents. In this case we have a zero-temperature superfluid interacting with an ideal gas of phonons. The collective modes of such a gas are invisible in this kinematical regime, and  no EFT description can be applied to it. However, the EFT of the underlying superfluid is still valid.
\item Microscopic distances, high frequencies: $\lambda \lesssim a$, $\omega \gtrsim c_s/a$. The superfluid EFT also breaks down in this regime, and one must resort to the microscopic theory.
\end{enumerate}
\begin{figure}[h]
    \centering
        \begin{tikzpicture}[baseline=-0.3em, decoration={
		      markings,
		      mark=at position 0.35 with {\arrow[line width=1pt]{>}},
		      mark=at position 0.85 with {\arrow[line width=1pt]{>}}
		}
	   ]
        \draw[gray, thick, ->] (0,0.8) -- (16.95*0.88,0.8);    
        \draw[gray, thick, <-] (0,-0.8) -- (16.95*0.88,-0.8); 
        \draw[gray] (5.25*0.88,0.65) -- (5.25*0.88,0.95);
        \draw[gray] (11.25*0.88,0.65) -- (11.25*0.88,0.95);
        \draw[gray] (5.25*0.88,-0.65) -- (5.25*0.88,-0.95);
        \draw[gray] (11.25*0.88,-0.65) -- (11.25*0.88,-0.95);
        \node[black] at (16.95*0.88+0.3,0.8) {$\lambda$};
        \node[black] at (-0.3,-0.8) {$\omega$};
        \node[black, above] at (5.25*0.88,0.98) {$a$};
        \node[black, above] at (11.25*0.88,0.98){$\ell_{\mathrm{mfp}}$};
        \node[black, below] at (5.25*0.88,-0.98) {$c_s/a$};
        \node[black, below] at (11.25*0.88,-0.98) {$c_s/\ell_{\mathrm{mfp}}$};
        \node[Maroon, align=center] at (2.625*0.88,0) {\small no EFT};
        \node[Maroon, align=center] at (8.25*0.88,0)
        {\small EFT for a superfluid\\ \small with thermal phonons};
        \node[Maroon, align=center] at (14.25*0.88,0)
        {\small EFT for first and\\ \small second sound};
    \end{tikzpicture}
\end{figure}
Notice that we are implicitly assuming that the second sound's propagation speed ($c_2$) is of the same order as the first sound's ($c_1 = c_s$). In fact, at very low temperatures their ratio is $c_2 \simeq c_1/\sqrt{3}$~\cite{Landau:1941lul}.

At higher temperatures this simple analysis gets modified at the quantitative level {\em i)} by the fact that interactions among thermal phonons become more and more important at higher and higher energies and {\em ii)} by the possible presence of gapped excitations, such the rotons of ${^4}{\rm He}$, which can also get thermally excited and thus contribute to the thermal properties of the normal fluid component. As a result, thinking of the normal fluid component as being made up of nearly free, thermally excited quasi-particles becomes less and less precise and useful. Related to this, although the two sound modes survive at high temperatures, their physical interpretation in terms of independent compressional modes of the normal and superfluid components becomes more subtle, due to their mixing. In particular the second sound's speed becomes smaller and smaller at higher temperatures. So, the simple estimates above for the relevant scales that define the different regimes of applicability of our effective theories must be modified.

Still, the qualitative picture remains the same: there is a normal, thermal fluid component interacting with a zero-temperature superfluid. Since, as we will now see, this picture can be phrased in terms of symmetries, the only phenomenon that can break it at higher temperatures is a phase transition. For liquid ${^4}{\rm He}$, this corresponds to the loss of superfluidity itself, at the so-called $\lambda$-point, at temperature $T \simeq 2.17 \text{ K}$.

\subsubsection{An Effective Theory for Finite-Temperature Superfluids} \label{sec:finiteTsuperfluids}

\noindent The two-fluid model we just described is usually implemented at the level of hydrodynamical equations of motion and conservation laws~\cite[e.g.,][]{Landau:1941lul,khalatnikov2018introduction}. Here we want to briefly describe a simple EFT approach \cite{Nicolis:2011cs}. Although, as we saw in detail, a more comprehensive analysis of the dynamics of finite temperature systems requires the Schwinger--Keldysh formulation of EFTs, for non-dissipative processes a standard EFT action suffices. In particular, since all hydrodynamical systems become less and less dissipative at longer and longer wavelengths, we can think of the EFT action that we will present here as the correct EFT description to lowest order in the derivative expansion.

We want to describe a superfluid interacting with a normal fluid. In EFT, as we saw above, a superfluid is characterized by a Lorentz-scalar $\psi(x)$, to be expanded about the equilibrium configuration \eqref{superfluid background}, and  whose dynamics are invariant by the shift symmetry \eqref{shift}. To describe a normal fluid, we adopts the Eulerian description with three fields, detailed already in Section~\ref{sec:Eulerianfluids}. This involves three Lorentz-scalars $\phi^I(x)$ ($I=1,2,3$), the comoving coordinates, to be expanded about the equilibrium configurations $\langle \phi^I \rangle \propto x^I$ (see Eq.~\eqref{perturbed fluid}). The dynamics of these scalars are invariant under internal volume-preserving diffeomorphisms,
\begin{align}
    \phi^I \to  \xi^I (\phi) \,, \qquad \det \frac{\partial \xi^I}{\partial \phi^J} = 1 \, .
\end{align}

Using these ingredients, the invariant building blocks the effective action can depend on, to lowest order in derivatives, are~\cite{Nicolis:2011cs}
\begin{align}
    X = \sqrt{-\partial_\mu \psi \partial^\mu \psi} \, , \qquad\; b = \sqrt{\det \partial_\mu \phi^I \partial^\mu \phi^J} \, , \qquad\; y = \frac{1}{3!} \frac{1}{b} \epsilon^{\mu\nu\lambda\sigma} \epsilon_{IJK}  \partial_\mu \psi  \partial_\nu \phi^I \partial_\lambda \phi^J \partial_\sigma \phi^K \, .
\end{align}
In the latter we recognize the coupling between the superfluid field, $\psi$, and the 4-velocity of the normal fluid component, as in Eq.~\eqref{eq:umu fluids}, which we refrain from calling ``$u$'' as this now indicates the 4-velocity of the superfluid component in Eq.~\eqref{eq:superfluidu}.
So, the effective action is
\begin{align} \label{finite-T superfluid}
    S = \int d^4 x \, F(X, b, y) + \mbox{higher $\partial$'s} \, ,
\end{align}
where $F$ is a generic function, which will depend on the specific superfluid under consideration. In particular, it can be thought as implementing the superfluid's equation of state. Note that this is more general than the action~\eqref{U(b,y)} for a fluid with conserved charged. There, we imposed invariance under the chemical shifts~\eqref{eq: chemical shift conserved charge}, which forbids the dependence on $X$, which is not invariant.

To make contact with the thermodynamics and hydrodynamics of the two-fluid model, it is enough to derive from the action above, via the Noether theorem,  the stress energy tensor and the  $U(1)$ current (associated with the shift symmetry). Their conservation laws are equivalent to the hydrodynamical equations, while the various thermodynamic quantities can be extracted from their form at equilibrium. Since this has now been done numerous times throughout this work, we spare the reader the details, which can be found in~\cite{Nicolis:2011cs}. A thorough study of this theory, its various limiting cases, and the corresponding thermodynamic identifications has been done in~\cite{Ballesteros:2016kdx}. For example, one finds that the chemical potential, entropy density, and temperature are related to our field theory variables and to $F$ by
\begin{align}
    \upmu = y \, , \qquad s = b \, , \qquad T = - \frac{\partial F}{\partial b} \, .
\end{align}
Note that the chemical potential in not $X$ anymore, which now receives an additional contribution coming from the relative velocities between the normal and superfluid component~\cite{Nicolis:2011cs}.

Different superfluids are characterized by different $F$ functions. However, there is a certain degree of universality when we go to very low temperatures, because, as we described above, in that case the normal fluid component is essentially made up of free thermal phonons, whose dynamics depend on a single quantity---their propagation speed $c_s$.
So, at very low temperatures, the thermal contributions to the effective action should be calculable just by knowing the sound speed of the underlying zero-temperature superfluid. Indeed, one finds \cite{Carter:1995if, Nicolis:2011cs, Kourkoulou:2022doz}
\begin{align} \label{low T sf F}
    F(X, b, y) \simeq P(X) - C \left[ \frac{b^4}{c_s(X)} \left(1 - \big( 1- c_s^2(X)\big)\frac{y^2}{X^2}\right)^2 \right]^{1/3} 
    \qquad \qquad (T \to 0) \,,
\end{align}
where $P(X)$ is the zero-temperature effective action,  $c_s(X)$ the corresponding sound speed \eqref{sf cs}, and $C$ is a dimensionless number that happens to be very close to one, $C = ( 1215/128\pi^2 )^{1/3} \simeq 0.99$.

As an application of this formalism, one can for instance study the free dynamics of excitations about equilibrium. This is done by expanding the Lagrangian to quadratic order in small perturbations of the fields. The general structure is
\begin{align} \label{quadratic finite T sf}
    {\cal L} \to  \frac{1}2 \left[ K_N \, \dot{\vec \pi}^{\, 2} - G_N {(\vec \nabla \cdot \vec \pi)}^2 \right] + \frac{1}2 \left[ K_S \, \dot \pi ^2 - G_S {(\vec \nabla \pi)}^2 \right] + M (\vec \nabla \cdot \vec \pi) \dot \pi \, ,
\end{align}
where the $\vec\pi$'s are the perturbations of the normal fluid's $\phi^I$'s, $\pi$ is the perturbation of the superlfluid's $\psi$, and the various coefficients\footnote{$K$, $G$, $M$, $N$, $S$ stand respectively for ``kinetic'', ``gradient'', ``mixing'', ``normal'', and ``superfluid''.} depend on the function $F$ and its derivatives evaluated at equilibrium~\cite{Nicolis:2011cs}. 

Such a quadratic Lagrangian describes the propagation of two independent `sound' modes, made up of different admixtures of $\pi$ and the longitudinal part of $\vec \pi$, whose propagation speeds one can  easily get by solving the corresponding secular equation.
Things are particularly clean in the low-temperature limit, where, by using in \eqref{quadratic finite T sf} the coefficients associated with the low-temperature effective action \eqref{low T sf F}, one gets the propagation speeds
\begin{align}
    c_1^2 = c^2_s(X) \, \, \qquad c_2^2 = \frac{c_1^2}{3} \, ,
\end{align}
in agreement with Landau's result~\cite{Landau:1941lul}.

\subsection{Vortex Lines}

\noindent A useful relativistic generalization of vorticity is given by $\omega^\mu =\epsilon^{\mu \nu \rho \sigma} u_\nu \partial_\rho u_\sigma$~\cite[e.g.,][]{Becattini:2015ska}. For a nonrelativistic superfluid, the temporal and spatial components reduce to $u_0 \to -1$ and $\vec u \to - \vec \nabla \psi/m$, with the latter being small, and corresponding to the standard superfluid velocity given by the gradient of the condensate wavefunction~\citep[e.g.,][]{pethick2002bose} (up to an overall, conventional minus sign). In this limit, the components of $\omega^\mu$ are dominated by their spatial part, $\vec\omega \to \vec\nabla \times \vec u$, as expected. It is simple to check that if $u_\mu$ is proportional to the derivative of a scalar function, as in Eq.~\eqref{eq:superfluidu}, then the superfluid angular velocity vanishes, $\omega^\mu = 0$.
Consequently, if the system is in a superfluid phase, its motion is irrotational. The only way to generate a nontrivial vorticity is if the scalar field $\psi$ develops a singularity such that $\partial_\mu \partial_\nu \psi \neq \partial_\nu \partial_\mu \psi$, and $\omega^\mu$ can become nonzero. Physically, this corresponds to small regions in the material where the symmetry is restored, and the system returns to its normal fluid phase. These are nothing but {\it superfluid vortices}, which are string-like objects, whose thickness is comparable with the typical inter-atomic separation, and which tend to nucleate naturally in response some angular momentum being injected into the superfluid. Indirect evidences of their existence were already obtained more than half a century ago~\cite{hall1956rotation,vinen1961detection}, later followed by direct observations~\cite[e.g.,][]{williams1974photographs,matthews1999vortices,anderson2000vortex,ku2014motion}. Indeed, superfluid vortices have attracted much attention in the standard condensed matter community, and for reviews we refer the reader to~\cite{RevModPhys.59.87,RevModPhys.80.885,RevModPhys.81.647,sonin2014tkachenko}.

A crucial property of superfluid vortices is that the circulation around them is quantized. Following the discussion of Section~\ref{sec:superfluids}, in fact, the scalar field $\psi$ can be considered as the phase of an order parameter which transforms linearly under the $U(1)$ symmetry. In the simplest instance, the order parameter has charge one, and can be taken to be $O(x) \propto e^{-i \psi(x)}$. In the standard interpretation in terms of Bose--Einstein condensation, the role of the order parameter is played precisely by the wave function, as also evident from the discussion of Sections~\ref{sec:weak coupling} and \ref{sec:GP}. From the single-valuedness of $O(x)$ follows the quantization of the superfluid circulation around a vortex line,
\begin{align} \label{eq:circulation}
	\begin{tikzpicture}[baseline=-0.3em, decoration={
			markings,
			mark=at position 0.35 with {\arrow[line width=1pt]{>}},
			mark=at position 0.85 with {\arrow[line width=1pt]{>}}
			}
		]
		\draw[gray, line width=6pt] (-0.6,-1.5) .. controls (-0.4,-0.65) ..  (0,0);
		\draw[thick,postaction=decorate] (0,0) ellipse (1.2cm and 0.5em);
		\draw[gray,fill=gray] (-0.60,-1.499) ellipse (2.815pt and 0.13em);
		\draw[gray, line width=6pt] (0,0) .. controls (0.4,0.65) ..  (0.6,1.5);
		\draw[gray,fill=gray] (0.60,1.499) ellipse (2.815pt and 0.13em);
		\draw[<->,thick] (0.43, 1.7) -- (0.77,1.7);
		\node at (0.6, 1.99) {$\sim \! \ell$};
		\node at (-0.78,-0.49) {$C$};
	\end{tikzpicture}  \qquad\quad \oint_C dx^\mu \, \partial_\mu \psi  = \Delta \psi = 2 \pi k \,, \qquad \text{ with } k \in \mathds{Z} \,,
\end{align}
where $C$ is {\it any} closed circuit which, in the relativistic setup, can enclose the vortex both in space {\it and} time.\footnote{In the relativistic context, a closed circuit in spacetime is any curve $x^\mu(\tau)$ parametrized by $\tau \in [\tau_i, \tau_f]$, such that $x^\mu(\tau_i) = x^\mu(\tau_f)$.} Moreover, $\Delta \psi$ is the change in the scalar field when moving around the circuit. The quantity $\partial_\mu \psi$ is sometimes called the superfluid four-momentum~\cite[e.g.,][]{carter1992momentum}. The standard nonrelativistic formulation is recovered when the circuit is purely spatial, in which case Eq.~\eqref{eq:circulation} corresponds to the well-known quantization of the circulation of the superfluid velocity,
\begin{align} \label{eq:NRcirculation}
	\Gamma \equiv \oint_{C_{\rm spatial}} d\vec x \cdot \vec u = \frac{2\pi k}{m} \,.
\end{align}

From this, we can estimate the energy stored within the superfluid vortex. In particular, at distances close to the vortex itself, one expects the energy to be dominated by the kinetic energy contribution. The superfluid velocity can then be estimated from Eq.~\eqref{eq:NRcirculation} as $u \sim \left(m \ell\right)^{-1}$, where $\ell$ is again the typical inter-atomic distance, comparable to the thickness of the vortex. The energy per unit vortex length associated with this is $d E_{\rm vortex} / dL \sim m u^2/\ell \sim  \frac{1}{m \ell^3}$, corresponding, not surprisingly, to the typical energy scale (per unit length) of the microscopic system~\citep[e.g.,][]{pethick2002bose}.\footnote{There are in fact logarithmic enhancements for the energy of a vortex, which are due to the behavior of the velocity field at large distances, and are thus calculable within our effective field theory \cite{Horn:2015zna}.}

\vspace{1em}

The fact that the size and energy of a vortex excitation are of typical microscopic magnitude, implies that the process of formation of a vortex cannot be described within the EFT presented in the previous sections. For example, the process of annihilation of a vortex would necessarily release to the system an energy and a momentum of the order of the cutoff, and the corresponding final state could be outside the regime of validity of the low energy EFT. From a more formal viewpoint, in the presence of a single straight vortex line, Eq.~\eqref{eq:circulation} implies that the equilibrium profile of the scalar field must be $\bar\psi(x) = \bar\upmu \, t + k \, \varphi$, with $\varphi$ being the azimuthal angle around the vortex line. It is clear that such a configuration cannot be generated by a long-wavelength modulation like that in Eq.~\eqref{eq:superfluidgoldstone}, $\psi(x) = \bar\upmu\left(t + \pi(x)\right)$. 
The azimuthal angle has a discontinuity originating from its multi-valuedness, and thus cannot be reproduced by a smooth Goldstone field $\pi(x)$.
Indeed, many of the effective descriptions of superfluid vortices in the literature are explicitly constructed from a microscopic theory, as done, for example, in~\cite{Watanabe:2013iia,hatsuda1994topological,Hatsuda:1995gh}.

Nonetheless, the very presence of a large separation between the typical vortex size and energy, and the wavelengths and frequencies of the phonons described by our EFT, suggests that, if we do not wish to describe the {\it formation or annihilation} of vortex lines, it is possible to include them as hard degrees of freedom, whose coupling to soft phonons will be completely dictated by symmetries. This is conceptually very much analogous to the coupling between soft photons and a heavy charged point-particle~\cite[e.g.,][]{Galley:2010es}, or between soft gluons and heavy quarks~\cite[e.g.,][]{Manohar:2000dt}.

Specifically, the vortex line can be described in terms of an effective {\it string} theory, where the interaction between the vortex and the long-wavelength phonons is encoded by a local action, with the usual conceptual and practical advantages that this entails. In the nonrelativistic limit, this program was carried out in~\cite{lund1991defect,Lucas:2014tka,Gubser:2014yma}, while in the relativistic formalism we adopt here, it was initiated in~\cite{Lund:1976ze,Endlich:2013dma} and completed in~\cite{Horn:2015zna}. 

\begin{figure}
	\centering
	\resizebox{0.55\textwidth}{!}{
        \begin{tikzpicture}
			\draw[->, gray] (0,0) -- (2,0);
			\draw[->, gray] (0,0) -- (0,2);
			\draw[->, gray] (0,0) -- (-1.15,-1.15);
			
			\node[scale=0.5,gray] at (2.19,0.02) {$X^2$};
			\node[scale=0.5,gray] at (-1.25,-1.28) {$X^1$};
			\node[scale=0.55,gray] at (0.04,2.14) {$X^0$};
									
			\draw[gray, fill = gray, opacity = 0.2] (-0.79, -1.0) .. controls (-0.3,-0.7) and (0.8,-0.9) .. (1.64,-0.36) .. controls (1.8, 0.5) and (1.4, 1.3) .. (1.65,1.74) .. controls (1.3,1.5) and (0.2,1.9) .. (-0.2,1.5) .. controls (-0.05,0.9) and (-1.0,-0.5) .. (-0.79,-1.0);

            \draw[line width = 0.1mm] (-0.6,-0.91) .. controls (-0.45, 0.4) and (-0.1, 0.7) .. (-0.12,1.57);

            \draw[line width = 0.1mm] (0.1,-0.81) .. controls (0.4, 0.4) and (0.25, 0.7) .. (0.33,1.68);
            
			\draw[line width = 0.1mm] (0.95,-0.65) .. controls (1.25, 0.5) and (0.75, 0.7) .. (0.87,1.67);

            \draw[line width = 0.1mm] (1.42,-0.48) .. controls (1.71, 0.5) and (1.25, 0.7) .. (1.33,1.66);

            \draw[line width = 0.1mm] (-0.82, -0.81) .. controls (-0.3,-0.5) and (0.8,-0.7) .. (1.67,-0.14);

            \draw[line width = 0.1mm] (-0.64, -0.10) .. controls (-0.3,0.2) and (0.8,-0.1) .. (1.67,0.44);

            \draw[line width = 0.1mm] (-0.34, 0.7) .. controls (-0.3,0.68) and (1.0,0.6) .. (1.59,1.04);

            \draw[line width = 0.1mm] (-0.18, 1.3) .. controls (0.2,1.55) and (1.2,1.15) .. (1.585,1.57);
			
			\draw[->, gray] plot [smooth] coordinates {(2.1,1.7) (1.7,1.6) (1.9, 1.4) (1.4,1.3)};
			
			\node[scale=0.5, gray, align=left] at (2.6, 1.73) {string \\[-0.25em] worldsheet};

            \node[scale=0.37, align=center, rotate = 7] at (0.59, 0.21) {\small $\tau = \text{fixed}$};

            \node[scale=0.37, align=center, rotate = -90] at (0.41, 1.03) {\small $\sigma = \text{fixed}$};

		\end{tikzpicture}
	}
	\caption{Parametrization of the vortex line configuration, $X^\mu(t,\sigma)$, in the $\tau = X^0 = t$ gauge. Every slice with $\tau=$~fixed represents a spatial configuration of the string at a given instant in time. Every $\sigma=$~fixed slice, instead, follows the motion of a given infinitesimal chunk of string through time.} \label{fig:vortex}
\end{figure}
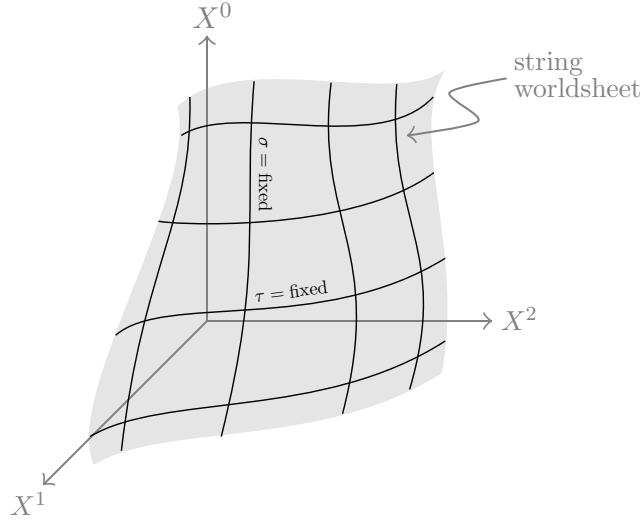

When considering only long-wavelength perturbation, the small vortex thickness can be neglected. The configuration of a vortex at a given instant in time can then be described by a 3-vector, $\vec X(t,\sigma)$, where $\sigma$ is a coordinate along the vortex. This represents the position of each element of the vortex line, labeled by $\sigma$, at each instant in time, $t$, as shown in Figure~\ref{fig:vortex}.
To formulate everything in a relativistic fashion, it is however convenient to introduce a redundancy, just like it happens for a relativistic string~\cite[e.g.,][]{Polchinski:1998rq}, and describe the vortex configuration by a 4-vector, $X^\mu(\tau,\sigma)$, where now $\tau$ is a generic worldsheet coordinate, not necessarily coinciding with physical time. The main reason to introduce this redundancy is that, by describing the vortex configuration in space and time via a 4-vector, we are setting the stage to make Lorentz invariance manifest, which is clearly of great advantage to our relativistic treatment. To go back to the more physical description in terms of $\vec X(t)$, one can choose the arbitrary worldsheet coordinate to coincide with the redundant component of the string worldsheet, i.e. set $\tau = X^0$~\cite{Horn:2015zna}. We will see this more explicitly a bit below, after presenting the effective string action.

\vspace{1em}

Now, the motion of our superfluid vortex is completely described by $X^\mu(\tau,\sigma)$, but how can one include its interaction to the phonon of the superfluid it lives in? Given the appearance of a Lorentz index in the string worldsheet, a na\"ive answer would be to construct manifestly Poincar\'e-invariant combinations of $X^\mu$ and $\partial^\mu \psi$. This is possible, but since in the presence of a vortex the scalar field is multi-valued, a construction of this type would require much care. It turns out that there is a more economic way of meeting our needs, by resorting to an alternative description of a zero-temperature superfluid. Indeed, when working in three spatial dimensions, a theory of a scalar field, which shifts under the $U(1)$ symmetry, $\psi \to \psi + a$, admits a dual description in terms of an anti-symmetric tensor---a 2-form, $\mathcal{A}_{\mu\nu}$---with a local gauge invariance~\cite{Cremmer:1973mg,buchbinder1992effective},
\begin{align} \label{eq:gauge2form}
	\mathcal{A}_{\mu \nu}(x) \to \mathcal{A}_{\mu\nu}(x)+ \partial_\mu \theta_\nu(x) - \partial_\nu \theta_\mu(x) \,.
\end{align}
In terms of this new tensor, the effective action for the bulk superfluid---i.e., the superfluid excluding the vortex line---dual to the one in Eq.~\eqref{superfluid action}, is given by~\cite{Horn:2015zna},
\begin{align}
	S_{\rm bulk} = \int d^4x \, G(Y) \,, \qquad \text{ with } \quad Y = \sqrt{- F_\mu F^\mu} \,, \quad \text{ and } \quad F^\mu = \frac{1}{2} \epsilon^{\mu \nu \rho \sigma} \partial_\nu \mathcal{A}_{\rho \sigma} \,,
\end{align}
where $F_\mu$ is the 2-form analogue of the gauge invariant electromagnetic tensor.
The conditions relating the EFT above to the thermodynamical quantities characterizing the superfluid are now, 
\begin{align} \label{eq:thermo2form}
	n = Y \,, \qquad \rho = - G(Y) \,, \qquad p = G(Y) - Y G^\prime(Y) \,, \qquad u_\mu = - \frac{F_\mu}{Y} \,.
\end{align}
As we can see, just like the quantity $X$ corresponds to the local chemical potential, $Y$ corresponds to the local number density. Finally, by comparing the expression for the superfluid four-velocity in Eqs.~\eqref{eq:superfluidu} and \eqref{eq:thermo2form}, one finds that the relation between the 2-form and the scalar field of the standard formulation is given by,
\begin{align} \label{eq:scalar2form}
	- \frac{\partial_\mu \psi}{X} = - \frac{F_\mu}{Y} \,.
\end{align}
By specializing the equation above to its different components, it is also possible to find what background value for $\mathcal{A}_{\mu\nu}$ implements the superfluid symmetry breaking pattern, $\langle \psi \rangle = \bar \upmu \, t$. In particular, one can show that a good equilibrium configuration is~\cite{Horn:2015zna},
\begin{align} \label{eq:background2form}
	\big\langle\mathcal{A}_{0i}\big\rangle = 0 \,, \qquad \big\langle\mathcal{A}_{ij}\big\rangle = - \frac{1}{3} \bar n \, \epsilon_{ijk} x^k \,.
\end{align}
The EFT presented just now is completely equivalent to the one presented around Eq.~\eqref{superfluid action}, as they describe the exact same physics.

Two comments are now in order,
\begin{itemize}
	\item Since in Eq.~\eqref{eq:scalar2form} both $\psi(x)$ and $\mathcal{A}_{\mu\nu}(x)$ appear acted on by derivatives, we deduce that the actual relationship between them is nonlocal, meaning that it involves an integration over the whole spacetime. As we will see below, this is the reason why, in this context, the 2-form formulation is more convenient: it allows to write the coupling between the vortex and the superfluid phonon in a local fashion.
	
	\item Although perfectly equivalent, the scalar field description is considerably more transparent. In fact, by just looking at the background values in Eq.~\eqref{eq:background2form}, one would hardly guess that they implement the symmetry breaking pattern of a superfluid as described previously, especially as far as the breaking of time translations is concerned. Indeed, time does not appear anywhere. It is only via the relation to the scalar field that the fact that we are indeed describing a superfluid, with the correct broken and unbroken symmetries, becomes evident.
\end{itemize}

We can now introduce small fluctuations around the equilibrium configuration. Since $\mathcal{A}_{\mu\nu}$ is anti-symmetric, it has six independent components, which can be parametrized in terms of two 3-vectors, as follows:
\begin{align}
	\mathcal{A}_{0i}(x) = \bar n \, A_i(x) \,, \qquad \mathcal{A}_{ij}(x) = \bar n \, \epsilon_{ijk} \left[ - \frac{1}{3} x^k + B^k(x) \right] \,.
\end{align}
Under the gauge transformation in Eq.~\eqref{eq:gauge2form}, $\vec A$ and $\vec B$ transform as
\be
\vec A \to \vec A +  \dot{\vec \theta} - \vec \nabla \theta_0\; , \qquad \vec B \to \vec B + \vec\nabla \times \vec \theta
\ee 
(upon redefining $\theta_\mu$ by a factor of $\bar n$). To grasp the physical meaning of these two vectors, we study their quadratic action. However, before doing that we must add a gauge fixing term. A particularly convenient one is~\cite{Horn:2015zna},
\begin{align}
	S_{\rm g.f.} = - \frac{1}{2 \xi} \int d^4x \, \left( \nabla_i \mathcal{A}^{i \mu} \right)^2 \,,
\end{align}
which is the analogue of Coulomb gauge in electromagnetism.
By expanding $S_{\rm bulk}$ and $S_{\rm g.f.}$ in small fluctuations, one finally obtains the quadratic action for $\vec A$ and $\vec B$:
\begin{align} \label{eq:SAB}
	\begin{split}
		S^{(2)}_{\rm bulk} + S^{(2)}_{\rm g.f.} ={}& \left( \bar \rho + \bar p \right) \int d^4x \bigg\{ \frac{1}{2} \big( \vec\nabla \times \vec A \big)^2 + \frac{1}{2} \dot{\vec B}^2 - \frac{c_s^2}{2} \big( \vec\nabla\cdot\vec B \big)^2 - \dot{\vec B} \cdot \big( \vec \nabla \times \vec A \big) \\
		& \qquad\qquad\qquad\;\; - \frac{1}{2\xi} \big( \vec\nabla \times \vec B \big)^2 + \frac{1}{2\xi} \big( \vec\nabla \cdot\vec A \big)^2 \bigg\} \,,
	\end{split}
\end{align}
where the speed of sound is related to thermodynamic variables as expected, $c_s^2 \equiv Y G^{\prime\prime}/G^\prime\big|_{Y = \bar n} = d\bar p/d\bar\rho$. We see that, working in the $\xi \to 0$ limit things get considerably simplified. First of all, the gauge fixing reduces to $\vec \nabla \cdot \vec A = \vec \nabla \times \vec B = 0$, i.e., $\vec A$ is purely transverse and $\vec B$ is purely longitudinal. With this at hand, one can easily show that, upon varying Eq.~\eqref{eq:SAB}, the linearized equations of motion for the two 3-vectors are,
\begin{align}
	\nabla^2 \vec A = 0 \,, \qquad \ddot{\vec B} - c_s^2 \nabla^2 \vec B = 0 \,.
\end{align}
From here we deduce that $\vec B$ represents a propagating mode, corresponding precisely to the superfluid {\it phonon}. The field $\vec A$, instead, is constrained and does not propagate any additional degree of freedom. In this sense, it is very much analogous to the Coulomb field in electromagnetism and, for this reason, it was dubbed {\it hydrophoton}~\cite{Endlich:2013dma}.

\vspace{1em}

At this point, we know how to describe a vortex line by itself, via its embedding coordinates $X^\mu(\tau,\sigma)$, and how to describe the bulk dynamics of the superfluid in term of the 2-form, $\mathcal{A}_{\mu\nu}(x)$. We are now ready to put these two things together, and finally include the interactions between the vortex and the superfluid it lives in. This happens via the action~\cite{Lund:1976ze,Zee:1994qw,Horn:2015zna}
\be
S_{\rm vortex}[X^\mu, {\cal A}_{\mu\nu}] = S_{\rm KR} + S_{\rm NG^\prime} \; ,
\ee
with
\begin{subequations} \label{eq:Sstring}
	\begin{align}
		S_{\rm KR} ={}& \lambda \int d\tau d \sigma \, \mathcal{A}_{\mu \nu}(X) \, \partial_\tau X^\mu \partial_\sigma X^\nu \,, \label{eq:SKR} \\
		S_{\rm NG^\prime} ={}& - \int d\tau d\sigma \, \sqrt{-{\rm det} g} \, \mathcal{T} \big( g^{\alpha \beta} h_{\alpha\beta} , \, Y \big) \,,
	\end{align}
\end{subequations}
where the indices $\alpha,\beta$ run over $(\tau,\sigma)$. The tensors $g_{\alpha\beta}$ and $h_{\alpha\beta}$ are so-called induced metric tensors, built out of the string embedding coordinates and the available objects that carry Lorentz indices. Specifically,
\begin{align}
	g_{\alpha \beta} = \eta_{\mu \nu} \partial_\alpha X^\mu \partial_\beta X^\nu \,, \qquad h_{\alpha \beta} = u_\mu(X) u _\nu(X) \, \partial_\alpha X^\mu \partial_\beta X^\nu \,.
\end{align}
Moreover, $g^{\alpha \beta}$ is the inverse of $g_{\alpha \beta}$.
The two terms in Eq.~\eqref{eq:Sstring} are the generalization of the so-called Kalb--Ramond and Nambu--Goto actions, well known in the description of relativistic fundamental strings~\cite[e.g.,][]{Polchinski:1998rq}, and every function of spacetime is evaluated on the vortex worldsheet, $x^\mu = X^\mu$. Finally, $\lambda$ is a coupling constant, while the $\mathcal{T}$ is a functional generalization of the string tension.

As already anticipated, both $S_{\rm KR}$ and $S_{\rm NG'}$ are {\it local} couplings between the 2-form without derivatives and the string worldsheet. Given the nonlocal relation between $\mathcal{A}_{\mu\nu}(x)$ and $\psi(x)$, the corresponding coupling between the worldsheet and the scalar field would be correspondingly nonlocal too. This is the reason why the 2-form description is, in this context, more convenient.

The criteria followed to write Eqs.~\eqref{eq:Sstring} are the usual derivative expansion, Poincar\'e invariance, gauge invariance for the 2-form, as well worldsheet reparametrization invariance, i.e. a generic relabeling of the coordinates describing the vortex configuration in time, $(\tau,\sigma) \to \big(\tau(\tau^\prime,\sigma^\prime), \sigma(\tau^\prime,\sigma^\prime)\big)$. Before moving on, it is interesting to spend a few words on this reparametrization invariance, which might be the least common in the standard condensed matter literature. First of all, the freedom to redefine freely the coordinate $\tau$ is precisely what allowed us to go from the more physical 3-vector vortex configuration, $X^i(t,\sigma)$, to the more redundant, but more convenient, 4-vector formulation, $X^\mu(\tau,\sigma)$. To go from one to another one exploits the freedom to set $\tau = t$. Reparametrization of the $\sigma$ coordinate, instead, has an interesting physical meaning: it means that excitations along the vortex are unphysical. This property is not guaranteed for any string. For example, for a guitar string, these excitations are very much physical, and correspond to compression of the string along its length. We thus stress that reparametrization invariance of the $\sigma$ coordinate along the string is an {\it assumption} about the nature of superfluid vortices \cite{Horn:2015zna}.

At this point, once the hard work of developing the EFT has been done, it is convenient to go back to the more physical formulation, by setting $\tau = t$ and thus $X^\mu = (t, \vec X)$. The action can then be expanded systematically, and the theory can be used for perturbative calculations. The criteria for this expansion are, once again, small derivatives and small amplitudes of the bulk field, $\vec A$ and $\vec B$, together with small string velocities, $\partial_t \vec X \ll c_s$, as demonstrated in~\cite{Horn:2015zna}. As an example, in the $\xi \to 0$ gauge, the first few terms in this expansion are given by~\cite{Horn:2015zna},\footnote{We point out a typo in the $\big(\vec{\nabla}\cdot\vec{B}\big)^3$ term reported in ~\cite{Horn:2015zna} (cf. Eq.~(5.2) there), which should have the opposite sign.}
\begin{subequations}
	\begin{align}
		\begin{split}
			S_{\rm bulk} \to{}& \left( \bar \rho + \bar p \right) \int d^4x \bigg[ \frac{1}{2}  \big( \vec\nabla \times \vec A \big)^2 + \frac{1}{2} \dot{\vec B}^2 - \frac{c_s^2}{2} \big( \vec\nabla\cdot\vec B \big)^2 \\
			& \qquad \qquad \qquad \; + \frac{1-c_s^2}{2} \vec \nabla \cdot \vec B \left( \dot{\vec B} - \vec \nabla \times \vec A \right)^2 + \frac{\kappa_3}{3!} \left( \vec \nabla \cdot \vec B \right)^3 + \dots \bigg] \,,
		\end{split} \\
		S_{\rm KR} \to{}& \bar n \lambda \int dtd\sigma \left[ \epsilon_{ijk} \left( - \frac{1}{3} X^k + B^k(X) \right) \partial_t X^i \partial_\sigma X^j + \vec A(X) \cdot \partial_\sigma \vec X + \dots \right] \,, \label{eq:SKRexpanded} \\
        \begin{split}
		    S_{\rm NG^\prime} \to {}& \int dt d\sigma \Big| \partial_\sigma \vec X \Big| \bigg[ -T_{(00)} + 2 T_{(01)} \vec \nabla \cdot \vec B(X) \\
            & \qquad \qquad \qquad \qquad + 2 T_{(10)} \left( \dot{\vec B}(X) - \vec \nabla \times \vec A(X) \right) \cdot \vec v_\perp + \dots \bigg] \,.
        \end{split}
	\end{align}
\end{subequations}
Here $\vec v_\perp$ is the component of the string velocity, $\vec v \equiv \partial_t \vec X$, locally perpendicular to the string itself. Moreover, the effective couplings are given by,
\begin{align}
	\kappa_n \equiv \frac{G^{(n)}(\bar n) \bar n^{n-1}}{G^\prime(\bar n)} \,, \qquad T_{(mn)} \equiv a^m b^n \frac{\partial^n}{\partial a^n} \frac{\partial^m}{\partial b^m} \mathcal{T}(a,b)\bigg|_{\rm bkg} \,,
\end{align}
both evaluated on the equilibrium configuration. From the presence of terms that are linear in the bulk fields, we understand what the role played by the vortex line is: it is a {\it source} for the phonon and hydrophoton modes. In~\ref{app:feynmanrules} we report the first few Feynman rules for the interaction between vortex lines and superfluid modes, as derived from the action above.

The physics described by this EFT is very rich, and it can be applied (possibly after generalizing it) to a variety of contexts---see Section~\ref{sec: further superfluids}. However, before moving on, let us mention without proof a few interesting properties of the theory we have just presented~\cite{Horn:2015zna}.
\begin{itemize}
	\item As one can see from Eqs.~\eqref{eq:SKR} and \eqref{eq:SKRexpanded}, the effective vortex action, and hence its equations of motion, feature a single time derivative. One must compare this to the standard behavior of the equations for a point-particle, which instead feature two time derivatives. This is at the origin of several peculiar properties of the dynamics of vortex lines as, for example, the fact that it is enough to provide the initial conditions for their {\it positions and shapes}, but {\it not} velocities, to completely determine their motion.
	
	\item The effective coupling $\lambda$ can be related to the circulation of the superfluid around the vortex, $\lambda = \left(\bar \rho + \bar p\right) \Gamma / \bar n$. As discussed in Eqs.~\eqref{eq:circulation} and \eqref{eq:NRcirculation}, in the microscopic theory the circulation is quantized. At long distances, this manifest itself in the fact that $\lambda$ does not get renormalized at any loop order.
	
	\item The other effective couplings of the EFT do get renormalized. Even more, they acquire a running with momentum/distance already at the {\it classical} level, i.e. without computing any loop corrections in the bulk fields.
\end{itemize}

\begin{figure}[t]
	\centering
	\includegraphics[width=0.5\textwidth]{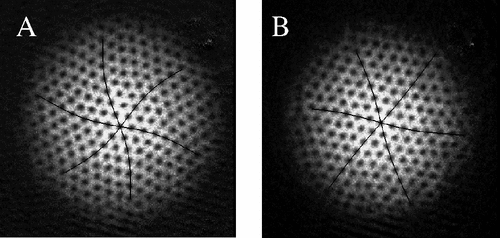}
	\caption{Experimental image of a vortex lattice featuring a Tkachenko mode (black lines), as reported in~\cite{coddington2003observation}. The two panels correspond to the same lattice, imaged a two different instant of times after the excitation of the mode.} \label{fig:tkachenko}
\end{figure}

We conclude this section by mentioning an interesting connection between the physics of superfluid vortices and that of solids. Specifically, in response to an external rotation of the superfluid, vortices are spontaneously created. When two or more vortices are present in the system, they can interact with one another through the exchange of bulk modes, phonons and hydrophotons, in a way that is attractive when the circulations have opposite sign, and repulsive otherwise. At high enough total angular momentum, the population of vortices is sufficient large that the mutual interaction allows them to organize themselves in a triangular lattice, essentially forming a solid structure within the superfluid, as also observed experimentally~\cite[e.g.,][]{abo2001observation}. The interplay between the superfluid phonon and the lattice vibrational modes give rise to a gapless excitation corresponding to an elastic mode in the direction transverse to the vortices, and characterized by a quadratic dispersion relation, $\omega_{\vec k} \propto k^2$ --- the so-called the Tkachenko mode~\cite[e.g.,][]{tkachenko1966vortex,tkachenko1966stability,baym2003tkachenko}. These excitations were first observed experimentally in~\cite{coddington2003observation}, as also shown in Figure~\ref{fig:tkachenko}.

An EFT for the systematic description of superfluid vortex lattices and their excitations has been developed in~\cite{Moroz:2018noc,Du:2022xys}. The authors use this EFT to compute, for example, transport coefficients in the vortex lattice, as well as the spectrum and decay rate of the Tkachenko modes. Their treatment share many similarities with the one presented here, with the difference that they work directly in the nonrelativistic limit, i.e. imposing Galilei rather than Lorentz invariance, and limit themselves to a two-dimensional system. In this case, the dual description of a superfluid is not given by a 2-form, but rather by a gauge field.

\subsection{Rotons} \label{sec:rotons}

\noindent The degrees of freedom we discussed so far---phonons and vortex lines---are universal: since they are mandated by the symmetries of the system, they are present in the spectrum of any superfluid. In this section we want to discuss the so-called {rotons}. Contrary to the two previous cases, not all superfluids feature rotons among their collective excitations, but they have been observed in trapped gases with weak dipolar intractions~\cite[e.g.,][]{blakie2012roton,chomaz2018observation} as well as, most notably, in superfluid ${^4}{\rm He}$~\cite[e.g.,][]{cowley1971inelastic,ma1998excitations,gibbs1999collective,keller2004roton,campbell2015dynamic}. The experimental dispersion relation of the latter is reported in Figure~\ref{fig:heliumspectrum}. As one can see, besides the phonon region at low momenta, the dispersion relation also shows a gapped minimum around a finite value $k_*$ of the momentum, 
\begin{align} \label{eq:omegap}
	\omega_{\vec k} = \Delta + \frac{\left(k - k_* \right)^2}{2m_*} + \dots \,,
\end{align}
where the dots stand for higher powers of $\left(k - k_*\right)$. Dimensional analysis tells us that, in terms of the typical microscopic mass and length scales, $m$ and $\ell$, the parameters above are roughly given by $\Delta \sim 1/(m \ell^2)$, $k_* \sim 1/\ell$, and $m_* \sim m$.
Collective excitations with energy around this minimum are exactly stable \cite{Nicolis:2017eqo}, and correspond to what one labels as rotons.

It is well known that, when present, rotons play a crucial role in several low temperature properties of the material at hand, such as specific heat and viscosity coefficients~\cite[e.g.,][]{landau1949theoryA,landau1949theoryB}. Remarkably, in superfluid ${^4}{\rm He}$, this is true even at temperatures as low as $T \simeq 1 \text{ K}$, despite the rotons' gap being large, $\Delta \simeq 0.75 \text{ meV} \simeq 9.1 \text{ K}$. This is due to the fact that, contrary to gapped excitations at zero momentum, the finite value of their typical momentum endows them with a large phase space, $d^3 k \sim k_*^2 \, dk$. It is then desirable to have a theoretical understanding of these degrees of freedom. The first works in this direction date back to Landau~\cite{Landau:1941lul,Landau:1947mij} and Feynman~\cite{feynman1954atomic}, and have been followed by several studies (for a brief review, see the Introduction of~\cite{glyde1993role}).

The typical energies and momenta of roton excitations are comparable to the typical microscopic energies and momenta characterizing the superfluid. Therefore, as it happens for vortex lines, we expect to be unable to describe their creation or annihilation within the EFT valid for phonons. However, since rotons with momenta near $k_*$ are stable, they can be considered as hard degrees of freedom interacting with soft phonons, again, just like vortex lines. Specifically, they can be described via an effective {\it point-particle} theory, as done in~\cite{Nicolis:2017eqo}. Physically, this corresponds to considering the limit where the roton wave packet is much more localized than the typical phonon wavelength. Since the most notable systems which feature rotons are nonrelativistic, we will restrict our discussion to this limit.

To write down the point-particle theory corresponding to rotons, we first need to determine their quantum numbers. Since phonons and rotons lie on the same dispersion relation, which connects them smoothly, we assume that they carry the same quantum numbers. In particular, our medium breaks boost, time translations and the internal $U(1)$ symmetry, and thus only preserves spatial translations and rotation. As usual, to characterize the quantum numbers of a given single particle state with momentum $\vec k$, one looks at how this transforms under the group left unbroken by the presence of the particle (the so-called ``little group''). In this case, this corresponds only to rotations around $\vec k$, and the corresponding quantum number is nothing but the particle's helicity. Now, since phonons in a superfluid are purely compressional modes, they carry zero helicity and, therefore, they are labeled by their 3-momentum alone, with no other degree of freedom. The same is then assumed to be true for rotons.

To build the lowest order action to describe a point-particle roton in a superfluid, we start by noticing that, following Eq.~\eqref{eq:omegap}, its Hamiltonian\footnote{Note that, in this section, when talking about ``Hamiltonian'' and ``Lagrangian'' we refer to the proper ``Hamiltonian'' and ``Lagrangian'', rather than their densities, as mostly done in the rest of this work.} around the minimum is given by,
\begin{align}
	H\big(\vec k\big) = \Delta + \frac{\big( |\vec k| - k_*\big)^2}{2m_*} + \dots \,,
\end{align}
The relation between the roton velocity and its momentum is given by one of Hamilton's equations,\footnote{From Eq.~\eqref{eq:xdot} we deduce that the kinematic of rotons is rather peculiar. Rotons on the left of their minimum, $|\vec k| < k_*$, have velocity and momentum pointing in opposite directions.}
\begin{align} \label{eq:xdot}
	\dot{\vec x} = \frac{\partial H}{\partial \vec k} = \big( |\vec k| - k_* \big) \frac{\hat k}{m_*} \qquad \Rightarrow \qquad |\vec k| = k_* + m_* \, \dot{\vec x} \cdot \hat k \,.
\end{align}
The equation above can be used to write the Lagrangian, and hence the action, describing rotons in the point-particle limit,
\begin{align} \label{eq:Spp}
	\begin{split}
		S_{\rm p.p.}\big[\vec x, \hat k\big] = \int dt \, L_{\rm p.p.} ={}& \int dt \left[ \dot{\vec x} \cdot \vec k - H\big(\vec k\big) \right] = \int dt \left[ \dot{\vec x} \cdot \hat k |\vec k| - \Delta -   \frac{\big( |\vec k| - k_*\big)^2}{2m_*}  \right] \\
		={}& \int dt \left[ - \Delta + k_* \dot{\vec x}\cdot \hat k  + \frac{m_*}{2} \big( \dot{\vec x} \cdot \hat k \big)^2 \right] \,.
	\end{split}
\end{align}
It is clear that the previous action is invariant under spatial translations and rotations. It is not invariant under boost, as it is expected, since they are broken by the medium. Nonetheless, we know by now, that boost invariance is only broken spontaneously, and it is restored by the long-wavelength fluctuations of the medium, the phonons, which transform nonlinearly under the corresponding symmetry. This means that, imposing Galilei boost invariance, the roton velocity can only appear in the combination $\dot{\vec x} - \vec u$, where $\vec u$ is the superfluid velocity defined by the nonrelativistic limit of Eq.~\eqref{eq:superfluidu}. Finally, all the parameters appearing in the action~\eqref{eq:Spp} are implicitly functions of the chemical potential which means that, when perturbations are turned on, they can depend on the superfluid invariant taken in the nonrelativistic limit, $\mathcal{X}$, as discussed in Section~\ref{sec:NRsuperfluid}. In conclusion, at low energies and momenta, the most general action describing the interaction between a localized roton and long-wavelength phonons is given by~\cite{Nicolis:2017eqo},
\begin{align}
	S_{\rm p.p.}\big[ \vec x, \hat k, \pi \big] = \int dt \left[ - \Delta(\mathcal{X}) + k_*(\mathcal{X}) \big( \dot{\vec x} - \vec u \big) \cdot \hat k + \frac{1}{2} m_*(\mathcal{X}) \left( \big( \dot{\vec x} - \vec u \big) \cdot \hat k \right)^2 \right] + \text{higher } \partial\text{'s} \,,
\end{align}
where every function of spacetime is evaluated on the point-particle's trajectory, $\mathcal{X} \equiv \mathcal{X} \big(t, \vec x(t)\big)$ and $\vec u \equiv \vec u\big(t, \vec x(t)\big)$. In this context, by ``higher $\partial$'s'', we refer to derivatives acting on the {\it bulk} fields characterizing the superfluid. As usual, these will be suppressed by powers of the typical microscopic scale $\ell$, in the case of spatial gradients, and by powers of $\ell/c_s \sim m \, \ell^2$, in the case of time derivatives.

\vspace{1em}

We can now apply the effective point-particle theory defined by Eq.~\eqref{eq:Spp} to a physical process. As an example, we study the roton-phonon scattering rate, which plays an important role in the determination of physical quantities such as the viscosity coefficients of ${^4}{\rm He}$~\cite{landau1949theoryB}. To do that, we need to expand the effective Lagrangian up to quadratic order in phonon field. Recalling that, in terms of the canonically normalized phonon field, we have (see Section~\ref{sec:NRsuperfluid}),
\begin{align}
	\mathcal{X} = \bar{\mathcal{X}} + \dot \pi_c - \frac{c_s}{2\sqrt{\bar\rho_m}} \big( \vec\nabla \pi_c \big)^2 \,, \qquad \vec u = - \frac{c_s}{\sqrt{\bar \rho_m}} \vec \nabla \pi_c \,,
\end{align}
the Lagrangian can be expanded as $L_{\rm p.p.} = L_{(0)} + L_{(1)} + L_{(2)} + \dots$, with,
\begin{subequations} \label{eq:Lrotphonon}
	\begin{align}
		L_{(0)} ={}& - \bar \Delta + \bar{k}_* \, \dot{\vec x} \cdot \hat k + \frac{1}{2} \bar m_* \big( \dot{\vec x} \cdot \hat k \big)^2 \,, \\
		L_{(1)} ={}& L_0^\prime \dot \pi_c + \frac{c_s}{\sqrt{\bar \rho_m}} \vec{\nabla} \pi_c \cdot \vec k \,, \\
		L_{(2)} ={}& \frac{1}{2} L_0^{\prime\prime} \dot{\pi}_c^2 - \frac{1}{2} \frac{c_s}{\sqrt{\bar\rho_m}} L_0^\prime \big( \vec \nabla \pi_c \big)^2 + \frac{c_s}{\sqrt{\bar \rho_m}} \dot\pi_c \, \vec\nabla \pi_c \cdot \vec k^{\,\prime} + \frac{1}{2} \frac{c_s^2}{\bar\rho_m} \bar m_* \big( \vec\nabla \pi_c \cdot \hat k \big)^2 \,, 
	\end{align}
\end{subequations}
where we defined the roton momentum as $\vec k \equiv \big( \bar k_* + \bar m_* \, \dot{\vec x} \cdot \hat k \big) \hat k$, primes denote derivatives with respect to $\mathcal{X}$, and barred quantities are evaluated on the background, $\bar{\mathcal{X}}$.

Na\"ively, one might be tempted to take the Lagrangian in Eqs.~\eqref{eq:Lrotphonon} at face value and use it to compute the amplitude for the roton-phonon scattering. However, this is yet not the end of the story. When computing a low energy process, we must include all gapless degrees of freedom, which can be excited with arbitrarily small energies and momenta. Within our point-particle theory, the bulk superfluid phonon is not the only excitation satisfying this criterion: we must also account for the fluctuations of the roton's trajectory. Indeed, the unperturbed roton's trajectory, $\vec x_0(t)$, spontaneously breaks a number of spacetime symmetries,\footnote{\label{footnote: unbroken H rotons}For example, a roton moving on a straight line with velocity $\vec v_0$, on top of the symmetries already broken by the surrounding medium, breaks all other spacetime symmetries except for rotations around $\vec v_0$ and the further modified time translations, generated by $\bar{\bar{H}} = \bar{H} - \vec{v}_0 \cdot \vec{P} = H - \bar\upmu \spacy Q - \vec v_0\cdot \vec P$.} and its fluctuations, $\vec x(t) = \vec x_0(t) + \delta \vec x(t)$, are the associated, gapless Goldstone bosons. The same is true for the fluctuations of the direction of the roton's momentum, $\hat k(t) = \hat k_0(t) + \delta \hat k(t)$, given that $\hat k$ is one of the canonical conjugate variables of $\vec x$.

We can then expand the Lagrangian in Eqs.~\eqref{eq:Lrotphonon} also around $\vec x_0(t)$ and $\hat k_0(t)$. From now on, we choose a roton moving in a straight line,
\begin{align}
	\dot{\vec x}_0(t) \equiv \vec v_0 = v_0 \, \hat k_0 \,, \qquad \hat k_0(t) \equiv \hat k_0 \,,
\end{align}
and remember that all the bulk fields are computed on its trajectory, $\pi_c \equiv \pi_c \left(t, \vec x_0(t) + \delta \vec x(t)\right)$. For the process we are considering, we only need terms that are up to quadratic in the phonon and/or the trajectory fluctuations. These are given by,
\begin{subequations}
	\begin{align}
		L_{(1,0)} ={}& L_{(0,0)}^\prime \dot \pi_c + \frac{c_s}{\sqrt{\bar \rho_m}} k_0 \hat k_0 \cdot \vec \nabla \pi_c \,, \\
		L_{(2,0)} ={}& \frac{1}{2} L_{(0,0)}^{\prime\prime} \dot \pi_c^2 - \frac{1}{2} \frac{c_s}{\sqrt{\bar \rho_m}} \left[ L_{(0,0)}^\prime \big( \vec \nabla \pi_c \big)^2 + 2 k_0^\prime \, \hat k_0 \cdot \vec\nabla\pi_c \dot\pi_c \right] + \frac{1}{2} \frac{c_s^2 \, \bar m_*}{\bar \rho_m} \big( \hat k_0 \cdot \vec \nabla\pi_c \big)^2 \,, \\
		\begin{split}
			L_{(1,1)} ={}& L_{(0,0)}^\prime \delta \vec x \cdot \vec\nabla\dot\pi_c + k_0^\prime \, \hat k_0 \cdot \delta \dot{\vec x} \, \dot \pi_c \\
			& + \frac{c_s}{\sqrt{\bar \rho_m}} \left[ \bar m_* \big( \hat k_0 \cdot \delta \dot{\vec x} \big) \big( \hat k_0 \cdot \vec\nabla\pi_c \big) + k_0 \big( \delta \vec x \cdot \vec \nabla \big) \big( \hat k_0 \cdot \vec \nabla \big) \pi_c + k_0 \, \delta \hat k \cdot \vec\nabla \pi_ c\right] \,,
		\end{split} \\
		L_{(0,2)} ={}& \frac{1}{2} \bar m_* \left( \hat k_0 \cdot \delta \dot{\vec x} \right)^2 + k_0 \, \delta \hat k \cdot \delta \dot{\vec x} - \frac{1}{2} v_0 k_0 \, \delta \hat k^2 \,,
	\end{align}
\end{subequations}
where in the last line we used that fact that, since $\hat k^2 = 1$, it must be $\hat k_0 \cdot \delta \hat k = - \frac{1}{2} \delta \hat k^2$.
Moreover, we defined $L_{(0,0)} \equiv - \bar \Delta + \bar k_* v_0 + \frac{1}{2} \bar m_* v_0^2$ and $k_0 \equiv \bar k_* + \bar m_* v_0$, and from now on $\pi_c \equiv \pi_c(t, \vec x_0(t))$. We also notice that by $\dot \pi_c$ and $\vec \nabla \pi_c$ we indicate the partial derivatives of the field with respect to its first and second argument. In particular, one should not confuse $\dot \pi_c$ with $d\pi_c/dt$.

From the equations above we see that, analogously to what happens for vortex lines, the role of the roton is that of a {\it source} for the phonon field, localized on its trajectory. Consequently, the derivation of the corresponding Feynman rules is a bit unusual, mostly due to the fact that part of the fields live in the bulk of the superfluid, while others live on the roton's trajectory. The correct procedure is detailed in~\ref{app:feynmanrules}, where we determine the phonon-roton vertices, as well as the propagator for the roton trajectory fluctuations. The resulting amplitude for the phonon-roton scattering is given by the following diagrams,
\vspace{0.5em}
\begin{figure}[h!]
	\centering
	\resizebox{0.9\textwidth}{!}{
		\begin{tikzpicture}[decoration={
			markings,
			mark=at position 0.5 with {\arrow{>}}}]
			\node[scale = 0.8] at (-4,0) {$i \mathcal{M}=$};
			
			\draw[->, snake it, semithick] (-3,-1) -- (-2.36,-0.36);
			\draw[->, snake it, semithick] (-1.64,-0.36) -- (-1,-1);

			\filldraw[color=gray!25, fill=gray!25, thin](-2,0) circle (0.5);
			
			\draw[gray, postaction={decorate}] (-3,1) -- (-2.36,0.36);
			\draw[gray, postaction={decorate}] (-1.64,0.36) -- (-1,1);
						
			\node[scale=0.82] at (-3,-1.35) {$(\omega_i, \vec k_i)$};
			\node[scale=0.82] at (-1,-1.35) {$(\omega_f, \vec k_f)$};
			
			\node[scale=0.8] at (-0.5,0) {$=$};
			
			\node[scale=0.6] at (1,0) {$\otimes$};
			\draw[gray, postaction={decorate}] (0,1) -- (0.94,0.06);
			\draw[gray, postaction={decorate}] (1.06,0.06) -- (2,1);
			\draw[->, snake it, semithick] (0,-1) -- (0.94,-0.06);
			\draw[->, snake it, semithick] (1.06,-0.06) -- (2,-1);
			
			\node[scale=0.8] at (2.5,0) {$+$};
			
			\node[scale=0.6] at (4,0.3) {$\otimes$};
			\draw[gray, postaction={decorate}] (3,1) -- (3.93,0.35);
			\draw[gray, postaction={decorate}] (4.07,0.35) -- (5,1);
			\draw[->, snake it, semithick] (4,0.21) -- (4,-0.26);
			\node at (4,-0.3) [circle,fill,inner sep=0.7pt]{};
			\draw[->, snake it, semithick] (3,-1) -- (3.97,-0.33);
			\draw[->, snake it, semithick] (4.03,-0.33) -- (5,-1);
			
			\node[scale=0.8] at (5.5,0) {$+$};
			
			\node[scale=0.6] at (6.5,0) {$\otimes$};
			\draw[gray, postaction={decorate}] (6,1) -- (6.46,0.08);
			\draw[->, snake it, semithick] (6,-1) -- (6.46,-0.08);
			\draw[->, thick, densely dotted] (6.58,0) -- (7,0);
			\draw[thick, densely dotted] (7,0) -- (7.42,0);
			\node[scale=0.6] at (7.5,0) {$\otimes$};
			\draw[gray, postaction={decorate}] (7.54,0.08) -- (8,1);
			\draw[->, snake it, semithick] (7.54,-0.08) -- (8,-1);
			
			\node[scale=0.8] at (8.5,0) {.};
		\end{tikzpicture}
	} 
\end{figure}

\noindent Here the dotted line represents the propagation of an intermediate trajectory fluctuation, while the crossed circle represents a source, corresponding to the unperturbed roton, $(\vec v_0 \,t , \hat k_0)$. We stress that the gray solid lines, representing the incoming and outgoing roton, are there simply for visual aid: they do not represent asymptotic states of our theory, and they are not created nor annihilated.
The contributions corresponding to the three diagrams above are,
\begin{subequations}
	\begin{align}
		\begin{tikzpicture}[baseline=-0.3em, decoration={
			markings,
			mark=at position 0.5 with {\arrow{>}}}]
			\node[scale=0.6] at (1,0) {$\otimes$};
			\draw[gray, postaction={decorate}] (0,1) -- (0.94,0.06);
			\draw[gray, postaction={decorate}] (1.06,0.06) -- (2,1);
			\draw[->, snake it, semithick] (0,-1) -- (0.94,-0.06);
			\draw[->, snake it, semithick] (1.06,-0.06) -- (2,-1);
		\end{tikzpicture} ={}& i \bigg\{ L_{(0,0)}^{\prime\prime} \omega_i \omega_f - \frac{c_s}{\sqrt{\bar \rho_m}} \left[ L_{(0,0)}^\prime \vec k_i \cdot \vec k_f + k_0^\prime \left( \omega_i k_f^{\paral} + \omega_f k_i^{\paral} \right)\right] + \frac{c_s^2 \, \bar m_*}{\bar \rho_m}  k_i^{\paral} k_f^{\paral} \bigg\} \,, \label{eq:seagull} \\[0.8em]
		\begin{tikzpicture}[baseline=-0.3em, decoration={
			markings,
			mark=at position 0.5 with {\arrow{>}}}]
			\node[scale=0.6] at (4,0.3) {$\otimes$};
			\draw[gray, postaction={decorate}] (3,1) -- (3.93,0.35);
			\draw[gray, postaction={decorate}] (4.07,0.35) -- (5,1);
			\draw[->, snake it, semithick] (4,0.21) -- (4,-0.26);
			\node at (4,-0.3) [circle,fill,inner sep=0.7pt]{};
			\draw[->, snake it, semithick] (3,-1) -- (3.97,-0.33);
			\draw[->, snake it, semithick] (4.03,-0.33) -- (5,-1);
		\end{tikzpicture} ={}& \frac{i}{\sqrt{\bar\rho_m}c_s} \left[ \frac{c_s k_0}{\sqrt{\bar\rho_m}} - L_{(0,0)}^\prime v_0 \right] v_0 \frac{\left( k_i^{\paral} - k_f^{\paral} \right)^2}{1 - \hat k_i \cdot \hat k_f} \left( \hat k_i \cdot \hat k_f + \bar \upmu c_s \frac{dc_s}{d \bar \upmu} \right) \,, \\[0.8em]
		\begin{tikzpicture}[baseline=-0.3em, decoration={
			markings,
			mark=at position 0.5 with {\arrow{>}}}]
			\node[scale=0.6] at (6.5,0) {$\otimes$};
			\draw[gray, postaction={decorate}] (6,1) -- (6.46,0.08);
			\draw[->, snake it, semithick] (6,-1) -- (6.46,-0.08);
			\draw[->, thick, densely dotted] (6.58,0) -- (7,0);
			\draw[thick, densely dotted] (7,0) -- (7.42,0);
			\node[scale=0.6] at (7.5,0) {$\otimes$};
			\draw[gray, postaction={decorate}] (8,1) -- (7.54,0.08);
			\draw[->, snake it, semithick] (7.54,-0.08) -- (8,-1);
		\end{tikzpicture} ={}& - \frac{i}{\bar \rho_m} \bigg\{ \vec k_i^{\perp} \cdot \vec k_f^{\perp} \bigg[ \frac{v_0}{k_0 \tilde \omega^2} \left( c_s k_0 k_i^{\paral} - \sqrt{\bar \rho_m} L_{(0,0)}^\prime \omega_i \right) \left( c_s k_0 k_f^{\paral} - \sqrt{\bar \rho_m} L_{(0,0)}^\prime \omega_f \right) \notag \\[-0.85em]
		\begin{split}
			& \qquad \quad + \frac{c_s}{\tilde \omega} \left( c_s k_0  \left( k_i^{\paral} + k_f^{\paral} \right) - \sqrt{\bar\rho_m} L_{(0,0)}^\prime \left( \omega_i + \omega_f \right) \right) \bigg] \\[0.15em]
			& \qquad \quad + \frac{1}{\bar m_* \tilde \omega^2} \left[ \left(c_s  k_0 k_i^{\paral} - \sqrt{\bar\rho_m} L_{(0,0)}^\prime \omega_i \right) k_i^{\paral} + \left( c_s \bar m_* k_i^{\paral} - \sqrt{\bar\rho_m} k_0^\prime \omega_i \right) \tilde \omega \right] 
			\end{split} \\[0.15em]
		& \qquad \quad \qquad \quad \! \times \left[ \left(c_s  k_0 k_f^{\paral} - \sqrt{\bar\rho_m} L_{(0,0)}^\prime \omega_f \right) k_f^{\paral} + \left( c_s \bar m_* k_f^{\paral} - \sqrt{\bar\rho_m} k_0^\prime \omega_f \right) \tilde \omega \right] \bigg\} \,. \notag
	\end{align}
\end{subequations}
In the above expressions we define the phonon momentum parallel and perpendicular to the roton's momentum, $k^{\paral} \equiv \vec k \cdot \hat k_0$ and $\vec k^\perp \equiv \vec k - k^{\paral} \hat k_0$, as well as $\tilde \omega = \omega_i - \vec{v}_0 \cdot \vec k_i = \omega_f - \vec{v}_0 \cdot \vec k_f$. In particular, as we show again in~\ref{app:feynmanrules}, we have used the fact that in this point-particle theory the only conservation law is given by $\omega_i - \vec{v}_0 \cdot \vec k_i = \omega_f - \vec{v}_0 \cdot \vec k_f$.

We are now ready to compute the scattering cross section. To make our lives easier, we will restrict ourself to the case of a roton at rest, $v_0=0$. In this limit, the conservation law imposes $k_i = k_f \equiv k$, and $\omega_i = \omega_f = c_s k$. Note that this is not the usual choice of reference frame that particle physicists use in their everyday life. The presence of the medium, in fact, spontaneously breaks boosts and, consequently, scattering off a roton with velocity $\vec v_0$ is a {\it physically} different process than scattering off a roton at rest. After some tedious work, the resulting amplitude is found to be,
\begin{align} \label{eq:iMrest}
	i\mathcal{M}_{\rm rest} ={}& - i \frac{c_s \bar k_* k^2}{\bar \rho_m} \bigg\{ \hat k_i \cdot \hat k_f \left[ \frac{\bar \rho_m}{c_s \bar k_*} \frac{d\bar \Delta}{d \bar \rho_m} + \hat k_i^{\paral} +\hat k_f^{\paral} \right] + \frac{\bar k_*}{\bar m_* c_s} \hat k_i^{\paral\,2} \hat k_f^{\paral\,2} \\
	& \qquad\qquad\; + \frac{\bar\rho_m^2}{c_s \bar k_*} \left[ \frac{d^2\bar \Delta}{d\bar\rho_m^2} + \frac{1}{\bar m_*} \left( \frac{d\bar k_*}{d \bar \rho_m} \right)^{\!2} \right] + B \, \frac{\bar \rho_m}{c_s \bar k_*} \frac{d\bar \Delta}{d\bar\rho_m} - \frac{\bar \rho_m}{\bar m_* c_s} \frac{d\bar k_*}{d\bar \rho_m} \left(  \hat k_i^{\paral\,2} + \hat k_f^{\paral\,2} \right)  \bigg\} \,, \notag
\end{align}
where we defined, 
\begin{align} \label{eq:B}
	\begin{split}
		B \equiv{}& 1 - 2\frac{\bar \rho_m}{c_s} \frac{dc_s}{d\bar\rho_m} + \frac{\bar k_*}{\bar m_* c_s} \hat k_i^{\paral} \hat k_f^{\paral} \left( \hat k_i^{\paral} + \hat k_f^{\paral} \right) + \frac{\bar \rho_m}{\bar m_* c_s^2} \frac{d\bar \Delta}{d\bar\rho_m} \hat k_i^{\paral} \hat k_f^{\paral} - \frac{\bar \rho_m}{\bar m_* c_s}  \frac{d\bar k_*}{d\bar \rho_m} \left( \hat k_i^{\paral} + \hat k_f^{\paral} \right) \,.
	\end{split}
\end{align}
In deriving the expressions above we have used the fact that, from Eqs.~\eqref{eq:lambdan} and \eqref{eq:Xderivatives}, we deduce that,
\begin{align} \label{eq:chain1}
	\frac{d^n}{d\bar{\mathcal{X}}^n} = \left( \frac{m c_s}{\sqrt{\bar\rho_m}} \right)^n \frac{d^n}{d\bar\upmu^n} \;,
\end{align}
while also using, 
\begin{align} \label{eq:chain2}
	\frac{d}{d\bar\upmu} = \frac{d\bar p}{d\bar\upmu} \frac{d\bar\rho_m}{d\bar p} \frac{d}{d\bar\rho_m} = \frac{\bar\rho_m}{mc_s^2} \frac{d}{d\bar\rho_m} \quad \text{ and } \quad \frac{d^2}{d\bar\upmu^2} = \frac{\bar\rho_m}{m^2 c_s^4} \left[ \left( 1 - 2 \frac{\bar\rho_m}{c_s} \frac{dc_s}{d\bar\rho_m} \right) \frac{d}{d\bar\rho_m} + \bar\rho_m \frac{d^2}{d\bar\rho_m^2} \right] \,.
\end{align}
Once the scattering amplitude has been computed, the corresponding cross section is simply given by~\cite{Endlich:2010hf,Nicolis:2017eqo},
\begin{align} \label{eq:sigmaroton}
	d\sigma = \frac{1}{2 \omega_i} \frac{1}{c_s} \big| \mathcal{M}_{\rm rest} \big|^2 (2\pi) \delta\big( \omega_i - \omega_f \big) \frac{d^3k_f}{(2\pi)^3 2\omega_f} \qquad \Rightarrow \qquad \frac{d\sigma}{d\Omega} = \frac{\big| \mathcal{M}_{\rm rest} \big|^2}{16\pi^2 c_s^4} \,.
\end{align}

Let us compare our Eqs.~\eqref{eq:iMrest} and \eqref{eq:sigmaroton} with the various studies appeared in the literature. Ideed, up to now, there was no agreement on the result given by different approaches. The roton-phonon scattering cross section for a roton at rest was first studied by Landau and Khalatnikov in~\cite{landau1949theoryA}. They worked in the limit of negligible $d\bar \Delta / d\bar\rho_m$, justified by its anomalously small value in low-pressure ${^4}{\rm He}$~\cite{godfrin2021dispersion}. The same result is also obtained in~\cite{Matchev:2021fuw} by means of an approach somewhat inspired by Heavy Quark Effective Theory~\cite{Manohar:2000dt}. In this regime, the result reported in Eq.~(3.19) of~\cite{landau1949theoryA} and Eq.~(4.7) of~\cite{Matchev:2021fuw} almost matches the result obtained from our Eqs.~\eqref{eq:iMrest} and \eqref{eq:sigmaroton}, with the exception of the very last term proportional to $d\bar k_*/d\bar\rho_m$. This term was simply forgotten in~\cite{landau1949theoryA}, and its presence was first pointed out in~\cite{Nicolis:2017eqo}. However, we find the expressions reported in Eqs.~(79b), (79c) and (82) of~\cite{Nicolis:2017eqo} to be incorrect.\footnote{2/3 of the present authors are responsible for \cite{Nicolis:2017eqo}. They thank the remaining 1/3 for clearing up the situation.}. The results presented here, instead, are in agreement with those obtained in~\cite{castin2017landau} by means of more traditional methods---namely, the so-called local density approximation. In particular, up to a conventional normalization factor, Eq.~(20) in~\cite{castin2017landau} (Eq.~(4) of the erratum in the editorial version) matches precisely our Eq.~\eqref{eq:iMrest}, after converting from relativistic to nonrelativistic normalization (see footnote 5).\footnote{It is also interesting to note that, in our treatment, the first two terms in Eq.~\eqref{eq:B} arise from the diagram in Eq.~\eqref{eq:seagull}, where the two phonon lines are directly attached to the roton trajectory. In~\cite{castin2017landau}, instead, the same terms arise from a diagram containing the phonon nonlinearity. This is due to the difference between the Lagrangian and Hamiltonian formalisms: the results obtained in the two frameworks should not be compared diagram-by-diagram.}
We stress again that the roton-phonon scattering cross section plays a relevant role in the determination of the viscosity coefficients of ${^4}{\rm He}$~\cite{landau1949theoryB} and, therefore, it is important to pin it down precisely.

\vspace{1em}

\subsection{Further Readings} \label{sec: further superfluids}

\noindent Together with hydrodynamics, the EFT of superfluids is arguably the one that has attracted the most attention, has been developed further, and has been applied to a variety of problems.

\paragraph{Further developments of the EFT for superfluids} Several different corners of the EFT presented in this section have been investigated in the literature, especially within the framework of the weakly coupled theory presented in Section~\ref{sec:weak coupling}. An interesting aspect is the investigation of the response of the superfluid to the presence of a body moving in it. This has been done in~\cite{Berezhiani:2019pzd,Berezhiani:2020umi}, where the authors studied friction phenomena (both standard friction, and the so-called dynamical friction). Among other things, they found that the lowest order EFT is almost always inadequate at describing instances where the body is moving at supersonic speeds. (For discussions of the same phenomenon, but within more traditional frameworks, see~\cite[e.g.,][]{kamenev2011field,Lancaster:2019mde,Hartman:2020fbg}.)
The properties of the ground state of the theory, discussed as a full-fledged quantum state, have instead been presented in~\cite{Berezhiani:2025tkp}.
Multi-component superfluids, i.e.~superfluids made of many species and featuring several $U(1)$ symmetries, have been studied in~\cite{Haber:2015exa,Haber:2017kth,Haber:2017oqb,Trabucco:2025duz}, including superconductors, which are obtained by gauging some of the $U(1)$ groups. This is especially relevant given the importance of these systems in extreme astrophysical conditions, such as neutron stars. 
From a more formal viewpoint, it has been noted that the EFT for superfluids presented here, besides the $U(1)$ symmetry we discussed, also features a 2-form symmetry, $U(1)^{(2)}$~\cite{Delacretaz:2019brr}.\footnote{This generalizes to a $(d-2)$-form symmetry in the case of $d$ spacetime dimensions.} The symmetry group $U(1) \times U(1)^{(2)}$, however, has a mixed anomaly. The authors have then recast the hydrodynamics of a superfluid by inverting the roles: they rederived it as the hydrodynamic theory of a system with an emergent, mixed-anomalous $U(1) 
\times U(1)^{(2)}$ symmetry. Moreover, it has been argued that, were the Universe in a finite temperature superfluid state, it might offer a solution to the long-standing cosmological constant problem~\cite{Khoury:2018vdv}. The authors of~\cite{Pajer:2018egx}, instead, have classified all possible additional symmetries that the EFT for superfluids can have at leading order in the derivative expansion, by suitably choosing specific forms of the $P(X)$ function.

\paragraph{More on superfluid vortices} Besides what we discussed in this section, the physics of superfluid vortices features many interesting aspects. One of their peculiarities is that, when their number is sufficiently high, they spatially organize themselves in a triangular lattice~\cite{tkachenko1966vortex,tkachenko1966stability,tkachenko1969elasticity}. As any other lattice, at long wavelengths, this features collective excitations, essentially corresponding to small displacements of the vortex positions with respect to their equilibrium values. The gapless excitations have a quadratic dispersion relation, and go under the name of Tkachenko modes~\cite{sonin2014tkachenko}. The effective field theory for such excitations has been developed in~\cite{Moroz:2018noc,Du:2022xys}, leveraging the vortex--boson duality, and in~\cite{Glodkowski:2025krf} using the coset construction. The inclusion of defects in the vortex lattice, their interaction with the Tkachenko modes, and the role they play in the melting of the lattice, are discussed in~\cite{Nguyen:2020yve,Nguyen:2023quf}. 
The existence and properties of vortex lattices, however, are not the only peculiarities of superfluid vortices. For example, they can coexist together with the so-called ``dark solitons'', i.e., localized nonlinear waves where the superfluid density locally drops to zero. An EFT similar to the one discussed here has been used in~\cite{Mateo:2016bai} to show the existence of configurations where the vortex lines are attached to a dark soliton, and to show that these configurations are stable under ordinary experimental conditions. Moreover, a vortex generated in a confined superfluid as, for example, a trapped atom cloud, exhibits a spontaneous precessional motion around the center of the cloud. The presence of this effect becomes particularly transparent when derived within our EFT~\cite{Esposito:2017xzg}, which also allows to derive a fully non-perturbative relation between the superfluid density and the trapping potential. Finally, the interaction between sound modes (the phonons) and vortex rings has been discussed in~\cite{Garcia-Saenz:2017wzf}. 

\paragraph{Superfluids in conformal field theories and holography} The EFT for relativistic superfluids has also played a central role in the study of certain conformal field theories. The literature on the topic is vast, and we do not attempt to cover all of it here, but we rather present a few examples. Thanks to the operator--state correspondence, it has been shown that superfluid states can be used to compute the scaling dimensions of operators in a $(2+1)$-dimensional CFT characterized by a large global $U(1)$ charge and spin~\cite[e.g.,][]{Hellerman:2015nra,Monin:2016jmo,Loukas:2016ckj,Banerjee:2017fcx,Cuomo:2017vzg}. For increasingly large spin, the corresponding superfluid states are those with a single phonon, two vortices, or several vortices. Extensions of these results have also been carried out for parity-violating CFTs~\cite{Cuomo:2021qws}, as well as $(4+1)$-dimensional CFTs~\cite{Cuomo:2019ejv}. Interestingly, vortex rings and Kelvin waves (fluctuations of the vortex line) also play a role. A number of studies have then been performed in the nonrelativistic limit~\cite[e.g.,][]{Favrod:2018xov,Kravec:2018qnu,Kravec:2019djc,Hellerman:2021qzz}. Implications for heavy CFT operators are discussed in~\cite{Delacretaz:2020nit}, while the constraining power that spontaneously broken boosts have on the CFT states is elucidated~\cite{Komargodski:2021zzy}. (For other similar studies see, for example,~\cite{Cuomo:2021ygt,Cuomo:2021cnb,Dondi:2022wli,Badel:2022fya,Cuomo:2022kio,Dondi:2024vua,Choi:2025tql}, while for a review on this and other similar topics see~\cite{Gaume:2020bmp}.) Superfluids have also provided a very fruitful test field for the so-called AdS/CMT program, which aims at studying old and new phases of matter by means of the celebrated holographic duality~\cite[e.g.,][]{Hartnoll:2009sz,Hartnoll:2016apf}. Specifically, the bulk theory dual to a superfluid state on the boundary is scalar QED coupled to gravity, as first argued in~\cite[e.g.,][]{Hartnoll:2008kx,Herzog:2008he}.\footnote{The original nomenclature has been that of ``holographic superconductors'', although the symmetry breaking pattern and corresponding spectrum seem to be more appropriately those of a superfluid.} The fact that such a bulk theory indeed matches to the correct superfluid EFT on the boundary, at quadratic order, has been proved in~\cite{Esposito:2016ria}.

\paragraph{Superfluids as and for dark matter} The theory of superfluidity, declined within the EFT presented in this work, has also often cross-pollinated with the theory of dark matter, incarnated as both possible dark matter components of the Universe, as well as proposed targets for laboratory searches. On the one hand, it has been argued that dark matter may be in a superfluid state. This hypothesis has received a great deal of attention, and has been thoroughly investigated~\cite[e.g.,][]{Goodman:2000tg,Berezhiani:2015bqa,Berezhiani:2017tth,Berezhiani:2018oxf,Berezhiani:2021rjs,Berezhiani:2025maf}. Were dark matter in a superfluid state, it would allow to overcome some difficulties faced by the standard cold dark matter paradigm, such as the ``core-cusp'' problem and the ``too-big-to-fail'' one~\cite[e.g.,][]{Bullock:2017xww}. On the other hand, superfluid ${^4}{\rm He}$ has a number of features that make it an interesting possible target to look for dark matter particles with masses as light as $m_\chi \sim \mathcal{O}({\rm keV})$. This was originally proposed in~\cite{Guo:2013dt,Schutz:2016tid,Knapen:2016cue}, where the study was carried out with more traditional methods. The use of a relativistic EFT to study the dark matter--phonon interaction was first introduced in~\cite{Acanfora:2019con}, and then further developed in~\cite{Caputo:2019cyg,Caputo:2019xum,Caputo:2019ywq}. The inclusion of 3-phonon final states, and their directional signature, has been done in~\cite{Caputo:2020sys}, which is the first instance where the EFT presented here allows to obtain results otherwise prohibitive within the traditional approaches. The first inclusion of roton degrees of freedom in this context has instead been done in~\cite{Matchev:2021fuw,You:2022pyn}.

\paragraph{Superfluids in cosmology} One can argue that cosmology is, to some extent, the theory of spontaneously broken spacetime symmetries in the presence of gravity. At early times, in particular, this is precisely the broad mechanism driving the inflationary expansion of the Universe. In this respect, it is not surprising that superfluids coupled to gravity have made their appearance in cosmology in a number of contexts. This connection is made clear in the so-called EFT of inflation~\cite[e.g.,][]{Cheung:2007st}, while explicit use of the EFT we discussed in this section, with the corresponding $P(X)$ Lagrangian, is done in~\cite[e.g.,][]{Finelli:2018upr}. Special superfluids coupled to gravity have also been proposed as possible infrared modifications of gravity itself, in the form of the so-called ``ghost condensate''~\cite[e.g.,][]{Arkani-Hamed:2003pdi,Arkani-Hamed:2003juy}.

%% file: classification.tex
\section{A Symmetry-based Classification of Media: Lessons Learned and some Speculations} \label{sec: classification}

\noindent The EFTs presented so far have all been constructed in an inductive way, starting from some known properties of the systems under consideration and, based on those, identifying their symmetry breaking pattern and the appropriate field content. Knowing what we know now, however, it is possible to look back at all the previous discussions and appreciate the emergence of a more global, symmetry-based picture. Ultimately, the systems considered in this work differ from one another only in which spacetime symmetries they spontaneously break, and which ones they preserve. Once this is specified, their low energy description is fixed by the requirement of the existence of some unbroken (effective) spacetime translations, which are {necessary} in order to be able to classify the energy levels of the system, and to define long-distance homogeneity, which is a general property of any medium realized in the lab~\cite{Esposito:2020wsn}. This can be supplemented with the requirement of long-distance isotropy, i.e., of the existence  unbroken continuous rotations as well. Notice however that, contrary to spacetime translations, this last condition is not mandatory. A perfect crystal, for example, never looks isotropic, regardless of how far away you look at it. From this viewpoint, the assumption of invariance under continuous rotations is just a simplifying one, which can be lifted if necessary~\cite[e.g.,][]{Kang:2015uha,Kang:2018bqc,Nicolis:2020rqz}.
In this Section, we revisit the previous EFTs, looking at them from this unifying viewpoint~\cite{Nicolis:2015sra}. By following this approach to the end, we will recognize the existence of other possible phases of matter, some of which are admittedly rather speculative.

As explained, we will always assume the existence of some unbroken generators playing the role of time translations, spatial translations and continuous rotations. These can be either the original spacetime generators (if unbroken), or a linear combination of the original (broken) spacetime generators together with the generators of some additional (broken) symmetry. We indicate with $P^\mu$ and $J^i$ the generators of the original spacetime translations and rotations, and with $\bar P^\mu$ and $\bar{J}^i$ the ones that remain unbroken at low energies. The ground state of a system, $|\Psi\rangle$, is identified as the state minimizing the generator of the effective unbroken time translations:
\begin{align}
    \bar P^0 |\Psi \rangle = 0 \,,
\end{align}
where we arbitrarily set the ground state eigenvalue to zero. Overall, $\bar P^\mu$ and $\bar{J}^i$ are the generators employed to define conservation of energy, momentum and angular momentum in the long-wavelength processes involving the Goldstones.

Finally, the symmetry breaking patterns presented here are the ones that are strictly necessary to define the corresponding media, based on the role played by spacetime symmetries. Additional symmetries---broken or not---can always be added on top of what we present here. Most of the times, this is a harmless complication of the theory. However, when the additional internal symmetries do not commute with the unbroken spacetime translations, this can lead to the emergence of new phenomena and interesting subtleties, such as the appearance of the so-called ``gapped Goldstones''~\cite[e.g.,][]{Morchio:1987aw, Nicolis:2012vf,Kapustin:2012cr,Watanabe:2013uya,Nicolis:2013sga,Cuomo:2020gyl}, and the already mentioned mismatch between the number of broken generators and the number of Goldstone modes~\cite[e.g.,][]{Low:2001bw,Watanabe:2013iia}. 

As far as the latter is concerned, it will be useful for what follows to briefly explain how to deduce the number of Goldstones from the symmetry breaking pattern. Let us label with $Q_a$ the set of broken generators. Whenever the commutator between the unbroken spacetime translations and some broken
generators, $Q_a$, contains other broken generators, $Q'_a$, i.e.,
\begin{align} \label{eq:inverseHiggscondition}
    [\bar P^\mu, Q_a] = i f^\mu {}_{a} {}^{b} \spacy Q_b' + \text{other generators} \,,
\end{align}
it is possible to impose the so-called ``inverse Higgs constraints''~\cite[e.g.,][]{Ivanov:1975zq,Low:2001bw,McArthur:2010zm,Nicolis:2013sga}. Their technical implementation requires the use of the coset construction~\cite[e.g.,][]{Coleman:1969sm,Callan:1969sn,ogievetsky1974nonlinear,Ivanov:1975zq,Delacretaz:2014oxa}, and it is beyond the scope of this section. What we care about now is that, at the end of the day, this allows to express the Goldstone fields associated with the $Q_a$'s in terms of derivatives of the Goldstone fields associated with the $Q'_a$'s, schematically
\begin{align} 
 \, \pi^b \sim  f^\mu {}_{a} {}^{b} \, \partial_\mu \pi^{\prime \, a} \, .
\end{align}
One can then eliminate the former from the theory by imposing these relations.

\subsection{The Classification}

\noindent Before proceeding, we remind the reader that the algebra of the Poincar\'e group---i.e., spacetime translations, rotations and Lorentz boosts---is given by,
\begin{align} \label{eq:Lorentzalgebra}
    \begin{split}
        \big[J^i, P^j\big] = i \epsilon^{ijk} P^k \,, \qquad \big[K^i, P^j\big] ={}& i \delta^{ij} P^0 \,, \qquad \big[K^i, P^0\big] = -i P^i \,, \\
        \big[J^i, J^j\big] = i \epsilon^{ijk} J^k \,, \qquad \big[J^i, K^j\big] ={}& i\epsilon^{ijk} K^k \,, \qquad \big[K^i,K^j\big] = - i \epsilon^{ijk} J^k \,,
    \end{split}
\end{align}
where $K^i$ are the generators of Lorentz boosts. Since the effective unbroken generators, $\bar P^\mu$ and $\bar{J}^i$, completely replace the original spacetime translations and rotations, they must necessarily satisfy their algebra among themselves, i.e.,
\begin{align} \label{eq:unbrokenalgebra}
    \big[ \bar J^i, \bar P^j \big] = i \epsilon^{ijk} \bar P^k \,, \qquad\text{ and }\qquad \big[ \bar J^i, \bar J^j \big] = i \epsilon^{ijk} \bar J^k \,,
\end{align}
where all other commutators vanish. On the other hand, as we discusssed in the Introduction, there is no unbroken version of the boost generators.  
With this at hand, we can rediscuss some aspects of solids, fluids, and superfluids~\cite{Nicolis:2015sra}. 

\subsubsection{Solids and Fluids: $P^0,\, \bar P^i,\, \bar J^i$}

\noindent As we saw in Section~\ref{sec:solids}, solids spontaneously break boosts, spatial translations and rotations. To recover homegeneity and isotropy at long distances, we must impose the existence of six additional internal generators, $Q^i$ and $\tilde{Q}^i$. 
These are employed to define the effective unbroken spatial translations and rotations as $\bar{P}^i = P^i - \alpha\spacy Q^i$ and $\bar{J}^i = J^i + \tilde{Q}^i$. 
From this viewpoint, $\alpha$ is a free parameter, which one has the freedom to introduce for abelian subgroups (such as spacetime translations) but not for non-abelian ones (such as rotations). Requiring for $\bar{P}^i$ and $\bar{J}^i$ to obey the standard algebra of the Lorentz group, Eq.~\eqref{eq:Lorentzalgebra}, we deduce that the new internal generators must satisfy,
\begin{align}
    \big[ Q^i, Q^j \big] = 0 \,, \qquad \big[ \tilde Q^i, Q^j \big] = i \epsilon^{ijk} Q^k \,, \qquad \big[ \tilde Q^i, \tilde Q^j \big] = i \epsilon^{ijk} \tilde Q^k \,.
\end{align}
This is indeed the algebra of the 3-dimensional Euclidean group, $ISO(3)$, made up of internal rotations and shifts. These act on the comoving coordinates as in Eqs.~\eqref{solid rotations} and \eqref{solid shifts}, i.e.,
\begin{align} \label{eq:ISO3}
    \phi^I(x) \to O^I {}_J \spacy \phi^J(x) + a^I \,,
\end{align}
with $O^I{}_J$ an $SO(3)$ matrix, and $a^I$ a constant.
    
As far as the counting of Goldstones is concerned, we have a total of nine broken generators ($K^i$, $P^i$, and $J^i$), but six relations of the kind~\eqref{eq:inverseHiggscondition}:
\begin{align} \label{eq:commsolids}
    \big[ \bar P^0, K^i \big] = i P^i \,, \qquad \text{ and } \qquad \big[ \bar P^i , J^j \big] = i \epsilon^{ijk} P^k \,.
\end{align}
We then can then impose six conditions, and only three Goldstones survive. These are precisely to the fluctuations of the comoving coordinates, $\phi^I(x) = \alpha \left( x^I + \pi^I(x) \right)$, corresponding to the three phonon modes sustained by a 3-dimensional solid: one longitudinal and two transverse.

In this language, the theory describing a perfect fluid is obtained as a special case of the EFT presented above. In fact, while for a solid one has the freedom to translate and rotate the volume elements, for a fluid one can do more. One can also freely deform them, as long as their  volumes remain the same~\cite[e.g.,][]{landau:1987bo}. The internal symmetry group that one postulates to describe a fluid is therefore that of all possible volume preserving diffeomorphisms (see Section~\ref{sec:Eulerianfluids}),
\begin{align}
    \phi^I(x) \to \xi^I\big(\phi(x)\big) \,, \qquad \text{with} \qquad \det \frac{\partial \xi^I(\phi)}{\partial \phi^J} = 1 \,.
    \end{align}
The transformations above include the $ISO(3)$ transformations in Eq.~\eqref{eq:ISO3}. Therefore, from a standpoint purely based on symmetries, a perfect fluid is nothing but a very special type of solid, one tuned to a point of enhanced symmetry. 
The group of volume preserving diffeomorphisms has infinitely many generators, but the  unbroken symmetries are still only those generated by $\bar P^i$ and $\bar J^i$, corresponding to the ground state's homogeneity and isotropy. The infinitely many Goldstone fields can all be expressed, through inverse-Higgs constraints, in terms of the phonon fields, along the lines of what discussed in~\cite{Goon:2014ika,Goon:2014paa}.

\subsubsection{Type-I Superfluids: $\bar P^0,\, P^i,\, J^i$}

\noindent Since the original time translations are broken, the existence of $\bar P^0$ mandates for the presence of an internal $U(1)$ symmetry. This must also be spontaneously broken, and its generator, $Q$, combines with $P^0$ to form $\bar P^0 = P^0 - \bar\upmu \spacy Q$. 
In Section~\ref{sec:superfluids}, we learned that $\bar\upmu$ is the equilibrium chemical potential. However, from the purely symmetry viewpoint we are now taking, $\bar\upmu$ is simply a dimensionful parameter characterizing the ground state of the system.

In this case, there is a total of four broken generators ($K^i$ and $P^0$), but three relations of the kind~\eqref{eq:inverseHiggscondition},
\begin{align} \label{eq:commPK}
    \big[\bar P^i, K^j\big] = \big[P^i, K^j\big] = - i \delta^{ij} P^0 \,.
\end{align}
Using this, one can eliminate the three Goldstones associated with $K^i$, in favor of the single Goldstone associated to $P^0$. This matches perfectly with the fact that, as we saw in Section~\ref{sec:superfluids}, the above mentioned symmetry breaking pattern can be implemented via a single scalar field, $\psi(x) = \bar\upmu \left( t + \pi(x) \right)$, with correspondingly only one degree of freedom: the superfluid phonon. For reasons that will be clear in Section~\ref{sec:morephases}, these systems have been dubbed ``type-I superfluids''~\cite{Nicolis:2015sra}.

\subsection{More Phases of Matter} \label{sec:morephases}

\noindent Looking at the above discussion one might have noticed that not all possible combinations of broken generators have been covered. Beside boosts, solids and fluids break spatial translations and rotations, while type-I superfluids only time translations. We will now discuss the missing pieces of the puzzle, starting from the ones that are actually realized in Nature, and then proceeding to some more speculative systems.

\subsubsection{Type-II Superfluids: $\bar P^0,\, P^i,\, \bar J^i$}

\noindent The so-called ``type-II superfluids''~\cite{Nicolis:2015sra}, also called ``p-wave'' superfluids, have the same symmetry breaking pattern as their type-I counterpart, with the addition of broken rotations. To recover long-distance isotropy as well, the most minimal internal symmetry group one can postulate is then $U(1) \times SO(3)$. If $\tilde{Q}^i$ are the generators of this internal $SO(3)$, the effective unbroken generators are obtained as $\bar P^0 = P^0 - \bar\upmu \spacy Q$ and $\bar{J}^i = J^i + \tilde{Q}^i$. In this case, there are seven broken generators ($K^i$, $P^0$, and $J^i$). However, besides the relations in Eq.~\eqref{eq:commPK}, no additional commutators of the kind~\eqref{eq:inverseHiggscondition} are present. The final number of Goldstone modes is then four. 

\begin{figure}[t!] 
\centering
\resizebox{0.31\textwidth}{!}{
	\begin{tikzpicture}
		\draw[->, thick, Maroon] (-1.1,0) -- (-0.6,0.7);
        \draw[->, thick, blue, dashed] (-1.1,0) -- (-0.3,0.1);
		\node at (-1.1,0) [circle,fill,inner sep=0.8pt]{};

        \draw[->, thick, Maroon] (0.2,0.5) -- (0.7,1.2);
        \draw[->, thick, blue, dashed] (0.2,0.5) -- (1,0.6);
		\node at (0.2,0.5) [circle,fill,inner sep=0.8pt]{};

        \draw[->, thick, Maroon] (-1.8,0.8) -- (-1.3,1.5);
        \draw[->, thick, blue, dashed] (-1.8,0.8) -- (-1,0.9);
		\node at (-1.8,0.8) [circle,fill,inner sep=0.8pt]{};

        \draw[->, thick, Maroon] (-0.4,1.2) -- (0.1,1.9);
        \draw[->, thick, blue, dashed] (-0.4,1.2) -- (0.4,1.3);
		\node at (-0.4,1.2) [circle,fill,inner sep=0.8pt]{};

        \draw[semithick] (-2,-0.2) rectangle (1.2,2.1);

        \node[scale=0.6] at (-0.5,0.85) {$\color{Maroon} \vec{S}$};
        \node[scale=0.6] at (-0.15,0.13) {$\color{blue} \vec{L}$};
	\end{tikzpicture}
} \hspace{3em}
\resizebox{0.31\textwidth}{!}{
	\begin{tikzpicture}
        \begin{scope}[shift={(-0.2,-0.05)}, rotate around={30:(-1.1,0)}]
		\draw[->, thick, Maroon] (-1.1,0) -- (-0.6,0.7);
        \draw[->, thick, blue, dashed] (-1.1,0) -- (-0.3,0.1);
        \node at (-1.1,0) [circle,fill,inner sep=0.8pt]{};
        \end{scope}

        \begin{scope}[rotate around={-15:(0.2,0.5)}]
        \draw[->, thick, Maroon] (0.2,0.5) -- (0.7,1.2);
        \draw[->, thick, blue, dashed] (0.2,0.5) -- (1,0.6);
		\node at (0.2,0.5) [circle,fill,inner sep=0.8pt]{};
        \end{scope}

        \begin{scope}[shift={(0.9,0.45)}, rotate around={120:(-1.8,0.8)}]
        \draw[->, thick, Maroon] (-1.8,0.8) -- (-1.3,1.5);
        \draw[->, thick, blue, dashed] (-1.8,0.8) -- (-1,0.9);
		\node at (-1.8,0.8) [circle,fill,inner sep=0.8pt]{};
        \end{scope}

        \begin{scope}[shift={(0.1,0.55)}, rotate around={-90:(-0.4,1.2)}]
        \draw[->, thick, Maroon] (-0.4,1.2) -- (0.1,1.9);
        \draw[->, thick, blue, dashed] (-0.4,1.2) -- (0.4,1.3);
		\node at (-0.4,1.2) [circle,fill,inner sep=0.8pt]{};
        \end{scope}

        \draw[semithick] (-2,-0.2) rectangle (1.2,2.1);

        \node[scale=0.6] at (-1.2,0.98) {$\color{Maroon} \vec{S}$};
        \node[scale=0.6] at (-0.5,0.55) {$\color{blue} \vec{L}$};
	\end{tikzpicture}
}
\caption{Different phases of superfluid ${^3}{\rm He}$. Red solid arrows represent the atomic spin, $\vec{S}$, while blue dashed arrows the atomic orbital angular momentum, $\vec{L}$. {\bf Left panel:} $A$-phase. The directions of both $\vec{S}$ and $\vec{L}$ are fixed. {\bf Right panel:} $B$-phase. The relative angle between $\vec{S}$ and $\vec{L}$ is fixed, but their overall orientation is random.} \label{fig:He3}
\end{figure}
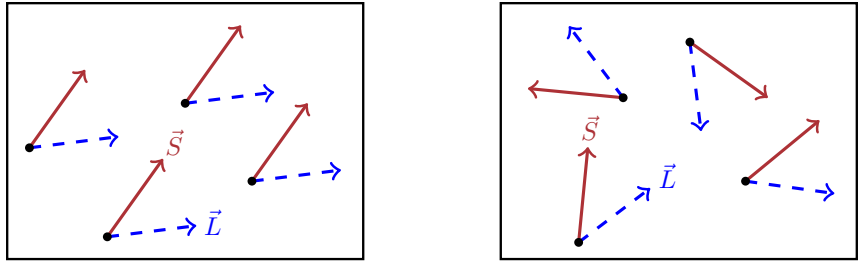

As anticipated, type-II superfluids do exist in Nature. A nonrelativistic realization of them is the $B$-phase of superfluid ${^3}{\rm He}$ at temperatures well below the critical one~\cite[e.g.,][]{vollhardt2013superfluid}. Indeed, since it is a nonrelativistic system, it has seven degrees of freedom: three components of the atomic spin, three components of the atomic angular momentum, and the phase of the order parameter. At zero temperature, its possible ground states look like those reported in Figure~\ref{fig:He3}. In the $B$-phase, the relative angle between spin and orbital angular momenta is fixed, but their overall orientation is random. This breaks both internal spin rotations and spatial rotations, but preserves a linear combination of two: the one that leaves unchanged their relative angle. The breaking of the internal $U(1)$, instead, is the same as for type-I superfluids. The EFT for the $B$-phase of superfluid ${^3}{\rm He}$ has been developed in~\cite{Fujii:2016mbc}, up to next-to-leading order in the derivative expansion. The $A$-phase of superfluid ${^3}{\rm He}$, instead, would require giving up the assumption of isotropy, as only a discrete subgroup of rotations is preserved at long distances.

In a relativistic context, the symmetry breaking pattern of a type-II superfluid can instead be implemented as done in~\cite{Endlich:2013vfa}, via a triplet of complex 4-vector fields, ${A}^a_\mu(x)$, with $a=1,2,3$, transforming linearly under both the internal $U(1)$ and the internal $SO(3)$:
\begin{align}
    A^a_\mu(x) \to e^{i \theta} \spacy O^a{}_b \spacy A^b_\mu(x) \,,
\end{align}
where $\theta$ is a constant and $O^a{}_b$ is an $SO(3)$ matrix.
In particular, one must require the following expectation value on the ground state,
\begin{align}
    \langle A_\mu^a(x) \rangle \propto e^{i \bar\upmu \spacy t} \, \delta_\mu^a \,.
\end{align}

\subsubsection{Supersolids: $\bar P^0,\, \bar P^i,\, \bar J^i$} \label{sec:supersolids}

\noindent As compared to solids and type-I superfluids, supersolids get the best (or worst) of both. Specifically, all spacetime symmetries are broken~\cite[e.g.,][]{andreev1969quantum,chester1970speculations,PhysRevLett.25.1543}. One is then lead to the introduction of an internal symmetry group given by $ISO(3) \times U(1)$, such that all unbroken generators can be reconstructed: $\bar P^0 = P^0 - \bar\upmu \spacy Q$, $\bar{P}^i = P^i - \alpha \spacy Q^i$, and $\bar{J}^i = J^i + \tilde{Q}^i$. The number of broken generators is now ten, but with the same two commutators as in Eq.~\eqref{eq:commsolids}. The number of physical Goldstone excitations is then four. Indeed, the symmetry breaking pattern can be implemented via four scalar fields, $\psi(x)$ and $\phi^I(x)$, acquiring  ground state expectation values that are that of a superfluid for the former, and those of a solid for the latter:
\begin{align}
    \langle \psi(x) \rangle = \bar{\upmu} \spacy t \,, \qquad \text{ and } \qquad \langle \phi^I(x) \rangle = \alpha \spacy x^I \,.
\end{align}
There are now three possible Lorentz invariants that one can build out of these fields, with the lowest possible number of derivatives:
\begin{align}
    X = \sqrt{-\partial_\mu \psi \spacy \partial^\mu \psi} \,, \qquad B^{IJ} = \partial_\mu \phi^I \spacy \partial^\mu \phi^J \,, \qquad V^I = \partial_\mu \psi \spacy \partial^\mu \phi^I \,,
\end{align}
where the last one is the only new ingredient as compared to superfluids and solids. The most general action for supersolids is then given by,
\begin{align}
    S = \int d^4x \spacy F\big(X, B^{IJ}, V^I\big) + \mbox{higher $\partial$'s} \,,
\end{align}
where the $I,J$ indices must be contracted into $SO(3)$-invariant combinations. Recently, a number of experiments have observed supersolid behavior~\cite[e.g.,][]{Li2017,Leonard:2016gtj,tanzi2019observation,guo2019low}, while the EFT describing such systems is discussed in~\cite{Son:2005ak}.

As for solids, one can again promote the internal $ISO(3)$ symmetry acting on the comoving coordinates to the larger group of volume preserving diffeomorphisms, without the need for additional Goldstone modes. The resulting EFT is nothing but that of a finite temperature superfluid, discussed in Section~\ref{sec:finiteTsuperfluids}.

\subsubsection{Galileids}

\noindent The first class of exotic phases of matter are the so-called ``galileids'', introduced in~\cite{Nicolis:2015sra}. These come in two types, defined by the following symmetry breaking patterns:
\begin{align*}
    \text{Type-I galileids: } & \qquad \bar P^0 = P^0\,, \quad \bar{P}^i = {P}^i - \beta \spacy Q^i \,, \quad \bar{J}^i = J^i \,, \\
    \text{Type-II galileids: } & \qquad \bar P^0 = P^0 - \gamma \spacy Q \,, \quad \bar{P}^i = {P}^i - \beta \spacy Q^i \,, \quad \bar{J}^i = J^i \,,
\end{align*}
where, as usual, $\beta$ and $\gamma$ are some a priori generic parameters, that will essentially characterize the ground state of the system.
Both these declinations share some common properties and obstructions, and we will then discuss them together.

First of all, in order for the unbroken generators to satisfy the algebra~\eqref{eq:unbrokenalgebra}, it must be,
\begin{align}
    \big[ J^i, Q^j \big] = i \epsilon^{ijk} Q^k \,.
\end{align}
The $Q^i$'s do not commute with spatial rotations, but rather transform as 3-vectors. Consequently, they cannot be the generators of an internal symmetry: they must generate a spacetime transformation. This alone makes the galileids stand out with respect to all the other systems presented so far. On the simple basis of spacetime covariance, we are led to conclude that there must also be an additional generator, $Q$, extending the previous ones to a full Lorentz 4-vector, $Q^\mu = (Q, Q^i)$. For a type-II galileid, the new generator is also broken, and it is precisely the one compensating for the breaking of the original time translations. For type-I galileids, instead, $Q$ must remain unbroken, just like $P^0$. As discussed in~\cite{Nicolis:2015sra}, this makes it harder to implement the corresponding symmetry breaking pattern at the level of fields and one must resort, for example, to a reducible representation of the Lorentz group. In this instance, however, a systematic classification of all possible invariants of the full symmetry group is a much less straightforward task. Nonetheless, one can employ the coset construction to show that the most general theory for both type-I and type-II galileids is inevitably plagued by instabilities~\cite{Nicolis:2015sra}.

These instabilities can be avoided in a minimal way, by adding a single internal generator, $D$, which commutes with all spacetime generators, but such that,
\begin{align}
    \big[ Q^\mu, P^\nu \big] = -i \eta^{\mu\nu} D \,.
\end{align}
This is the algebra of galileon theories~\cite[e.g.,][]{Nicolis:2008in,Trodden:2011xh,deRham:2012az,Deffayet:2013lga}, which were originally proposed as possible modifications of gravity. From this, the dubbing of these systems. At the field level, this is implemented by a single real scalar transforming as,
\begin{align}
    \phi(x) \to \phi(x) + c + b_\mu x^\mu \,,
\end{align}
where the $c$ shift is generated by $D$, while the $b_\mu x^\mu$ one by $Q^\mu$. 

Now, in the case for type-I galileids, where time translations are unbroken, the desired symmetry breaking pattern is implemented by the following background configuration for the scalar field,
\begin{align} \label{eq:bkggalileidI}
    \langle \phi(x) \rangle = \frac{1}{2} \beta {| \vec{x} |}^2 \,.
\end{align}
Phrased this way, the system clearly features a single excitation mode, i.e., the fluctuation of the scalar field on top of the above background profile. Indeed, we now break seven generators ($K^i$, $P^i$ and $D$) but we have six commutators of the kind~\eqref{eq:inverseHiggscondition},
\begin{align} \label{eq:inverseHiggsgalileid}
    \big[ \bar P^0, K^i \big] = i P^i \,, \quad \text{ and } \quad \big[ \bar P^i, P^j \big] = i \spacy \beta \spacy \delta^{ij} D \,,
\end{align}
thus leaving us with only one Goldstone.

A very similar situation is true for type-II galileids. Since we must now break time translations as well, the correct background configuration that does that is,
\begin{align} \label{eq:bkggalileidII}
    \langle \phi(x) \rangle = \frac{1}{2} \spacy \beta \spacy {|\vec{x}|}^2 + \frac{1}{2} \spacy \gamma \spacy t^2 \,.
\end{align}
In this case the number of broken generators is eight ($K^i$, $P^0$, $P^i$ and $D$). However, on top of the commutators already presented in Eq.~\eqref{eq:inverseHiggsgalileid}, we now have an additional one,
\begin{align}
    \big[ \bar P^0, P^0 \big] = i \spacy \gamma \spacy D \,.
\end{align}
This allows to eliminate the Goldstone associated to the additional broken generator, keeping the final number down to just one.

So, can galileids exist in Nature? Clearly, even if they could in principle exist, they have not been observed in any of the systems realized in lab or observed indirectly so far. Moreover, they can be free of instabilities, for the same reasons that apply to a generic galileon theory~\cite[e.g.,][]{Nicolis:2008in,Deffayet:2009mn,DeFelice:2010nf,Andrews:2010km}. Nonetheless, were they to be found, they would feature a rather exotic property: their background stress--energy tensor would be inhomogeneous. Indeed, it is possible to show that at low energies their stress--energy tensor takes the following schematic form~\cite{Nicolis:2015sra}:
\begin{align}
    T_{\mu\nu} \sim \partial \phi \spacy \partial \phi \left( \partial\partial \phi \right)^{n-2} \,,
\end{align}
where $n=2,3,\dots$, and we have been cavalier about the possible contractions of the 4-vector indices coming with each derivative. Now, when evaluated on the backgrounds in Eqs.~\eqref{eq:bkggalileidI} and \eqref{eq:bkggalileidII}, this leads to,
\begin{align}
    \left\langle T_{\mu\nu} \right\rangle \sim x^2 \,,
\end{align}
where $x$ can be either time or space. This is clearly not translationally invariant. Thus, were they to exist, galileids would be characterized by very peculiar energies and pressures.

\subsubsection{Framids}

\noindent We conclude this section with a speculation. Looking back, the attentive reader might have noticed that two symmetry breaking patterns were left out from our previous discussions. Specifically, given our broad definition of ``condensed matter'' as systems who break boosts, the last possibilities are,
\begin{align*}
    \text{Type-I framids: } & \qquad \bar{P}^0 = P^0 \,, \quad \bar{P}^i = P^i \,, \quad \bar{J}^i = J^i \,, \\
    \text{Type-II framids: } & \qquad \bar{P}^0 = P^0 \,, \quad \bar{P}^i = P^i \,, \quad \bar{J}^i = J^i + \tilde{Q}^i \,.
\end{align*}
At the field level, the first instance is implement by a single Lorentz 4-vector field, acquiring the expectation value $\langle A_\mu(x) \rangle = \delta_\mu^0$. The second instance is instead realized by a triplet of Lorentz 4-vectors, with $\langle A^a_\mu(x)\rangle = \delta_\mu^a$~\cite{Nicolis:2015sra,Rothstein:2017twg}.

We can see immediately that these correspond to some very special systems. Let's consider, for example, type-I framids. These break no other spacetime symmetry but boosts. This means that it doesn't even make sense to talk about the idea of displacing whatever ``stuff'' makes them up, in neither time nor space. This is because spacetime translations are unbroken symmetries, and thus there is no physical difference between different points in space and different instants in time. The same is true for rotations. The only physically sensible thing one can do to a type-I framid is to change its local {\it velocity}, as boosts are indeed broken. This amounts to a change in the local reference frame, hence the name ``framids''. The same thing applies to type-II framids, with the addition that one can also change their local orientation, due to the breaking of rotations. 

Does this make framids unphysical systems? Not necessarily. Of course, an almost {\it classical} condensed matter system could never feature the properties described above. It would necessarily be made of some constituents which, when collected in small volumes, can clearly be displaced in both time and space as a physically meaningful action. However, this might not be true for systems where {\it quantum} effects are important. For instance, let's consider a superfluid, like ${^4}{\rm He}$. On the one hand, it is fundamentally made up of atoms which can be individually displaced in space and time. On the other hand, after condensation, the collective behavior of these atoms is described by a single degree of freedom: the scalar field $\psi$. Indeed, at long distances spatial translations are unbroken: they are not a physical action anymore, only time translations are. Therefore, the symmetry breaking pattern of framids could be present, at least in principle, in systems whose ground state is determined predominantly by quantum effects.

What really makes framids hardly realizable in Nature (though not rigorously excluded) is their intrinsically relativistic nature, and their absence of thermodynamics, as we now explain. As shown explicitly in~\cite{Nicolis:2015sra,Kourkoulou:2021ksw}, the low energy equilibrium stress--energy tensor of framids takes necessarily the form of a cosmological constant, i.e.,
\begin{align}
    \left\langle T_{\mu\nu} (x) \right\rangle = \Lambda \spacy \eta_{\mu\nu} \,.
\end{align}
Therefore, their background configuration must be characterized by a highly relativistic pressure,
\begin{align} \label{eq:framidpressure}
    \bar p = - \bar \rho = \Lambda \,.
\end{align}
This alone makes framids hardly realizable in lab, where one has only access to nonrelativistic constituents, implying pressures that are much smaller than energy densities. It does not, however, exclude the possibility of realizing them in genuinely relativistic contexts, such as those offered by astrophysics and/or early time cosmology.

Nonetheless, the really peculiar feature of framids is that they do not allow for thermodynamics.\footnote{As explained in~\cite{Nicolis:2015sra}, the lack of themodynamics is actually closely related to the relation~\eqref{eq:framidpressure}.} Indeed, their ground state does not feature any free parameter. In all phases of matter considered in this work, one can manipulate the boundary conditions to change these parameters, which results in different, physically inequivalent states of the same system. For example, one such free parameter is the equilibrium chemical potential of a superfluid, $\bar{\upmu}$. Different values of the chemical potential correspond to different values of energy and pressure, while still maintaining the superfluid nature of the system. Formally, this happens in two ways (often, but not necessarily related):
\begin{itemize}
    \item From the symmetry breaking pattern viewpoint, any time a spacetime translation is broken, a free parameter appears in the definition of the unbroken $\bar{P}^\mu$. This is the case for the chemical potential, $\bar{\upmu}$, in superfluids and supersolids (see Sections~\ref{sec:superfluids} and \ref{sec:supersolids}). But it is also the case for the parameter $\alpha$ appearing in solids, fluids and supersolids (see Sections~\ref{sec:solids}, \ref{fluids} and \ref{sec:supersolids}). This characterizes the level of compression of the mechanical degrees of freedom of the system or, if properly generalized, the level of shear (see Section~\ref{sec:solidphonons}).\footnote{Note that these free parameters do not appear in the unbroken rotations, $\bar{J}^i$. This is due to their non-Abelian nature, which forces a unique combination of the original spacetime generators, $J^i$, and additional generators, $\tilde{Q}^i$.}

    \item From a field perspective, when such free parameters are present, they appear in the background value taken by the relevant fields---see, e.g., Eqs.~\eqref{solid equilibrium}, \eqref{superfluid background}, and \eqref{eq:bkggalileidII}. More specifically, they label a family of physically inequivalent homogeneous and isotropic solutions to the background equations of motion. Different solutions can be connected by the introduction of Goldstone field profiles with a nontrivial spacetime dependence~\cite[e.g.,][]{Nicolis:2015sra}. For example, starting from a superfluid state characterized by a chemical potential $\bar\upmu$, one can obtain a different superfluid state, characterized by a chemical potential $\bar\upmu + \delta\bar\upmu$, by shifting the original Goldstone field by a term linear in time (see, e.g., Eq.~\eqref{eq:superfluidgoldstone}),
    \begin{align}
        \pi(x) \to \pi(x) + \frac{\delta\bar\upmu}{\bar\upmu} \left( t + \pi(x) \right) \,.
    \end{align}
\end{itemize}
In the case of framids, no such free parameters appear, neither in the symmetry breaking pattern, nor in the background field configurations. This seems to imply that their state cannot be changed by a tuning of the boundary conditions, for example, squeezing them, shearing them, and so on. They essentially posses only one possible equilibrium state, as opposed to all other known condensed matter systems, which posses a continuum of them. In light of all this, it would be quite surprising to see framids realized as a physical system, although it would not be the first time that Nature surprises us.

\subsection{Further Readings}

\noindent The more formal developments discussed in this section, beside highlighting a unifying, symmetry-based viewpoint on different material media, offer ideas for possible exotic field theories. These have been used in a number of fields, not directly related to condensed matter.

\paragraph{Gapped Goldstones} Quite some effort has gone towards understanding the physics of the so-called gapped Goldstones, which we only briefly mentioned in this section. These are {\it exact} Goldstone, as they realize nonlinearly an exact symmetry of the system. Their energy, however, does not vanish in the $\vec{k} \to 0$ limit. The possibility of their existence was first pointed out in~\cite{Morchio:1987aw, Nicolis:2012vf}, where the authors identified the necessity of being at finite density for a broken non-Abelian charge, together with the fact that the Goldstones' gap is completely universal, and fixed by the chemical potential and by the symmetry algebra. The picture was further expanded in~\cite{Nicolis:2013sga}, where the coset construction was employed to unravel the existence of additional types of gapped Goldstone, whose gap is not universal anymore, but rather depends on the details of the symmetry breaking mechanism. Counting rules applying to the number of gapped Goldstones have been proved in~\cite{Kapustin:2012cr,Watanabe:2013uya,Hayata:2014yga}, where the authors have also analyzed the relation between gapped Goldstones and the more traditional gapless ones, with both linear and quadratic dispersion relations. A number of examples where gapped Goldstones arise have instead been presented in~\cite{Son:2000xc,Miransky:2001tw,Watanabe:2013uya,Catinari:2024qap}, though not always framing them as gapped Goldstones.\footnote{In fact, some of these works predate~\cite{Nicolis:2012vf}.} These examples include QCD at finite isospin density, and (anti-)ferromagnets in presence of an external magnetic field. An EFT for soft gapless Goldstones and slow gapped ones have been, instead, developed in~\cite{Cuomo:2020gyl}. This describes gapped Goldstones even when their gap is of order of the  EFT cutoff, allowing not to integrate them out, which would result in losing all the information about the original symmetry group, which they realize nonlinearly. The study of the properties of scattering amplitudes involving gapped Goldstones has been done in~\cite{Brauner:2017gkr}.

\paragraph{Chiral superfluids} A good deal of EFT attention has gone to ``chiral'' superfluids. These are superfluids that, besides the usual $U(1)$ symmetry, also break parity and time-reversal. These are made, for example, of Cooper pairs bound in a state of angular momentum $L=1$. In three spatial dimensions, many of them are also anisotropic, such as the $A$-phase of bulk ${^3}{\rm He}$, as already mentioned. The EFT for this has been developed in~\cite{Furusawa:2020pjq}, with special emphasis on the determination of the corresponding Hall viscosity. The finite temperature case, where the normal component is characterized by a macroscopic motion with respect to the superfluid one, has instead been discussed in~\cite{Selch:2024yls}. In the case of two spatial dimensions, instead, isotropy is preserved, with the symmetry breaking pattern being $U(1) \times SO(2) \to U(1)$. Consequently, there exists a single Goldstone mode, and the corresponding EFT has been studied in~\cite{Hoyos:2013eha,Moroz:2014ska,Golan:2019svj}, while the instance where the two-dimensional space is curved has been discussed in~\cite{Moroz:2015cft,Jiang:2021grx}. The authors of~\cite{Brauner:2018xhh} have instead studied the ground state of the $A$-phase of a thin film of ${^3}{\rm He}$ under an external magnetic field perpendicular to the film itself. Interestingly, the spin degrees of freedom develop a non-uniform helical structure. The relation between p-wave chiral superfluids and standard spinless bosonic superfluids in the presence of a strong magnetic field is, instead, discussed in~\cite{Hsiao:2024rbg}.

\paragraph{Other media in cosmology} Some of the exotic phases of matter described here (or their generalizations) have been proposed as possible states coupled to gravity. One such proposal involves the so-called gaugids~\cite{Piazza:2017bsd}. These bear a number of similarities with both solids and framids, and allow for an uncommonly large tensor-to-scalar ration. Another possibility is the so-called Einstein--aether theories~\cite{Jacobson:2000xp,Jacobson:2004ts,Carroll:2004ai,Armendariz-Picon:2010nss}, which are nothing but framids coupled to gravity. (In fact, they were proposed before framids were discussed as a possible phase of matter.) Inflation driven by a solid or supersolid phase of the Universe has, instead, been investigated in~\cite{Endlich:2012pz, Ricciardone:2016lym,Celoria:2020diz,Celoria:2021cxq}. Among other peculiarities, these models allow for a blue-tilted tensor power spectrum---essentially, gravitational waves---, with more power expected at small scales. Moreover, compared to more standard single field inflationary models, primordial non-Gaussianities can be strongly enhanced. The systematic work of coupling the phases of matter discussed in this section to gravity, determine whether or not they allow for a period of quasi-de Sitter expansion, and studying the associated cosmological perturbations, has instead been done in~\cite{Cabass:2021iii,Celoria:2017bbh}, while the study . Finally, the classification of shift symmetric, boost-breaking scalar theories in cosmology has been performed in~\cite{Grall:2020ibl}, where the authors also identify a new class of galileid states.

\paragraph{Holography} Contrary to solids, fluids and superfluids, the literature on the holographic principle applied to the other media presented in this section is rather small. To the best of our knowledge, this has only been done for supersolids. Holography has been employed to study their dynamical and thermodynamical properties~\cite{Baggioli:2022aft}, as well as their generalization to spacetime supersolids via the introduction of a time-dependent chemical potential~\cite{Yang:2023dvk} (much analogous to time crystals~\cite[e.g.,][]{Shapere:2012nq,Wilczek:2012jt}). Other holographic theories with properties resembling those of supersolids have been identified~\cite{Kiritsis:2015oxa,Yang:2023vxz}, although their interpretation as such is still not completely clear.

%% file: education.tex
\section{Effective Field Theories in the Classroom} \label{sec:edu}

\noindent It is interesting to ponder whether the viewpoints, techniques, and results reviewed here can suggest novel ways to teach some topics in physics. In fact, it is fair to say that there is a lot of inertia in the way we teach physics, especially at the undergraduate level, where we tend to perpetuate subfield-specific traditional derivations of classic results, which, their importance notwithstanding, tend to hide the unity of physics---the overarching structure that connects all subfields as well as  contemporary approaches to physics research. From this viewpoint, we feel that a field-theoretic, action-based approach to the physics of ``matter", like the one we have emphasized here, can offer useful alternative ways to think about standard properties of, for instance, nonrelativistic, classical matter. To make the point concrete, we want to offer two examples.


\subsection{Nearly Incompressible Perfect Fluids}

\noindent In many situations in nonrelativistic hydrodynamics, one takes the incompressible  limit. Incompressibility is usually presented as an approximate property of a given fluid, such as water. In reality, it is a property of a given kinematical regime for any fluid, as we now discuss. An incompressible fluid, as the name honestly advertises, does not like to be compressed, meaning that it requires large changes in pressure for even tiny changes in density. A measure of incompressibility is, in fact, the speed of sound, $c_s^2 = dp/ d \rho$, which is a dimensionful quantity.\footnote{This is obscured by natural units, which should be used with care, especially when Lorentz invariance is broken.} As such, it can only be small or large in comparison with other physical quantities with the same units. 
The precise statement is then that any fluid behaves as nearly incompressible, as long as its fluid flows---i.e., the local velocities of its volume elements---are slow compared to the speed of sound, $v \ll c_s$. 
While this condition can be satisfied by both transverse and longitudinal configurations, it constrains these two sectors very differently. Vortical, transverse modes can remain fully nonlinear while satisfying it, since their velocity is unconstrained by the size of their spatial gradients. Indeed, we saw in Section~\ref{sec: hydro modes fluid} that there is a linearized solution to the equations of motion for which the transverse hydrodynamical modes have a time-independent velocity profile $\vec v_T(\vec x) = -\dot{\vec \pi}_T(t,\vec x) = -\vec h(\vec x)$---see Eqs.~\eqref{transverse mode solutions} and \eqref{eq: u hydro modes}. This can be made arbitrarily small, and in particular much smaller than $c_s$, without having to require anything neither about the size of its gradients, nor about the amplitude of the field which, by virtue of Eq.~\eqref{transverse mode solutions}, can be made arbitrarily large. Longitudinal, compressional modes, by contrast, obey the dispersion relation $\omega_{\vec k} = c_s k$, which ties their velocity directly to their gradients, $\vec v_L(\vec x,t) = -\dot{\vec \pi}_L(\vec x,t) \sim c_s \vec\nabla \vec \pi_L(\vec x,t)$. Making $v_L$ much smaller than $c_s$ therefore forces the profile $\vec \pi_L$ itself to have small gradients, and hence small time derivatives, i.e., an almost trivial profile.
From our field theory viewpoint, this suggests the following interpretation of the incompressible limit: there are some kinematical regimes characterized by slow fluid flows, $v \ll c_s$, where vortices can have rich, nonlinear, complicated dynamics, while compressional modes are fully in the linear regime. Any nontrivial compressional mode has finite energy, and can thus be integrated out in first approximation. This also suggests how to  systematically compute corrections to the incompressible fluid limit: one should deal with the linear compressional modes in perturbation theory, applying the perturbative expansion of field theory. This approach was initiated in~\cite{Endlich:2010hf}, and applied to a number of physical situations in~\cite{Endlich:2013dma}. 

Consider for example a nonrelativistic fluid in Lagrangian formulation:
\begin{align} \label{eq: S[X] incompressible fluids}
    S[X] = \int d^3 \phi \spacy dt \,  \rho_0 \left[ \frac12 \dot{\vec X}\, ^2 - V \big( \big|\partial X / \partial \phi \big| \big) \right] \,,
    \qquad \text{ with } \qquad  \vec X \equiv \vec X(\phi, t) \,,
\end{align}
where the $\phi^I$'s are the comoving (i.e., Lagrangian) coordinates, $X^i$ are the physical (i.e., Eulerian) ones, $\rho_0 = {\rm constant}$ is the comoving mass density, and $V$ is the fluid's internal energy per unit mass. This action can be derived by taking the nonrelativistic limit of the effective action in Eq.~\eqref{action LF3fields}, i.e., the limit $\dot{\vec{X}} \ll 1$ and $c_s \ll 1$.\footnote{In light of the discussion in Section~\ref{sec: nonrelativistic fluid}, this action describes a fluid in the isentropic regime with a constant ratio~$\sigma = s/n$.} Alternatively, it can be obtained directly by thinking of a nonrelativistic fluid as a collection of infinitesimal volume elements of mass $\rho_0 \, d^3 \phi$, with a potential energy that depends only on the local compression level, which is measured by the Jacobian of the mapping $\phi^I \leftrightarrow X^i$.
The volume-preserving diffeomorphisms symmetry acting on $\phi^I$ is manifest, since $d^3 \phi$, $\dot {\vec X}$, and $\big|\partial X / \partial \phi \big|$ are all invariant. 

As a check, let us show that the action above reproduces the known equation of motion. Indeed, the equation of motion obtained by varying the action \eqref{eq: S[X] incompressible fluids} with respect to the fields $X^i$ can, after some massaging, be put in the form
\begin{align}
    \ddot { X}^i - \det \mathcal{J} \, (\mathcal{J}^{-1})_i {}^I \frac{\partial}{\partial \phi^I}  V'(\det \mathcal{J}) = 0 \,, \qquad  \text{ with } \qquad  \mathcal{J}_I {}^i \equiv \frac{\partial X^i}{\partial \phi^I} \,.
\end{align}
Notice that the matrix $\mathcal{J}$ is the inverse of the matrix $J$ introduced in Section~\ref{sec:solids}. Upon using the standard relationship between time derivatives at constant $\phi^I$ and time derivatives at constant $\vec x = \vec X(\phi, t)$, and noticing that the first law of thermodynamics implies that $V$ is related to the pressure by $p = - \rho_0 V'(\det \mathcal{J})$, the equation of motion above can be rewritten as Euler's equation,
\begin{align}
    \partial_t \vec v + \big(\vec v \cdot \vec \nabla \, \big) \vec v = - \frac{1}{\rho_m} \vec \nabla p \,, 
\end{align}
where $\vec v $ and $p$ are to be thought of as functions of $\vec x$ and $t$. The mass density $\rho_m$ is related to the comoving mass density by $\rho_0 = \rho_m \spacy \det \mathcal{J}$.

Now, just by looking at the action~\eqref{eq: S[X] incompressible fluids}, it is clear that there is a qualitative difference between fluid configurations that do not change the local density of the fluid and those that do. To be more precise, consider   a fluid that at infinity, or at the boundary of the volume it occupies, has constant mass density, say  $\det \mathcal{J} =1$. We can ask what is the energy of a generic fluid configuration with these boundary conditions. We notice that the first variation of $V$ about a configuration with  $\det \mathcal{J} =1$ is a total spatial derivative, which doesn't contribute to the action:
\begin{align}
    d^3 \phi \, \delta V \simeq d^3 \phi  \, V'(1) \delta (\det \mathcal{J}) = d^3 \phi \, V'(1) \big(\mathcal{J}^{-1}\big)_i {}^I \frac{\partial \delta X^i}{\partial \phi^I} = d^3 x \, V'(1) \, \frac{\partial \delta x^i}{\partial x^i} \,,
\end{align}
where we used the facts that $\delta\left(\det \mathcal{J}\right) = \det \mathcal{J} (\mathcal{J}^{-1})_i{}^I\left(\delta \mathcal{J}\right)_I{}^i$, $d^3x = d^3\phi \det \mathcal{J}$, together with $\det \mathcal{J} = 1$.
And so, with fixed boundary conditions, the contribution of compressional modes to the total internal energy is sign-definite, at least for perturbative compressions or dilations, because it starts at second order:
\begin{align}
    \int d^3 \phi \, \delta V \simeq \int d^3 \phi  \, \frac 12 V''(1) (\delta \det \mathcal{J})^2 \,.
\end{align}
In particular, if the fluid is stable, this must be positive definite. On the other hand, fluid configurations that do not change the local density contribute {\em zero} to the internal energy, and only carry kinetic energy, which can be made arbitrarily small by taking slower and slower flows, regardless of how large and complicated a displacement $\vec X(\phi, t)$ can build up over time. This is nothing but the explicit verification of what we discussed heuristically at the beginning of this section: volume-preserving vorticose solutions can be arbitrarily complicated while still contributing to the fluid's incompressible configurations. Nontrivial compressional solutions, instead, always carry finite energy. 
The equations of motion couple these two kinds of motion but, as we will see below, the coupling involves directly the velocity of the volume-preserving flow, because static volume-preserving deformations do not contribute to the action, thanks to the volume-preserving diffeomorphism symmetry.

The coefficient $V''(1)$ above happens to be precisely the squared speed of sound, $c_s^2$, and the solutions for the compressional modes are, of course, sound. We thus reach the conclusion that any fluid can sustain near-incompressible flows as long as their typical speed is much smaller than the speed of sound, and corrections to this behavior
can be systematically computed treating sound waves in perturbation theory. Explicitly, one can formally decompose any fluid configuration as
\begin{align}
    \vec X(\phi, t) = \vec  x_0(\phi, t) + \vec \psi\big(\vec x_0(\phi, t), t\big) \,,
\end{align}
where, at any given time, $x_0^i$ is a volume preserving diffeomorphism of $\phi^I$, and $\vec \psi$---the compressional mode---is longitudinal with respect to~$\vec x_0$~\cite{Endlich:2013dma}: 
\begin{align}
\det \frac {\partial x_0^i}{\partial \phi^I} = 1 \,, \qquad \vec \nabla_{x_0}  \times \vec \psi(\vec x_0, t) = 0 \,.
\end{align}
Plugging into the Lagrangian, changing integration variables when convenient, and integrating by parts, one finds
\begin{align}
    S[X] = S_0[x_0] + S_{\rm sound}[\psi] + S_{\rm int}[ x_0, \psi ] \,,
\end{align}
where
\begin{itemize}
    \item $S_0$ describes the incompressible flow:
    \begin{align}
        S_0[x_0] = \int d^3\phi \spacy dt \,   \frac{1}{2} \spacy \rho_0 \spacy \dot{\vec x}_0^{\spacy 2} \, .
    \end{align}
    The simplicity of this action is deceiving, in that one has to remember that $\vec x_0$ is a volume-preserving diffeomorphism of $\phi^I$. To make this explicit, one can introduce a Lagrangian multiplier $p(\phi, t)$, and add
    \begin{align}
        p(\phi, t) \left(\!\det \frac {\partial x_0^i}{\partial \phi^I} - 1 \right)
    \end{align}
    to the Lagrangian. The equation of motion one obtains in this way is exactly Euler's equation, with $p$ playing the role of pressure (hence the name). Solving such equations of motion in general is the difficult part of fluid dynamics.
    \item $S_{\rm sound}$ describes the dynamics of sound waves, including their self interactions:
    \begin{align}
        \begin{split}
        S_{\rm sound}[\psi] & = \rho_0 \int d^3 x_0 dt \left[ \frac{1}{2} \dot{\vec \psi} \,^2 - V\big(\det(1 + \partial_i \psi^j) \big) \right] \\
        & = \rho_0 \int d^3 x_0 dt \,   \frac{1}{2} \Big( \dot{\vec \psi} \,^2 -  c_s^2 {(\vec \nabla \cdot \vec \psi \,)}^2 \Big)+ \dots \,,
        \end{split}
    \end{align}
    where the spatial derivatives are taken with respect to~$\vec x_0$, and the dots stand for terms with higher power of  $\partial \psi$.
    Here one must remember that $\vec \psi$ is purely longitudinal. This is most easily implemented by rewriting it as the gradient of a scalar,
    \begin{align}
        \vec \psi (\vec x_0, t) = \vec \nabla \Psi(\vec x_0, t) \,,
    \end{align}
    and working with the field $\Psi$.
    \item $S_{\rm int}$ couples the vorticose and compressional sectors:
    \begin{align}
        S_{\rm int}[ x_0, \Psi ]  = \rho_0 \int d^3 x_0 dt \, v_0^i v_0^j \, \partial_i \partial_j \Psi + \dots \,,
    \end{align}
    where $\vec v_0 = \dot{\vec x}_0 (\phi, t)$ is the vorticose part of the velocity field, which has zero divergence with respect to~$\vec x_0$, and the dots stand for terms that are quadratic in $\Psi$.
\end{itemize}
This reorganization of the action is an {\em exact} rewriting of perfect fluid dynamics---it just comes from a field redefinition---but, as we emphasized, it allows for straightforward perturbative computations of small effects involving the compressional modes for slow vorticose fluid flows. 

For example, one can easily compute the power emitted in sound waves by a vorticose source with $v_0 \ll c_s$~\cite{Endlich:2013dma}, which matches Lighthill's classic result \cite{lighthill1954sound}---see also Eq.~\eqref{eq: vortex radiated power}:
\begin{align}
    P \simeq \frac{\rho_0}{2 \pi \, c_s^5} \bigg[ \frac{1}{2} \big\la \ddot Q \,^2 \big\ra + \frac{1}{15} \big\la \ddot Q_{ij} {}^2 \big\ra\bigg] \,,
\end{align}
where the angular brackets denote time-average, and $Q$ and $Q_{ij}$ are suitable ``kinetic multipoles'', 
\begin{align}
    Q = \int d^3 x_0  \, \frac13 v_0^2 \,, \qquad   Q_{ij} = \int d^3 x_0  \Big(v_0^i v_0^j - \frac13 v_0^2 \Big)  \,.
\end{align}

As a more original application of these techniques, one can integrate out the compressional mode $\Psi$, and end up with a sound-mediated long-range interaction between vortices, which should be added to the purely incompressible action $S_0$. 
For two well separated vorticose regions, to lowest order in the multipole expansion one finds~\cite{Endlich:2013dma}
\begin{align}
    \Delta S_0 = - \int dt \, \frac{\rho_0}{8\pi \, c_s^2} \, K_1^{ij} \, K_2^{kl} \, \partial_i \partial_j \partial_k \partial_l \spacy r  \,,
\end{align}
where $r$ is the distance between the two vortices, and for each of them
\begin{align}
    K_{ij} = Q_{ij} + Q \spacy \delta_{ij}  =  \int d^3 x \, v_0^i v_0^j \,.
\end{align}
It is interesting to estimate the importance of such a correction, which also shows the regime of validity of the approximations involved: for two vortices of typical size $\ell$  and typical speed $v$, we have
\be
\frac{\Delta S_0}{S_0 } \sim \left(\frac{\ell}{r}\right)^3 \left(\frac{v}{c_s}\right)^2 \; .
\ee
The first suppression factor has to do with the multipole expansion, which is under control for $\ell \ll r$; the second suppression factor has to do with the near incompressible approximation, which, as expected, is under control for $v \ll c_s$.


\subsection{The Archimedean Principle and Generalizations thereof}

\noindent We are all familiar with Archimedes' principle: any object immersed in a fluid is buoyed by a force equal to the weight of the displaced fluid. In this context, ``object'' usually means an ordinary solid object, impermeable to the fluid; also note that Archimedes' principle is a property of hydrostatics: the fluid and the object must be at rest in the same reference frame. 

Can one find generalizations of Archimedes' principle that apply to more general objects and in more dynamical situations? For instance, there are certain generalized ``objects'' that exist in all fluids and just cannot stay put: sound waves. They maximally violate the framework where Archimedes' principle works: {i)} They lack a sharp boundary with the surrounding fluid, so it is not obvious what their intrinsic volume and weight are; {ii)} They always travel at the speed of sound relative to the surrounding fluid; {iii)} They do not obey an equation of motion of the form $\vec F = m \spacy \vec a$, so it is not clear how the concept of a buoyant force can apply. 
Nonetheless, when one considers sound wave-packets much smaller than the typical scales over which physical properties like pressure and velocity vary, sound waves do follow well-defined trajectories. One can then ask: does a sound wave float or sink? 

One could approach this question by solving the wave equation for a fluid under gravity, while accounting for the fact that the local equilibrium pressure of the fluid depends on the local value of the gravitational field, which makes the pressure different at different points in space. Consequently, the sound speed must also depend on the local value of the gravitational field, and one expects to find modifications to the wave equation leading to refraction phenomena. From this viewpoint, any tendency of a sound wave to ``float'' or ``sink''---i.e., for its trajectory to bend upward or downward---would be unrelated to buoyancy. It would simply be standard refraction.
We find it more revealing, however, to approach the question using an effective point-particle theory for our sound wave-packet. This will allow us to recast refraction as the manifestation of a direct interaction between sound waves and gravity with, as we will see, far-reaching implications.

As a warmup, let us see how to implement this program in the case of an ordinary object in a fluid subject to a gravitational field.
First, one writes the most general Lagrangian or Hamiltonian for a point-particle compatible with the spacetime symmetries that are unbroken by the surrounding fluid, in the absence of external fields or perturbations in the fluid \cite{Nicolis:2017eqo}. For a homogenous fluid at rest at a given mass density $\rho_m$, the unbroken symmetries are spatial rotations and spacetime translations.\footnote{As we know by now, this is to be intended in the sense of the unbroken effective generators.} Galilei boosts are broken---the fluid picks out a reference frame. Thus, for a point-particle moving slowly relative to the fluid, the most general Hamiltonian is
\begin{align} \label{object}
    H(\vec p , \vec x) = E_0(\rho_m) + \frac{p^2}{2 m_{\rm eff}(\rho_m)} + {\cal O}\big(p^4\big) \, ,
\end{align}
where $E_0$ is a (nonrelativistic) rest-energy of the object, whose role will become clear in a moment, and $m_{\rm eff}$ is an effective mass, which generally differs from the mass the object would have outside the fluid: if the object moves in the fluid, it drags along a fraction of the fluid's mass, of order of the mass of the displaced volume, which effectively renormalizes the mass parameter entering the kinetic energy term~\cite{landau:1987bo}. 
Crucially, note that the parameters $E_0$ and $m_{\rm eff}$ generally depend on the density $\rho_m$ of the surrounding fluid.

Next, let us consider what happens in the presence of long-wavelength perturbations of the fluid's physical properties. 
For our purposes, we will still restrict to a static fluid configuration, but now in the presence of an external gravitational field, $\Phi(\vec x)$. To lowest order in its size, our object simply perceives a static, homogeneous fluid in its immediate surroundings, with a density that depends on the object's position. In hydrostatics, for fluids with a barotropic equation of state (pressure $P \equiv P (\rho_m)$), the local value of the density is related to the local value of the gravitational potential by the condition for hydrostatic equilibrium:
\begin{align} \label{Phi}
    c_s^2(\rho ) \, d \log \rho_m + d \Phi = 0 \, ,
\end{align}
which, up to an irrelevant integration constant for $\Phi$, can be integrated to yield $\rho_m \equiv \rho_m\spacy(\Phi)$.

Thus, the Hamiltonian of the object is still given by Eq.~\eqref{object}, with the understanding that $\rho_m$ now depends on the value of $\Phi$ at the object's position.
Then, the force acting on an object with negligible momentum immersed in a static fluid is
\begin{align}
    \vec F = \frac{1}{c_s^2} \frac{dE_0}{d\log \rho_m} \vec \nabla \Phi \, .
\end{align}
This expression should be interpreted as the {\em net} force acting on the object, given by the vector sum of the gravitational force and the buoyant force.
In other words, the object's net gravitational mass is
\begin{equation} \label{mg for object}
m_{\rm g} = - \frac{1}{c_s^2} \frac{dE_0}{d\log \rho_m} \, .
\end{equation}

We can now go back to the physical meaning of the rest energy $E_0(\rho)$: to place an object of small volume $V$ at a point in the fluid with a given $\rho$, one must do work $p(\rho)\, V$ against the pressure; moreover, if the object has mass $m$, it has a gravitational energy $m \, \Phi$. Combining the two contributions, one has
\begin{align}
    \frac{dE_0}{d \log \rho_m} = \frac{d p}{d \log \rho_m} V + m \frac{d {\Phi}}{d \log\rho_m} = c_s^2 \big[\rho_m V - m ] \, . 
\end{align}
Then, we see that the object's net gravitational mass, Eq.~\eqref{mg for object}, is nothing but its mass minus the mass of displaced fluid, in agreement with the Archimedean principle.

The generalization of this effective point-particle formalism to a sound wave-packet follows immediately. The Hamiltonian of a wave-packet of momentum $\vec p$ traveling in an unperturbed fluid at density $\rho_m$ is 
\begin{align} \label{eq:Hsound}
    H (\vec p, \vec x) = c_s(\rho_m) \spacy |\vec p \,|  \, .
\end{align}
Notice that unlike for the quantities $E_0$ and $m_{\rm eff}$ above, which depend on the object, the dependence of the sound speed on the density $\rho_m$ is uniquely determined by the fluid's equation of state.
In the presence of an external gravitational field $\Phi(\vec x)$, again applying Eq.~\eqref{Phi}, Hamilton's equations read
\begin{align} \label{eom}
    \dot{\vec x} = c_s \spacy \hat p \, , \qquad \dot {\vec p} = \frac{d \log c_s}{d \log \rho_m} \, \frac{E}{c_s^2} \, \vec \nabla \Phi \, ,
\end{align}
where $E = c_s |\vec p \,|$ is the energy carried by the wave. 
The first equation implies that, unsurprisingly, the instantaneous velocity of the wave-packet is aligned with $\vec p$, and its magnitude is the local value of the sound speed. The second equation tells us how the trajectory of the wave-packet bends under gravity. For ordinary equations of state the derivative on the right hand side is positive, as the speed of sound is larger for denser media. Thus, the trajectory tends to bend upwards, towards regions of higher gravitational potential. It is in this sense that we assert that sound waves tend to ``float''.

\begin{figure}[t]
\centering
\includegraphics[width=0.65\textwidth]{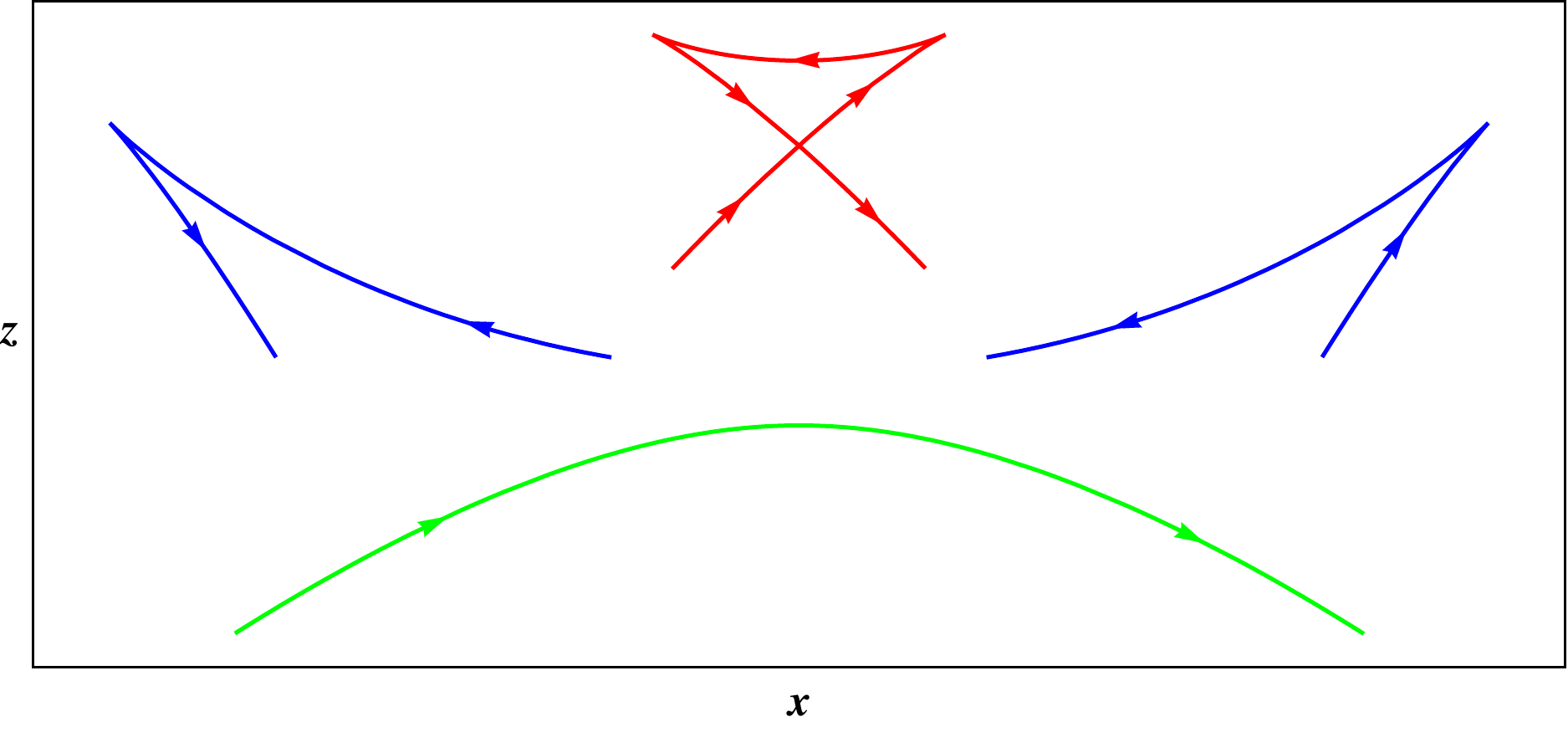} 
\caption{Possible trajectories of rotons under gravitational acceleration $\vec g=-g\hat z$, with horizontal momentum component directed rightwards. Depending on the value and orientation of their initial momentum, rotons can sink in peculiar ways~\cite{Nicolis:2017eqo}.} \label{rotons}
\end{figure}

The same framework can be applied to all types of collective excitations, including genuinely quantum mechanical ones, for which a straightforward application of Archimedes' principle initially seems even more outrageous (or hopeless). For instance, \cite{Nicolis:2017eqo}  considered rotons in a superfluid. Upon introducing gravity in the same manner as above, one finds that rotons generally tend to sink, but, given their relationship between momentum and velocity, they can follow very erratic trajectories, as illustrated in Figure~\ref{rotons}. 

Returning now to classical sound waves, Eqs.~\eqref{eom} are completely equivalent to what one could have derived from a more standard analysis of refraction~\cite{Nicolis:2017eqo}. 
However, our formalism shows that refraction can be thought of as the product of an {\em interaction} between a wave and {gravity}. Therefore, as for all interactions, not only is the wave affected by gravity, but {\em gravity} is affected in turn by the wave: associated with our wave-packet is a point-like source for gravitational fields. Its gravitational mass can be derived by varying our Hamiltonian in Eq.~\eqref{eq:Hsound} with respect to the gravitational potential. It is nothing other than the coefficient of $-\vec \nabla \Phi$ in Eq.~\eqref{eom},
\begin{align} \label{mg}
    m_{\rm g} = - \frac{d \log c_s}{d \log \rho_m} \spacy \frac{E}{c_s^2} \, ,
\end{align}
and it is generally negative. Notice that we never invoked relativistic mass--energy equivalence and this phenomenon is purely nonrelativistic. The implication is that, contrary to the received wisdom, sound waves carry with them a small fraction of the mass of the surrounding medium.

This claim can be verified, and generalized, in a number of ways \cite{Esposito:2018sdc}. 
For instance, one can consider the EFTs for fluids, superfluids, and solids discussed in this work, and couple them to a nontrivial spacetime metric, $g_{\mu\nu}(x)$. One can then derive the energy--momentum tensor and compute the spatial integral of the perturbation of $T^{00}(x)$ associated with a sound wave-packet. This gives the total energy of the wave-packet. 
To linear order, such an integral averages to zero over a few periods. However, taking into account quadratic corrections, the nonrelativistic limit yields precisely Eq.~\eqref{mg}. For solids, which admit both longitudinal and transverse waves, the correct expression is a minor generalization of \eqref{mg}~\cite{Esposito:2018sdc}.

We conclude that \emph{sound carries mass} in a very physical sense: it is a source of gravity, traveling at the speed of sound. Although never observed, there are instances where this tiny effect might become sufficiently enhanced to be accessible to measurement~\cite{Esposito:2018sdc}. 

%% file: conclusions.tex
\section{Final Thoughts and Acknowledgments} 

\noindent 
A few closing comments are in order.
First, we have presented a limited, partial set of viewpoints and results on a vast subject. Our review is not meant to be  exhaustive, even remotely.   
We have mostly focused on topics and examples we have personally worked on, not because we think that those are the most important ones, but because they are the ones we know and understand better, and so they are the only ones we are qualified to write about. 
Related to this, we strove for completeness in our list of references, but we only know a fraction of the literature, and so there are probably egregious omissions, for which we apologize.
Finally, some of the research described here is very much work in progress, and we did our best to share our present understanding of the subject matter. Things might evolve in unexpected directions, and viewpoints that sound compelling now might turn out to be na\"ive. Our review should thus be viewed not as the final word on these subjects, but rather as a status report.

Having gotten apologies out of the way, we also want to express our enthusiasm for the physics and the approaches described here: we find it fascinating that the language of relativistic quantum field theory, spontaneous symmetry breaking, and effective field theory lends itself so well to describing physical systems that are traditionally quite removed from the realm of particle physics. We hope that some of our readers  share our enthusiasm and  find our review interesting or useful.

We want to express our gratitude to our editor, Claudia~de~Rham, for inviting us to
write this review, and for her patience as we drifted past one deadline after another. The work presented here is the result of countless collaborations and discussions with several colleagues over the years, exception made for what is imprecise or even
wrong, which is our full responsibility. We then wish to thank Can~Onur~Akyuz, Lasma~Alberte, Tom\'a\v{s}~Brauner, Andrea~Caputo, Pier~Giuseppe~Catinari, Paolo~Creminelli, Gabriel~Cuomo, Luca~V.~Delacr\'etaz, Sergei~Dubovsky, Solomon~Endlich, Eren~Firat, Sebastian~Garcia-Saenz, Emma~Geoffray, Andrew~Gomes, Garrett~Goon, Walter~Goldberger, Thomas~Gre\'goire, Sean~A.~Hartnoll, Bart~Horn, Lam~Hui, Austin~Joyce, Jonghee~Kang, Andrei~Khmelnitsky, Ilia~Komissarov, Ioanna~Kourkoulou, Rafael~Krichevsky, Michael~Landry, Hong~Liu, Tom~Melia, Ermis~Mitsou, Daniel~McLoughlin, Alexander~Monin, Filippo~Nardi, Klaas~Parmentier, Shashin~Pavaskar, Federico~Piazza, Fulvio~Piccinini, Alessandro~Podo, Antonio~D.~Polosa, Rafael~Porto, Massimiliano~Riva, Riccardo~Rattazzi, Rachel~A.~Rosen, Ira~Z.~Rothstein, Luca~Santoni, Sergey~Sibiryakov, Dam~T.~Son, John~Staunton, Franco~Strocchi, Guanhao~Sun, Junpu~Wang, Sebastian~Will, and Shengjia~Zhou.

AE acknowledges support by the Italian Ministero dell’Universit\`a e della Ricerca through the FIS2 Starting Grant project LEAPS (CUP: B53C25003020001). AN was partially supported by the US Department of Energy (award no. DE-SC0011941). RP is supported by the US Department of Energy grant DE-SC0010118, and part of his work was carried out at the Kavli Institute for Theoretical Physics (KITP) and at the Aspen Center for Physics, which are supported respectively by the National Science Foundation grants PHY-2309135 and PHY-2210452.

%% file: app_strain_tensor.tex
\appendix

\section{The Strain Tensor and an Improvement thereof}\label{sec: strain}

\noindent For solids, the strain tensor is usually introduced in the following way \cite{LL_elasticity}. One starts by considering an {\em unstrained} solid, that is, a static solid at equilibrium at vanishing applied stresses, and introduces a generic deformation thereof. This can be parametrized with a displacememt field $\vec u(\vec x \,)$,  with $\vec x$ being a generic position in the unstrained solid, and $\vec x + \vec u(\vec x \,)$ being the new position of the corresponding infinitesimal volume element in the deformed (or {\em strained}) solid.

The infinitesimal squared distance between two  nearby points $\vec x$ and $\vec x + d \vec x$ in the unstrained solid,
\begin{align}
    d \ell ^2= d\vec x \cdot d \vec x \, ,
\end{align}
after the deformation becomes
\begin{align} \label{dl'}
    d\ell' {}^2= d\ell ^2+ 2 \, u_{ij} (\vec x) \, dx^i dx^j \, ,
\end{align}
with
\begin{align} \label{strain}
    u_{ij} \equiv \frac12 \bigg( \frac{\partial u^j}{\partial x^i} + \frac{\partial u^i}{\partial x^j} + \frac{\partial u^k}{\partial x^i} \frac{\partial u^k}{\partial x^j} \bigg) \, .
\end{align}
This is the strain tensor and, as clear from how it is introduced, it parametrizes how infinitesimal distances change under a  deformation of the solid.

Notice that nowhere in the derivation above is it assumed that the displacement field $\vec u$ and its derivatives are small. So, in principle one can use the strain tensor to measure infinitesimal distances in a deformed solid no matter how large the deformation is. In practice, however, one usually only considers small deformations and restricts the analysis to linear order, in which case the strain tensor reduces to
\begin{align} \label{linear strain}
    u_{ij}  \simeq \frac12 \big( \partial_i u_j + \partial_j u_i \big) \,, \quad \text{ with } \quad \partial_i u_j \ll 1 \, .
\end{align}
There is some degree of confusion in the literature about this point. For example, when Ref.~\cite{LL_elasticity} derives the first law of thermodynamics for solids as $dE = T dS -V \,  T_{ij} \, du_{ij}$, that $d u_{ij}$ is an infinitesimal change in the {\em linear} strain tensor \eqref{linear strain}, even though the linear approximation is never used in the derivation. This suggests that the nonlinear strain \eqref{strain} only plays a limited role for the dynamics of solids and, indeed, it is usually ignored in the literature.

In fact, one of the problems that we see with the definition above, is that such a definition confuses the different symmetries at play in a solid. There is the fundamental $SO(3)$ symmetry of physical space, under which the deformed positions $x^i + u^i(\vec x)$ must transform as vectors, and there is the symmetry group $G \subset SO(3)$ of the unstrained solid, under which the unperturbed positions $x^i$ transform as vectors. 
In a crystalline solid, which is anisotropic already in its unstrained equilibrium state, these two symmetries are different.

To clarify the issue and make progress on it, it is useful to insist on distinguishing the {\em comoving} coordinates---the $\vec x \,$'s of the unstrained solid---from the physical coordinates of the points of the deformed solid, $\vec x + \vec u(\vec x)$.
To this end, we revert to our notation:
\begin{align*}
    \phi^I : \quad & \mbox{comoving coordinates}\,, \\
    x^i : \quad & \mbox{physical coordinates of the deformed solid} \,.
\end{align*}
The advantage of this notation, with different indices on $\phi$ and $x$, is that implementing the symmetries is now straightforward: $x^\mu = (t, \vec x)$ transforms in the usual way under Poincar\'e, while the $\phi^I$ are Poincar\'e scalars. The rotation symmetry group of the solid's equilibrium state, $G \subset SO(3)$, acts on the comoving $I$ index, not on the spatial index $i$.

We already saw that, to lowest order in the derivative expansion, the matrix
\begin{align}
    B^{IJ} (x)= \partial_\mu \phi^I(x) \, \partial^\mu \phi^J(x)
\end{align}
is the most general Lorentz scalar that is also invariant under shifts of $\phi$ (which implement the continuum limit homogeneity of the solid's unstrained ground state). It transforms as a tensor under the internal rotation symmetry group $G$. It  can thus be taken as {\em the} fundamental building block to construct an action that is invariant under all the symmetries. What is its relationship to the strain tensor defined above?

Consider for simplicity a static configuration, $\phi^I \equiv \phi^I(\vec x)$, so that,
\begin{align}
    B^{IJ} = J_i{}^I J_{i}{}^J = \left(J^t \cdot J\right)^{IJ} \, , \quad \text{ with } \quad J_i{}^I \equiv \partial_i \phi^I \, .
\end{align}
Notice that $J$ is simply the Jacobian matrix for the $x^i \to \phi^I$ mapping. For the inverse mapping, we have,
\begin{align}
    \frac{\partial x^i}{\partial \phi^I} = \big(J^{-1} \big)_I{}^i \, .
\end{align}
Given that in physical space the only two-index tensor that is invariant under the spatial $SO(3)$ rotational symmetry is $\delta_{ij}$, we can unambiguously define the metric in comoving space as the induced metric
\begin{align} \label{eq:gIJ}
    g_{IJ} = \delta_{ij} \, \frac{\partial x^i}{\partial \phi^I} \frac{\partial x^j}{\partial \phi^J} = \left[ J^{-1} \cdot \left(J^t\right)^{-1} \right]_{IJ} = \big(B^{-1}\big)_{IJ} \,.
\end{align}
This is invariant under the spatial $SO(3)$, and transforms covariantly under $G$, like our $B^{IJ}$, of which it is simply the inverse.

Now, the unstrained configuration corresponds to $x^i = \phi^i$, and so a generic deformation of the solid corresponds to
\begin{align}
    x^i(\phi) = \phi^i + u^i(\phi) \, .
\end{align}
In terms of $\vec u(\phi)$, the induced metric becomes
\begin{align} \label{eq:gofu}
    g_{IJ} = \delta_{IJ}  + 2 \, u_{IJ} \, ,
\end{align}
where $u_{IJ}$ is the same as the nonlinear strain in Eq.~\eqref{strain}, but where all the derivatives are now explicitly with respect to the $\phi^I$ comoving coordinates.
We thus recover the same  infinitesimal distance as in Eq.~\eqref{dl'}, but now in a notation that makes the separation between spatial ($x$) and comoving ($\phi$) coordinates explicit:
\begin{align}
    ds^2 = d \vec x \cdot d \vec x = (\delta _{IJ}+ 2 \, u_{IJ}) \, d\phi^I d\phi^J \, .
\end{align}

It is also interesting to consider  a further infinitesimal deformation of an already (nonlinearly) deformed solid, i.e.,
\begin{align}
    \phi^I(\vec x) \to \phi^I(\vec x) + \delta \phi^I(\vec x) \, ,
\end{align}
or, equivalently,
\begin{align}
    x^i(\phi) \to x^i(\phi) + \delta x^i(\phi) \, ,
\end{align}
for instance to derive thermodynamic relationships.
In matrix notation, $B^{IJ}$ changes by
\begin{align}
\delta B  = - B \cdot  \delta \! \left(B^{-1}\right) \cdot B  
 = - J^t \cdot \big( J  \cdot \delta g \cdot J^t \big)\cdot J  \, .
\end{align}
The Jacobian factors outside the parentheses simply convert tensors from physical space to comoving space. The combination in parentheses is, instead, given by,
\begin{align} \label{eq:JdgJ}
    \big( J  \cdot \delta g \cdot J^t  \big)_{ij} & = \frac{\partial \phi^I}{\partial x^i} \frac{\partial \phi^J}{\partial x^j} \, \delta \left(  \frac{\partial x^k}{\partial \phi^I} \frac{\partial x^k}{\partial \phi^J} \right) = \frac{\partial (\delta x^i)}{\partial x^j} + \frac{\partial (\delta x^j)}{\partial x^i} \equiv 2 \spacy \delta u_{ij} \,,
\end{align}
where the definition at the end is simply based on a superficial analogy with the definition of the {\em linear} strain \eqref{linear strain}. One should remember however, that the (nonlinear) strain $u_{IJ}$ is naturally a tensor with comoving indices. Upon computing its infinitesimal variation $\delta u_{IJ}$, if desired one can convert those indices to spatial ones by multiplying with the appropriate Jacobian factors. The $\delta u_{ij}$ defined above is  precisely the outcome of that procedure:
\begin{align}
    \delta u_{ij} \equiv (J \cdot \delta u \cdot J^t)_{ij} \, ,
\end{align}
where we used Eq.~\eqref{eq:JdgJ} together with the fact that, from Eq.~\eqref{eq:gofu}, $\delta g = 2 \spacy \delta u$.
Combining everything,  we thus get,
\begin{align}
    \delta B^{IJ} = - 2 \spacy J_i{}^{I} J_j{}^{J} \spacy  \delta u_{ij} \,.
\end{align}

The $\delta u_{ij}$ we just defined is what matters when computing the infinitesimal work needed to deform a solid at equilibrium in the presence of nontrivial stresses: 
\begin{align}
    \delta W = -\oint_{S} dS \; T_{ij} \spacy \hat n^i \spacy \delta x^j = -\int_V dV \; \partial_i \big( T_{ij} \, \delta x^j  \big) = - \int_V dV \;  T_{ij} \, \partial_i  \delta x^j  = -  \int_V dV \;  T_{ij} \, \delta u_{ij} \, ,
\end{align}
where we used the fact that $T_{ij}$ is symmetric and that, at equilibrium and in the static limit, $\partial_i T_{ij} = 0$. Moreover, $\delta u_{ij}$ also characterizes the local change in {\em physical} volume of a given comoving volume element:
\begin{align}
    \delta\left( {d^3 x} \right)= \delta {\sqrt{\det g}} \, d^3 \phi = \sqrt{\det g} \, \left(g^{-1}\right)^{IJ} \spacy \frac{\delta g_{JI}}{2} \, d^3 \phi = \delta u_{ii} \, d^3 x \, ,
\end{align}
where we used the known expression for the variation of the determinant of a matrix, $\delta \spacy \det g = \det g \left( g^{-1} \right)^{IJ} \delta g_{JI}$, together with Eqs.~\eqref{eq:gIJ} and \eqref{eq:JdgJ}. The result is in agreement with the standard result $\delta V = \delta u_{ii}  V$~\cite{LL_elasticity}.

Notice that, in our derivations above, we never linearized in the pre-existing deformation,  only in its infinitesimal variation $\delta \vec x$. And so, we see that for static solids the descriptions in terms of our $B^{IJ}$ or in terms of a nonlinear strain tensor $u_{IJ}$ are completely equivalent, at nonlinear level as well, provided one insists on differentiating explicitly the comoving coordinates from the spatial ones. 

However, when  considering dynamical situations, it is much more convenient to stick with our $B^{IJ}$ description, that is, to parametrize the dynamics in terms of our three scalar fields $\phi^I(t, \vec x)$. This is because the space-{\em time} symmetries---Lorentz invariance, or Galilei invariance in the nonrelativistic limit---are easy to implement in this description but much more cumbersome in the alternative $\vec x(t, \phi^I)$ one.

%% file: app_nonlinear_Lorentz.tex
\section{Nonlinear Realization of the Lorentz Symmetry} \label{app:nonlinearLorentz}

\noindent As discussed in several occasions throughout this work, the spontaneously broken spacetime symmetries are realized nonlinearly at the level of the Goldstone modes. Once the theory is expressed in terms of the latter, these original spacetime symmetries imply nontrivial relations among the effective coefficients. Examples include the relations holding between the cubic and quadratic coefficients of the EFT for solids,  Eq.~\eqref{eq:solidcoefficients}, or the relations between the quadratic, cubic and quartic couplings in the EFT for zero-temperature superfluids,  Eqs.~\eqref{eq:S3and4superfluid}.

It is instructive to make this concrete, by showing in practice how such nonlinear symmetries are implemented and how they imply constraints between the effective coefficients. We consider the EFT for solids, whose Lagrangian up to cubic order is reported in Eq.~\eqref{eq: cubic action phonon solids}. Now, a generic infinitesimal Lorentz transformation acting on the spacetime coordinates is given by,
\begin{align}
    x^\prime {}^\mu = \Lambda^\mu {}_\nu \spacy x^\nu \simeq x^\mu + \omega^\mu {}_\nu \spacy x^\nu \,,
\end{align}
with $\omega_{\mu\nu}$ antisymmetric. The comoving coordinates, $\phi^I(x)$, are three scalar fields, implying that they transform under Lorentz as,
\begin{align}
    \phi^I(x) \to \phi^I(x - \omega \cdot x) \,.
\end{align}
Given the parametrization in Eq.~\eqref{eq: solid phonon field def}, and expanding in small $\omega$, we can read off how the Goldstone fields transforms under Lorentz,
\begin{align}
    \pi^I(x) \to \pi^I(x) - \omega^\mu{}_\nu \spacy x^\nu \spacy \partial_\mu \pi^I(x) - \omega^I{}_\mu \spacy x^\mu \,.
\end{align}
As anticipated, the second term is proportional to the field itself, and it is a {\it linear} variation. The third term, instead, is independent on the field: it is the {\it nonlinear} variation.
The time derivatives and spatial gradients of the phonon field then transform as,
\begin{subequations} \label{eq:Lorentzvariations}
\begin{align}
    \dot \pi^I \to{}& \dot\pi^I + \vec{\beta}\cdot\vec{x} \spacy\spacy \ddot \pi^I + t \spacy \vec{\beta}\cdot\vec{\nabla} \spacy \dot\pi^I + \vec{\beta}\cdot\vec{\nabla}\pi^I + \big( \vec{\theta}\times\vec{x} \big) \cdot \vec{\nabla}\dot\pi^I  + \beta^I \,, \\
    \begin{split}
        \nabla^J \pi^I \to{}& \nabla^J\pi^I + \bm\beta\cdot\bm x \spacy \nabla^J \dot\pi^I + t \spacy \vec{\beta}\cdot\vec{\nabla} \spacy \nabla^J \pi^I + \beta^J \dot\pi^I + \big( \vec{\theta}\times\vec{x} \big)\cdot\vec{\nabla} \spacy \nabla^J\pi^I - \big( \vec{\theta}\times\vec{\nabla} \big)^J \pi^I \\
        & - \epsilon^{IJK} \theta^K \,,
    \end{split}
\end{align}
\end{subequations}
where we used the fact that $\omega^{I0} = \beta^I$ and $\omega^{IJ} = \epsilon^{IJK}\theta^K$, with  $\vec{\beta}$ and $\vec{\theta}$ being the small velocity and angle parametrizing, respectively, the infinitesimal boost and rotation.
Given this, and with a bit of tedious algebra, the Lorentz transformations of the quadratic operators appearing in the Lagrangian~\eqref{eq: cubic action phonon solids} read,
\begin{subequations}
\begin{align}
    \dot{\vec{\pi}}^2 \to{}& \dot{\vec{\pi}}^2 + 2 \spacy \vec{\beta}\cdot\nabla\pi\cdot\dot{\vec{\pi}} + \text{total derivatives} \,, \\
    \left[ \nabla\pi \right]^2 \to{}& \left[ \nabla\pi \right]^2 + 2 \left[ \nabla\pi \right] \vec{\beta}\cdot\dot{\vec{\pi}} - 2 \left[ \nabla\pi \right] \vec{\theta}\cdot\big(\vec{\nabla}\times\vec{\pi}\big) + \text{total derivatives} \,, \\
    \left[ \nabla\pi\nabla\pi^t \right] \to{}& \left[ \nabla\pi\nabla\pi^t \right] + 2 \spacy \vec{\beta}\cdot\nabla\pi\cdot\dot{\vec{\pi}} + \text{total derivatives} \,.
\end{align}
\end{subequations}
Disregarding the total derivatives, in order for the terms above not to spoil Lorentz invariance, they must cancel against terms coming from the transformation of higher-order operators. Specifically, since the variations above are quadratic in the fields, one must look at the nonlinear variations of the cubic operators, which will also be quadratic in the fields. The linear variations of the cubic operators will be compensated by the nonlinear variations of the quartic operators, and so on. The nonlinear transformations of the cubic operators, deduced again from Eqs.~\eqref{eq:Lorentzvariations}, are given by,
\begin{subequations}
\begin{align}
    \left[ \nabla\pi \right]^3 \xrightarrow{\text{ nonlinear }}{}& \left[ \nabla\pi \right]^3 \,, \\
    \left[ \nabla\pi \right] \left[ \nabla\pi^2 \right] \xrightarrow{\text{ nonlinear }}{}& \left[ \nabla\pi \right] \left[ \nabla\pi^2 \right] - 2 \left[\nabla\pi\right] \vec{\theta}\cdot\big(\vec{\nabla}\times\vec{\pi}\big) \,, \\
    \left[ \nabla\pi \right]\left[ \nabla\pi\nabla\pi^t \right] \xrightarrow{\text{ nonlinear }}{}& \left[ \nabla\pi \right]\left[ \nabla\pi\nabla\pi^t \right] + 2 \left[\nabla\pi\right] \vec{\theta}\cdot\big(\vec{\nabla}\times\vec{\pi}\big) \,, \\
    \left[ \nabla\pi^2\nabla\pi^t \right] \xrightarrow{\text{ nonlinear }}{}& \left[ \nabla\pi^2\nabla\pi^t \right] + \left[\nabla\pi\right] \vec{\theta}\cdot\big(\vec{\nabla}\times\vec{\pi}\big) + \text{total derivatives} \,, \\
    \left[ \nabla\pi \right] \dot{\vec{\pi}}^2 \xrightarrow{\text{ nonlinear }}{}& \left[ \nabla\pi \right] \dot{\vec{\pi}}^2 + 2 \left[ \nabla\pi \right]\vec{\beta}\cdot\dot{\vec{\pi}} \,, \\
    \dot{\vec{\pi}}\cdot\nabla\pi\cdot\dot{\vec{\pi}} \xrightarrow{\text{ nonlinear }}{}& \dot{\vec{\pi}}\cdot\nabla\pi\cdot\dot{\vec{\pi}} + \vec{\beta}\cdot\nabla\pi\cdot\dot{\vec{\pi}} + \left[ \nabla\pi \right] \vec{\beta}\cdot\dot{\vec{\pi}} + \text{total derivatives} \,.
\end{align}
\end{subequations}
Putting everything together, the quadratic variation of the Lagrangian up to cubic order reads,
\begin{align}
    \begin{split}
        \mathcal{L} \to \mathcal{L} &+ \left( 1 - c_T^2 + g_6 \right) \vec{\beta}\cdot\nabla\pi\cdot\dot{\vec{\pi}} + \left( c_T^2 - c_L^2 + 2 g_5 + g_6 \right) \dot{\vec{\pi}} \cdot \nabla\pi \cdot\vec{\beta} \\
        &+ \left( c_L^2 - c_T^2 - 2g_2 + 2g_3 + g_4 \right) \left[ \nabla\pi \right]\vec{\theta}\cdot\big(\vec{\nabla}\times\vec{\pi}\big) \,.
    \end{split}
\end{align}
Thus, Lorentz invariance is ensured provided that the cubic couplings are related to the sounds speeds, precisely as in Eq.~\eqref{eq: solid cubic constraints}. The advantage of formulating the theory in a Lorentz invariant fashion and then introducing some Lorentz-breaking expectation values for the fields, rather than working directly at the level of the Goldstones, is that these constraints are automatically implemented, provided that Lorentz indices are properly contracted in the original theory.

%% file: Routhian.tex
\section{Integrating out and the Routhian} \label{app: routhian}

\noindent When we integrate out a gapped field (say, $\phi_2$) to get an EFT for the remaining fields (say, $\phi_1$), at tree level we simply solve the equation of motion for $\phi_2$ and plug the solution back into the action,
\begin{align}
    S_{\rm eff}[\phi_1] = S\big[\phi_1, \bar \phi_2[\phi_1]\big] \, ,
\end{align}
where $\bar \phi_2[\phi_1]$ is the solution for $\phi_2$ for a given configuration $\phi_1(x)$. This procedure works because the equation of motion for $\phi_1$ one gets from the effective action $S_{\rm eff}$ are equivalent to those associated with the original action:
\begin{align}
    \frac{\delta S_{\rm eff}}{\delta \phi_1(x)} = \frac{\delta S}{\delta \phi_1(x)} + \int d^4 y \frac{\delta S}{\delta \phi_2(y) } \bigg|_{\bar \phi_2} \frac{\delta \bar \phi_2(y)}{\delta \phi_1(x) } = \frac{\delta S}{\delta \phi_1(x)} \, ,
\end{align}
where we used the fact that $\bar \phi_2$ solves the $\phi_2$ eom for generic $\phi_1(x)$.

There is a hidden subtlety in this argument, which has to do with boundary conditions. The original variational problem has fixed boundary conditions for $\phi_1$ {\em and} $\phi_2$. The variational problem associated with $S_{\rm eff}$ only has fixed boundary conditions for $\phi_1$, because that's the only field that is left. One must then check whether an infinitesimal variation of $\phi_1$ that vanishes at the boundary induces an infinitesimal variation of $\phi_2$,
\begin{align}
    \delta \bar \phi_2(x) = \int d^4 y   \frac{\delta \bar \phi_2(x)}{\delta \phi_1(y) } \, \delta \phi_1(y)\, ,
\end{align}
that also vanishes at the boundary. For a gapped $\phi_2$ field, this is ensured by the exponential decaying of its Green's functions at infinity. For a gapless $\phi_2$, by contrast, there are subtleties that we will now explore.

In special circumstances like the ones explored in the main text, we can still get a local dynamics for $\phi_1$ even if we integrate out a {\em gapless} $\phi_2$ field with a shift symmetry acting on it,
\begin{align}
    \phi_2 (x) \to \phi_2(x) + {\rm constant} \, .
\end{align}
In this case, however, the vanishing of $\delta \bar \phi_2$ at infinity for any $\delta \phi_1$ that decays at infinity is not guaranteed anymore, and when varying the effective action with respect to $\phi_1$ we might be left with a boundary term:
\begin{align}
    \begin{split}
        \delta S_{\rm eff} &= \int d^4 x \, \frac{\delta S}{\delta \phi_1(x)}\delta \phi_1(x)
        + \int d^4 x \, \frac{\partial {\cal L}}{\partial(\partial_\mu \phi_2(x))} \bigg|_{\bar \phi_2} {\delta \partial_\mu \bar \phi_2}(x) \\
        &= \int d^4 x \, \frac{\delta S}{\delta \phi_1(x)}\delta \phi_1(x)
        + \oint dS_\mu \frac{\partial {\cal L}}{\partial(\partial_\mu \phi_2(x))} \bigg|_{\bar \phi_2} {\delta \bar \phi_2}(x) \, ,
    \end{split}
\end{align}
where we went from the first to the second line using $\delta \partial_\mu \bar \phi_2 =  \partial_\mu \delta \bar \phi_2 $, performing an integration by parts, and using the fact that $\bar \phi_2$ satisfies the Euler--Lagrange equation. The latter, thanks to the postulated shift symmetry on $\phi_2$, is nothing but the conservation of the current
\begin{align}
    J^\mu = \frac{\partial {\cal L}}{\partial(\partial_\mu \phi_2(x))} \, .
\end{align}
In other words, the boundary term in the variation above is
\begin{align} \label{boundary term}
    \oint dS_\mu \, \bar J^\mu(x) \, \delta\bar \phi_2(x)  \, ,
\end{align}
where $\bar J^\mu$ is the current evaluated on the solution for $\phi_2$, $\bar J^\mu[\phi_1] = J^\mu\big[\phi_1, \bar\phi_2[\phi_1]\big]$, and so is {\em identically} conserved, that is, conserved for any, generically off-shell configuration $\phi_1(x)$.

\emph{If} $\bar J^\mu$ is a (possibly trivial) local function of $\phi_1(x)$ and its derivatives, as in the explicit examples analyzed in the text,  the boundary term  \eqref{boundary term} can further be rewritten as the variation of a boundary term,
\begin{align} \label{boundary term 2}
    \delta\bigg(\oint d S_\mu \, \bar J^\mu (x) \,\bar \phi_2(x) \bigg) \, ,
\end{align}
because the variation of $\bar J^\mu$ induced by a $\delta \phi_1$ that vanishes at infinity also vanishes at infinity. In this case, putting everything together and using the fact that $\bar J^\mu$ is identically conserved, we see that the original variational problem is equivalent to that associated with the new effective action
\begin{align}
    \tilde S_{\rm eff}[\phi_1] = \int d^4 x \, \Big\{ {\cal L}\big[\phi_1, \bar \phi_2[\phi_1]\big] - \bar J^\mu[\phi_1] \, \partial_\mu \bar \phi_2[\phi_1]  \Big\} \, .
\end{align}
The integrand is precisely the Routhian \cite{landau1976mechanics}. Note that the boundary term in Eq. \eqref{boundary term 2} does affect the equation of motion for $\phi_1$ because $\bar \phi_2$ is a nonlocal function of $\phi_1$. Its derivative $\partial_\mu \bar \phi_2$, on the other hand, is not, and therefore the Routhian is local.

It is worth making explicit how this construction realizes the trade between fixing the boundary value of the field $\phi_2$ or its conjugate charge alluded to in footnote \ref{foot: routhian}. In going from Eq. \eqref{boundary term} to \eqref{boundary term 2}, we used the fact that $\delta \phi_1 = 0$ implies $\delta\bar J^\mu = 0$ on the boundary, i.e.\ it implies that the conserved
current---and not the eliminated field $\bar\phi_2$---is held fixed there. Restricting to the initial  and final time slices, on which $dS_\mu = \mp\,\delta_\mu^0\, d^3x$, this reduces to the statement that the charge density is fixed at the boundary, $\delta\bar J^0 = 0$. Because $J^0 = \partial{\cal L}/\partial\dot\phi_2$ is also equal to the momentum density conjugate to $\phi_2$, this is the field-theory counterpart of the fact that a partial Legendre transform for a cyclic coordinate trades $\delta q = 0$ for $\delta p = 0$ at the endpoints~\cite{landau1976mechanics}.

For the perfect fluid of Section~\ref{LF3fields}, $\phi_2 = \tau$ enters ${\cal L}$ only through $\dot\tau$, so the current has a single nonvanishing component,
$\bar J^0 = \pi_\tau$ and $\bar J^I = 0$. Moreover, solving the $\tau$ equation of motion---which is just $\partial_t \pi_\tau = 0$---with the gauge choice
$\pi_\tau|_{t=0} = 1$ gives $\pi_\tau \equiv 1$ for \emph{any} configuration
$\vec X(t,\phi)$ (which plays the role of $\phi_1$), so $\delta \bar J^\mu = 0$ holds identically and the charge is
automatically fixed. In this frame the covariant Routhian density
${\cal L} - \bar J^\mu \partial_\mu \bar\phi_2$ collapses to the elementary
Routhian ${\cal L} - \pi_\tau \dot\tau$ of Eq.~\eqref{eq: routhian}, and the
boundary term lives entirely on the equal-time surfaces.

%% file: app_feynman_rules.tex
\section{Feynman Rules} \label{app:feynmanrules}

\noindent Feynman rules are the basic building blocks for standard amplitude calculations both in relativistic quantum field theory as well as in many-body theory. While this is likely standard knowledge for the reader, we find it useful to briefly review how to derive them, in order to homogenize the differences that one often encounters in the presentation of the topic between the two fields~\citep[e.g.,][]{Srednicki:2007qs,coleman2015introduction}, as well as to establish the notation which we use throughout the text.

In general, the nontrivial part of the $S$-matrix element for any process between a given initial state, $|\alpha\rangle$, and a given final state, $|\beta\rangle$, can be written as,
\begin{align}
    \langle \beta | S - \mathds{1} | \alpha \rangle = i \mathcal{M} \spacy (2\pi)^4 \delta^{(4)}(k_{\alpha,{\rm tot}} - k_{\beta,{\rm tot}}) \,,
\end{align}
where the $\delta^{(4)}$ stands for the $\delta$-functions implementing conservation of energy and momentum, $k^\mu_{\alpha,{\rm tot}}=(\omega_{\alpha,{\rm tot}}, \vec{k}_{\alpha,{\rm tot}})$ denotes the total energy and momentum of the initial state, and $k^\mu_{\beta,{\rm tot}}$ does the same for the final state. Finally, $i\mathcal{M}$ is the matrix element one is interested in computing. In a theory where not all components of $k^\mu$ are conserved (as for example our roton EFT, Section~\ref{sec:rotons}), there will be fewer $\delta$-functions (with their ubiquitous factors of $(2\pi)$). 

Now, a possible way to compute this $S$-matrix element, common to both high energy and many-body physics, is via time-dependent perturbation theory and the Dyson's series. However, the application of Dyson's series to our context is subtle. This is because the fundamental building block of time-dependent perturbation theory is the interaction Hamiltonian. As discussed at length in the main text, local effective field theories are best described in the Lagrangian language. Due to the Goldstone nature of our fields, all their interactions involve derivatives, and thus modify the definition of the conjugate momenta with respect to that of the free theory. When going from Lagrangians to Hamiltonians, this introduces a number of subtleties involving non-covariant terms. These eventually cancel out, but in a rather nontrivial way~\cite[e.g.,][]{Duncan:2012aja}.

We thus find it more transparent to stick with Lagrangians.
We will also limit our treatment to bosonic fields, $\phi_a$, where the $a$ can be a collection of indices, which can be internal, spacetime ones, or both. The extension to fermions is straightforward, provided one introduces anti-commutation rules and an ordering convention. Now, consider a certain interaction Lagrangian, which is a function of spacetime through its dependence on the fields,
\begin{align}
    \mathcal{L}_{\rm int}(x) \equiv \mathcal{L}_{\rm int}\big( \phi(x), \partial \phi(x) \big) \,.
\end{align}
For practical purposes, it is convenient to always choose the fields $\phi_a(x)$ as the ones with {\it canonically normalized kinetic terms}.
We work with a convention where the Fourier transform of the fields and their corresponding functional derivatives are
\begin{align}
    \phi_a(x) = \int \frac{d^4k}{(2\pi)^4} \spacy \hat{\phi}_a(k) \spacy e^{i k \cdot x} \,, \qquad \text{ and } \qquad \frac{\delta \hat{\phi}_a(k)}{\delta\hat{\phi}_b(q)} = (2\pi)^4 \delta^{(4)}(k-q) \delta_{ab} \,.
\end{align}
Here $\delta_{ab}$ is a collection of Kronecker $\delta$'s for the internal indices, and metric tensors for spacetime ones.
The prescription to obtain the Feynman rule associated to $\mathcal{L}_{\rm int}$ is simple. First, take $\mathcal{L}_{\rm int}$ and express it in terms of $\hat\phi_a(k)$. Then, if $\mathcal{L}_{\rm int}$ depends on $n$ fields with indices $a_1, a_2, \dots, a_n$ (not necessarily all different), the corresponding Feynman rule is simply,
\begin{align}
    i \mathcal{M}_{a_1,a_2,\dots,a_n}(k_1, k_2, \dots, k_n) = i \spacy \frac{\delta}{\delta \hat{\phi}_{a_1}(-k_1)} \spacy \frac{\delta}{\delta \hat{\phi}_{a_2}(-k_2)} \cdots \frac{\delta}{\delta \hat{\phi}_{a_n}(-k_n)} \spacy \mathcal{L}_{\rm int}(x = 0) \,.
\end{align}
The recipe above produces the vertex Feynman rule with {\it all outgoing particles}. To turn the $i$-th particle from outgoing to incoming, one must simply change the sign of its 4-momentum, $k_i^\mu \to -k_i^\mu$. 

The vertex Feynman rule must then be paired with a rule for external legs. Specifically, if the $i$-th external leg corresponds to a physical degree of freedom labeled by $\lambda_i$, created and annihilated by the field $\phi_{a_i}$, then for that external leg one must introduce whatever function connects the field indices with the label of physical states, $\varepsilon_{a_i}^{\lambda_i}$: the generalization of standard polarization vectors.

We now report the Feynman rules for a two prototypical theories discussed in this work: relativistic superfluid phonons and their self-interactions, and nonrelativistic phonon--roton interactions.

First, however, let us briefly comment about conservation of energy and momentum. Within the EFT for phonons in a superfluid, it is possible to define both energy and momentum conservation. Specifically, given than time translations are broken, the energy of a phonon is given by the eigenvalue of the one-phonon state with respect to the unbroken time translations, $\bar{H} = H - \bar{\upmu} \spacy Q$. Since spatial translations are unbroken, instead, the phonon momentum is defined in the usual way, as the eigenvalue of the $\bar{P}^i = P^i$ operator. 
Consider then the interactions between nonrelativistic superfluid phonons and a roton moving with constant velocity, $\dot{\vec{x}}_0(t) = v_0 \spacy \hat{k}_0$. As we discussed in footnote~\ref{footnote: unbroken H rotons}, the roton trajectory breaks time translations (which were already broken by the superfluid state), as well as spatial translations. The only unbroken combination that is left is $\bar{\bar{H}} = \bar{H} - \vec{v}_0 \cdot \vec{P}$. This is what defines conservation of energy in processing involving phonons and rotons moving in a straight line. Specifically, since $\bar{H}$ and $\vec{P}$ define the phonon energy, $\omega$, and momentum, $\vec{k}$, the conserved quantity is the combination $\omega - \vec{v}_0 \cdot \vec{k}$.

\vspace{2em}

\begin{table}[h]
	\begin{tabular}{C{0.2\textwidth}|C{0.75\textwidth}}
		\Xhline{6\arrayrulewidth}
		\multicolumn{2}{c}{Relativistic superfluid phonons} \\
		\Xhline{6\arrayrulewidth}
		Energy conservation & $(2\pi) \delta(\omega_{\rm in} - \omega_{\rm out})$ \\
		Momentum conservation & $(2\pi)^3 \delta^{(3)}(\vec k_{\rm in} - \vec k_{\rm out})$ \\\Xhline{3\arrayrulewidth}
		Feynman diagram & Rule \\
		\hline
		\begin{tikzpicture}
			\draw[snake it, thick] (-1,0) -- (1,0);
			\node at (-1,0) [circle,fill,inner sep=1.1pt]{};
			\node at (1,0) [circle,fill,inner sep=1.1pt]{};
			\node at (0,0.4) {\footnotesize $k$};
			\node at (0,-0.4) {\footnotesize phonon propagator};
		\end{tikzpicture} & $\frac{i}{\omega^2 - c_s^2 \vec k^2 + i \epsilon}$ \\
		\begin{tikzpicture}
			\draw[->, snake it, thick] (0,0) -- (-1,0);
			\draw[->, snake it, thick] (0,0) -- (0.5, {sqrt(3)/2});
			\draw[->, snake it, thick] (0,0) -- (0.5, {-sqrt(3)/2});
			\node at (0,0) [circle,fill,inner sep=1.1pt]{};
			\node at (-1.25,0) {\footnotesize $k_1$};
			\node at (0.7, 1.1) {\footnotesize $k_2$};
			\node at (0.7, -1.1) {\footnotesize $k_3$};
		\end{tikzpicture} & $\frac{1}{\sqrt{Z}} \bigg\{ g_3 \, \omega_1 \omega_2 \omega_3 - \left(1-c_2^2\right) \left[ \omega_1 {\vec k}_2 \cdot {\vec k}_3 + \omega_2 {\vec k}_1 \cdot {\vec k}_3 + \omega_3 {\vec k}_1 \cdot {\vec k}_2 \right] \bigg\}$ \\[-2em]
		\begin{tikzpicture}
			\draw[->, snake it, thick] (0,0) -- (-1,1);
			\draw[->, snake it, thick] (0,0) -- (-1,-1);
			\draw[->, snake it, thick] (0,0) -- (1,1);
			\draw[->, snake it, thick] (0,0) -- (1,-1);
			\node at (0,0) [circle,fill,inner sep=1.1pt]{};
			\node at (-1.25,1.25) {\footnotesize $k_1$};
			\node at (1.25,1.25) {\footnotesize $k_2$};
			\node at (1.25,-1.25) {\footnotesize $k_3$};
			\node at (-1.25,-1.25) {\footnotesize $k_4$};
		\end{tikzpicture} & {\begin{align*} \frac{i}{Z} \bigg\{& g_4 \, \omega_1 \omega_2 \omega_3 \omega_4 \\
	& + \left[ 2\left( 1- c_s^2\right) - g_3 \right] \Big( \omega_1 \omega_2 \vec k_3 \cdot \vec k_4 + \omega_1 \omega_3 \vec k_2 \cdot \vec k_4 + \omega_1 \omega_4 \vec k_2 \cdot \vec k_3 \\
	& \quad\qquad \qquad \qquad \qquad + \omega_2 \omega_3 \vec k_1 \cdot \vec k_4 + \omega_2 \omega_4 \vec k_1 \cdot \vec k_3 + \omega_3 \omega_4 \vec k_1 \cdot \vec k_2 \Big) \\
	& + \left(1 - c_s^2 \right) \left[ \vec k_1 \cdot \vec k_2 \vec k_3 \cdot \vec k_4 + \vec k_1 \cdot \vec k_3 \vec k_2 \cdot \vec k_4 + \vec k_1 \cdot \vec k_4 \vec k_2 \cdot \vec k_3 \right] \bigg\} \end{align*}} \\
    \Xhline{6\arrayrulewidth}
	\end{tabular}
	\caption{Feynman rules for phonons in a bulk superfluid. To switch a given external leg from outgoing to incoming it is sufficient to change sign to its energy/momentum. The effective coefficient, $g_n$, are defined in Eq.~\eqref{eq:gn}. Details on the theory are found in Section~\ref{sec:phonon}.} \label{tab:Feynman1}
\end{table}

\newpage

\begin{table}[t]
	\begin{tabular}{C{0.2\textwidth}|C{0.75\textwidth}}
		\Xhline{6\arrayrulewidth}
		\multicolumn{2}{c}{nonrelativistic rotons and superfluid phonons} \\
		\Xhline{6\arrayrulewidth}
		Energy conservation & $(2\pi) \delta(\omega_{\rm in} - \omega_{\rm out} - \vec{v}_0 \cdot \vec k_{\rm in} + \vec{v}_0 \cdot \vec k_{\rm out})$ \\
		Momentum conservation & None \\\Xhline{3\arrayrulewidth}
		Feynman diagram & Rule \\\hline
		\multicolumn{2}{l}{The roton unperturbed trajectory cannot be created nor destroyed, and it plays the role of a} \\[-0.6em]
		\multicolumn{2}{l}{source (crossed circle) for the phonon (wavy line) and trajectory fluctuations (dotted line).} \\\hline
		\begin{tikzpicture}[decoration={markings,mark=at position 0.5 with {\arrow{>}}}]			
			\node[scale=0.9] at (-1,0) {$\otimes$};
			\draw[->, thick, densely dotted] (-0.88,0) -- (0,0);
			\draw[thick, densely dotted] (0,0) -- (0.89,0);
			\node[scale=0.9] at (1,0) {$\otimes$};
			\node at (0,0.4) {\footnotesize $\tilde\omega$};
			\node at (0,-0.4) {\footnotesize trajectory fluctuation};
			\node at (0,-0.8) {\footnotesize propagator};
		\end{tikzpicture} & $\begin{pmatrix} \frac{i v_0}{k_0 \tilde\omega^2} P^{ij} + \frac{i}{\bar m_* \tilde\omega^2} \hat k_0^i \spacy \hat k_0^j & -\frac{P^{ij}}{k_0 \tilde\omega} \\ \frac{P^{ij}}{k_0 \tilde\omega} & 0 \end{pmatrix} \;, \quad \text{ with } \quad P^{ij} \equiv \delta^{ij} - \hat k_0^i \hat k_0^j$ \\
        \rule{0pt}{2.5em}
		\begin{tikzpicture}[decoration={markings,mark=at position 0.5 with {\arrow{>}}}]			
			\node[scale=0.9] at (-1,0) {$\otimes$};
			\draw[->, thick, snake it] (-0.88,0) -- (0.9,0);
			\node at (1.2,0) {\footnotesize $k$};
		\end{tikzpicture} & $\frac{c_s}{\sqrt{\rho_m}} k_0 \hat k_0 \cdot \vec k - L_{(0,0)}^\prime \omega$ \rule{0pt}{2em} \\
		\begin{tikzpicture}[decoration={markings,mark=at position 0.5 with {\arrow{>}}}]			
			\node[scale=0.9] at (-1,0) {$\otimes$};
			\draw[->, thick, snake it] (-0.88,0.07) -- (0.7,0.9);
			\draw[->, thick, snake it] (-0.88,-0.07) -- (0.7,-0.9);
			\node at (1,0.9) {\footnotesize $k_1$};
			\node at (1,-0.9) {\footnotesize $k_2$};
		\end{tikzpicture} & {\begin{align*} -i \bigg\{& L_{(0,0)}^{\prime\prime} \omega_1\omega_2 - \frac{c_s}{\sqrt{\bar\rho_m}} \left[ L_{(0,0)}^\prime \vec k_1 \cdot \vec k_2 + k_0^\prime \left( \omega_1 \hat k_0 \cdot \vec k_2 + \omega_2 \hat k_0 \cdot \vec k_1 \right) \right] \\ &+ \frac{c_s^2 \bar m_*}{\bar \rho_m} \hat k_0 \cdot \vec k_1 \, \hat k_0 \cdot \vec k_2  \bigg\} \end{align*}} \\
		\begin{tikzpicture}[decoration={markings,mark=at position 0.5 with {\arrow{>}}},baseline = -1em]			
			\node[scale=0.9] at (-1,0) {$\otimes$};
			\draw[->, thick, snake it] (-0.88,0.07) -- (0.7,0.9);
			\draw[->, thick, densely dotted] (-0.88,-0.07) -- (0.7,-0.9);
			\node at (1,0.9) {\footnotesize $k$};
			\node at (1,-0.9) {\footnotesize $\tilde \omega$};
		\end{tikzpicture} & $i \begin{pmatrix} L_{(0,0)}^\prime \omega k^i - \frac{c_s}{\sqrt{\bar\rho_m}} k_0 \, \hat k_0 \cdot \vec k \, k^i - \tilde \omega \left( \omega k_0^\prime - \frac{c_s \bar m_*}{\sqrt{\bar\rho_m}} \hat k_0 \cdot \vec k \right) \hat k_0^i  \\ -i \frac{c_s}{\sqrt{\bar \rho_m}} k_0 k^i  \end{pmatrix}$ \\[-1.4em] 
        \\
        \Xhline{6\arrayrulewidth}
	\end{tabular}
	\caption{Feynman rules for superfluid phonons interacting with a roton moving with constant velocity, $\dot{\vec x}_0(t) = v_0 \hat k_0$ and constant direction, $\hat k_0(t) = {\rm constant}$. Moreover, $L_{(0,0)} \equiv - \bar \Delta + \bar k_* v_0 + \frac{1}{2} \bar m_* v_0^2$ and $k_0 \equiv \bar k_* + \bar m_* v_0$, and the primes indicate derivatives with respect to $\bar{\mathcal{X}}$---see Eqs.~\eqref{eq:chain1} and \eqref{eq:chain2}. To switch a given external leg from outgoing to incoming it is sufficient to change sign to its energy/momentum. Details on the theory are found in Section~\ref{sec:rotons}.}
\end{table}